\documentclass[12pt,a4paper,openright]{report} 
\usepackage{amsmath}
\usepackage[utf8]{inputenc}
\usepackage{graphicx}
\usepackage[hmargin=2cm,vmargin=2cm]{geometry}
\usepackage[french]{babel}
\usepackage{color}
\usepackage{tikz}
\usepackage{here}
\usepackage[Conny]{fncychap} 
\usepackage{epigraph}

\newcommand{\quotes}[1]{``#1''}

\definecolor{quotemark}{gray}{0.7}
\makeatletter
\newlength\origparskip

\newcommand{\fquote}{%
	\@ifnextchar[{\fquote@i}{\fquote@i[]}
}

\def\fquote@i[#1]{%
	\@ifnextchar[{\fquote@ii{#1}}{\fquote@ii{#1}[]}
}%

\def\fquote@ii#1[#2]{%
	\def\pqm@tempa{#1}%
	\def\pqm@tempb{#2}%
	\noindent
	\list
	{}
	{\setlength{\leftmargin}{0.3\textwidth}%
		\setlength{\rightmargin}{0.1\textwidth}%
		\setlength{\origparskip}{\parskip}}%
	\item[]%
	\begin{picture}(0,0)%
		\put(-15,-8){\makebox(0,0){\scalebox{4}{%
					\textcolor{quotemark}{\textquotedblright}}}}%
	\end{picture}%
	\begingroup
	\itshape
	\ignorespaces}%

\def\endfquote{%
	\endgroup
	\par
	\raggedleft
	\ifx\pqm@tempa\empty
	\else
	{\bfseries --- \pqm@tempa\par}%
	\setlength{\parskip}{\origparskip}%
	\ifx\pqm@tempb\empty
	\else
	(\pqm@tempb)%
	\fi
	\fi
	\par
	\endlist}
\makeatother
\usepackage{hyperref}
\usepackage{pifont}
\usepackage{pdfpages}
\usepackage{enumitem}
\usepackage{amssymb}
\usepackage[all,cmtip]{xy}
\usepackage{amsmath,amscd}
\usepackage[T1]{fontenc}
\usepackage{graphicx}
\usepackage{array,multirow,makecell}
\setcellgapes{1pt}
\makegapedcells
\usepackage{colortbl}
\newcolumntype{R}[1]{>{\raggedleft\arraybackslash }b{#1}}
\newcolumntype{L}[1]{>{\raggedright\arraybackslash }b{#1}}
\newcolumntype{C}[1]{>{\centering\arraybackslash }b{#1}}
\definecolor{Zgris}{rgb}{0.87,0.85,0.85}
\newsavebox{\BBbox}

\newcommand{\chaptertoc}[1]{\chapter*{#1}
	\addcontentsline{toc}{chapter}{#1}
	\markboth{\slshape\MakeUppercase{#1}}{\slshape\MakeUppercase{#1}}}
\providecommand{\openone}{\leavevmode\hbox{\small1\kern-4.3pt\normalsize1}}

\begin{document}
\includepdf[pages=-]{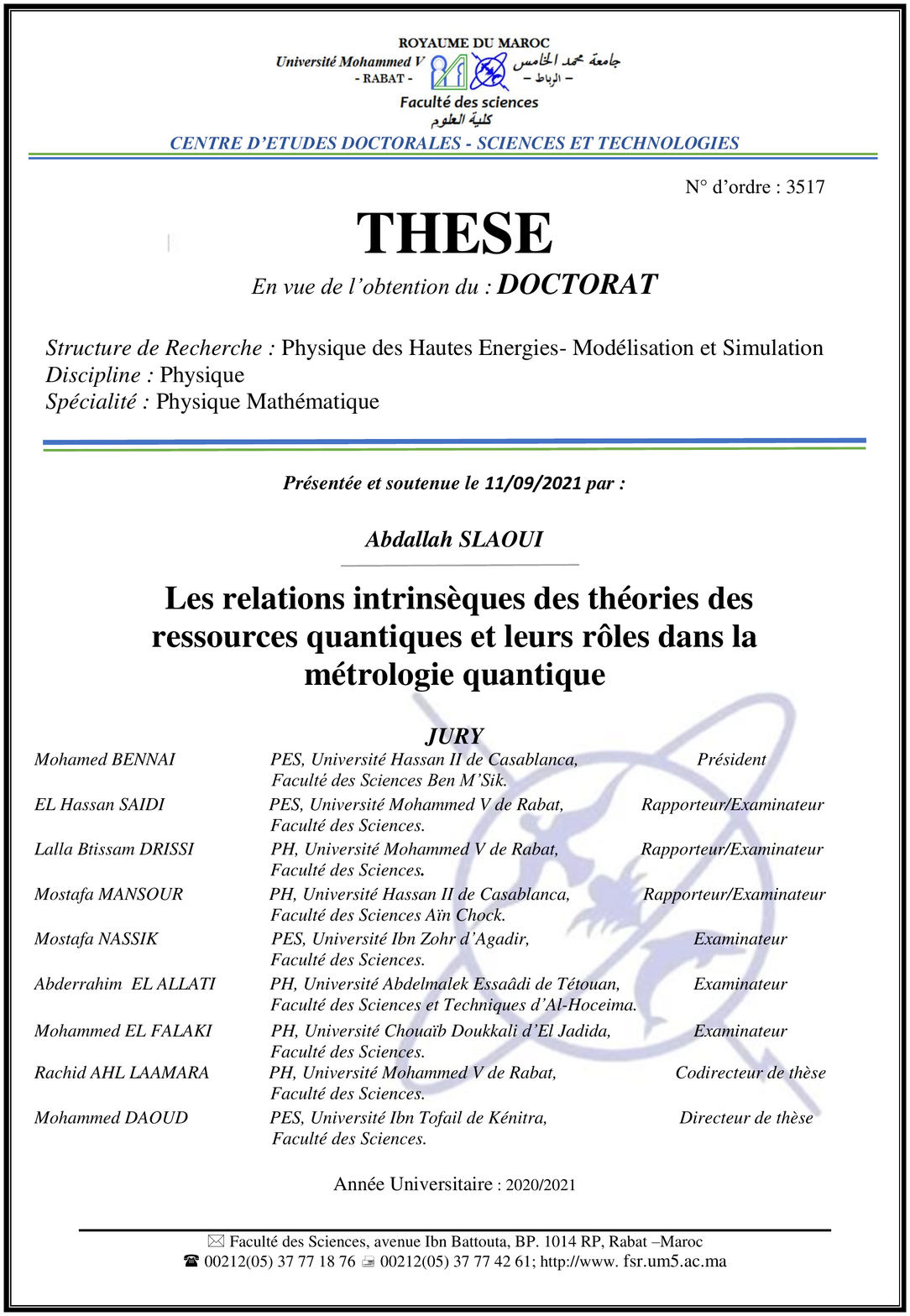}
\pagenumbering{roman} 
\include{page-de-garde}
\include{chapitre1}
\include{chapitre2}
\include{chapitre3}
\chaptertoc{Remerciements}
\vspace{-2cm}Les travaux présentés dans cette thèse ont été effectués au sein du Laboratoire de Physique des Hautes Energies-Modélisation et Simulation de l'université Mohammed V-Rabat, Faculté des Sciences, sous la direction et l'encadrement de Monsieur {\bf Mohammed DAOUD}, Professeur de l'enseignement supérieur à la Faculté des Sciences de Kénitra, et le co-encadrement de Monsieur {\bf Rachid AHL LAAMARA}, Professeur Habilité à la Faculté des Sciences de Rabat.\par 

La rédaction de cette thèse et la réalisation de ce programme de doctorat n'auraient pas été possibles sans les opportunités, les conseils et l'aide que m'ont offerts plusieurs autres personnes impliquées dans ce travail. D'abord et avant tout, j'ai beaucoup de chance et je suis profondément reconnaissant à mon parrain scientifique et mon directeur de thèse, Monsieur {\bf Mohammed DAOUD}, Professeur de l'enseignement supérieur à la Faculté des Sciences Ibn Tofail de Kénitra, pour tous ses encouragements, son soutien continu et ses conseils qui m'ont permis de terminer ma thèse avec succès. Je peux dire que c'est lui qui m'a formé à la rédaction d'articles de recherche. Je lui suis très reconnaissant pour ses enseignements, qui m'ont permis d'apprendre à étudier des problèmes de recherche et à noter des idées. Il m'a laissé prendre le temps de réfléchir aux idées et aux problèmes sur lesquels la thèse se développe. Il m'a donné la liberté de les poursuivre à mon propre niveau et m'a fourni les commentaires, les directives et l'aide nécessaires. Je le remercie vivement de m'avoir inspiré de nouveaux travaux, d'avoir organisé des discussions enthousiastes, de maintenir l'esprit de groupe et, surtout, de m'avoir toujours apporté son soutien total par tous les moyens possibles. C'était une grande opportunité de travailler sous sa direction et ses connaissances profondes ainsi que sa patience inégalée ont été inestimables dans ma croissance en tant que chercheur.\par

Je voudrais également exprimer mes remerciements à mon co-encadrant et mon conseiller pédagogique, Monsieur {\bf Rachid AHL LAAMARA}, Professeur Habilité à la Faculté des Sciences de Rabat, pour ses conseils et ses encouragements tout au long de ce travail, pour sa disponibilité et pour m'avoir fait bénéficier de ses compétences, merci également pour les efforts qu'il n'a cessé de déployer pour la réussite de plusieurs activités de notre laboratoire. Aussi, je le remercie chaleureusement pour m'avoir donné l'opportunité d'enseigner le cours de théorie de l'information quantique pour les étudiants du Master Physique Mathématique.\par

Je tiens à remercier Monsieur {\bf Mohamed BENNAI},  Professeur de l'enseignement supérieur à la Faculté des Sciences Ben M'Sik de Casablanca, pour avoir accepté de présider mon jury de thèse. Veuillez recevoir Monsieur l’expression de mon respect et de ma profonde gratitude.\par

J'exprime mes profonds remerciements au Monsieur {\bf El Hassan SAIDI}, Professeur de l'enseignement supérieur et directeur du Laboratoire de Physique des Hautes Energies-Modélisation et Simulation de département de physique de la Faculté des Sciences, Université Mohammed V de Rabat, où les travaux de cette thèse ont été préparés. Je lui adresse mes plus vifs remerciements pour m'avoir accueilli dans ce laboratoire et pour la confiance qu'il m'a accordée depuis mes premiers pas jusqu'à la rédaction de ce manuscrit, ainsi pour avoir accepté d'être rapporteur de
ma thèse. Ses encouragements continus, qui m'ont donné la force et le courage de réaliser cette thèse, ainsi que d'avoir accepté de juger ce travail. Je le remercie également pour son enseignement durant les années du Master de Physique Mathématique avec une très grande compétence.\par

Je remercie vivement Madame {\bf Lalla Btissam DRISSI}, Professeur Habilitée à la Faculté des Sciences de Rabat, de m'avoir fait l'honneur de participer au jury de ma soutenance en tant que rapporteur ainsi pour son enseignement de bonne qualité durant notre formation dans le Master Physique Mathématique. Je la remercie pour ses conseils avisés.\par

Mes remerciements chaleureuses s'adressent à Monsieur {\bf Mostafa MANSOUR}, Professeur Habilité à la Faculté des Sciences Ain Chock de Casablanca, pour avoir accepté de rapporter mon travail de recherche et de faire partie du jury de thèse. Veuillez accepter, Professeur, l'expression de ma plus profonde gratitude.\par

Par la même occasion, je suis également reconnaissant à Monsieur {\bf Mostafa NASSIK}, Professeur d'enseignement supérieur à la Faculté des Sciences Ibn Zohr d'Agadir pour avoir accepté d'être examinateur de ma thèse.\par 

Mes remerciements chaleureuses s'adressent à Monsieur {\bf Mohammed EL FALAKI}, Professeur Habilité à la Faculté des Sciences d'El Jadida, pour avoir accepté d'être membres du jury en tant que examinateur et pour avoir évalué ce travail.\par

Un grand merci à Monsieur {\bf Abderrahim EL ALLATI}, Professeur Habilité à la Faculté des Sciences et Techniques d'Al Hoceima d'avoir accepté d'être examinateur. \par

Je suis profondément reconnaissant à tous les membres du jury qui ont bien voulu me faire l’honneur d’assister à ma soutenance afin de juger la qualité de ce travail. Je leur suis redevable pour leurs précieux commentaires.\par

Je suis redevable à Monsieur {\bf Muzzamal SHAUKAT}, Professeur assistant à l'Instituto Superior Técnico (IST), l'Université de Lisbonne, Portugal. Shaukat a suivi de près mes activités de recherche depuis mes premiers jours de doctorat et une grande partie de mon travail est influencée par ses précieuses remarques. Il m'a également prodigué des conseils utiles pour ma carrière. Merci Shaukat!\par

J'ai la chance d'avoir rencontré Monsieur {\bf Ahmed SALAH}, professeur à l'Autorité de l'énergie atomique du Caire, Égypte. Salah m'a fait découvrir la théorie nouvellement développée de l'optique quantique, m'a aidé avec sa littérature et a généreusement partagé ses idées. Certaines de ces idées se sont matérialisées par des résultats concrets.\par

Au cours de mes recherches de thèse, j'ai bénéficié de visites prolongées à \quotes{\emph{l'Institut Henri Poincaré}} et au \quotes{\emph{Collège de France}} à Paris, au \quotes{\emph{Centre International de Rencontres Mathématiques}} à Marseille, en France, au \quotes{\emph{Centre International Abdus Salam de Physique Théorique}} à Trieste, en Italie, et à \quotes{\emph{l'Institut des Sciences Mathématiques de Madrid}}, en Espagne. Je voudrais remercier de nombreuses personnes qui m'ont accueilli dans ces établissements, et grâce à ces opportunités, cette période a été la plus stimulante et la plus productive de mon doctorat. Et en parlant de gens, je ne peux pas être assez reconnaissant envers tous ceux qui ont partagé une partie du voyage dans tous ces groupes et les régions environnantes. Il y a beaucoup de gens avec qui j'ai partagé des discussions, des cafés et des matchs de football, et cela fait beaucoup de noms à retenir!\par

Bien que mon doctorat se soit déroulé loin de chez moi, ma famille a toujours été là pour moi et m'a encouragée. Je remercie ma mère Amina, mes sœurs et mes frères pour leur amour et leur soutien. J'espère que je les ai rendus fiers. 

J'ai eu la chance d'avoir des discussions intéressantes sur le sujet de la thèse avec de nombreux collègues et amis spécialistes de la physique quantique au LPHE-MS, et chacun d'entre eux a contribué d'une manière ou d'une autre à mon travail, parmi lesquels H. El Hadfi, B. Amghar, L. Bakmou et F. Z. Ramadan.\par

Enfin, je voudrais profiter de cette occasion pour remercier tous mes professeurs, collègues et amis chercheurs qui m'ont apporté l'aide et le soutien tout au long de ma carrière académique. Je tiens à remercier toutes les personnes qui ont contribué à la réalisation de ce travail.

\chaptertoc{Résumé de la Thèse}

Les théories des ressources quantiques permettent de quantifier un effet quantique utile, de développer de nouveaux protocoles pour sa détection et de déterminer les processus exacts qui maximisent son utilisation pour des tâches pratiques. Ces théories visent à transformer les phénomènes physiques, telles que l'intrication et la cohérence quantique, en propriétés utiles à l'exécution de tâches concrètes liées à l'information quantique. D'autre part, la théorie de l'estimation des paramètres quantiques vise à déterminer la précision ultime de tous les paramètres contenus dans l'état d'un système quantique donné. Son principal problème est de trouver un schéma de mesure ultime permettant de dépasser la limite quantique standard qui n'est pas atteinte par les stratégies classiques. Pour déterminer cette précision ultime, il faut trouver l'information quantique de Fisher maximale indispensable pour saturer la borne de Cramér-Rao quantique.\par

Dans cette thèse, nous nous focalisons sur les théories des ressources de l'intrication, des corrélations quantiques de type discorde et de la cohérence quantique, les phénomènes quantiques les plus intrigants exploités jusqu'à présent dans la théorie de l'information quantique. Nous commençons par présenter en détail les outils théoriques de ces ressources quantiques en mettant en évidence les techniques et les problèmes de calcul les plus remarquables. Dans ce sens, nous abordons plusieurs méthodes mathématiques qui permettent de résoudre certains problèmes liés à leurs quantifications, et quelques résultats analytiques pour les systèmes bipartites sont donnés. Nous examinons également les connexions intrinsèques entre ces ressources quantiques en extrayant les liens qui unissent les mesures correspondantes.\par

En revanche, la révolution de la technologie quantique a suscité un intérêt croissant en faveur de la métrologie quantique, et l'intrication quantique a été employée pour surmonter la limite classique dans plusieurs protocoles d'estimation quantique. Dans le présent document, nous analysons le rôle des corrélations quantiques au-delà de l'intrication en vue d'améliorer la précision d'un paramètre inconnu. Selon nos résultats, les corrélations peuvent être capturées à l'aide de l'information quantique de Fisher, et les corrélations quantiques de type discorde peuvent être exploitées pour garantir la précision des protocoles d'estimation de phase.\par

Cette thèse comprend également des contributions sur la dynamique de ces ressources quantiques dans différents modèles de systèmes quantiques ouverts. Parmi nos objectifs, est d'étudier les effets de l'environnement sur ces ressources quantiques et d'obtenir des techniques pour les protéger contre les effets de la décohérence intrinsèque.\par

\underline{\bf Mots-clés:} Intrication quantique, discorde quantique, cohérence quantique, systèmes quantiques ouverts, décohérence, métrologie quantique, corrélations quantiques, théories des ressources quantiques, incertitude quantique, information de Fisher quantique, théorie de l'estimation quantique 

\chaptertoc{Abstract}
Quantum resource theories allow us to quantify a useful quantum phenomenon, to develop new protocols for its detection and determine the exact processes that maximize its use for practical tasks. These theories aim at transforming physical phenomena, such as entanglement and quantum coherence, into useful properties for the execution of concrete tasks related to quantum information. On the other hand, the theory of quantum parameter estimation aims at determining the ultimate accuracy of all parameters contained in the state of a given quantum system. Its main problem is to find an ultimate measurement scheme allowing to exceed the standard quantum limit which is not reached by classical strategies. To determine this ultimate precision, one must find the maximum Fisher quantum information required to saturate the quantum Cramér-Rao bound.\par

In this thesis, we focus on the resource theories of entanglement, discord-like quantum correlations, and quantum coherence, the most intriguing quantum phenomena exploited so far in quantum information theory. We begin by presenting in detail the theoretical tools of these quantum resources, focusing on the most remarkable techniques and computational problems. In this sense, we discuss several mathematical methods that solve some problems related to their quantifications, and some analytical results for bipartite quantum systems are given. We also examine the intrinsic connections between these quantum resources by extracting the links that unite the corresponding measures.\par

In contrast, the revolution of quantum technology has led to a growing interest in quantum metrology, and quantum entanglement has been employed to overcome the classical limit in several quantum estimation protocols. In this work, we analyze the role of quantum correlations beyond entanglement in improving the accuracy of an unknown parameter. According to our results, correlations can be captured using quantum Fisher information, and quantum discord correlations can be exploited to ensure the accuracy of phase estimation protocols.\par

This thesis includes also the contributions on the dynamics of these quantum resources in various models of open quantum systems. Among our objectives, is to study the effects of the environment on these quantum resources and to obtain techniques to protect them from the effects of intrinsic decoherence.\par

\tableofcontents
\addcontentsline{toc}{chapter}{Table des figures}
\listoffigures
\chaptertoc{Introduction Générale}
\pagenumbering{arabic}
\begin{fquote}[Leonardo da Vinci (1452-1519)]
	He who loves practice without theory is like the sailor who boards ship without a rudder and compass and never knows where he may cast.
\end{fquote}
La beauté de la nature est inexprimable. Il est donc d'un intérêt fondamental de comprendre les phénomènes naturels. Tous les systèmes et processus physiques sont régis par les lois de la nature. La physique quantique, formalisée dans les premières décennies du 20e siècle, est la théorie la plus fondamentale dont nous disposons pour décrire la nature, y compris les particules et les forces élémentaires qui composent l'univers. Elle s'est révélée être une théorie très réussie pour expliquer les phénomènes microscopiques et ses prédictions correspondent aux expériences avec une grande précision. Cette théorie contient des éléments qui sont fondamentalement différents de ceux exigés dans la description de la nature par la physique classique \cite{Dirac1981}. En effet, dans la physique classique, notamment la mécanique newtonienne et la théorie électromagnétique de Maxwell, les particules se déplacent selon des trajectoires, et leurs vitesses et accélérations peuvent donc être déterminées à partir de leurs positions. Les choses se compliquent toutefois lorsque l'on introduit des objets trop petits, tels que des atomes ou des électrons. Cela a mené à la formulation de la théorie de la mécanique quantique et à la densité de probabilité trouvée à partir de la fonction d'onde permettant de se substituer aux trajectoires des corps pour les systèmes quantiques. Plus important encore, la physique quantique est à l'origine de développements technologiques qui ont contribué à façonner la société actuelle, tels que le transistor, l'énergie nucléaire, le laser, l'imagerie médicale par résonance magnétique et le microscope électronique. En outre, il s'agit d'une discipline essentielle pour comprendre la physique des particules, la théorie quantique des champs, la physique statistique quantique, en passant par la physique des matériaux et d'autres branches de la physique moderne, qui sont à la base de toute l'électronique moderne. De surcroît, la science quantique a dépassé le cadre de la physique pour transformer notre compréhension de l'information et ouvrir la voie à des technologies de l'information nouvelles et révolutionnaires, telles que les ordinateurs quantiques et l'internet quantique. Cependant, les fondements de la physique quantique restent une source de débat scientifique et philosophique.\par

L'une des caractéristiques les plus marquantes de la physique quantique est l'existence de corrélations quantiques entre différents systèmes quantiques. Dans un monde classique, si un système à l'état pur peut être divisé en deux sous-systèmes, alors la somme des informations des sous-systèmes constitue l'information complète du système global. Ceci n'est plus vrai dans le formalisme quantique. En particulier, il existe des états quantiques constitués de deux ou plusieurs systèmes physiques pour lesquels l'information complète de l'ensemble n'est pas disponible, même lorsque les sous-systèmes sont complètement aléatoires. Erwin Schrödinger \cite{Schrodinger1935} a lancé le terme \quotes{intrication quantique} \cite{Horodecki2009} pour décrire cette caractéristique quantique.\par

Le stockage, la transmission et le traitement de l'information constituent les éléments de base de la théorie de l'information et ces éléments sont basés sur des porteurs physiques. Lorsque ces porteurs physiques sont des systèmes quantiques tels que des atomes ou des photons, ils constituent la base de la théorie quantique de l'information. Grâce à la théorie quantique, de nouvelles possibilités s'offrent aux tâches de traitement de l'information et de communication. Les performances accrues des systèmes quantiques dans ces tâches seraient principalement dues à la superposition et à l'intrication quantiques présentes dans les systèmes quantique. Parmi les applications notables, citons la téléportation quantique \cite{Bouwmeester1997,Pirandola2015}, cryptographie quantique ou la distribution quantique de clés \cite{Gisin2002,Bennett1992}, la correction d'erreurs quantiques \cite{Scott2004}, la factorisation de grands nombres \cite{Shor1994} et la recherche d'une grande base de données sur un ordinateur quantique \cite{Grover1996}. Évidemment, la certification des corrélations quantiques est essentielle pour le développement plus rapide des technologies quantiques. Au-delà de cette demande pratique, une compréhension fondamentale des corrélations quantiques, incluant leur caractérisation et leur quantification, joue un rôle clé dans l'exploration de la frontière entre les théories classiques et les caractéristiques uniques de la physique quantique. La question demeure cependant de savoir quels types de corrélations sont véritablement quantiques. \par 

L'intrication est à la base de nombreuses tâches quantiques fondamentales \cite{Plenio2014,Guhne2009}, et est souvent considérée comme un synonyme de corrélations quantiques dans les premières études, bien qu'il soit maintenant reconnu que la notion de corrélations quantiques a une portée beaucoup plus large, et que l'intrication est un type particulier, bien que le plus important de corrélations quantiques, c'est-à-dire que l'intrication peut être identifiée comme des corrélations quantiques non locales \cite{Werner1989}. L'étude de l'intrication remonte explicitement aux travaux fondamentaux d'Einstein, Podolsky et Rosen \cite{Einstein1935}, et de Schrödinger \cite{Schrodinger1936,Schrodinger21935}, dès les années 1930. Aujourd'hui, l'intrication est considérée comme une ressource clé de l'information quantique et est souvent liée à la non-localité quantique \cite{Bell1987,Brunner2014}. Sa théorie peut être grossièrement divisée en trois parties. La première est de type qualitative, c'est-à-dire qu'elle répond à la question \quotes{Est-ce que cet état est intriqué ou non?}. La deuxième partie de type comparative pose la question \quotes{Cet état est-il plus intriqué que cet autre?}, et enfin de type quantitative, elle pose la question \quotes{Quel est le degré d'intrication de cet état?}, et donne ses réponses sous la forme de mesures d'intrication attribuant un chiffre à chaque état. Des problèmes de quantification se posent naturellement lorsque l'intrication est utilisée comme ressource pour des tâches de traitement de l'information quantique. Par exemple, il est bien connu que les états intriqués sont, d'une certaine manière, le carburant des processus de téléportation quantique, et pour chaque étape de la transmission, un système maximalement intriqué est nécessaire. Le processus fonctionne également avec des états moins que maximalement intriqués, mais ils deviennent moins efficaces. En ce sens, l'une des interrogations centrales de la théorie de l'information quantique est l'étude de l'intrication dans divers systèmes quantiques et, en particulier, la manière dont elle peut être quantifiée.\par

En 2002, après avoir étudié la corrélation entre l'appareil et le système lors d'une mesure, Ollivier et Zurek ont réalisé \cite{Ollivier2001} que les états séparables, tels que définis par Werner \cite{Werner1989}, peuvent encore avoir une certaine corrélation dans le sens où ils peuvent être perturbés par des mesures locales. Cette mesure nouvelle des corrélations quantiques au-delà de l'intrication est appelée \quotes{discode quantique}. Ils ont constaté que l'intrication n'est pas la seule corrélation quantique qui n'a pas de contrepartie classique. D'autres types de corrélations quantiques, comme la discorde quantique, peuvent également être responsables de l'accélération de certains algorithmes quantiques, alors que l'intrication peut disparaître ou être négligeable \cite{Modi2012}. Depuis les premiers énoncés de ce concept, de nombreux efforts de recherche ont été consacrés à la compréhension des propriétés mathématiques et des significations physiques de la discorde et de quantités similaires. Des études complètes sur les propriétés de la discorde sont disponibles dans \cite{Adesso2016}, et les références \cite{Brodutch2017,Adesso22016} fournissent des perspectives récentes dans ce domaine.\par

Curieusement, il est connu que la discorde et l'intrication quantique ne constituent pas des caractéristiques équivalentes. L'intrication se caractérise par le fait qu'une information complète sur les sous-systèmes ne reproduit pas l'information complète du système global. Il n'est pas surprenant que ce ne soit pas la seule caractérisation d'un système quantique. Par exemple, la disparition d'une partie d'un sous-système après la mesure d'une autre est une autre caractéristique propre aux systèmes quantiques et n'a pas d'analogue classique. La quantité qui tente de capturer cette caractéristique unique est la discorde quantique.\par

L'étude de la discorde quantique présente de nombreux défis et questions ouvertes. Une difficulté majeure avec les quantités de type discorde est qu'elles sont difficiles à calculer ou à analyser. Formellement, la discorde quantique est définie comme étant la différence entre l'information mutuelle quantique (corrélation totale) et la quantité maximale d'information mutuelle accessible à une mesure quantique (corrélation classique). Pour quantifier les corrélations (classiques) localement accessibles, cette quantification implique une optimisation sur toutes les mesures locales possibles. Cette mesure est choisie pour maximiser les corrélations classiques. En raison de la complexité du processus d'optimisation, le calcul de la discorde quantique et de ses variantes (comme la discorde géométrique) n'est pas une tâche aisée et les résultats analytiques ne sont connus que pour certaines familles restreintes d'états. Pour pallier ce problème, plusieurs autres versions de la discorde ont également été proposées, notamment la discorde quantique linéaire \cite{Ma2015}, l'incertitude quantique locale \cite{Girolami2013} et l'information quantique de Fisher locale \cite{Kim2018}. La plupart de ces quantificateurs de corrélation quantique pour les états quantiques purs peuvent coïncider, et peuvent parfois présenter des comportements similaires pour les états mixtes. Cependant, il existe également des différences subtiles pour ces corrélations quantiques et leurs interprétations physiques sont également différentes. Tout cela indique que les propriétés des corrélations quantiques d'un système sont complexes et que leur caractérisation sous différents aspects est nécessaire.\par

La possibilité pour les systèmes quantiques d'exister dans des \quotes{états de superposition} révèle la nature ondulatoire de la matière et constitue une grande différence avec la physique classique. De fait, l'observation d'interférences en optique classique nécessite généralement une cohérence (il existe des versions spatiales et temporelles), ce qui implique que les différentes parties d'une onde ont une relation de phase fixe. Le même principe s'applique aux interférences entre les particules quantiques, bien que la phase soit désormais une propriété intrinsèque de la fonction d'onde, plutôt qu'un champ classique \cite{Theurer2017,Biswas2017}. Dans cette image, si la cohérence quantique existe entre deux points de la fonction d'onde d'une particule, celle-ci peut être comprise comme étant dans une superposition des deux positions simultanément \cite{Wang2017}. Voilà pourquoi on dit souvent que les systèmes dans de tels états de superposition possèdent une cohérence quantique. L'application de la cohérence quantique comme ressource pour les protocoles d'information quantique a récemment fait l'objet d'une grande attention car elle relie les questions fondamentales sur la nature physique aux aspects pratiques des technologies quantiques à venir (voir la référence \cite{Chitambar2019} pour une introduction). La construction d'un cadre mathématiquement rigoureux et physiquement significatif pour sa caractérisation et sa quantification a été l'un des principaux objectifs des chercheurs de la communauté quantique, car cela est non seulement essentiel pour les fondations quantiques, mais peut également fournir la base de ses applications potentielles dans une grande variété de sujets prometteurs. Nous nous focalisons dans cette thèse sur les développements récents concernant la caractérisation quantitative de la cohérence, dont l'essence est l'adoption du point de vue de traiter la cohérence quantique comme une ressource physique, tout comme la théorie des ressources de l'intrication.\par

Dans le cadre général, la mécanique quantique fournit des modèles exacts ou arbitrairement précis pour la description de nombreux systèmes simples, et des extensions numériques avec des approximations contrôlées rendent l'étude de phénomènes naturels plus complexes faisable et quantitativement satisfaisante. Cependant, lorsque les systèmes physiques concernés ne sont pas en eux-mêmes particulièrement complexes mais qu'ils interagissent avec des degrés de liberté externes échappant à notre contrôle, là où ces interactions ont le moins de chances d'être négligeables, la dynamique globale de ces systèmes est difficile à prendre en compte quantitativement. Ce problème fait l'objet de la théorie des systèmes quantiques ouverts \cite{Davies1976,Rivas2012}, un domaine de recherche actuel riche et en pleine expansion, qui constitue l'un des contextes généraux du travail présenté dans cette thèse. Brièvement, l'interaction entre un système quantique et son environnement peut entraîner la dissipation ou la perte d'informations contenues dans le système au bénéfice de son environnement. Pour obtenir une description complète d'un système quantique, il faut intégrer l'effet de l'environnement à l'hamiltonien du système original. C'est la philosophie qui a présidé à l'étude des systèmes quantiques ouverts. L'une des principales tâches de cette théorie est de résoudre l'équation dite \quotes{équation maîtresse} des systèmes quantiques ouverts. En général, la résolution d'une telle équation peut être très compliquée et nécessite de nombreux calculs fastidieux, en particulier dans le cas non-Markovien dans lequel le processus d'évolution présente le comportement des effets de mémoire \cite{Breuer2016}.\par

Par ailleurs, la décohérence quantique constitue un obstacle majeur à la mise en œuvre des nouvelles propositions de la science et de la technologie de l'information quantique, telles que le calcul quantique \cite{DiVincenzo2000}, la métrologie quantique \cite{Caves1981,Wineland1992} et la simulation quantique \cite{Britton2012}. En raison de son omniprésence dans tous les types de dispositifs physiques, l'étude, la réduction et le contrôle de la décohérence constituent depuis longtemps un sujet brûlant dans ces domaines. Dans cette optique, le comportement dynamique des corrélations et de la cohérence quantique présentes dans un système quantique ouvert composite dépend fortement du bruit produit par le milieu environnant. En conséquence, l'un des aspects les plus importants de l'environnement est de savoir s'il peut être décrit comme sans mémoire (markovien) ou avec mémoire (non markovien). En réalité, la dynamique de l'intrication quantique dans les systèmes quantiques ouverts a été largement étudiée dans la littérature. Néanmoins, peu de travaux ont traité de l'effet de l'environnement sur la cohérence et d'autres quantificateurs des corrélations quantiques de type discorde.\par

Plusieurs quantités d'intérêt dans la théorie de l'information quantique ne correspondent pas à des observables quantiques et ne peuvent pas être évaluées directement par des mesures. Il s'agit par exemple de la pureté d'un état quantique, d'une phase quantique ou de corrélations quantiques. Dans tous ces cas, il faut recourir à la mesure indirecte et déduire la valeur de la quantité d'intérêt à partir de son influence sur une source donnée. La façon canonique de traiter ce problème est d'utiliser les outils de la théorie de l'estimation quantique locale \cite{Giovannetti2006,Giovannetti2011}. Les problèmes d'estimation consistent à attribuer des valeurs appropriées à des quantités qui sont inconnues parce qu'elles ne sont pas entièrement accessibles par observation. Compte tenu des informations disponibles, on peut élaborer une stratégie optimale pour extraire des estimateurs appropriés des inconnues. Une procédure d'estimation a pour objectif de trouver la meilleure stratégie pour déduire la valeur d'un paramètre inconnu avec la plus grande précision possible. Ceci est réalisé en effectuant des mesures indirectes sur le système quantique, c'est-à-dire en déduisant la valeur du paramètre en traitant l'ensemble des résultats de mesure d'une autre observable, ou d'un ensemble de variables. Cette théorie donne la limite ultime de la précision avec laquelle les paramètres d'un système physique peuvent être mesurés par une source quantique. Cette limite, appelée limite quantique de Cramér-Rao \cite{cramer1946}, n'est peut-être pas réalisable en général, soit en raison de limitations pratiques, soit même pas en principe \cite{Paris2009}. Cependant, elle donne une indication de la faisabilité d'une certaine conception d'un nouveau détecteur quantique et de la possibilité d'obtenir un meilleur résultat que la technologie actuelle.\par

La métrologie quantique est un mécanisme qui utilise les ressources distinctives de la mécanique quantique, telle que l'intrication, pour augmenter la précision de l'estimation des paramètres par des mesures quantiques au-delà de la limite de sa contrepartie classique. Elle est enracinée dans la théorie de l'estimation quantique, initiée par Helstrom \cite{Helstrom1976} et Holevo \cite{Holevo1982}, qui ont proposé la relation d'incertitude basée sur les paramètres. Braunstein et ses collaborateurs \cite{Braunstein1994} ont développé cette théorie du point de vue de la borne de Cramér-Rao \cite{cramer1946}, qui caractérise la manière dont un paramètre peut être estimé à partir d'une distribution de probabilité, et ont obtenu l'information optimale de Fisher sur différents schémas de mesure quantique pour un état quantique donné en fonction d'un paramètre. Cette information est souvent appelée \quotes{l'information quantique de Fisher}. En raison de l'importance des mesures de précision dans différents domaines de la physique, l'information quantique de Fisher a suscité un grand intérêt de la part des chercheurs. Giovannetti et son équipe \cite{Giovannetti2004} ont découvert que la mise à l'échelle de l'information de Fisher quantique présente une amélioration de $N^{\frac{1}{2}}$ par rapport à sa contrepartie classique si un état maximalement intriqué de $N$-qubits est utilisé. Cette découverte a stimulé l'émergence de la métrologie quantique, qui a été appliquée à différents systèmes quantiques pour améliorer la précision des mesures.\par

A ce jour, les ressources quantiques sont très importantes dans tous les protocoles de la métrologie quantique. Dans cette thèse, nos préoccupations sont les corrélations quantiques, la cohérence quantique, leurs connexions intrinsèques et leurs rôles en vue d'améliorer la précision des paramètres dans l'estimation quantique. A cette fin, cette thèse est présentée d'une manière similaire à un ouvrage avec de nombreux détails techniques, ce qui pourrait aider les chercheurs à suivre et mieux comprendre les résultats correspondants. En plus de présenter une vue globale des principaux développements des corrélations quantiques et de la cohérence quantique, nous essayons de résumer et reformuler certains calculs éparpillés dans un grand nombre de littérature, en combinaison bien sûr avec nos propres résultats.\par

Au premier chapitre, nous passons en revue les éléments de base et la terminologie de la théorie de Claude Shannon, de la mécanique quantique et de la théorie de l'information quantique. Nous clarifions la signification de ces termes et fixons les notations et la terminologie adoptées dans le reste de l'ouvrage. Ensuite, nous abordons la théorie des ressources de l'intrication quantique, nous expliquons comment elle peut être quantifiée et nous présentons les expressions analytiques correspondantes.\par

Dans le deuxième chapitre, l’intérêt est porté au corrélations quantiques au-delà de l'intrication quantique. Nous vous ferons part de plusieurs mesures du type discorde et de ses propriétés, ainsi que des méthodes mathématiques employées dans les calculs analytiques. De plus, nous étudions l'interaction entre les effets quantiques globaux, comme la corrélation quantique, et les effets locaux, en particulier l'incertitude quantique sur les observables uniques.\par

Dans le troisième chapitre, nous présentons la description générale des systèmes quantiques ouverts. Nous traitons leur évolution, avec un accent particulier sur les systèmes quantiques markoviens, en utilisant l'équation maîtresse. L'objectif est d'introduire la théorie de la décohérence quantique et la façon dont elle peut expliquer la transition du mode quantique au mode classique. Dans la deuxième partie, nous passons en revue les aspects fondamentaux liés à la théorie des ressources de la cohérence quantique, en particulier la structure des états libres et les opérations libres correspondantes. Nous présentons ensuite une revue détaillée des développements récents de la quantification de la cohérence quantique. De plus, certaines de leurs extensions valables pour les systèmes de dimension finie, par exemple d'autres mesures de cohérence basées sur l'intrication quantique et la discorde quantique, seront également examinées.\par

Dans le quatrième chapitre, nous introduisons les notions de base dont nous avons besoin pour procéder à l'estimation de l'état quantique. Nous abordons également le problème de l'estimation multiparamétrique. Nous donnons la forme explicite de l'information quantique de Fisher (dans le cas de l'estimation à un seul paramètre) et de la matrice d'information quantique de Fisher (dans le cas de l'estimation multiparamétrique). Le rôle des corrélations quantiques, contenues dans les états quantiques, en métrologie quantique a également été examiné. Nos résultats suggèrent que la corrélation quantique peut être exploitée pour garantir la précision des protocoles d'estimation de phase.\par

Le cinquième chapitre présente nos contributions dans le domaine. Nous terminons ce manuscrit de thèse par une conclusion générale évoquant les idées principales développées tout au long du document. Certaines de nos perspectives sont également mentionnées.
\section*{Liste des contributions de l'auteur}
\underline{\bf Contributions et Publications:}\\
	La plupart des informations présentées ici ont été rapportées sous diverses formes dans ces articles:
	\begin{itemize}[font=\color{black} \Large, label=\ding{42}]
	\item The dynamics of local quantum uncertainty and trace distance discord for two-qubit $X$ states under decoherence: A comparative study, \underline{\bf A. Slaoui}, M. Daoud and R. Ahl Laamara, Quantum Information Processing, {\bf17} (2018) 178; arXiv: 1904.04382.
	
	\item Universal evolution of non-classical correlations due to collective spontaneous emission, \underline{\bf A. Slaoui}, M. I. Shaukat, M. Daoud and R. Ahl Laamara, The European Physical Journal Plus, {\bf133} (2018) 413; arXiv: 1902.02224.
	
	\item Phonon-mediated quantum discord in dark solitons,
	M. I. Shaukat, \underline{\bf A. Slaoui}, H. Terças and M. Daoud, The European Physical Journal Plus, {\bf135} (2020) 357; arXiv: 1903.06627.
	
	\item Quantum Fisher information matrix in Heisenberg $XY$ model, L. Bakmou, \underline{\bf A. Slaoui}, M. Daoud and R. Ahl Laamara. Quantum Information Processing, {\bf18} (2019) 163; arXiv: 1904.07507.
	
	\item A comparative study of local quantum Fisher information and local quantum uncertainty in Heisenberg $XY$ model,
	\underline{\bf A. Slaoui}, L. Bakmou, M. Daoud and R. Ahl Laamara. Physics Letters A, {\bf383} (2019) 2241-2247; arXiv:1904.12132.
	
	\item The dynamic behaviors of local quantum uncertainty for three qubit $X$ states under decoherence channels,
	\underline{\bf A. Slaoui}, M. Daoud and R. Ahl Laamara, Quantum Information Processing, {\bf18} (2019) 250; arXiv: 1907.00696.
	
	\item Quantum discord based on linear entropy and thermal negativity of qutrit-qubit mixed spin chain under the influence of external magnetic field, F. B. Abdallah, \underline{\bf A. Slaoui} and M. Daoud. Quantum Information Processing, {\bf19} (2020) 252.
	
	\item Influence of Stark-shift on quantum coherence and non-classical correlations for two two-level atoms with a single-mode cavity field, \underline{\bf A. Slaoui}, A. Salah and M. Daoud, Physica A: statistical mechanics and its applications, {\bf558} (2020) 124946; arXiv: 2003.11338.
\end{itemize}
\underline{\bf Communications orales:}
\begin{enumerate}[font=\color{black} \Large, label=\ding{229}]
	\item Évolution des corrélations quantiques et processus de décohérence pour les systèmes bipartites, Journées Nationales des Doctorants et des Jeunes Chercheurs 2017. l'ENSET- Rabat.
	
	\item La dynamique des corrélations quantiques mesurée par l'incertitude quantique locale et la discorde géométrique pour les systèmes bipartites, La Quatrième Ecole sur la Théorie de l'Information et la Cryptographie Quantiques, Juin (2017). Faculté des Sciences de Rabat.
	
	\item Les comportements dynamiques de l'incertitude quantique locale pour les états $X$ tripartites sous les canaux de decohérence, RNDJC-2018. Faculté des Sciences de Rabat.
	
	\item Expression analytique de l'incertitude quantique locale pour les systémes tripartites et leur dynamique sous les canaux de décohérence, RNJCP9-2018. Université Hassan II de Casablanca, Maroc.
	
	\item La dynamique des correlations quantiques pour les systémes tripartites sous les canaux de décohérence, 1CNPFA-2019. Faculté Polydisciplinaire Beni Mellal, Maroc.
	
	\item Local quantum Fisher information vs local quantum uncertainty, RNICQ-2019. FS, Université Ibn Tofail. Kenitra, Maroc.
	\item Characterizing nonclassical correlations via quantum Fisher information. Sciences de l'Information Quantique: Fondements et applications. RNIQ-2021 (événement virtuel).
 \end{enumerate}
\underline{\bf Communications par affiche:}
\begin{enumerate}[font=\color{black} \Large, label=\ding{42}]
	\item Comparison of quantum discord and local quantum uncertainty, and their dynamics under decoherence, Conference on Quantum Information Theory, December (2017), Institut Henri Poincare, Paris, France.
	
	\item The various measurements of quantum correlations and their dynamics in open systems, October (2017), Rencontre des Jeunes Physiciens, Collége de France, Paris, France.
	
	\item Quantifiers of nonclassical correlations by quantum uncertainty; Local quantum Fisher information vs local quantum uncertainty, Centre International de Rencontres Mathématiques. Janvier (2020), Luminy, Marseille. France.
\end{enumerate}

\chapter{Principes de la théorie quantique de l'information}
La théorie de l'information classique a été fondée par Claude Shannon peu après la seconde guerre mondiale et suppose que l'information est codée dans des systèmes physiques évoluant selon les lois de la physique classique \cite{Shannon1948,Jaynes1957}. Comme fondement de sa théorie, Shannon a développé un modèle de communication abstrait très simple dans le contexte de l'envoi d'informations sur un canal tel qu'un fil téléphonique. Ce modèle fourni un moyen de quantifier l'information et applicable dans de nombreuses situations et pour tous les types de communication et de traitement de l'information. Il nous permet également de tirer des limites sur la complexité ou les coûts des tâches telles que le stockage d'informations ou l'envoi des données sur des canaux bruyants.\par 

Avec l'idée que l'information est associée à une représentation physique, et que le stockage, le traitement et la transmission d'informations sont tous régis par des lois physiques, une grande attention a été consacrée à l'étude de nombreux phénomènes des systèmes physiques d'un point de vue théorique dans le but d'exploiter leurs propriétés physiques dans le codage de l'information \cite{Keyl2002}. D'autre part, les lois de la mécanique classique peuvent être obtenues à partir des lois de la mécanique quantique en faisant des choix particuliers pour un processus quantique et pour l'état d'un système quantique. Cela signifie que la théorie classique émerge de la théorie quantique comme une approximation. Dans ce sens, il semble plus convenable de poser une question sur la façon dont ces lois de la physique quantique peuvent être utilisées pour quantifier, communiquer et partager de l'information dans les structures de la matière à l'échelle microscopique à partir de son état qui est représenté par un ensemble de grandeurs physiques \cite{Wilde2011}.\par

La théorie de l'information quantique a été développée pour explorer la nature de l'information dans le monde quantique et à révéler de nouvelles capacités de traitement de l'information \cite{Timpson2004}. Cette nouvelle théorie revêt une importance capitale pour améliorer la sécurité et pour transmettre des informations de manière fiable en préservant l’information et des corrélations allant au-delà de ce qui est possible dans les modèles traditionnels, tout en ajoutant l'ingrédient magique de la mécanique quantique, notamment le principe de superposition et l’intrication quantique.\par

Avant d'étudier les nouveaux aspects que la mécanique quantique ajoute à la théorie de l'information, il est utile de connaître quelques bases de la théorie de l'information classique pour étudier comment elle doit être modifiée dans des situations où les effets quantiques sont importants. Pour cette raison, nous commençons dans ce chapitre par présenter les bases de la théorie de l'information classique. En particulier, nous allons discuter comment l'information peuvent être quantifiée et comment nous pouvons comparer les informations dans deux ensembles de données. Nous présentons les différents types de mesures informationnelles telles que l'entropie de Shannon, l'entropie relative, l'information mutuelle, l’entropie conjointe et conditionnelle. Cela fournit les bases pour la deuxième partie de ce chapitre, où nous étudions la théorie quantique de l’information, dans laquelle l'information est représentée par des états quantiques et laissée évoluer selon les lois de la mécanique quantique. Dans cette partie, nous nous intéressons, successivement, aux quelques formalismes de la mécanique quantique ainsi qu'au formalisme mathématique des différentes mesures de l'intrication quantique en tant que type spécial de corrélations quantiques. Nous commencerons par définir l'opérateur densité, l'entropie de von Neumann et le bit quantique. Ensuite, nous allons introduire quelques critères de séparabilité permettant de distinguer les états quantiques intriqués et non intriqués. Dans le même esprit, nous introduisons quelques mesures de l’intrication quantique pour les systèmes quantique bipartites et nous donnons l'expression analytique de chaque mesure discutée.
\section{Théorie classique de l'information}

Les systèmes de communication sont constitués d'une source d'information, d'un émetteur ou d'un encodeur, d'un canal bruyant et d'un récepteur (décodeur). Dans ce schéma, le contenu de l'information d'un événement ne joue aucun rôle dans la théorie de l'information mais l'incertitude associée à son occurrence \cite{Nyquist1928,Hartley1928}. Par exemple, dans le cas simple d'un appareil téléphonique, ce qui est transmis n'est pas ce qui est dit sur le téléphone, mais plutôt un signal analogique qui enregistre les ondes sonores émises par le locuteur, puis ce signal analogique est envoyé numériquement après un codage. Il est intéressant de souligner que l'unité de mesure fondamentale en théorie de l'information de tous les supports d'information, textes, signaux téléphoniques, ondes radio et tous les modes de communication peuvent être codés en "bit", un objet qui représente une distinction entre deux possibilités soit la valeur "0" soit "1" et qui est représenté physiquement (par exemple par un interrupteur) et une séquence de bits nous permet de transmettre un message \cite{Watrous2018}.\par

Selon Shannon, le concept d'information est décrit quantitativement dans un cadre probabiliste et que l'information est quantifiée en termes d'incertitude. Il considère que l'incertitude associée au système physique est la quantité d'informations qu'il transporte. En effet, la probabilité concerne ce que nous savons, les informations que nous avons ou pourrions obtenir sur un événement. Nous pouvons donc dire que les informations dépendent de l'observateur et avec le gain d'informations, il y a une réduction de l'incertitude sur un événement associé. En ce sens, moins l'occurrence d'un résultat de l'événement est favorable (c-à-d, la probabilité de l’événement est faible), plus d'information est apprise sur son occurrence. De façon similaire, plus l'occurrence d'un résultat de l'événement est favorable, moins d'information est apprise sur son occurrence \cite{Verdu1998}. Cette première section tente de présenter l'étude formelle de la théorie de l’information classique qui a été introduite par Claude Shannon en 1948. Nous allons passer en revue les définitions ainsi que certaines de ses propriétés importantes permettant de jeter la base en vue d’introduire et aborder la théorie de l’information quantique.

\subsection{Mesure de l'information: Entropie de Shannon}
 
La probabilité et l'information sont liées par la notion d'entropie, qui est introduite par Boltzmann pour décrire les mélanges statistiques en thermodynamique statistique \cite{Clausius1865,Boltzmann1896}. Dans ce cas, l'entropie est une mesure du désordre qui exprime l’incertitude ou le caractère aléatoire du système, et donc de l’information manquante sur un système donné. A partir de cette définition et de considérations axiomatiques sur l'information, Shannon a identifié une autre forme d'entropie statistique pour la source de l'information et a démontré le sens de sa formule comme une mesure d'information. Ce concept est connu sous le nom de l'entropie de Shannon, et il peut être identifiée comme l'information portée par un état donné.\par

Considérons un événement, qui est une source d'information, décrit par une variable aléatoire $X$ de taille finie. Supposons que cette source d'information choisit au hasard $n$ symboles $X\equiv\{x_1, x_2,..., x_n\}$ et chaque symbole $x_i$ correspond à un résultat de l’événement avec la probabilité $p\left(x_i\right)$. D'après notre discussion précédente, la fonction $\zeta\left(x\right)$ qui mesure la quantité d’information fournie par un événement $x$ doit être nécessairement décroissante avec sa probabilité $p\left(x\right)$, c-à-d, $\zeta\left(x\right)=f\left( 1/p\left(x\right)\right)$ où $f$ est une fonction croissante. Aussi, la fonction $f$ est nulle pour un événement qui n'apporte aucune information (c-à-d, $p\left(x\right)=1$ et $f\left(1\right)=0$). De plus, pour une source d'informations qui produit deux symboles indépendants $x_{1}$ et $x_{2}$ avec la probabilité $p\left(x_{1}, x_{2}\right)$ (dans ce cas, $p\left(x_{1},x_{2}\right)=p\left(x_{1}\right)p\left(x_{2}\right)$), la quantité d’information des deux symboles $x_{1}$ et $x_{2}$ est égale à la somme de leurs quantités d’information individuelles, c-à-d,
\begin{align}
\zeta\left(x_{1},x_{2}\right)&=f\left(\frac{1}{p\left(x_{1}, x_{2}\right)}\right)=f\left( \frac{1}{p\left(x_{1}\right)p\left(x_{2}\right)}\right)= f\left(\frac{1}{p\left(x_{1}\right)}\right)+f\left(\frac{1}{p\left(x_{2}\right)}\right)=\zeta\left(x_{1}\right)+\zeta\left(x_{2}\right).
\end{align}
D'après les axiomes ci-dessus, Nous pouvons conclure que la fonction $f$ est une fonction logarithmique. Conformément à ce choix, la base du logarithme ici égale à $2$ car elle détermine l'unité d'information (bit classique). Dans cette direction, le contenu d'information émanant de la source d'information s'appelle son entropie de Shannon \cite{Shannon1948}. Elle est définie par
\begin{equation}
{\cal H}\left( X\right)=\sum\limits_{x_{i}} p\left(x_{i}\right)\zeta\left(x_{i}\right)=-\sum\limits_{x_{i}}p\left(x_{i}\right)\log_{2}\left(p\left(x_{i}\right) \right),\label{H(X)}
\end{equation}
avec la convention que $0\log_{2}0\equiv0$. Notez que l'entropie d'une variable aléatoire $X$ est une quantité positive et ne dépend pas de ses valeurs, c’est-à-dire les éléments de l’alphabet $x_{i}$, mais dépend uniquement des probabilités $p\left(x_{i}\right)$ que ces valeurs prennent. Dans le cas d'un ensemble de variables aléatoires, l'entropie de Shannon (\ref{H(X)}) de chaque variable peut être utilisée pour spécifier d'autres mesures qui contiennent des informations sur ses relations avec d'autres variables. Ce sera l'objet des sous-sections suivantes.
\subsection{L'entropie relative}

Considérons deux variables aléatoires classiques $X$ et $Y$ donnant les résultats $\{x_i\}$ et $\{y_i\}$ avec des probabilités $\{p_x\}$ et $\{p_y\}$, respectivement. L’entropie relative, également connue sous le nom de divergence de Kullback Leibler, est l’incertitude relative à la distribution de probabilité $p_x$ par rapport à celle de $p_y$. Autrement dit, il s'agit d'une mesure de la similarité entre deux variables aléatoires et de la proximité des probabilités associées. L'entropie relative donnée par ${\cal H}\left(X\|Y\right)$ est définie comme suit
\begin{equation}
	{\cal H}\left(X\|Y\right)=-\sum\limits_{x,y}p_{x}\log_{2}\left(\frac{p_{y}}{p_{x}}\right)=-\sum\limits_{x,y}p_{x}\log_{2}\left(p_{y} \right)-{\cal H}\left( X\right).  \label{ER}
\end{equation}
Notez que l'entropie relative est toujours positive, ${\cal H}\left(X\|Y\right)\geq 0$ , avec condition d'égalité est déduite si seulement si $p_x=p_y$ pour toutes les valeurs de $x$ et de $y$, c'est-à-dire pour des distributions identiques. Pour prouver cette propriété, nous devons utiliser l'égalité mathématique suivante; $\log_{2}\left( x\right)\ln2 =\ln\left(x\right)\leq x-1$ pour tout $x$ positif et avec égalité si seulement si $x=1$. Donc nous avons
\begin{align}
{\cal H}\left(X\|Y\right)&=-\sum\limits_{x,y}p_{x}\log_{2}\left(\frac{p_{y}}{p_{x}} \right)\notag\\&\geq \frac{1}{\ln 2}\sum\limits_{x,y}p_{x}\left(1- \frac{p_{y}}{p_{x}}\right)\notag\\&= \frac{1}{\ln 2}\sum\limits_{x,y}\left( p_{x}-p_{y}\right)\notag\\&= \frac{1}{\ln 2}( \underbrace {\sum\limits_x {{p_x}} }_{=1}-\underbrace {\sum\limits_y {{p_y}} }_{=1})=0. 
\end{align}
Cette propriété de l'entropie relative explique pourquoi elle est considérée comme une mesure de distance entre deux distributions en théorie de l'information.
\subsection{L'entropie conjointe et conditionnelle}

Pensons aux mêmes couples des variables aléatoires $\left(X,Y\right)$. La mesure d'information conjointe de $X$ et $Y$ est simplement l'entropie de la distribution conjointe de $X$ et $Y$, qui est équivalente à l’information combinée de ces deux variables aléatoires. En d'autres termes, c'est l'incertitude qui reste sur $Y$ une fois que nous connaissons la valeur de $X$. Elle est définie comme suit
\begin{equation}
{\cal H}\left( X,Y\right)=-\sum\limits_{x,y}p\left( x,y\right) \log_{2} p\left( x,y\right), 
\end{equation}
où $p\left( x,y\right)$ est la probabilité conjointe. Notez que l'entropie conjointe est symétrique, c-à-d ${\cal H}\left( X,Y\right)={\cal H}\left( Y,X\right)$. En outre, si $X$ et $Y$ sont indépendants, et grâce à la propriété additive de l'entropie de Shannon, l'entropie conjointe
peut s'écrire sous la forme d'une somme des entropies de Shannon de chaque variable
\begin{equation}
{\cal H}\left( X,Y\right)={\cal H}\left( X\right)+{\cal H}\left(Y\right).
\end{equation}
De façon analogue, on peut définir une entropie conditionnelle de ces deux variables comme le degré d'incertitude de la réalisation d'une variable aléatoire $X$ si la valeur d'une autre variable aléatoire $Y$ est connue. C'est une mesure de la moyenne du manque d'information sur la valeur $X$ si l’on connait parfaitement les informations de la seconde variable aléatoire $Y$. Elle est définie par
\begin{equation}
{\cal H}\left( X/Y\right)=-\sum\limits_{x,y}p\left( x/y\right) \log_{2} p\left( x/y\right), 
\end{equation}
où la probabilité conditionnelle est donnée par la loi de Bayer
\begin{equation}
p\left( x/y\right)=\frac{p\left( x,y\right) }{p\left(y\right)}=\frac{p\left(y/x \right)p\left( x\right)  }{p\left( y\right) }.\label{Bayer}
\end{equation}
Une propriété particulière de l'entropie conjointe est qu'elle est inférieure ou égale aux entropies singulières, c-à-d ${\cal H}\left(X/Y\right)\leq {\cal H}\left(X\right)$. Cela signifie que la connaissance de l'information d'une variable supplémentaire $Y$ ne peut pas augmenter l'entropie de la variable $X$. Pour prouver cette égalité, nous devons utiliser l'inégalité de Jensen; pour toute fonction concave $f$, les valeurs $t_{1},...,t_{n}$ et $\lambda_{1},...,\lambda_{n}\in[0,1]$ avec $\sum\limits_{i}\lambda_{i}=1$ ($i=1,...,n$), il s'ensuit que $\sum\limits_{i}\lambda_{i}f\left(t_{i} \right)\leq f( \sum\limits_{i}\lambda_{i}t_{i})$. Nous avons donc
\begin{align}
	{\cal H}\left( X/Y\right)&=-\sum\limits_{x,y}\frac{p\left(x,y\right)}{p\left(y\right)} \log_{2}\frac{p\left(x,y\right)}{p\left(y\right)}\notag\\&=\sum\limits_{x,y}p\left(x,y\right) \log_{2}\frac{p\left(y\right)}{p\left(x,y\right)}\notag\\&=\sum\limits_{x,y}p\left(x\right)\frac{p\left(x,y\right)}{p\left(x\right)} \log_{2}\frac{p\left(y\right)}{p\left(x,y\right)}\notag\\&=\sum\limits_{x}p\left(x\right)\sum\limits_{y}\frac{p\left(x,y\right)}{p\left(x\right)} \log_{2}\frac{p\left(y\right)}{p\left(x,y\right)}\notag\\&\leq \sum\limits_{x}p\left(x\right) \log_{2}\sum\limits_{y}\frac{p\left(x,y\right)}{p\left(x\right)}\frac{p\left(y\right)}{p\left(x,y\right)}\notag\\&=\sum\limits_{x}p\left(x\right)\log_{2}\frac{1}{p\left(x\right)}={\cal H}\left( X\right).
\end{align}
En utilisant l'équation (\ref{Bayer}), nous pouvons également prouver que l'entropie conditionnelle est liée à l'entropie conjointe par cette équation
\begin{equation}
{\cal H}\left( X,Y\right)={\cal H}\left( X\right)+{\cal H}\left( Y/X\right)={\cal H}\left( Y\right)+{\cal H}\left( X/Y\right).
\end{equation}
Dans la sous-section suivante, nous montrons comment ces définitions se rapportent au concept d'information mutuelle, qui est le concept fondamental qui nous traitons tout au long de cette thèse.
\subsection{Information mutuelle}

Pour deux variables discrètes $X$ et $Y$ dont la distribution de probabilité conjointe est $p\left(x, y\right)$, l'information mutuelle mesure la quantité d'informations communes entre elles. Elle est principalement utilisée pour quantifier les corrélations totales entre ces deux variables aléatoires. Elle est définie comme un cas particulier de l'entropie relative, car elle permet de distinguer entre une distribution de probabilité conjointe $p\left(x, y\right)$ et le produit des distributions de probabilité de chaque variable $p\left(x\right)p\left(y\right)$. Par conséquent, l'application de la formule (\ref{ER}) nous permet d'obtenir plus facilement l'information mutuelle comme suit
\begin{align}
	{\cal I}\left( X:Y\right)&={\cal H}\left( p\left(x,y \right) \|p\left(x\right)p\left(y\right) \right)=\sum\limits_{x,y}p\left(x,y \right)\log_{2}\left(\frac{p\left(x,y \right)}{p\left(x\right)p\left(y\right)} \right). \label{IMC}
\end{align}
L'équation ci-dessus (\ref{IMC}) peut également être exprimée en termes d'entropie conjointe et d'entropie conditionnelle comme
\begin{align}
{\cal I}\left( X:Y\right)&=-\sum\limits_{x}\sum\limits_{y}p\left(x,y \right)\log_{2}p\left(x\right)+\sum\limits_{x}\sum\limits_{y}p\left(x,y \right)\log_{2}p\left(x/y\right)\notag\\&={\cal H}\left( X\right)-{\cal H}\left(X/Y \right)\notag\\&= {\cal H}\left( Y\right)-{\cal H}\left(Y/X \right)\notag\\& ={\cal H}\left( X\right)+{\cal H}\left( Y\right)-{\cal H}\left(X,Y \right).\label{IMM}
\end{align}
A partir de l'équation (\ref{IMM}), l'information mutuelle peut être considérée comme la réduction de l'incertitude sur une variable aléatoire étant donné la connaissance d'une autre. Une information mutuelle élevée entre deux variables aléatoires indique une forte réduction de l'incertitude, une faible information mutuelle indique une petite réduction et une information mutuelle nulle signifie que les variables sont indépendantes.\par

\begin{figure}[H]
	\begin{minipage}[b]{.50\linewidth}
		\centering
		\vfill $\left(a\right)$
		
		\includegraphics[width=8.5cm]{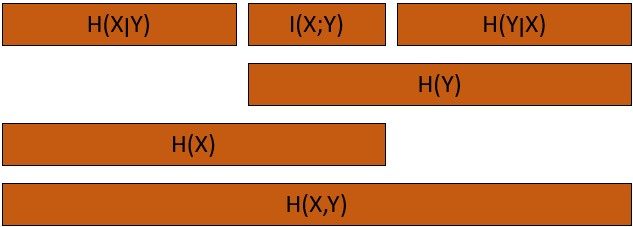}
	\end{minipage}
	\begin{minipage}[b]{.50\linewidth}
		\centering
		\vfill $\left(b\right)$
		
		\includegraphics[width=8cm]{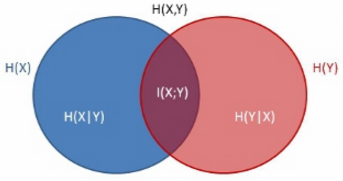}
	\end{minipage}
	\caption[Représentations intuitives des différents types d'entropies]{(a) Une représentation intuitive des différents types d'entropies (b) Une représentation traditionnelle à l'aide du diagramme de Venn.
	}\label{veen}	
\end{figure}

Comme le montre la Fig.(\ref{veen}), nous pouvons représenter toutes les différentes quantités citées précédemment sous la forme d'un diagramme appelé diagramme de Venn. Ce diagramme illustre les relations entre l'information mutuelle ${\cal I}\left( X:Y\right)$, l'entropie conjointe ${\cal H}\left( X,Y\right)$, l'entropie conditionnelle ${\cal H}\left( X/Y\right)$ et les entropies de Shannon ${\cal H}\left(X\right)$ et ${\cal H}\left(Y\right)$ des variables aléatoires $X$ et $Y$.
\subsection{La capacité du canal classique}

L'un des objectifs de la théorie de l'information est de connaître les limites supérieures et inférieures de la capacité des canaux bruyants. Physiquement, un canal classique peut être décrit comme le transfert d'un système classique de l'émetteur au récepteur. Si le transfert est intact et non perturbé, le canal est sans bruit. Contrairement, si le système classique interagit en cours de route avec un autre système, un canal classique bruyant se produit \cite{Bennett1997}. Supposons qu'un émetteur, Alice, souhaite transmettre des informations classiques à un récepteur, Bob, en utilisant un canal de communication classique avec un alphabet d’entrée associé $X$. Alice représentera les messages possibles en préparant le canal dans divers états classiques. Bob récupérera les informations en soumettant le canal à une mesure avec un alphabet de sortie associé $Y$. Par conséquent, le canal de communication possède une information mutuelle ${\cal I}\left( X:Y\right)$ entre le signal d'entrée $X$ et la sortie reçue $Y$. Il est intéressant de souligner que l'information mutuelle est une fonction concave de la distribution $p\left(x\right)$ lorsque la distribution conditionnelle $p\left(x/y\right)$ est fixée (voir l'équation (\ref{IMM})). Cette concavité implique qu'il existe une distribution $p\left(x\right)$ qui maximise l'information mutuelle. Alors, la capacité de Bob à récupérer le message d'Alice sans erreur est l'information mutuelle maximale entre l'entrée $X$ et la sortie $Y$ du canal et nous la désignons comme
\begin{equation}
{\cal C}=\max\limits_{p\left(x\right)}{\cal I}\left( X:Y\right)=\max\limits_{p\left(x\right)}\left[ {\cal H}\left( X\right)-{\cal H}\left(X/Y \right)\right],
\end{equation}
où la maximisation porte sur toutes les distributions de probabilités d'entrée $p\left(x\right)$ de $X$.

\section{Théorie quantique de l'information}

La physique fournit un moyen de comprendre le monde sur la base des lois fondamentales, et les phénomènes éventuellement exposés dans le monde sont conformes à ces lois fondamentales des théories en physique. Au niveau microscopique, les lois qui régissent les interactions physiques sont décrits par la mécanique quantique. D'un autre côté, la théorie quantique est une théorie probabiliste, et il était donc inévitable qu'une théorie de l'information quantique soit développée pour unifier la mécanique quantique et la théorie de l’information. Dans cette nouvelle théorie, les informations quantiques sont codées comme une caractéristique des systèmes quantiques (par exemple, la polarisation des photons ou le spin des particules) et qui sont associées à un état quantique décrit par l'opérateur densité agissant sur un espace de Hilbert \cite{CohenTannoudji1977,von-Neuman1932}. Comme nous l'avons dit, les principaux outils mathématiques utilisés par l'information quantique appartiennent à la mécanique quantique. À cet égard, cette partie passe en revue les concepts de base de la mécanique quantique ainsi que le processus de mesure pour présenter les états et les opérateurs des systèmes quantiques. Ensuite, nous allons présenter quelques mesures qui nous permettent de quantifier le degré de l'intrication dans un état quantique.

\subsection{Opérateur ou matrice densité}

En mécanique quantique, un espace de Hilbert ${\cal H}$ (espace des états) est associé à chaque système. C'est un espace vectoriel à $n$ dimension munie d'une base orthonormée $\{\left| {\cal U}_i \right\rangle, i=0,...,n-1\}$. Un élément quelconque de l'espace ${\cal H}$ est appelé vecteur-Ket et nous le désignons comme $\left| \psi \right\rangle$ et sa dimension dépend du système considéré. Son vecteur dual est noté $\left\langle \psi \right| $. N'importe quel état représenté par le vecteur $\left| \psi \right\rangle$ dans l'espace ${\cal H}$ peut être développé dans une base orthonormée comme une combinaison linéaire
\begin{equation}
	\left| \psi \right\rangle=\sum_{i=0}^{n-1}v_i \left| {\cal U}_i \right\rangle, \label{vu}
\end{equation}
que les physiciens quantiques appellent souvent une superposition et les amplitudes de probabilité $v_i$ sont des coefficients complexes. La probabilité de trouver le système dans l'état $\left| {\cal U}_i \right\rangle$ n'est autre que $|v_i|^{2}$ avec $\sum_{i=0}^{n-1}|v_i|^{2}=1$. Maintenant, si nous avons deux systèmes quantiques avec les espaces de Hilbert ${\cal H}_1$ et ${\cal H}_2$, respectivement, l'espace de Hilbert associé au système quantique combiné est ${\cal H}={\cal H}_1\otimes{\cal H}_2$. Il est important de noter que ces deux espaces de Hilbert pourraient représenter des systèmes physiques entièrement différents, par exemple, le premier pourrait être l'espace de spins d'électrons, tandis que le second pourrait être l'espace de polarisations de photons.\par

L'opérateur densité ou la matrice densité est l'outil qui permet de décrire tous les états quantiques possibles d’un système physique donné en une seule matrice à un instant donné. Il nous permet également de réduire les calculs en théorie de l'information quantique. Dans ce qui suit, nous discuterons la forme de cette matrice densité dans des états quantiques purs et mixtes et de leurs propriétés correspondantes.
\subsubsection{Cas d’un état pur}

Un état quantique général est représenté par un opérateur densité $\rho$ qui est un opérateur hermitien positif avec trace unitaire. C'est-à-dire qu'il doit remplir les trois conditions
\begin{equation}
{\rm Tr}\left(\rho \right)=1, \hspace{1cm}\rho=\rho^\dag, \hspace{1cm} \left\langle {\cal U}_i \right|\rho\left| {\cal U}_i \right\rangle\geqslant0,
\end{equation}
pour tous les vecteurs ${\cal U}_i \in {\cal H}$. L'état d'un système quantique est dit pur s'il n'a qu'une seul valeur propre non nulle. Il se caractérise par un maximum de connaissances et leur matrice densité correspond au projecteur sur l'état $\left| \psi \right\rangle$. Il s'écrire comme $\rho=\left| \psi \right\rangle\left\langle  \psi \right| $ et leur éléments valent $\rho_{ij}=\left\langle {\cal U}_i \right|\rho\left| {\cal U}_j \right\rangle$. Dans ce cas, les connaissances sur la préparation du système ne manquent pas et la préparation génère avec certitude l'état souhaité. En utilisant les principes de la mécanique quantique, nous pouvons obtenir les propriétés importantes suivantes:
\begin{enumerate}
	\item Si on effectue une mesure de la quantité physique $A$ qui correspondant à l'observable $\hat A$, la probabilité d'obtenir la valeur propre $v_i$ est
	\begin{equation}
	P\left( v_i\right)=|\left\langle {\cal U}_i|\psi\right\rangle |^{2}=\left\langle \psi \right| P_i\left| \psi \right\rangle, \label{P}
	\end{equation}
où $\hat P_i$ représente le projecteur 	$\left| {\cal U}_i \right\rangle \left\langle {\cal U}_i \right|$. La valeur moyenne d'un opérateur $\hat A$ est exprimée en fonction de la matrice densité comme
	\begin{equation}
	\left\langle \hat A \right\rangle=\sum_{i,j}v_{i}^{\ast}v_{j}A_{ij}=\sum_{i,j}\rho_{ij} \left\langle  {\cal U}_i \right| \hat A\left| {\cal U}_j \right\rangle =\sum_{i}\left\langle {\cal U}_i\right| \rho \hat A\left| {\cal U}_i\right\rangle ={\rm Tr}\left(\hat A\rho\right). \label{A}
	\end{equation}
	D'après les équations (\ref{P}) et (\ref{A}), nous pouvons écrire donc	$P\left( v_i\right)={\rm Tr} \left(P_i \rho \right)$. De plus, il est facile de vérifier que
	\begin{equation}
		{\rm Tr}\left( \rho\right)=\sum_{i=0}^{n-1}\left\langle {\cal U}_i\right| \psi \left\rangle \right\langle \psi \left| {\cal U}_i\right\rangle =\sum_{i=0}^{n-1}v_{i}^{\ast}v_{i}=\sum_{i=0}^{n-1}|v_{i}|^{2}=1.
	\end{equation}
	\item Supposons que le résultat de la mesure donne la valeur $v_i$, et l’état du système après la mesure est $\left| \psi' \right\rangle=\frac{P_i \left| \psi \right\rangle}{\sqrt{\left\langle \psi \right| P_i\left| \psi \right\rangle}}$. On peut vérifier facilement que l'opérateur densité associé à la mesure $v_i$ est donné par
	\begin{equation}
	\rho_{v_i}=\frac{P_{i}\left| \psi \right\rangle \left\langle \psi \right| P_{i}}{\left\langle \psi \right|P_{i} \left| \psi \right\rangle}=\frac{P_{i}\rho P_{i}}{P\left( v_i\right)}.
	\end{equation}
	\item la matrice densité $\rho$ est un opérateur de projection, Alors, nous avons $\rho^2=\rho$ et ${\rm Tr}\left( \rho^2\right)=1$.
	\item Si le hamiltonien qui décrit le système est noté par $\hat H\left( t\right)$, l'évolution temporelle de l'état $\left| \psi\right\rangle$ est
	\begin{equation}
		\left| \psi \left( t\right) \right\rangle=U\left(t \right)\left| \psi \left( 0\right) \right\rangle \hspace{1cm} \Leftrightarrow \hspace{1cm} \frac{{\partial \left|  \psi\left( t\right) \right\rangle}}{{\partial t}}=-i {\hat H}\left( t\right)\left| \psi \left( t\right) \right\rangle,
	\end{equation}
où $U\left(t \right)$ et ${\hat H}\left(t \right)$ sont liés par $U\left(t \right)=\exp\left[ -i\int {\hat H}\left(t \right) t dt\right]$. L'évolution temporelle de ce système décrit par l'équation de Schrödinger peut s'écrire en fonction de l'opérateur densité comme
\begin{align}
	 \dot{\rho}\left( t\right)+i[ {\hat H}\left(t \right),\rho\left( t\right) ]=0, \label{EV}
\end{align}
où $[.,.]$ représente le commutateur des deux opérateurs.
\end{enumerate}
\subsubsection{Cas d’un état mixte}

Nous considérons maintenant le résultat de la mesure d'un état quantique mixte. Supposons que nous avons un mélange des états quantiques purs $\left| \psi_i\right\rangle$ avec une probabilité $p_{i}$, et chaque $\left| \psi_i\right\rangle$ peut être représenté par un vecteur dans l'espace ${\cal H}$ comme (\ref{vu}). La matrice densité d'un état mixte est définie comme un mélange statistique d'un ensemble d'états purs, et après la mesure, le système quantique susceptible de prendre un état pur parmi $n$ états possibles. Par conséquent, la matrice densité qui décrit le système global est donnée par
\begin{equation}
	\rho=\sum_{i}p_{i}\rho_i=\sum_{i}p_{i}\left| \psi_i\right\rangle \left\langle \psi_i\right|, 
\end{equation}
où $p_{i}$ est la probabilité que le système se trouve dans l'état normalisé $\left| \psi_i\right\rangle$ et la somme est prise sur tous les états accessibles au système. La probabilité $p_{i}$ satisfait évidemment
\begin{equation}
	0\leq p_i\leq 1,\hspace{2cm}\sum_{i}p_i=1, \hspace{2cm} p_{i}^{2} \leq 1.
\end{equation}
Pour un état pur, il n'y a qu'un seul $p_i$ qui est égal à l'unité et tous les autres sont nuls. Les propriétés de l'état pur sont aussi valables dans le cas de mélange statistique sauf que $\rho^{2}=\rho$. Alors, l'opérateur densité n’est pas un projecteur ($\rho^{2}\neq\rho$) et il est aussi facile de vérifier que ${\rm Tr}\left(\rho^{2}\right) <1$.
 
\subsubsection{Opérateur densité reduit}

Parfois, on ne s'intéresse qu'à l'un des sous-systèmes d'un système quantique composite. Cette situation est décrite à l'aide du concept de l'opérateur densité réduite. Considérons un système quantique $AB$ formé de deux sous-systèmes $A$ et $B$ dans les bases $\{\left|k\right\rangle, k=0,...,d_{A}-1 \} \in {\cal H}_A$ et $\{\left|m\right\rangle, m=0,...,d_{B}-1 \} \in {\cal H}_B$, respectivement. Si le système total est décrit par l'opérateur densité $\rho_{AB}$, alors le sous-système $A$ est décrit par l'opérateur densité réduite $\rho_{A}$. Il est défini par
\begin{equation}
	\rho_{A}={\rm Tr}_{B}\left( \rho_{AB}\right)=\sum_{m=0}^{d_{A}-1}\left\langle m\right| \rho_{AB}\left| m\right\rangle,
\end{equation}
où ${\rm Tr}_{B}$ est appelé trace partielle sur le sous-système $B$. De même pour le sous-système $B$, nous avons
\begin{equation}
\rho_{B}={\rm Tr}_{A}\left( \rho_{AB}\right)=\sum_{k=0}^{d_{B}-1}\left\langle k\right| \rho_{AB}\left| k\right\rangle.
\end{equation}
Les matrices densité réduite satisfont aux propriétés d'une matrice densité. Pour voir cela, considérons par exemple le cas dans lequel $\rho_{AB}$ est une matrice densité de l'état pur
\begin{equation}
	\rho_{AB}=\left| \psi_{AB}\right\rangle\left\langle \psi_{AB} \right|.
\end{equation}
Évidemment, si $\{\left| k\right\rangle_{A} \}$ et $\{\left| m\right\rangle_{B} \}$ sont des vecteurs de base orthonormés de ${\cal H}_A$ et ${\cal H}_B$, alors $\{\left| k_{A}\right\rangle\otimes\left| m_{B}\right\rangle\}$ est une base orthonormée pour ${\cal H}_A\otimes{\cal H}_B$ et l'état $\left| \psi_{AB}\right\rangle $ devient
\begin{equation}
	\left| \psi_{AB}\right\rangle =\sum_{k,m}\alpha_{km}\left| k_{A}\right\rangle \otimes \left| m_{B}\right\rangle,
\end{equation}
et la matrice densité réduite $\rho_{A}$ est écrite sous la forme suivante
\begin{align}
	\rho_{A}=\sum_{k,k',m}\alpha_{km}\alpha_{k'm}^{\ast}\left| k_{A}\right\rangle\left\langle k_{A}'\right|. \label{rhoAA}
\end{align}
En utilisant l'expression ci-dessus (\ref{rhoAA}), il est facile de voir que $\rho_{A}$ satisfait les propriétés d'une matrice densité:
\begin{list}{}{}
	\item $(i)$ $\rho_{A}=\rho_{A}^{\dagger}$;
	\item $(ii)$ $\rho_{A}\geq 0$, c'est-à-dire pour tout $\left| \phi\right\rangle \in {\cal H}_{A}$, nous avons $\left\langle \phi\right| \rho_{A}\left| \phi\right\rangle\geq 0$;
	\item $(iii)$ ${\rm Tr}\rho_{A}=1$.
\end{list}
De plus, en utilisant la relation générale de l'évolution temporelle de l'opérateur densité (\ref{EV}), nous obtenons
\begin{equation}
	\frac{{\partial \rho_{A}\left( t\right)}}{{\partial t}}=-i[ {\hat H}\left(t \right),\rho_{A}\left( t\right) ], \hspace{2cm}\frac{{\partial \rho_{B}\left( t\right)}}{{\partial t}}=-i[ {\hat H}\left(t \right),\rho_{B}\left( t\right) ].
\end{equation}
\subsection{Qubit et sphère de Bloch}

Dans les ordinateurs ordinaires, les informations sont stockées et traitées sous forme de bits, et donc, théoriquement, le bit est la plus petite unité qui transporte ou transmet des informations. Le terme bit fait référence aux nombres dans le système de chiffres binaires, qui est l'unité de base de la quantité d'informations dans les ordinateurs et les communications numériques. Cette unité ne peut contenir qu'une seule des deux valeurs et est donc physiquement appliquée dans une machine à deux états. Ces deux conditions sont représentées par “$0$” ou “$1$”. Il peut également être représenté comme “vrai” ou “faux”, “oui” ou “non” ou toute autre caractéristique avec deux valeurs.\par
Par analogie avec le bit classique, un bit quantique ou un qubit \cite{Schumacher1995} est la plus petite unité fondamentale de stockage d'information quantique. Contrairement au système classique où un bit devrait être dans un état ou dans l'autre, la mécanique quantique permet au qubit d'être dans une superposition des deux états simultanément. Cela signifie que les qubits peuvent exister dans plusieurs états à la fois, ce qui permet un calcul très rapide et la possibilité d'effectuer une multitude de calculs en même temps. Plusieurs systèmes quantiques peuvent stocker dans un qubit, nous citons ici par exemple un spin de l'électron dans l'expérience de Stern et Gerlach, dans lequel la séparation en deux faisceaux de l’atome d’argent révèle qu'il existe deux états possibles de spin et les deux niveaux peuvent être pris comme spin up et spin down \cite{Friedrich2003}. Un autre exemple concerne la polarisation d'un photon unique dans lequel les deux états peuvent être considérés comme étant la polarisation verticale et la polarisation horizontale.\par
Autrement dit, un qubit est l'état d'un système quantique à deux niveaux $\{\left| 0\right\rangle,\left| 1\right\rangle\}$ évoluant dans un espace de Hilbert à deux dimensions. N'importe lequel de ces états s'écrit mathématiquement sous la forme

\begin{equation}
	\left| \psi\right\rangle=\alpha\left| 0\right\rangle+\beta\left| 1\right\rangle, \label{qubit}
\end{equation}
où $\alpha$ et $\beta$ sont des nombres complexes qui satisfont la condition de normalisation $|\alpha|^{2}+|\beta|^{2}=1$ et les états $\left| 0\right\rangle$ et $\left| 1\right\rangle$ doivent être orthogonaux. Le qubit aura donc deux états jusqu'à ce que nous le mesurions. Une fois que nous avons implémenté une mesure sur le qubit, nous obtiendrons directement l'une des deux possibilités; soit l'état $\left| 0\right\rangle$ avec probabilité $|\alpha|^{2}$, soit l'état $\left| 1\right\rangle$ avec probabilité $|\beta|^{2}$.\par
\begin{figure}[H]
	\centerline{\includegraphics[width=7cm]{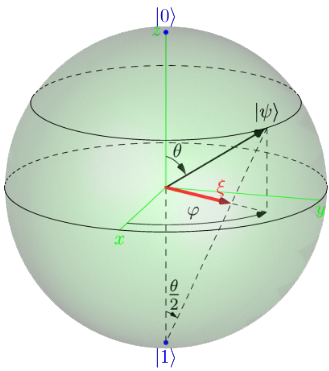}}
	\caption{Représentation géométrique d'un qubit sur la sphére de Bloch.}
	\label{Bloch}
\end{figure}

 
Puisque les facteurs de phase globaux des états quantiques en mécanique quantique n'affectent pas l'état physique, et seule la différence de phase entre $\alpha$ et $\beta$ est importante, nous pouvons prendre la représentation pour que le coefficient de l'état $\left|0 \right\rangle $ soit réel et non négatif. Donc, à partir de la condition de normalisation, un qubit (\ref{qubit}) peut être réécrit en fonction de deux angles ($\theta,\varphi$) comme
\begin{equation}
	\left|\psi \right\rangle =\cos\left( \frac{\theta}{2}\right)\left|0 \right\rangle+\exp^{i\varphi} \sin\left(\frac{\theta}{2} \right)\left| 1\right\rangle,
\end{equation}
avec les paramètres $\theta\in[0,\pi]$ et $\varphi\in[0,2\pi]$ exprimant l'amplitude relative et la phase relative des états de base, respectivement. Un état de ce type peut représenter géométriquement en tant qu'un point sur une sphère unitaire appelée la sphère de Bloch, où $\theta$ représente l'angle azimutal et $\varphi$ représente l'angle polaire (voir Fig.(\ref{Bloch})). Sa surface représente tous les états purs et son intérieur représente tous les états mixtes, et les états mixtes au maximum sont au centre de cette sphère. Il est également intéressant de noter que plusieurs opérations sur des qubits uniques qui sont couramment utilisés dans le traitement de l'information quantique peuvent être parfaitement décrites dans l'image de la sphère Bloch.
\subsection{Mesures quantiques}

\begin{figure}[H]
	\centerline{\includegraphics[width=8cm]{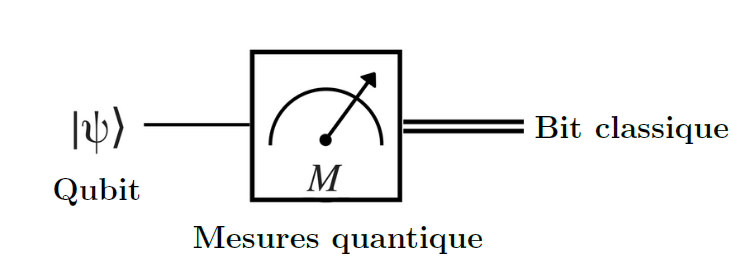}}
	\caption{Représentation schématique du processus de mesure}
	\label{mesure}
\end{figure}

L'un des problèmes les plus difficiles et controversés en mécanique quantique est le problème de la mesure et les opinions varient considérablement quant à son importance. Certaines personnes considèrent qu'en réalité il n'y a pas de problème du tout, tandis qu'à l'autre bout, d'autres paradoxes considèrent que le problème de la mesure est l'un des grands puzzles qui n'ont pas été résolus par la mécanique quantique. Le problème est que la mécanique quantique ne fournit que les possibilités de différents résultats possibles dans une expérience et elle ne fournit aucun mécanisme par lequel le résultat réel, qui est observé à la fin, se produit. Généralement, la mesure quantique est essentiellement ce que la Fig.(\ref{mesure}) représente; une opération sur un qubit, de sorte que l'état quantique est dans une superposition des vecteurs de base, pour obtenir un bit classique \cite{Braginsky1995,Wiseman2009}. Selon la théorie quantique, chaque grandeur physiquement mesurable qui est associée à une observable (c-à-d un opérateur hermitien) est caractérisée par un ensemble d'opérateurs $\{M_{k}\}_{k=1}^{N}$ satisfaisant la contrainte $\sum_{k=1}^{N}M_{k}^{\dagger}M_{k}=\openone$. Ces opérateurs agissant sur l'espace d'état du système et l'état $\left|\psi\right\rangle \in {\cal H}$ après la mesure devient
\begin{equation}
	\left|\psi\right\rangle\mapsto\left|\psi_{k}\right\rangle\equiv\frac{M_{k}\left| \psi\right\rangle}{\sqrt{p_{k}}},
\end{equation}
avec la probabilité
\begin{equation}
p_{k}=\left\langle \psi\right|M_{k}^{\dagger}M_{k}\left|\psi\right\rangle=\|M_{k}\left|\psi\right\rangle\|^{2} \geq 0.
\end{equation}

\subsubsection{Mesures projectives}

Les mesures projectives sont un cas particulier des mesures généralisées, dans lesquelles les opérateurs de mesure $\{M_{k}\}_{k=1}^{N}$ sont des opérateurs hermitiens appelés projecteurs. Le nombre de ces opérateurs est égal à la dimension de l'espace de Hilbert et ils sont reproductibles dans le sens que si nous effectuons une mesure projective une fois et obtenons le résultat $v_k$, la répétition de la mesure donne à nouveau le résultat $v_k$ et ne change pas l'état. Par conséquent, $M_{k}=P_{k}$ où $P_{k}P_{l}=\delta_{kl}P_{k}$ et $P_{k}^{2}=P_{k}$ avec $P_{k}$ représente un opérateur projecteur et $\delta_{kl}$ est le symbole de Kronecker. En utilisant cela, la probabilité d'observer le résultat avec l'indice $k$ lors de la mesure d'un système quantique dans l'état $\left| \psi\right\rangle$ est
\begin{equation}
p_{k}=\left\langle \psi\right|M_{k}^{\dagger}M_{k}\left|\psi\right\rangle=\left\langle \psi\right|P_{k}\left|\psi\right\rangle,
\end{equation}
et l'état immédiatement après la mesure $\left|\psi_{k}\right\rangle$ est
\begin{equation}
\left|\psi\right\rangle\mapsto\left|\psi_{k}\right\rangle\equiv\frac{P_{k}\left| \psi\right\rangle}{\sqrt{p_{k}}}.
\end{equation}
Si nous appliquons à nouveau l'opérateur $P_{k}$, cette fois à l'état $\left|\psi_{k}\right\rangle$, il n'y a pas de changement d'état. En effet, l'état après avoir appliqué $P_{k}$ pour la deuxième fois est
\begin{equation}
\left|\psi_k\right\rangle\mapsto\left|\psi_{k}'\right\rangle\equiv\frac{P_{k}\left| \psi_{k}\right\rangle}{\sqrt{p_{k}'}},
\end{equation}
où $p_{k}'=\left\langle \psi_{k}\right|P_{k}\left|\psi_{k}\right\rangle$. Nous pouvons maintenant dériver l'expression de $p_{k}'$ comme

\begin{equation}
p_{k}'=\left[\left\langle \psi\right| \frac{P_{k}^{\dagger}}{\sqrt{p_{k}}}\right] P_{k}\left[\frac{P_{k}}{\sqrt{p_{k}}} \left|\psi \right\rangle\right]=\frac{\left\langle \psi\right|P_{k}\left|\psi\right\rangle}{p_{k}}.
\end{equation}
Alors,
\begin{equation}
	\sqrt{p_{k}'p_{k}}=\sqrt{\left\langle \psi\right|P_{k}\left|\psi\right\rangle}=\sqrt{p_{k}}.
\end{equation}
Finalement, nous voyons que l'état n'a pas été affecté par la deuxième mesure
\begin{equation}
	\left| \psi_{k}'\right\rangle=\frac{P_{k}}{\sqrt{p_{k}'}}\frac{P_{k}\left| \psi\right\rangle }{\sqrt{p_{k}}}=\frac{P_{k}\left| \psi\right\rangle }{\sqrt{p_{k}}}=\left|\psi_{k}\right\rangle.
\end{equation}
Par conséquent, on peut dire que le résultat observé à la suite d'une mesure projective est déterministe.
\subsubsection{Opérateurs de mesure à valeur positive (POVMs)}

En général, nous détruisons un système quantique au cours du processus de mesure, ce qui signifier que les mesures projectives sont restrictives et ne sont pas toujours possibles. Aussi, la répétabilité d'une mesure projective est violée dans les cas où le système n'est mesuré qu'une seule fois. Cependant, il est possible d'envisager une notion de mesure plus généralisée en relâchant la contrainte d'orthogonalité. Cela conduit au concept d'opérateurs de mesure à valeur positive (POVM: Positive Operator-Valued Measure). Les opérateurs POVM ne sont pas nécessairement orthogonaux ou commutatifs et permettent la possibilité de résultats de mesure associés à des états non orthogonaux. Contrairement aux mesures projectives, le nombre d'éléments dans les mesures POVMs peut être supérieur à la dimension de l'espace de Hilbert. De plus, ils jouent un rôle important dans de nombreux domaines de la science de l'information quantique \cite{Brandt1999,Hamieh2004}. \par

Considérons une mesure généralisée avec des opérateurs de mesure $\{M_{k}\}$, les éléments d'une mesure POVM sont définis via
\begin{equation}
	{E}_k= M_{k}^{\dagger} M_{k},
\end{equation}
et la condition de normalisation devient $\sum_{k}{E}_k=\openone$. Il est clair que, $E_{k}^{\dagger}=M_{k}^{\dagger} \left(M_{k}^{\dagger}\right)^{\dagger}=E_{k}$, alors les éléments POVM sont hermitiens. Il est également facile de montrer que les $E_{k}$ sont des opérateurs positifs, c'est-à-dire que $\left\langle \psi \right| E_{k} \left|\psi \right\rangle \geq 0$ est vrai pour tous $\left|\psi \right\rangle \in {\cal H}$. De plus, la probabilité du résultat $k$ est simplement $p_{k}=\left\langle \psi \right| E_{k} \left|\psi \right\rangle$. En utilisant la décomposition polaire d'un opérateur, où pour tout opérateur $A$ on peut toujours trouver un transformation unitaire $U$ et un opérateur positif $Q$ de sorte que $A=UQ$ avec $Q=\sqrt{A^{\dagger}A}$. Dans le cas le plus simple, où l'opérateur $A$ est inversible et $U=AQ^{-1}$, les opérateurs $M_{k}$ deviennent
\begin{equation}
	M_{k}=U_{k}\sqrt{E_{k}},
\end{equation}
avec les $U_{k}$ sont des transformations unitaires arbitraires. Par conséquent, nous pouvons maintenant écrire l'état après la mesure comme
\begin{equation}
\left|\psi\right\rangle\mapsto\left|\psi_{k}\right\rangle\equiv\frac{U_{k}\sqrt{E_{k}}\left| \psi\right\rangle}{\sqrt{p_{k}}}.
\end{equation}

\subsection{Entropie de von Neumann et l'information quantique}

Par analogie avec l'entropie de Shannon, John von Neumann a introduit une définition de l'entropie via l'opérateur densité de la mécanique quantique $\rho$. Ici, nous donnerons une brève introduction à cette entropie et à sa version relative. Une explication plus approfondie est donnée dans le livre de Nielsen et Chuang \cite{Nielsen2000}. L'entropie de von Neumann d'un état quantique $\rho$ est définie par la formule
\begin{equation}
	S\left( \rho\right) =-{\rm Tr}\left(\rho\log_{2}\rho \right). \label{vn}
\end{equation}
Dans les calculs analytiques, cette entropie est calculée en diagonalisant la matrice $\rho$ comme
\begin{equation}
	\rho={\cal W}\Lambda {\cal W}^{-1}, \label{diarho}
\end{equation}
avec
\begin{equation}
	{\cal W}=\left( w_{1},...,w_{n}\right)^{T}, \hspace{1cm} \Lambda={\rm diag}\left( \lambda_{1},...,\lambda_{n}\right),
\end{equation}
où les vecteurs $w_{i}$ sont des vecteurs propres de $\rho$ et chacun avec une valeur propre $\lambda_{i}$. La matrice $\Lambda$ contient les valeurs propres dans la diagonale et sinon les zéros. Le logarithme de $\rho$ peut alors être calculé comme
\begin{equation}
	\log_{2}\rho={\cal W}\log_{2}\left(\Lambda \right){\cal W}^{-1}. \label{logeho}
\end{equation}
L'utilisation des équations (\ref{vn}), (\ref{diarho}) et (\ref{logeho}), et de la caractéristique cyclique de la trace ${\rm Tr}\left(AB \right)={\rm Tr}\left(BA \right) $, conduit à un calcul de l'entropie de von Neumann comme
\begin{equation}
	S\left( \rho\right) =-\sum_{i}\lambda_{i}\log_{2}\lambda_{i}={\cal H}\left( \lambda_{i}\right),
\end{equation}
qui est la formule pratique utilisée pour calculer l'entropie de von Neumann d'une matrice densité de dimension arbitraire où ${\cal H}\left( \lambda_{i}\right)$ est l'entropie de Shannon. L'entropie de von Neumann est donc équivalente à l'entropie de Shannon de la distribution de probabilité associée à la diagonalisation de $\rho$. Si $\left| \psi \right\rangle$ est un état pur, alors sa matrice densité $\rho$ n'a qu'une seule valeur propre égale à $1$, sa trace est égale à $1$ (alors $\log_{2}\rho\equiv0$), et son entropie de von Neumann est nulle. D'autre part, si l'on prépare le système dans un état mixte maximal $1/N$, c'est-à-dire un état mixte où tous les $N$ états purs correspondants ont les mêmes probabilités, alors l'état mixte représente l'incertitude maximale avec l'entropie de von Neumann $\log_{2}N$ en accord avec l'entropie de Shannon. De plus, l'entropie de von Neumann satisfait les propriétés suivantes:
\begin{itemize}
	\item Comme pour l'entropie de Shannon, l'entropie de von Neumann est toujours positive et nous avons $0\leq S\left(\rho \right)\leq \log_{2}N$.
	\item L'entropie de von Neumann est sous-additive:
	$S\left(\rho_{AB} \right)\leq S\left(\rho_{A} \right)+S\left(\rho_{B}\right)$. L'égalité a lieu si $\rho_{AB}=\rho_{A}\otimes\rho_{B}$.
	 \item Pour un état pur $\rho_{AB}$, nous avons $S\left(\rho_{A}\right)=S\left(\rho_{B}\right)$. 
	 \item Soit $p_i$ une distribution de probabilité et $\rho_{i}$ une famille d'états appartenant à des sous-espaces orthogonaux, alors nous pouvons obtenir
	 \begin{equation}
	 	S\left(\sum_{i}p_{i}\rho_{i} \right)={\cal H}\left( p_{i}\right) + \sum_{i}p_{i}S\left(\rho_{i}\right).
	 \end{equation}
	 \item $S\left(\rho_{AB} \right)$ est une fonction concave, c'est-à-dire si  $\lambda_{1},...,\lambda_{n}\geq0$ avec $\sum_{i=1}^{n}\lambda_{i}=1$, alors
	\begin{equation}
		S\left(\sum_{i=1}^{n}\lambda_{i}\rho_{i}\right)\geq \sum_{i=1}^{n}\lambda_{i}S\left( \rho_{i}\right). 
	\end{equation}
	\item $S\left( \rho\right)$ est invariante sous une transformation unitaire $U$: $S\left( \rho\right)=S\left( U\rho U^{\dagger}\right)$.
	 
\end{itemize}
\subsubsection{Entropie quantique conjointe, conditionnelle et relative}

Comme dans la théorie de l'information classique, il est extrêmement utile de définir une version quantique de l'entropie conjointe, conditionnelle et relative. Considérons un système quantique à deux parties dont l'état total est donné par l'opérateur densité $\rho_{AB}$ et on note par les opérateurs $\rho_{A}$ et $\rho_{B}$ les matrices densité réduites. L'entropie conjointe de von Neumann des sous-systèmes $A$ et $B$ est définie comme l'entropie de von Neumann de l'état du système total:
\begin{equation}
	S\left( A,B\right)=S\left( \rho_{AB}\right).
\end{equation}
Ainsi, l'entropie conditionnelle de von Neumann du système $A$ conditionnée par le système $B$ est définie comme
\begin{equation}
	S\left(\rho_{A}/\rho_{B}\right)=S\left( \rho_{AB}\right) -\rho_{B}.
\end{equation}
Pour deux matrices densité $\rho$ et $\sigma$, l'entropie relative quantique de $\rho$ par rapport à $\sigma$ est définie par
\begin{equation}
	S\left( \rho\|\sigma\right)={\rm Tr}\left( \rho\log_{2}\rho\right)-{\rm Tr}\left( \rho\log_{2}\sigma\right). \label{ERQ}
\end{equation}
Cette entropie (\ref{ERQ}) est calculée en obtenant d'abord les valeurs propres et les vecteurs propres des matrices $\rho$ et $\sigma$. Nous désignons les valeurs propres de $\rho$ et $\sigma$ comme $\lambda_i$ et $\mu_j$ respectivement, et leurs vecteurs propres correspondants comme $w_i$ et $u_j$. Si la base des vecteurs propres de $\rho$ est une base orthonormée, la matrice inverse ${\cal W}^{-1}$ dans l'équation (\ref{logeho})  peuvent être remplacés par la matrice de transposition ${\cal W}^{T}$, alors le deuxième terme de l'équation (\ref{ERQ}) peut être calculé comme

\begin{equation}
	{\rm Tr}\left(\rho\log_{2}\sigma \right)=\sum_{i}w_{i}^{T}\left(\rho\log_{2}\sigma\right) w_{i}=\sum_{i}\lambda_{i} w_{i}^{T}\log_{2}\sigma w_{i}.
\end{equation}
De même, nous avons
\begin{equation}
	\log_{2}\sigma=\sum_{j}\log_{2}\left(\mu_j \right) u_{j}u_{j}^{T}.
\end{equation}
Par conséquent, la formule finale de l'entropie relative (\ref{ERQ}) devient
\begin{equation}
		S\left(\rho\|\sigma\right)=\sum_{i=1}^{n}\lambda_{i}\left[ \log_{2}\lambda_{i}-\sum_{j=1}^{n}\left(w_{i}^{T}\mu_j \right)^{2} \log_{2}\mu_j\right].
\end{equation}

\subsubsection{L'information mutuelle quantique}

L'information mutuelle quantique est une généralisation au cas quantique du concept d'information mutuelle rencontrée en théorie classique de l'information. Elle porte sur l'interprétation des corrélations totales entre deux sous-systèmes $A$ et $B$ d'un système quantique bipartite $AB$. Elle est considérée comme un cas particulier de l'entropie relative entre l'état $\rho_{AB}$ et l'état $\rho_{A}\otimes\rho_{B}$. En utilisant l'équation (\ref{ERQ}) et la relation $\log_{2}\left(\rho_{A}\otimes\rho_{B}\right)=\log_{2}\rho_{A}\otimes \openone_{B}+\openone_{A}\otimes \log_{2}\rho_{B}$, nous trouvons l'extension quantique de l'information mutuelle (\ref{IMC}) comme
\begin{align}
	{\cal I}\left( \rho_{AB}\right)&=S\left( \rho_{AB}\|\rho_{A}\otimes\rho_{B}\right)\notag\\&={\rm Tr}\left[\rho_{AB}\log_{2}\rho_{AB} \right]-{\rm Tr}\left[\rho_{AB}\log_{2}\left( \rho_{A}\otimes\rho_{B}\right) \right]\notag\\&=S\left( \rho_{A}\right)+S\left( \rho_{B}\right)-S\left( \rho_{AB}\right). \label{IMQ}
\end{align}
Notez que l'information mutuelle quantique est non négative en raison de la sous-additivité de l'entropie de von Neumann, et nulle uniquement pour un état de produit $\rho_{AB}=\rho_{A}\otimes\rho_{B}$. Elle peut également être écrite en fonction de l'entropie conditionnelle quantique en faisant une généralisation directe de l'équation (\ref{IMM}). Nous trouvons
\begin{equation}
	{\cal I}\left( \rho_{AB}\right)=S\left( \rho_{A}\right)-S\left( \rho_{AB}/\rho_{B}\right).
\end{equation}
Malgré que l'entropie de Shannon et l'entropie de von Neumann présentent des similitudes, elles ont de nombreuses propriétés nettement différentes. Cependant, en raison de la relation intime entre ces entropies et les propriétés opérationnelles des informations, ces propriétés nous renseignent sur la nature de l'information.
\subsection{Entropie de Rényi}

L'une des lignes de recherche les plus fructueuses en théorie de l'information quantique consiste à généraliser les entropies quantiques au-delà de celles qui sont données par des combinaisons linéaires d'entropie de von Neumann. L'idée générale est de chercher une généralisation de l'entropie de von Neumann et à réécrire toutes les définitions susmentionnées en termes de nouvelles entropies. Cela peut être utile dans les applications de l'information quantique d'un point de vue théorique. Dans ce sens, plusieurs grandeurs entropiques ont été introduites et étudiées. En particulier, les entropies de Rényi-$\alpha$ sont une famille intéressante d'entropies additives qui ont trouvé des applications spéciales dans l'étude des capacités des canaux \cite{Alicki} ainsi que pour les quantifications des corrélations dans les systèmes quantiques \cite{Lennert2013,Wilde2014}. Considérant une variable aléatoire $X$ avec une distribution de probabilité $p_{x}$, où $p_{x}\geq 0$ et $\sum_{x}p_{x}=1$, son entropie de Rényi d'ordre $\alpha$ est définie comme

\begin{equation}
	{\cal H}_\alpha\left(X \right)=\frac{1}{1-\alpha}\log_{2}\sum_{x}p_{x}^{\alpha},
\end{equation}
où $0<\alpha<\infty$ et lorsque le paramètre $\alpha$ s'approche de 1, l'entropie de Rényi est réduite à l'entropie de Shannon. En effet
\begin{equation}
{\cal H}_1\left(X \right)=\lim_{\alpha\longrightarrow 1}{\cal H}_\alpha\left(X \right)\equiv-\sum_{x}p_{x}\log_{2}p_{x}={\cal H}\left(X \right).
\end{equation}
La version quantique de l'entropie de Rényi est définie comme
\begin{equation}
	S_{\alpha}\left(\rho \right)=\frac{1}{1-\alpha}\log_{2}\left[{\rm Tr}\left(\rho^{\alpha} \right) \right],
\end{equation}
et pour $\alpha=1$, l'entropie de Rényi quantique est égale à l'entropie de von Neumann
\begin{equation}
	S_{1}\left(\rho \right)=\lim_{\alpha\longrightarrow 1}S_{\alpha}\left(\rho \right)\equiv-{\rm Tr}\left(\rho\log_{2}\rho \right) =S\left( \rho\right).
\end{equation}
\section{Quantification et caractérisation de l'intrication quantique}

L'intrication quantique, comme un type spécial de corrélations quantiques, est une ressource précieuse pour effectuer et de façon plus performante plusieurs tâches de traitement de l'information quantique et a fait l'objet d'études intensives au cours des dernières décennies. Elle décrit les corrélations entre les systèmes quantiques qui n'ont pas d'analogues classiques \cite{Schrodinger1935,Horodecki2009}. En effet, les corrélations quantiques jouent un rôle crucial dans la plupart des applications de la théorie de l'information quantique, car elles sont à l'origine de plusieurs processus qui ne seraient pas possibles dans un contexte purement classique \cite{Hill1997,Braunstein2005,Ekert1991}. En conséquence, l'étude de l'intrication et en particulier comment elle peut être quantifiée est un sujet central dans ce champ en plein essor.\par

La description quantitative de l'intrication a commencé avec les fameuses inégalités de John Bell en 1964. Pendant plusieurs années, l'intrication était considérée comme synonyme de la violation de ces inégalités. Les prochaines étapes, du point de vue théorique, était le travail de Werner sur la caractérisation mathématique précise de l'intrication dans les états quantiques mixtes où il a décrit des états mixtes qui ne violent aucune inégalité de Bell \cite{Werner1989}. Depuis lors, un effort théorique remarquable a été consacré à la compréhension des corrélations quantiques. Cependant, il peut sembler surprenant qu'à ce jour il n'existe pas de théorie quantitative globale de l'intrication. En effet, la théorie de l'intrication pour l'état mixte est plus compliquée et moins bien comprise que celle de l'intrication pour l'état pur. Pire encore, il existe une variété de classifications et de mesures d'intrication basées sur différents concepts et méthodes sans liens clairs les uns avec les autres. \par

Comme nous l'avons mentionné précédemment, la dernière moitié de ce chapitre est consacrée à rappeler certains des résultats de base de la théorie de l'intrication et ses différents critères de détection. Dans ce but, cette partie est une introduction à nos connaissances actuelles sur la façon de décider si un état donné est séparable ou intriqué, et comment quantifier l'intrication avec des mesures appropriées. Mais d'abord, nous avons besoin d'une définition mathématique de l'intrication versus la séparabilité. Considérons un système quantique bipartite $AB$ qui est composé de deux sous-systèmes $A$ et $B$. Un état quantique pur $\left| \psi_{AB}\right\rangle$ est appelé séparable s'il peut s'écrire comme
\begin{equation}
	\left| \psi_{AB}\right\rangle=\left|\phi_{A} \right\rangle \otimes \left| \chi_{B}\right\rangle, \label{ES}
\end{equation}
où $\left|\phi_{A} \right\rangle\in {\cal H}_A$ et $\left| \chi_{B}\right\rangle\in {\cal H}_B$. Dans ce cas, les matrices densités réduites des deux sous-systèmes sont des états purs, alors
\begin{equation}
	\rho_{A}=\left|\phi_{A} \right\rangle\left\langle\phi_{A} \right|,\hspace{3cm}	\rho_{B}=\left|\chi_{B} \right\rangle\left\langle\chi_{B} \right|.
\end{equation}
D'un autre côté, si $\left| \psi_{AB}\right\rangle$ ne peut pas être exprimé comme un produit tensoriel des états de deux sous-systèmes (\ref{ES}), on dit que les sous-systèmes $A$ et $B$ sont dans un état intriqué et $\left| \psi_{AB}\right\rangle$ est dit être un état pur intriqué. Dans ce cas, les opérateurs densité réduite $\rho_{A}$ et $\rho_{B}$ correspondent à des états mixtes. A titre d’exemple, $\left|\psi \right\rangle =\left|00 \right\rangle$ est un état pur séparable, et les états de Bell,
\begin{equation}
	\left|\phi^{\pm} \right\rangle=\frac{1}{\sqrt{2}}\left(\left|00 \right\rangle\pm\left|11 \right\rangle\right),\hspace{1.5cm}{\rm et}\hspace{1.5cm}\left|\psi^{\pm} \right\rangle=\frac{1}{\sqrt{2}}\left(\left|01 \right\rangle\pm\left|10 \right\rangle\right),
\end{equation}
sont une classe spéciale d'états purs intriqués et qui sont maximalement intriqués.\par

Un état mixte est dit séparable, s'il peut être préparé par les deux parties de manière classique, et leur matrice densité ne peut contenir que des corrélations classiques. Mathématiquement, un état mixte est dit séparable s'il peut être écrit comme
\begin{equation}
	\rho_{AB}=\sum_{i}p_{i}\left|\phi_{i} \right\rangle\left\langle \phi_{i}\right|\otimes\left|\chi_{i} \right\rangle\left\langle \chi_{i}\right|,\label{SEM}
\end{equation}
sinon, $\rho_{AB}$ est un état intriqué. Dans le cadre de la généralisation de l'équation (\ref{SEM}) pour un état multipartite, un état $N$-partite agissant sur ${\cal H}={\cal H}^{d_1}\otimes{\cal H}^{d_2}...\otimes{\cal H}^{d_N}$ est séparable s'il peut être écrit comme une somme convexe de produits tensoriels d'états des sous-systèmes
\begin{equation}
	\rho_{N}=\sum_{i}p_{i}\rho_{i}^{(1)}\otimes\rho_{i}^{(2)}...\otimes\rho_{i}^{(N)}=\sum_{i}p_{i}\otimes_{j=1}^{N}\rho_{i}^{(j)}, \hspace{0.5cm}p_{i}\geq 0,\hspace{0.5cm}\sum_{i}p_{i}=1, 
\end{equation}
et l'état $\rho_{N}$ est dit k-séparable si nous pouvons l'écrire comme
\begin{equation}
	\rho_{N}=\sum_{i}p_{i}\rho_{i}^{(a_1)}\otimes\rho_{i}^{(a_2)}...\otimes\rho_{i}^{(a_k)},
\end{equation}
avec $a_l$ (où $l=1,2,...,k$) sont les sous-ensembles disjoints du l'ensemble globale $\{1,2,...,N\}$ et $\rho^{(a_l)}$ agit sur l'espace du produit tensoriel constitué par les sous-espaces de ${\cal H}$ marqués par les membres de $a_l$.
\subsection{Critéres de séparabilité}

Comme le montre l'équation (\ref{ES}), la définition de l'intrication d'un état bipartite ne fournit aucun critère constructif pour décider si un état donné à une représentation du produit tensoriel ou non. Par conséquent, la connaissance des états intriqués et non intriqués est une question fondamentale de la théorie de l'intrication quantique, et seulement dans quelques cas, cette question a une réponse simple. Dans le cas des états purs bipartites, la décomposition de Schmidt est une méthode efficace pour détecter l'intrication quantique. Cependant, pour les états mixtes bipartites, trouver une décomposition comme dans l'équation (\ref{ES}) ou prouver qu'il n'en existe pas est une tâche difficile. Ces dernières années, des efforts considérables ont été déployés pour analyser la séparabilité des états mixtes. Ceci a conduit à l'identification de plusieurs critères de séparabilité opérationnelle, mais la situation est moins simple et ces critères ne sont efficaces que pour les basses dimensions.

\subsubsection{La décomposition de Schmidt}

La décomposition de Schmidt est l'un des outils les plus importants pour analyser les états purs bipartites en théorie de l'information quantique \cite{Ekert1995,Radtke2006}. Il montre qu'il est possible de décomposer tout état pur bipartite en superposition d'états orthonormés. Soit ${\cal H}$ un espace de Hilbert défini comme un produit tensoriel de deux espaces de Hilbert ${\cal H}_A$ et ${\cal H}_B$. Un état pur $\left|\psi_{AB} \right\rangle$ du système bipartite composite $AB$ s'écrit comme $\left|\psi_{AB} \right\rangle=\sum_{i,j}\chi_{ij}\left| i\right\rangle_{A}\otimes\left| j\right\rangle_{B}$, où $\left| i\right\rangle_{A}$ et $\left| j\right\rangle_{B}$ sont respectivement des bases en ${\cal H}_A$ et ${\cal H}_B$. Selon le théorème de Schmidt, il existe des bases $\left|u_{i} \right\rangle_{A}\in {\cal H}_{A}$ et $\left|v_{i} \right\rangle_{B}\in {\cal H}_{B}$ dans lesquelles, pour chaque état pur $\left|\psi_{AB} \right\rangle$, peut être exprimé comme
\begin{equation}
	\left|\psi_{AB} \right\rangle=\sum_{i=1}^{n}\sqrt{\tilde{\chi}}_{i}\left|u_{i} \right\rangle_{A}\otimes\left|v_{i} \right\rangle_{B},
\end{equation}
où $n\leq\min\{{\rm dim}\left({\cal H}_{A} \right),{\rm dim}\left({\cal H}_{B} \right)\}$ est appelé le nombre de Schmidt et $\tilde{\chi}_{i}$ sont des nombres réels non négatifs, appelés coefficients de Schmidt, qui satisfont $\sum_{i=1}^{n}\tilde{\chi}_{i}=1$. En outre, les coefficients de Schmidt $\tilde{\chi}_{i}$ correspondent aux valeurs propres de l'une des matrices densités réduites
\begin{equation}
	\rho_{A}={\rm Tr}_{B}\left[\rho_{AB}\right]=\sum_{i=1}^{n}\tilde{\chi}_{i}\left|u_{i}\right\rangle_{A}\left\langle u_{i}\right|, \hspace{1cm} {\rm ou}\hspace{1cm} \rho_{B}={\rm Tr}_{A}\left[\rho_{AB}\right]=\sum_{i=1}^{n}\tilde{\chi}_{i}\left|v_{i}\right\rangle_{B}\left\langle v_{i}\right|.
\end{equation}
L'état $\left|\psi_{AB} \right\rangle$ est séparable si et seulement s'il n'y a qu'un seul coefficient de Schmidt non nul, c'est-à-dire le nombre de Schmidt $n=1$. S'il existe plusieurs coefficients de Schmidt non nuls, l'état est intriqué. De plus, si tous les coefficients de Schmidt sont non nuls et égaux, l'état est dit maximalement intriqué.\par

A titre d'exemple trivial du théorème de décomposition de Schmidt, considérons le vecteur $\left|\psi_{AB} \right\rangle=\left|11 \right\rangle$ dans un espace de Hilbert à $4$-dimensions, dans ce cas, pour chacun espace de Hilbert ${\cal H}_{A}$ et ${\cal H}_{B}$, la base de Schmidt est la base de calcul, alors
\begin{equation}
\left|u_{1} \right\rangle_{A}=\left|0\right\rangle_{A},\hspace{1cm}\left|u_{2} \right\rangle_{A}=\left|1 \right\rangle_{A}, \hspace{1cm}\left|v_{1} \right\rangle_{B}=\left|0 \right\rangle_{B}, \hspace{1cm}\left|v_{2} \right\rangle_{B}=\left|1\right\rangle_{B},
\end{equation}
et les coefficients de Schmidt sont $\tilde{\chi}_{1}=0$ et $\tilde{\chi}_{2}=1$. Par conséquent, l'état $\left|\psi_{AB} \right\rangle$ est séparable.\par
Comme exemple un peu moins trivial, considérons l'état suivant sur le même espace à $4$-dimensions que dans l'exemple précédent
\begin{align}
	\left|\psi_{AB} \right\rangle&=\frac{1}{2}\left( \left|00 \right\rangle+\left|01 \right\rangle+\left|10 \right\rangle+\left|11 \right\rangle\right)\notag\\&=\left[\frac{1}{\sqrt{2}} \left(\left|0 \right\rangle_{A}+\left|1 \right\rangle_{A} \right) \right]\left[\frac{1}{\sqrt{2}} \left(\left|0 \right\rangle_{B}+\left|1 \right\rangle_{B} \right) \right].
\end{align}
Dans cet exemple, la base de calcul n'est pas une base de Schmidt, nous choisissons donc les bases de Schmidt pour être
\begin{equation}
	\left|u_{1} \right\rangle_{A}=\frac{1}{\sqrt{2}} \left(\left|0 \right\rangle_{A}+\left|1 \right\rangle_{A} \right),\hspace{1cm}\left|u_{2} \right\rangle_{A}=\frac{1}{\sqrt{2}} \left(\left|0 \right\rangle_{A}-\left|1 \right\rangle_{A} \right),
\end{equation}
\begin{equation}
\left|v_{1} \right\rangle_{B}=\frac{1}{\sqrt{2}} \left(\left|0 \right\rangle_{B}+\left|1 \right\rangle_{B} \right),\hspace{1cm}\left|v_{2} \right\rangle_{B}=\frac{1}{\sqrt{2}} \left(\left|0 \right\rangle_{B}-\left|1 \right\rangle_{B} \right),
\end{equation}
et les coefficients de Schmidt sont $\tilde{\chi}_{1}=1$ et $\tilde{\chi}_{2}=0$.
\subsubsection{La purifucation}

Généralement, un état quantique mixte peut être compris soit comme un mélange statistique d'états quantiques purs, soit comme faisant partie d'un état pur de dimensions supérieures. Alors, il est possible d'introduire un autre système sur un état mixte, que nous appelons un système de référence E, de telle manière que l'état total soit un état pur. Cette technique est appelée la purification. En d’autre terme, une purification d'un état quantique mixte est un état pur dans un espace de Hilbert de dimension supérieure, dont la matrice densité réduite est identique à l'état d'origine.\par

Supposons une matrice densité mixte donnée par $\rho_{AB}$. La première étape pour purifier le système $AB$ consiste à diagonaliser la matrice, en calculant les valeurs propres et les vecteurs propres. Supposons que nous obtenions comme valeurs propres $\left\lbrace \lambda_1, \lambda_2,\lambda_3,\lambda_4 \right\rbrace$ et comme vecteurs propres $\left\lbrace \phi_1, \phi_2,\phi_3,\phi_4 \right\rbrace$. Pour calculer l'état pur $\left|ABE \right\rangle$, où $\rho_{ABE}=\left|ABE \right\rangle\left\langle ABE\right|$ et $\rho_{AB}={\rm Tr}_{E}\left[\rho_{ABE} \right]$, nous définissons un ensemble de vecteurs comme une base de Schmidt. Nous devons définir ces vecteurs comme
\begin{equation}
	\left|1 \right\rangle=\left( {\begin{array}{*{20}{c}}
		1\\
		0\\
		0\\
		0\end{array}} \right), \hspace{1cm}\left|2 \right\rangle=\left( {\begin{array}{*{20}{c}}
		0\\
		1\\
		0\\
		0\end{array}} \right), \hspace{1cm}\left|3 \right\rangle=\left( {\begin{array}{*{20}{c}}
		0\\
		0\\
		1\\
		0\end{array}} \right), \hspace{1cm}\left|4 \right\rangle=\left( {\begin{array}{*{20}{c}}
		0\\
		0\\
		0\\
		1\end{array}} \right). 
\end{equation}
En adaptant cette notation, l'état pur $\left|\psi_{ABE} \right\rangle$ est donc donné par
\begin{equation}
	\left|\psi_{ABE} \right\rangle=\sqrt{\lambda_1}\left|\phi_1 \right\rangle \otimes\left| 1\right\rangle+ \sqrt{\lambda_2}\left|\phi_2 \right\rangle \otimes\left| 2\right\rangle+ \sqrt{\lambda_3}\left|\phi_3 \right\rangle \otimes\left| 3\right\rangle+\sqrt{\lambda_4}\left|\phi_4 \right\rangle \otimes\left| 4\right\rangle, 
\end{equation}
Il est facile de vérifier que $\rho_{AB}={\rm Tr}_{E}\left[\rho_{ABE} \right]$. Il est également important de noter que le rang du système définit la dimension du système de référence qui le purifie. En effet, par exemple, si nous n'avons que deux valeurs propres non nulles, nous n'aurons besoin que de deux bases et ce sera la dimension du système de référence. Une autre observation importante est que l'état de référence est défini sauf s'il s'agit d'une opération unitaire et nous pouvons définir les bases par n'importe quel ensemble d'états orthonormaux.
\subsubsection{Critére de Peres-Horodecki}

Comme nous l'avons vu dans le paragraphe précédent, la décomposition de Schmidt est un indicateur nécessaire pour décider si un état pur donné est intriqué ou non, mais il s'avère qu'il est beaucoup plus difficile de définir la séparabilité pour les états mixtes. Par conséquent, nous devons passer à d'autres façons d'évaluer les états mixtes. Dans ce sens, l'étude des critères de séparabilité dans les états mixtes a attiré beaucoup d'attention et plusieurs critères ont été formulés. Commençons par l'un des critères les plus anciens et pourtant les plus importants pour la détection de l'intrication appelé le critère de la transposition partielle positive (PPT) ou le critère de Peres-Horodecki.\par
Le critère de la transposition partielle positive établit une condition nécessaire à la séparabilité dans le cas d'un état bipartite général. Il a été proposé pour la première fois dans le cadre de la théorie de l'intrication par Peres \cite{Peres1996}. Il implique un calcul algébrique simple sans aucune optimisation. Soit $\rho_{AB}$ un état quantique mixte d'un système bipartite, décrit par l'espace de Hilbert ${\cal H}={\cal H}_A\otimes{\cal H}_B$, alors un élément de la matrice d’un opérateur $\rho_{AB}$ est donnée par
\begin{equation}
	\rho_{m\mu,n\nu}=\left\langle m\right|\left\langle \mu\right|\rho\left| n\right\rangle \left| \nu\right\rangle=\left\langle m\mu\right|\rho\left| n\nu\right\rangle,
\end{equation}
où les indices $\left\lbrace m,\mu \right\rbrace$ décrivent le premier sous-système $A$ et les indices $\left\lbrace n,\nu \right\rbrace$ décrivent le second sous
système $B$. La transposition partielle de la matrice densité $\rho_{AB}$ par rapport au premier sous-système $A$ est obtenue en effectuant la transposition par rapport aux degrés de liberté de sous-système $B$, et vice versa. Ensuite, nous obtenons 
\begin{equation}
\rho_{m\mu,n\nu}^{T_A}=\rho_{n\mu,m\nu} \hspace{1cm} ou\hspace{1cm}\rho_{m\mu,n\nu}^{T_{B}}=\rho_{m\nu,n\mu}. \label{pptresults}
\end{equation}
Si un état bipartite $\rho_{AB}$ est séparable, sa transposition partielle $\rho^{T_A}$ est une matrice densité semi-définie positive, c-à-d $\rho^{T_A}\geq 0$. S'il y a au moins une seule valeur propre négative, l'état $\rho_{AB}$ est intriqué.
\subsection{Quantification de l'intrication quantique}

Dans les protocoles de l'information quantique, les états intriqués sont utilisés pour implementer de nombreux processus et ils sont consommés pendant l'exécution du protocole quantique. Par conséquent, il est nécessaire d'avoir des mesures qui nous disent si l'intrication contenue dans les systèmes quantiques est suffisante pour effectuer une certaine tâche. La quantification de l'intrication dans un état général s'avère être une tâche très difficile. En fait, il est clair qu'elle est plus complexe que le problème de la séparabilité. De plus, en raison des effets de décohérence inévitables, les applications réalistes doivent traiter des systèmes quantiques dans des états mixtes, et exactement dans cette situation, la question de la quantité d'intrication disponible est intéressante.\par
Une mesure d'intrication est une fonction mathématique qui rend-compte des caractéristiques cruciales associées à l'intrication et qui convertit tout état quantique $\rho$ en un nombre positif $E\left( \rho\right)$ avec les propriétés axiomatiques suivantes \cite{Vedral1997,Eltschka2014}:
\begin{enumerate}
	\item Premièrement, une exigence naturelle est que la mesure d'intrication $E\left( \rho\right)$ s'annulle si $\rho$ est séparable.
	\item Deuxièmement, une mesure d'intrication devrait être invariante en cas de changement local de la base. Cela signifie qu'il devrait être invariante sous les transformations unitaires locales.
	\begin{equation}
E\left( \rho\right)=E\left( U_{A}\otimes U_{B}\rho U_{A}^{\dagger}\otimes U_{B}^{\dagger}\right).
	\end{equation}
	\item Pour toute mesure d'intrication $E\left( \rho\right)$, il est nécessaire que $0\leq E\left( \rho\right)\leq 1$ avec $E\left( \rho\right)=0$ si et seulement si l'état est séparable, et $E\left( \rho\right)=1$ si $\rho$ est l'opérateur densité d'un état maximalement intriqué.
	\item Comme l'intrication ne peut pas être créée par des opérations locales et de communications classiques (LOCC), c'est à dire pour toute opération de la forme $A\otimes B$, il est raisonnable d'exiger que $E\left( \rho\right)$ n'augmente pas sous de telles transformations. Autrement dit, si $\wedge_{\rm LOCC}$ est une application positive qui peut être implémentée par LOCC, alors
	\begin{equation}
\rho \longmapsto \wedge_{\rm LOCC}\left( \rho\right) =\frac{A\otimes B \rho A^{\dagger}\otimes B^{\dagger}}{{\rm Tr}\left[ A\otimes B \rho A^{\dagger}\otimes B^{\dagger}\right]}.
	\end{equation}
	Et
	\begin{equation}
E\left[\wedge_{\rm LOCC}\left( \rho\right)\right]\leq E\left( \rho\right).
	\end{equation}
	En d'autres termes, si l'état quantique $\rho$ peut être transformé en un autre état $\wedge_{\rm LOCC}\left( \rho\right)$ via LOCC, alors $\rho$ est au moins aussi intriqué que $\wedge_{\rm LOCC}\left( \rho\right)$. Il est très intéressant de souligner que les opérations locales sont des opérations qui peuvent être décrites par des opérateurs appartenant à l'espace de Hilbert d'un seul sous-système. Si l'un des sous-systèmes peut décider de faire une mesure arbitraire de son espace, alors les opérations locales ne peuvent pas augmenter l'intrication de l'état. De plus, Alice et Bob ne peuvent pas utiliser les communications classiques pour construire un ensemble d'opérations locales de sorte qu'ils se retrouvent, en moyenne, avec une intrication plus élevée qu'au début.
	\item Une autre propriété qui est satisfaite par la plupart des mesures d'intrication est la convexité. Autrement dit, il faut que l'intrication diminue sous le mélange de deux ou plusieurs états;
	\begin{equation}
E\left( \sum_{k}p_{k}\rho_{k}\right)\leq \sum_{k}p_{k} E\left(\rho_{k}\right),
	\end{equation}
\item D'autres questions se posent si plus de deux exemplaires d'un état sont disponibles. Par exemple, si Alice et Bob partagent $n$ copies du même état $\rho$, il peut être raisonnable d'exiger une additivité, c'est-à-dire
\begin{equation}
E\left(\rho^{\otimes n} \right)=nE\left(\rho\right).
\end{equation}
Encore plus fort, on peut exiger une additivité complète. Cela signifie que si Alice et Bob partagent deux états différents $\rho_{1}$ et $\rho_{2}$, alors
\begin{equation}
E\left(\rho_{1}\otimes\rho_{2}\right)=E\left(\rho_{1}\right)+E\left(\rho_{2}\right).
\end{equation}		
\end{enumerate}
Toute mesure d'intrication $E\left(\rho\right)$ vérifiant les exigences ci-dessus est appelée une intrication monotone. Aussi, une bonne mesure devrait satisfaire au moins les trois premières conditions.
\subsubsection{Entropie d'intrication}

Pour un état pur bipartite $\rho_{AB}$, l'entropie d'intrication est une mesure de l'intrication conceptuellement très importante, qui est unique et qui a une signification opérationnelle \cite{Kabat1995}. Elle est définie comme l'entropie de von Neumann de la matrice densité réduite de l'un des sous-systèmes $\rho_{A}$ et $\rho_{B}$. Elle est donnée par
\begin{equation}
E\left( \rho_{AB}\right)=-{\rm Tr}\left( \rho_{A}\log_{2}\rho_{A}\right)=-{\rm Tr}\left( \rho_{B}\log_{2}\rho_{B}\right), \label{EI}
\end{equation}
où $\rho_{A}={\rm Tr}_{B}\left( \rho_{AB}\right)$. Pour illustrer la façon dont l'entropie de l'intrication est calculée, nous considérons l'état de Bell $\left|\phi^{+} \right\rangle_{AB}=\frac{1}{\sqrt{2}}\left(\left|00 \right\rangle_{AB}+\left|11 \right\rangle_{AB}\right)$, la matrice densité qui décrit cet état est donnée par
\begin{align}
	\rho_{AB}&=\left|\phi^{+} \right\rangle_{AB}\left\langle\phi^{+}\right|\notag\\&=\frac{1}{2}\left( \left|0 \right\rangle_{A}\left\langle 0\right|\otimes\left|0 \right\rangle_{B}\left\langle 0\right|+\left|0 \right\rangle_{A}\left\langle 1\right|\otimes\left|0 \right\rangle_{B}\left\langle 1\right|+\left|1 \right\rangle_{A}\left\langle 0\right|\otimes\left|1 \right\rangle_{B}\left\langle 0\right|+\left|1 \right\rangle_{A}\left\langle 1\right|\otimes\left|1 \right\rangle_{B}\left\langle 1\right|\right). 
\end{align}
Alors, l'opérateur densité réduite du sous-système $A$ est
\begin{align}
\rho_{A}=\frac{1}{2}\left[ \left|0 \right\rangle_{A}\left\langle 0\right|+\left|1 \right\rangle_{A}\left\langle 1\right|\right]=\left( {\begin{array}{*{20}{c}}
	\frac{1}{2}&0\\
	0&\frac{1}{2}\end{array}} \right).
\end{align}
On remarque que $\rho_{A}$ est une matrice diagonale. De ce fait, les entrées diagonales représentent les valeurs propres. En utilisant (\ref{EI}), nous trouvons l'entropie d'intrication du système $AB$ comme
\begin{equation}
E\left( \rho_{AB}\right)=S\left( \rho_{A}\right)={\cal H}\left( \frac{1}{2}\right)=-\frac{1}{2}\log_{2}\left(\frac{1}{2} \right)-\frac{1}{2}\log_{2}\left(\frac{1}{2} \right)=\log_{2}2=1. 
\end{equation}
D'après cet exemple, il est clair que l'entropie d'intrication prend sa valeur maximale pour les états intriqués au maximum et elle est égale à zéro pour les états séparables. Aussi, on peut facilement prouver que l'intrication du sous-système $B$ prend toujours la même valeur que celle du sous-système $A$. De plus, pour un état pur maximalement intriqué $\left|\psi \right\rangle$, l'entropie d'intrication prend la valeur
\begin{equation}
E\left( \left|\psi \right\rangle\right)=\log_{2}d,
\end{equation}
où $d$ est la dimension de la matrice densité réduite.
\subsubsection{Intrication de formation}

Pour le cas d'un système préparé dans l'état mixte, une mesure qui répond aux exigences d'intrication ci-dessus est l'intrication de formation $E_{f}$. Elle est définie comme l'intrication minimale que l'on obtient sur toutes les décompositions d'états purs d'un état mixte bipartite \cite{Bennett1996}. Elle s'écrit
\begin{equation}
E_{f}\left(\rho\right)=\min_{\{p_i,\left|\psi_i \right\rangle\}}\sum_{i}p_i E\left( \left|\psi_i \right\rangle\right),\label{EF}
\end{equation}
où la minimisation porte sur toutes les décompositions spectrales purs possibles de l'état mixte $\rho$. Il convient de noter que l'intrication de formation a une interprétation simple, car elle caractérise la décomposition de l'état qui fournit les entropies locales minimales. De manière équivalente, vue que l'intrication des états purs bipartites est mieux quantifiée via leurs entropies locales (\ref{EI}), l'intrication de formation est caractérisée par la décomposition générale de l'état qui fournit l'intrication à l'état pur minimale. D'un point de vue analytique, il est très difficile de mesurer ou de calculer cette mesure. En effet, il est extrêmement difficile d'obtenir toutes les décompositions possibles d'un état quantique multipartite dans le cas général. La solution analytique a été proposée dans un article de Wootters \cite{Wootters1998} qui a déterminé une formule simple permettant de calculer cette entropie de formation, dans le cadre d'une nouvelle mesure d'intrication que nous discutons dans ce qui suit. Il a défini l'intrication de formation pour tout état à deux qubits par la formule suivante
\begin{equation}
E_{f}\left(\rho\right)=h\left( \frac{1+\sqrt{1-|{\cal C}\left(\rho \right)|^{2}}}{2}\right), 
\end{equation}
où $h\left( x\right)$ est la fonction de l'entropie binaire définit comme $h\left( x\right)= -x\log_{2}x-\left(1-x\right)\log_{2}\left(1-x\right)$ et ${\cal C}\left(\rho \right)$ est la concurrence de l'état $\rho$ que nous allons définir dans la sous section suivante.
\subsubsection{Concurrence et les relations de Wootters}

Nous allons maintenant discuter la mesure de la concurrence et passer en revue certaines de ses propriétés. Nous adoptons l'approche historique et discutons d'abord la concurrence de deux qubits, puis discutons brièvement de ses généralisations à d'autres états bipartites. Pour un état pur ${\left| \psi  \right\rangle }$ d'un système à deux qubits, la concurrence ${\cal C}\left( {\left| \psi  \right\rangle } \right)$ est définie par
\begin{equation}
{\cal C}\left( {\left| \psi  \right\rangle } \right) = \left| {\left\langle \psi  \right|\left| {\tilde \psi } \right\rangle } \right| = \left| {\left\langle \psi  \right|\left( {{\sigma _y} \otimes {\sigma _y}} \right)\left| {{\psi ^*}} \right\rangle } \right|, 
\end{equation}
où $\left| {\tilde \psi } \right\rangle= \left( {{\sigma _y} \otimes {\sigma _y}} \right)\left| {{\psi ^*}} \right\rangle $, et $\left| {\tilde \psi } \right\rangle $ est appelé l'état "spin-flip" de l'état ${\left| \psi  \right\rangle }$, ${\sigma _y}$ désigne la matrice de Pauli et $\left| {{\psi ^*}} \right\rangle $ est le conjugué complexe de $\left| \psi  \right\rangle $. De plus, il existe des classes d'états pour lesquelles la concurrence prend une forme très simple. Le cas le plus simple est celui d'un état pur à deux qubits qui peut toujours s'écrire comme 
\begin{equation}
	\left|\psi \right\rangle=\alpha\left|00 \right\rangle+\beta\left|01 \right\rangle+\gamma\left|10 \right\rangle+\delta\left|11 \right\rangle,
\end{equation}
avec $\alpha$, $\beta$, $\gamma$ et $\delta$ sont des nombres complexes qui vérifient la condition de normalisation suivante
\begin{equation}
|\alpha|^{2}+|\beta|^{2}+|\gamma|^{2}+|\delta|^{2}=1.
\end{equation}
Ainsi, l'expression explicite de la concurrence d'un état pur à deux qubits est donnée par
\begin{equation}
{\cal C}\left( \left| \psi  \right\rangle \right)=2|\alpha\delta-\beta\gamma|\geq 0.
\end{equation}
Cela signifie que l'état $\left| \psi  \right\rangle$ est séparable si $\alpha\delta=\beta\gamma$. D'autre part, la concurrence peut également être généralisée à des systèmes bipartites de dimensions supérieures. Dans ce sens, nous notons que la concurrence pour un état pur peut être écrit comme suit
\begin{equation}
{\cal C}\left(\left| \psi  \right\rangle \right) =\sqrt{2\left(1-{\rm Tr}\left(\rho_{A}^{2}\right)\right)},
\end{equation}
où $\rho_{A}$ est l'une des matrices densité réduite que l'on peut définir pour un état pur bipartite. L'extension de la concurrence aux états mixtes est définie comme la concurrence moyenne d'un ensemble d'états purs représentant $\rho$, obtenue en faisant une minimisation sur toutes les décompositions à l'état pur
\begin{equation}
{\cal C}\left(\rho\right)=\min_{\{p_i,\left|\psi_i \right\rangle\}}\sum_{i}p_i C\left( {\left| \psi  \right\rangle } \right),\label{Concu}
\end{equation}
où $\rho= \sum\limits_i {{p_i}\left| {{\psi _i}} \right\rangle \left\langle {{\psi _i}} \right|} $	et ${\left| {{\psi _i}} \right\rangle }$ sont les états purs normalisée d'un système bipartite. Cette généralisation a d'abord été proposée par Rungta et al.\cite{Rungta2001} à travers la généralisation de l'opérateur "spin-flip" à des dimensions supérieures. Ils ont également montré que cette fonction remplit la condition de convexité et d'invariance sous les transformations unitaires et donc sa minimisation sur toutes les décompositions à l'état pur donne une mesure appropriée de l'intrication. Il a été montré dans la référence \cite{Wootters1998} que la quantité ci-dessus peut être calculée à partir de
\begin{equation}
{\cal C}\left( \rho  \right) = \max \left\{ {0,\sqrt{\lambda _1} - \sqrt{\lambda _2} - \sqrt{\lambda _3} - \sqrt{\lambda _4}} \right\}, \label{Concurrence}
\end{equation}
où les $\lambda _j$ sont les valeurs propres, dans l'ordre décroissant, de la matrice suivante
\begin{equation}
\tilde \rho  = \left( {{\sigma _y} \otimes {\sigma _y}} \right){\rho ^*}\left( {{\sigma _y} \otimes {\sigma _y}} \right).
\end{equation}
À titre d'exemple, considérons une famille d'états $X$ à deux qubits, qui est écrite dans une base de produits orthonormés $\{\left|00 \right\rangle,\left|01 \right\rangle,\left|10 \right\rangle,\left|11 \right\rangle\}$ avec des éléments de la matrice densité non nulle uniquement le long de la diagonale et de l'anti-diagonale. Ils sont appelés les états $X$ en raison de son apparence semblable à la lettre X et prennent la forme suivante
\begin{equation}
\rho_{X}=\left( {\begin{array}{*{20}{c}}
	\rho_{11}&0&0&\rho_{14}\\
	0&\rho_{22}&\rho_{23}&0\\
	0&\rho_{32}&\rho_{33}&0\\
	\rho_{41}&0&0&\rho_{44}\end{array}} \right). \label{matrixX}
\end{equation}
En utilisant la formule (\ref{Concurrence}), nous pouvons facilement trouver la concurrence de la matrice densité $\rho_{X}$ comme
\begin{equation}
	{\cal C}\left(\rho_{X}\right)=2\max\{0,{\cal C}_{1}\left(\rho_{X}\right),{\cal C}_{2}\left(\rho_{X}\right)\},\label{ConcurrenceX}
\end{equation}
avec
\begin{equation}
{\cal C}_{1}\left(\rho_{X}\right)=|\rho_{14}|-\sqrt{\rho_{22}\rho_{33}}, \hspace{1cm} {\rm et} \hspace{1cm} {\cal C}_{2}\left(\rho_{X}\right)=|\rho_{23}|-\sqrt{\rho_{11}\rho_{44}}.
\end{equation}
La formule de la concurrence simple, donnée ci-dessus, explique pourquoi ces états à deux qubits ont été largement utilisés pour étudier la dynamique de l'intrication dans de nombreux scénarios.
\subsubsection{Entropie relative de l'intrication}

Une autre idée pour quantifier l'intrication consiste à mesurer la distance qui sépare un état $\rho$ (intriqué) de l'ensemble des états séparables $S$. Il a été trouvé que parmi toutes les fonctions de distance possibles, l'entropie relative est physiquement la plus raisonnable \cite{Vedral1997}. Par conséquent, nous définissons l'entropie relative de l'intrication comme
\begin{equation}
E_{r}\left(\rho\right)=\min_{\sigma\in S}S\left(\rho\|\sigma \right)=\min_{\sigma\in S}{\rm Tr}\left(\rho\log_{2}\rho-\rho\log_{2}\sigma \right), \label{Er}
\end{equation}
où la minimisation est repris sur tous les états séparables et l'équation (\ref{Er}) donne la distance entre l’état $\rho$ et l'état séparable le plus proche.
\subsubsection{Négativité et Négativité logarithmique}

Les expressions analytiques de la plupart des mesures discutées ci-dessus ont des formules compactes uniquement pour le système quantique qubit-qubit, et en général, il n'est pas facile de les calculer pour un état quantique mixte arbitraire. En outre, chacune de ces mesures révèle un aspect spécifique de l'intrication quantique. D'autre part, la négativité est une mesure particulièrement utile et facilement calculable pour tous les états mixtes d'un système bipartite arbitraire. Elle a été introduite par Vidal et Werner \cite{Vidal2002} comme étant la version quantitative du critère PPT. Elle est définie comme la somme des valeurs absolues des valeurs propres négatives de la matrice densité partiellement transposée. Par conséquent, la négativité est définie comme
\begin{equation}
	{\cal N}\left(\rho \right)=\sum_{i}|\vartheta_i|,\label{neg}
\end{equation}
où les $\{\vartheta_i\}$ sont les valeurs propres négatives de l'opérateur densité partiellement transposé $\rho^{T_{A}}$ et $T_{A}$ désigne la transposition partielle par rapport au sous-système $A$. La négativité ${\cal N}$ est également liée à la norme trace de $\rho^{T_{A}}$ via
\begin{equation}
{\cal N}\left(\rho \right)=\frac{\|\rho^{T_{A}}\|_{1}-1}{2},
\end{equation}
où la norme trace est définie par $\|\rho^{T_{A}}\|_{1}={\rm Tr}\left(\sqrt{\rho^{T_{A}}{\rho^{T_{A}}}^{\dagger}}\right)$. Pour les systèmes quantiques tripartites avec la matrice densité $\rho_{ABC}$, la négativité (\ref{neg}) devient
\begin{equation}
{{\cal N}^{\left(3 \right)}}\left(\rho_{ABC}\right):= \sqrt[3]{{\cal N}(\rho_{ABC}^{T_A}){\cal N}(\rho_{ABC}^{T_B}){\cal N}(\rho_{ABC}^{T_C})},
\end{equation}
où ${\cal N}(\rho_{ABC}^{T_A})$ représente la négativité bipartite entre le qubit $A$ et le sous-système $B$ et $C$. ${\cal N}(\rho_{ABC}^{T_A})= \sum\limits_i {\left| {{\lambda
			_i}\left( {{\rho_{ABC}^{{T_A}}}} \right)} \right|} - 1$ et les ${\lambda
		_i}\left( {{\rho_{ABC}^{T_A}}} \right)$ sont les valeurs propres de $\rho_{ABC}^{T_A}$. Des définitions similaires valent pour ${\cal N}(\rho_{ABC}^{T_B})$ et ${\cal N}(\rho_{ABC}^{T_C})$. Pour les systèmes quantiques tripartites invariants sous la symétrie de permutations, c'est-à-dire le changement de qubits, ${\cal N}(\rho_{ABC})$ se réduit à la négativité bipartite de toute bipartition du système. Cela écrit comme
\begin{equation}
{{\cal N}^{\left( 3 \right)}}(\rho_{ABC}) = {\cal N}(\rho_{ABC}^{T_A}) = {\cal N}(\rho_{ABC}^{T_B}) = {\cal N}(\rho_{ABC}^{T_C}).
\end{equation}
Notez que la négativité est une intrication monotone sous l'action des opérations locales et de communications classiques. Ainsi, elle peut être considérée comme une bonne mesure pour quantifier le degré d'intrication dans les systèmes composites. Il existe une autre intrication monotone qui est basée sur la définition de la négativité appelée la négativité logarithmique et qui est également plus facile à calculer analytiquement. Elle est définie par
\begin{equation}
\xi_{N}\left( \rho\right)=\log_{2}\|\rho^{T_{A}}\|_{1}=\log_{2}\left[ 2{\cal N}\left(\rho \right)+1\right].\label{neglog}
\end{equation}
La fonction (\ref{neglog}) a la caractéristique intéressante d'être additive, mais elle n'est pas convexe.
\subsubsection{Notion de tangle}

Le tangle est une autre mesure de l'intrication dans les systèmes tripartites, qui n'est ni une généralisation directe ni une combinaison directe de mesures bipartites \cite{Coffman2000}. Il quantifie le degré d'intrication entre les trois sous-systèmes qui ne peuvent pas être décrits et quantifiés par l'intrication bipartite. De plus, lorsque l'un des sous-systèmes (disons $A$) est intriqué avec les deux autres sous-systèmes combinés ($BC$ dans ce cas), l'intrication peut être comprise comme une intrication bipartite entre deux parties $A|BC$ et elle est mesurée comme telle par l'une des mesures bien connues de l'intrication bipartite. Dans ce cas, le tangle $\tau_{g}$ des systèmes à trois qubits est exprimé comme
\begin{equation}
	\tau_{g}=\max_{\{i,j,k\}=\{A,B,C\}}\{{\cal C}_{i|jk}^{2}-{\cal C}_{i|j}^{2}-{\cal C}_{i|k}^{2}\},\label{tangel3}
\end{equation}
où ${\cal C}_{i|jk}$ est la concurrence bipartite entre les deux sous-systèmes $i$ et $jk$, donnée par
\begin{equation}
{\cal C}_{i|jk}=\sqrt{2\left( 1-{\rm Tr}\left(\rho_{i}^{2}\right) \right) },
\end{equation}
et ${\cal C}_{i|j}$ est la concurrence entre les paires des qubits $i$ et $j$, qui a été déterminée à l'aide de l'équation (\ref{Concurrence}).

\chapter{Les quantificateurs des corrélations quantiques de type discorde}
Dans le premier chapitre, nous avons souligné que le rôle central dans le succès de l'information quantique est dû à l'exploitation de l'intrication, c'est-à-dire à des corrélations particulières décrites par des lois quantiques et qui ne peuvent être décrites dans une image classique. Cela est motivé par le fait que l'intrication peut être utilisée comme une ressource utile pour l'étude fondamentale en mécanique quantique et pour des applications dans la téléportation quantique \cite{Braunstein1998,Bouwmeester1997}, le calcul quantique \cite{Sorensen2000}, la distribution des clés quantiques \cite{Ekert1991} et beaucoup plus. Cependant, selon diverses études \cite{Datta2005,Datta2007}, il a été prouvé expérimentalement qu'il existe des états quantiques séparables qui sont très utiles dans la technologie quantique pratique et que l'intrication quantique n'est pas la forme la plus générale de corrélations quantiques. Par exemple, l'augmentation de la vitesse d'exécution de certains protocoles quantiques avec des états séparables \cite{Braunstein1999,Meyer2000}, la non-localité quantique sans l'intrication quantique \cite{Bennett1999,Niset2006} et l'efficacité du calcul quantique \cite{Lanyon2008}. En ce sens, des efforts théoriques considérables ont été consacrés au cours des dernières années pour caractériser et quantifier les corrélations quantiques au-delà de l'intrication dans les systèmes quantiques multipartites. \par

Ici, nous suivons le chemin historique qui a conduit à définir une classe de corrélations quantiques plus générale que l'intrication. Historiquement, la discorde quantique entropique, introduite pour la première fois par Ollivier et Zurek \cite{Ollivier2001} puis proposée par Henderson et Vedral \cite{Henderson2001}, est le quantificateur de corrélations quantiques largement utilisé et plusieurs travaux ont été consacrés à cette classe de mesures. Ce type de mesure peut rendre compte efficacement de toutes les corrélations quantiques existant dans les systèmes quantiques. Il est défini comme la différence entre l'information mutuelle quantique et les corrélations classiques mesurées à l'aide d'une mesure projective effectuée sur l'un des sous-systèmes. Malheureusement, en raison d'une procédure d'optimisation des corrélations classiques sur toutes les mesures projectives locales, la discorde quantique n'a été calculée explicitement que pour les cas particuliers des systèmes à deux qubits, et une solution analytique n'existe pas encore.\par

Ces difficultés ont incité Dakić et ses collaborateurs \cite{Dakic2010} à proposer une formulation alternative de la discorde quantique appelée "la discorde quantique géométrique", qui mesure les corrélations quantiques d'un état en utilisant la norme minimale de Hilbert-Schmidt (ou la norme $2$ de Schatten) entre un état donné et les états classiques les plus proches. D'un point de vue analytique, le calcul de cette mesure géométrique nécessite un processus de minimisation plus simple que la discorde quantique entropique, mais plus récemment, il a été démontré \cite{Piani2012} que la discorde quantique géométrique peut augmenter sous les opérations quantiques locales agissant sur le sous-système non mesuré. Cela est dû à la distance de Hilbert-Schmidt, qui ne satisfait pas la propriété de contractivité sous les applications préservant les traces. Par conséquent, la discorde géométrique basée sur la norme de Hilbert-Schmidt n'est pas une bonne mesure des corrélations quantiques. Plusieurs variantes géométriques ont été introduites dans la littérature, mais la norme trace (la norme de Bures ou la norme $1$ de Schatten) s'est avérée être la seule norme de Schatten que l'on peut utiliser pour faire une mesure géométrique des corrélations quantiques \cite{Paula2013}. De nouvelles mesures liées à l'incertitude quantique ont été proposées pour surmonter ces difficultés et problèmes, telles que l'incertitude quantique locale \cite{Girolami2013}, l'information de Fisher quantique locale \cite{Kim2018} et le pouvoir interférométrique quantique \cite{Girolami2014}.\par

La première partie de ce chapitre est consacrée à la présentation de la discorde quantique et sa variante géométrique; ses définitions et ses propriétés importantes qui sont nécessaires pour discuter des résultats introduits dans les chapitres ultérieurs. De plus, à l'aide d'un exemple, nous montrons comment calculer analytiquement ces mesures dans les systèmes quantiques bipartites. Dans la deuxième partie, nous allons introduire le concept de l'incertitude quantique locale comme un autre indicateur de la discorde quantique dans les systèmes bipartites. Nous adoptons l'information d'interchange (Skew) pour mesurer l'incertitude quantique des observables locales, et nous dérivons analytiquement l'expression de l'incertitude quantique locale pour les états à deux qubits et son extension aux états à plusieurs qubits. Enfin, dans la dernière section de ce chapitre, nous discutons la notion de monogamie pour savoir comment les corrélations quantiques peuvent être partagées entre de nombreux sous-systèmes.

\section{La discorde quantique entropique}
\subsection{Définition}

Comme nous l'avons mentionné précédemment, des recherches récentes révèlent qu'il existe des corrélations autres que l'intrication, et la discorde quantique quantifie les corrélations totales non classiques dans un état quantique. La discorde quantique pourrait également être non nulle pour certains états mixtes séparables. Cela signifie que la séparabilité ne signifie pas l'absence des corrélations quantiques. En général, la définition de l'intrication contient une interprétation opérationnelle indiquant que les états séparables peuvent être préparés par des opérations locales et de la communication classique entre les sous-systèmes, contrairement aux états intriqués. Cependant, il a été démontré que ce n'est pas le cas pour la discorde quantique car les communications classiques peuvent donner lieu à des corrélations quantiques. Cela peut être compris en considérant que les états réduits des sous-systèmes peuvent être physiquement mélanges, c'est-à-dire qu'ils ne sont pas orthogonaux. Par conséquent, toutes les informations qui s'y rapportent ne peuvent pas être récupérées localement. Ce phénomène n'a pas d'analogue classique, ce qui explique la quantité des corrélations dans l'état séparable avec la discorde non nulle. Au-delà de son importance en tant que quantificateur des corrélations purement quantique, la discorde quantique est devenue de plus en plus importante dans plusieurs domaines ainsi que dans les applications en science de l'information quantique. En physique de la matière condensée, les corrélations quantiques dans les états séparables caractérisent, même à température finie, une transition de phase quantique \cite{Maziero2012}. Concernant les systèmes biologiques, il a été suggéré que les corrélations quantiques de type discorde pourraient jouer un rôle important dans la photosynthèse \cite{Bradler2010}. Dans le cadre de la théorie quantique des champs, il a été démontré que l'effet Unruh détruit le comportement des corrélations quantiques \cite{Celeri2010}. En outre, il a été suggéré que la discorde quantique, plutôt que l'intrication, est responsable de l'efficacité d'un ordinateur quantique, ce qui est confirmé à la fois théoriquement \cite{Datta2008} et expérimentalement \cite{Lanyon2008}. Werlang et ses collègues ont démontré que la discorde quantique est plus robuste que l'intrication dans les mêmes conditions de l'environnement markovien pour les systèmes quantiques ouverts \cite{Werlang2009}. De plus, Streltsov et Zurek ont montré que les résultats de mesure ne peuvent pas être parfaitement communiqués par des canaux classiques si l’appareil de mesure est dans un état non classique (un état qui ne contient que des corrélations quantiques). Cela signifie que la perte d'informations se produit même lorsque l'appareil de mesure n'est pas intriqué avec le système, et que la quantité de ces informations perdues s'avère être exactement la discorde quantique \cite{Streltsov2013}. Selon la réf \cite{Ollivier2001}, la discorde quantique est définie comme la différence entre deux expressions classiquement équivalentes de l'information mutuelle; c'est-à-dire l'information mutuelle quantique d'origine (\ref{IMQ}) et l'information mutuelle quantique induite par la mesure locale. Donc, pour un système composite bipartite $\rho _{AB}$, nous avons  
\begin{equation}
{\cal Q}\left( {{\rho _{AB}}} \right): = {\cal I}\left( {{\rho _{AB}}} \right) -{\cal J}_{B}\left( {{\rho _{AB}}} \right), \label{QDD}
\end{equation}
où les corrélations totales sont quantifiées par l'information mutuelle quantique ${\cal I}\left( \rho_{AB}\right)=S\left( \rho_{A}\right)+S\left( \rho_{B}\right)-S\left( \rho_{AB}\right)$, et la quantité ${\cal J}_{B}\left( {{\rho _{AB}}} \right)$ est définie comme une mesure des corrélations classiques
\begin{equation}
{\cal J}_{B}\left( {{\rho _{AB}}} \right)=\mathop {\max }\limits_{{\pi _B}^j} \left( {S\left( {{\rho _B}} \right) - \sum\limits_j {{p_{B,j}}S\left( {{\rho _{^{B,j}}}} \right)} } \right), \label{QCD}
\end{equation}
où le minimum est pris sur l'ensemble des opérateurs de mesure à valeur positive ${\pi _B}^j$ sur le sous-système $B$ qui satisfont $\sum_{j}{{\pi _B}^j}^{\dagger}{\pi _B}^j=\openone$, $S\left(\rho\right)$ est l'entropie de von Neumann et $S\left( {{\rho _{^{B,j}}}} \right)$ est l'entropie conditionnelle,
\begin{equation}
	{p_{B,j}} = {\rm Tr}_{AB}\left[\left( {I \otimes {\pi _B}^j} \right){{\rho _{AB}}\left({I \otimes {\pi _B}^{j^{\dagger}}} \right)}\right] \hspace{0.5cm}{\rm et}\hspace{0.5cm} {\rho _{B,j}} =\frac{{{\rm Tr}_A\left[ {\left( {I \otimes {\pi _B}^j} \right){\rho _{AB}}\left( {I \otimes {\pi _B}^{j^{\dagger}}}\right)} \right]}}{p_{B,j}},
\end{equation}
sont la probabilité et l'état conditionnel du sous-système $B$ associés au résultat $j$. Il convient de noter que la discorde quantique est mesurée par de fortes mesures qui conduisent à la perte de sa cohérence, mais elle révèle plus de corrélations quantiques que l'intrication. De plus, l'idée principale du calcul de la discorde quantique est de quantifier la quantité d'informations qui ne sont pas accessibles par une mesure locale, en extrayant certaines informations sur le sous-système $A$ en mesurant l'état du sous-système $B$ sans perturber l'état $A$. Semblable à l’intrication, la discorde quantique satisfait les propriétés suivantes
\begin{description}
	\item[Pr$_1$:] La discorde quantique est toujours définie positive ${\cal Q}\left( {{\rho _{AB}}} \right)\geq 0$. Ceci est le résultat direct de la concavité de l'entropie conditionnelle.
	\item[Pr$_2$:]La discorde quantique est inférieure ou égale à l'entropie de von Neumann du sous-système non mesuré (dans notre cas le sous-système $A$), alors ${\cal Q}\left( {{\rho _{AB}}} \right)\leq S\left(\rho_{A} \right)$.
	
	\item[Pr$_3$:]la discorde quantique ${\cal Q}\left( {{\rho _{AB}}} \right)$ s'annulent si et seulement si l'état $\rho_{AB}$ est classiquement corrélé ($\rho_{A}$ et $\rho_{B}$ sont composés d'états orthogonaux,), c'est-à-dire
	\begin{equation}
 \rho_{AB}=\sum_{i,j}p_{ij}\left|i \right\rangle\left\langle i\right|\otimes\left|j \right\rangle\left\langle j\right|.
	\end{equation}
	\item[Pr$_4$:]La discorde quantique n'est pas symétrique car l'entropie conditionnelle n'est pas symétrique. Cela signifie que nous ne trouvons pas la même valeur de la discorde quantique si nous changeons le sous-système mesuré $B$ par le sous-système $A$. Donc ${\cal Q}_{A}\left( {{\rho _{AB}}} \right)\neq {\cal Q}_{B}\left( {{\rho _{AB}}} \right)$.
	\item[Pr$_5$:]Pour tous les états bipartites purs, la discorde quantique se réduit à l'entropie d'intrication
	\begin{equation}
{\cal Q}\left( {{\rho _{AB}}} \right)=E\left( {{\rho _{AB}}} \right)=S\left( {{\rho _{A}}} \right)=S\left( {{\rho _{B}}} \right).
	\end{equation}
\item[Pr$_6$:]La discorde quantique est invariante sous les transformations unitaires locales. Nous avons donc
\begin{equation}
{\cal Q}\left( {{\rho _{AB}}} \right)={\cal Q}\left(\left( U_{A}\otimes U_{B}\right) \rho _{AB}\left( U_{A}^{\dagger}\otimes U_{B}^{\dagger}\right)\right).
\end{equation}	
\end{description}
La principale difficulté lors du calcul de la discorde quantique, pour plusieurs systèmes quantiques bipartites, réside dans la minimisation de l'entropie conditionnelle. Cela explique pourquoi il n'existe pas de méthode simple pour calculer explicitement la discorde quantique pour des états mixtes arbitraires. Par la suite, nous discuterons certains résultats liés aux expressions analytiques de la discorde quantique obtenues pour certaines formes spéciales des états de type qubit-qubit.
\subsection{La discorde quantique pour les états bipartites de forme $X$}
Ici, nous expliquons brièvement comment calculer analytiquement la discorde quantique pour les états $X$ à deux qubits en utilisant la méthode proposée par Wang dans la réf \cite{Wang2010}. Tout d'abord, la matrice densité (\ref{matrixX}) possède les valeurs propres suivantes
\begin{equation*}
{\eta _{1,2}} = \frac{1}{2}\left[ {{\rho _{11}} + {\rho _{44}} \pm \sqrt {{{\left( {{\rho _{11}} - {\rho _{44}}} \right)}^2} + 4{{\left| {{\rho _{14}}} \right|}^2}} } \right], \label{VPX1}
\end{equation*}
et
\begin{equation*}
{\eta _{3,4}} = \frac{1}{2}\left[ {{\rho _{22}} + {\rho _{33}} \pm \sqrt {{{\left( {{\rho _{22}} - {\rho _{33}}} \right)}^2} + 4{{\left| {{\rho _{23}}} \right|}^2}} } \right].\label{VPX2}
\end{equation*}
Dans ce cas, l'entropie de von Neumann des matrices réduites ${\rho _A}$ et ${\rho _B}$ est donnée par
\begin{align}
S\left( {{\rho _A}} \right) =  - \left( {{\rho _{11}} + {\rho _{22}}} \right){\log _2}\left( {{\rho _{11}} + {\rho _{22}}} \right) - \left( {{\rho _{33}} + {\rho _{44}}} \right){\log _2}\left( {{\rho _{33}} + {\rho _{44}}} \right),
\end{align}
et
\begin{align}
S\left( {{\rho _B}} \right) =  - \left( {{\rho _{11}} + {\rho _{33}}} \right){\log _2}\left( {{\rho _{11}} + {\rho _{33}}} \right) - \left( {{\rho _{22}} + {\rho _{44}}} \right){\log _2}\left( {{\rho _{22}} + {\rho _{44}}} \right).
\end{align}
Pour minimiser les corrélations classiques (\ref{QCD}), nous prenons un ensemble complet de mesures projectives sur le sous-système B qui sont représentées par les opérateurs $\{{\pi _B}^j=\left| {{B_j}} \right\rangle\left\langle {B_j}\right|\}$ ($j=1,2$) avec
\begin{equation}
\begin{array}{l}
\left| {{B_1}} \right\rangle  = \cos \left( {\frac{\theta }{2}} \right)\left| 1 \right\rangle  + {e^{i\varphi }}\sin \left( {\frac{\theta }{2}} \right)\left| 0 \right\rangle \\
\left| {{B_2}} \right\rangle  = \sin \left( {\frac{\theta }{2}} \right)\left| 1 \right\rangle  - {e^{i\varphi }}\cos \left( {\frac{\theta }{2}} \right)\left| 0 \right\rangle,
\end{array}
\end{equation}
où $0 \le \theta  \le \frac{\pi }{2}$ et $0 \le \varphi  \textless 2\pi$. La probabilité ${p_{B,j}}$ correspondant au résultat $j$ et les deux valeurs propres de $\rho _{B,j}$ après la mesure locale sont données par
\begin{equation}
{p_{B,j}} = \frac{1}{2}\left[ {1 + {{\left( { - 1} \right)}^j}\cos \theta \left( {1 - 2{\rho _{11}} - 2{\rho _{33}}} \right)} \right]
\end{equation}
et
\begin{equation}
{\eta_ \pm }\left( {{\rho _{B,j}}} \right) = \frac{1}{2}\left( {1 \pm \frac{{\sqrt {{\upsilon_j}} }}{{{p_{B,j}}}}} \right),
\end{equation}
avec
\begin{align}
{\upsilon _j} = &\frac{1}{4}{\left[ {1 - 2\left( {{\rho _{33}} + {\rho _{44}}} \right) + {{\left( { - 1} \right)}^j}\cos \theta \left( {1 - 2{\rho _{11}} - 2{\rho _{44}}} \right)} \right]^2}  \notag\\&+{\sin ^2}\theta \left[ {{{\left| {{\rho _{14}}} \right|}^2} + {{\left| {{\rho _{23}}} \right|}^2} - 2\left| {{\rho _{14}}} \right|\left| {{\rho _{23}}} \right|\sin \left( {2\varphi  + \phi } \right)} \right].
\end{align}
Alors, l'entropie de $\rho _{B,j}$ peut être écrite en termes de valeurs propres comme suit
\begin{equation}
S\left( {{\rho _{B,j}}} \right) = h\left( {{\eta_+ }\left( {{\rho _{B,j}}} \right)} \right),
\end{equation}
où $h\left(x \right)$ est l'entropie binaire de Shannon. Ainsi, l'entropie conditionnelle est
\begin{equation}
S\left( {{\rho _{A\left| B \right.}}} \right) = \sum\limits_{j = 1}^2 {{p_{B,j}}S\left( {{\rho _{B,j}}} \right)}  = {p_{B,1}}S\left( {{\rho _{B,1}}} \right) + {p_{B,2}}S\left( {{\rho _{B,2}}} \right).
\end{equation}
En définissant des dérivées partielles de cette entropie conditionnelle par rapport à $\theta $ et $\varphi$ égales à zéro. Après un calcul simple, il est facile de montrer que l'entropie conditionnelle est minimale lorsque $\theta  = \frac{\pi }{2}$. Dans ce cas, ${p_{B,1}} = {p_{B,2}}$ et $S\left( {{\rho _{B,1}}} \right) = S\left( {{\rho _{B,2}}} \right)$. Par conséquent, la valeur minimale de l'entropie conditionnelle est
\begin{equation}
{\zeta _1} = h\left( {\frac{1 + \sqrt {{{\left( {1 - 2\left( {{\rho _{33}} + {\rho _{44}}} \right)} \right)}^2} + 4{{\left( {\left| {{\rho _{14}}} \right| + \left| {{\rho _{23}}} \right|} \right)}^2}}}{2}} \right).
\end{equation}
La deuxième valeur extrémale est obtenue lorsque $\theta=0$ pour toute valeur arbitraire de $\varphi$ et la deuxième valeur minimale de l'entropie conditionnelle est
\begin{equation}
{\zeta_2} = - \sum\limits_{i = 1}^4 {{\rho _{ii}}{{\log }_2}} {\rho _{ii}} - h\left( {{\rho _{11}} + {\rho _{33}}} \right).
\end{equation}
Alors, l'expression explicite de corrélation classique dans l'état (\ref{matrixX})  prend la forme suivante
\begin{equation}
{\cal J}_{B}\left( {{\rho _{AB}}} \right) = \max \left( {{{\cal J}_1},{{\cal J}_2}} \right),
\end{equation}
avec
\begin{equation}
{{\cal J}_i} = h\left( {{\rho _{11}} + {\rho _{22}}} \right) - {\zeta _i}.
\end{equation}
Enfin, l'expression analytique de la discorde quantique peut être déterminée à partir de l'équation (\ref{QDD}) comme
\begin{equation}
{\cal Q}\left( {{\rho _{AB}}} \right) = \min \left( {{{\cal Q}_1},{{\cal Q}_2}} \right),\label{QDANLY}
\end{equation}
avec
\begin{equation}
{{\cal Q}_i} = h\left( {{\rho _{11}} + {\rho _{33}}} \right) + \sum\limits_{i = 1}^4 {{\eta_i}{{\log }_2}} {\eta_i} + {\zeta _i}.
\end{equation}
\subsection{Relation de Koashi-Winter}
Considérons un système bipartite $\rho_{AB}\in {\cal H}_{A}\otimes {\cal H}_{B}$ et sa purification $\left| \psi\right\rangle_{ABE}\in {\cal H}_{A}\otimes {\cal H}_{B}\otimes {\cal H}_{E}$. La dimension de l'espace global est ${\rm dim}\left({\cal H}_{ABE}\right)={\rm dim}\left({\cal H}_{A}\right).{\rm dim}\left({\cal H}_{B}\right).{\rm Rang}\left(\rho_{AB}\right)$. Il est important de souligner que le système de purification $E$ (l'environnement) est constitué par l'univers moins les sous-systèmes $A$ et $B$, puisque, dans ce cas, $\rho_{ABE}$ est une matrice densité pur. Cette purification créant des corrélations quantiques entre le système $AB$ et le système de purification $E$ et la distribution des corrélations d'une purification tripartite est régie par la relation de Koashi-Winter \cite{Koashi2004}. Cette relation est introduite par Koashi et Winter pour comprendre la distribution de l'intrication entre les deux sous-systèmes. Ils ont montré l'existence d'une relation monogame importante entre l'intrication de formation $E_{f}\left(\rho\right)$ (\ref{EF}) et la corrélation classique ${\cal J}\left( {{\rho}} \right)$ (\ref{QCD}). Elle est donnée par
\begin{equation}
E_{f}\left(\rho_{AB}\right)+{\cal J}_{B}\left( \rho_{BE} \right)=S\left(\rho_{B} \right), \label{RKW1}
\end{equation}
où $E_{f}\left(\rho_{AB}\right)$ est l'intrication de formation entre les sous-systèmes $A$ et $B$, ${\cal J}_{B}\left( \rho_{BE} \right)$ représente les corrélations classiques entre les sous-systèmes $B$ et $E$. L'équation (\ref{RKW1}) est valide si nous effectuons la mesure locale sur le sous-système $B$. De même, si nous effectuons une mesure sur le sous-système $A$, nous avons
\begin{equation}
E_{f}\left(\rho_{AB}\right)+{\cal J}_{A}\left( \rho_{AE} \right)=S\left(\rho_{A} \right).\label{RKW2}
\end{equation}
Notez que l'entropie de von Neumann $S\left(\rho_{A} \right)$ mesure la quantité de corrélation (classique et/ou quantique) entre le sous-système $A$ avec le monde extérieur. Si nous divisons le monde extérieur en deux parties, le sous-système $B$ et le système de purification $E$, alors la quantité de corrélation quantique entre $A$ et $B$, plus la quantité de corrélation classique entre $A$ et la partie de purification $E$, doit être égale à $S\left(\rho_{A} \right)$. Grâce au changement symétrique des qubits, c'est-à-dire que le sous-système $A$ devient $B$, $B$ devient $E$ et $E$ devient $A$, les équations (\ref{RKW1}) et (\ref{RKW2}), respectivement, deviennent comme suit
\begin{equation}
E_{f}\left(\rho_{EA}\right)+{\cal J}_{A}\left( \rho_{AB} \right)=S\left(\rho_{A} \right), 
\end{equation}
et
\begin{equation}
E_{f}\left(\rho_{BE}\right)+{\cal J}_{B}\left( \rho_{BA} \right)=S\left(\rho_{B} \right). \label{II}
\end{equation}
On remplace l'expression de corrélation classique en fonction de l'intrication de formation dans l'équation (\ref{QDD}), la discorde quantique devient alors
\begin{equation}
{\cal Q}\left( {{\rho _{AB}}} \right) = S\left( {{\rho_{A}}} \right)- S\left( {{\rho_{AB}}} \right)- E_{f}\left(\rho_{BE}\right).\label{QDrang2}
\end{equation}
En utilisant l'équation (\ref{II}), nous pouvons trouver une autre expression explicite de la discorde quantique pour les systèmes qubit-qubit de type $X$ et de rang-$2$. Pour ce type d'état, la matrice densité (\ref{matrixX}) peut également être écrite dans la base de calcul $\{\left|00 \right\rangle,\left|11 \right\rangle,\left|01 \right\rangle,\left|10 \right\rangle\}$ sous la forme suivante:
\begin{equation}
\rho_{X}=\left( {\begin{array}{*{20}{c}}
	\rho_{11}&\rho_{14}&0&0\\
	\rho_{41}&\rho_{44}&0&0\\
	0&0&\rho_{22}&\rho_{23}\\
	0&0&\rho_{32}&\rho_{33}\end{array}} \right)=\rho_{X_{1}}+\rho_{X_{2}},
\end{equation}
avec les matrices $\rho_{X_{1}}$ et $\rho_{X_{2}}$ sont données par
\begin{equation}
\rho_{X_{1}}=\left( {\begin{array}{*{20}{c}}
	\rho_{11}&\rho_{14}\\
	\rho_{41}&\rho_{44}\end{array}} \right),\hspace{1cm}{\rm et}\hspace{1cm}\rho_{X_{2}}=\left( {\begin{array}{*{20}{c}}
	\rho_{22}&\rho_{23}\\
	\rho_{32}&\rho_{33}\end{array}} \right).
\end{equation}
Dans cette écriture, les valeurs propres (\ref{VPX1}) et (\ref{VPX2}) deviennent
\begin{equation}
	{\eta _{1,2}} =\frac{1}{2}{\rm Tr}\left(\rho_{X_{1}} \right)\pm \sqrt{\left( \frac{1}{2}{\rm Tr}\left(\rho_{X_{1}}\right)\right)^{2} -{\rm det}\left(\rho_{X_{1}} \right)},
\end{equation}
et
\begin{equation}
{\eta _{3,4}} =\frac{1}{2}{\rm Tr}\left(\rho_{X_{2}} \right)\pm \sqrt{\left( \frac{1}{2}{\rm Tr}\left(\rho_{X_{2}}\right)\right)^{2} -{\rm det}\left(\rho_{X_{2}} \right)},
\end{equation}
où
\begin{equation}
{\rm det}\left(\rho_{X_{1}}\right)=\rho_{11}\rho_{44}-\rho_{14}\rho_{41}, \hspace{1cm}{\rm et}\hspace{1cm}{\rm det}\left(\rho_{X_{2}} \right)=\rho_{22}\rho_{33}-\rho_{23}\rho_{32}.
\end{equation}
Il est claire que pour $\rho_{X}$ soit de rang $2$, il suffit que ${\rm det}\left(\rho_{X_{1}}\right)=0$ et ${\rm det}\left(\rho_{X_{2}}\right)=0$. Par conséquent, les valeurs propres non nulles de la matrice densité $\rho_{X}$ sont
\begin{equation}
{\eta _{1}}=\rho_{11}+\rho_{44}\hspace{1cm}{\rm et}\hspace{1cm}{\eta _{3}}=\rho_{22}+\rho_{33}.
\end{equation}
Par la décomposition spectrale de la matrice densité $\rho_{X}$, nous trouvons
\begin{equation}
	\rho_{X}={\eta _{1}}\left|\psi_{1} \right\rangle_{AB}\left\langle \psi_{1}\right|  +{\eta _{3}}\left|\psi_{2} \right\rangle_{AB}\left\langle \psi_{2}\right|,
\end{equation}
où les vecteurs propres $\left|\psi_{1} \right\rangle_{AB}$ et $\left|\psi_{2} \right\rangle_{AB}$ propres sont donnés par
\begin{equation}
	\left|\psi_{1} \right\rangle_{AB}=\frac{1}{\sqrt{|\rho_{41}|^{2}+|\rho_{44}|^{2}}}\left(\rho_{41}\left| 00\right\rangle + \rho_{44}\left| 11\right\rangle \right),
\end{equation}
\begin{equation}
\left|\psi_{2} \right\rangle_{AB}=\frac{1}{\sqrt{|\rho_{23}|^{2}+|\rho_{33}|^{2}}}\left(\rho_{23}\left| 00\right\rangle + \rho_{33}\left| 11\right\rangle \right).
\end{equation}
En utilisant la méthode de purification (voir le premier chapitre), l'état pur $\left|\psi \right\rangle_{ABE}$ est donné par
\begin{equation}
	\left|\psi \right\rangle_{ABE}=\sqrt{\eta _{1}}\left|\psi_{1} \right\rangle_{AB}\otimes\left|0 \right\rangle_{E}+\sqrt{\eta _{3}}\left|\psi_{2} \right\rangle_{AB}\otimes\left|1 \right\rangle_{E}.
\end{equation}
Pour appliquer la relation Kaochi-Winter dans ce cas, nous devons calculer l'intrication de formation de l'état $\rho_{BE}$, Après un calcul simple, nous avons
\begin{equation}
E_{f}\left(\rho_{BE}\right)=h\left( \frac{1+\sqrt{1-|{\cal C}\left(\rho_{BE} \right)|^{2}}}{2}\right), 
\end{equation}
avec
\begin{equation}
|{\cal C}\left(\rho_{BE} \right)|^{2}=\frac{4\left(\rho_{11}+\rho_{44} \right) \left(\rho_{22}+\rho_{33}  \right)\left( \rho_{41}\rho_{23}-\rho_{33}\rho_{44} \right)  }{\left(|\rho_{41}|^{2} +|\rho_{44}|^{2}\right) \left( |\rho_{23}|^{2} +|\rho_{33}|^{2}\right)}.
\end{equation}
Enfin, nous constatons que l'expression explicite de la discorde quantique s'écrit comme suit
\begin{equation}
	{\cal Q}\left( {{\rho _{AB}}} \right)=h\left(\rho_{11}+\rho_{22} \right)-h\left(\rho_{11}+\rho_{44} \right)-h\left(\frac{1}{2}+\frac{1}{2}\sqrt{1-\frac{4\left(\rho_{11}+\rho_{44} \right) \left(\rho_{22}+\rho_{33}  \right)\left( \rho_{41}\rho_{23}-\rho_{33}\rho_{44} \right)  }{\left(|\rho_{41}|^{2} +|\rho_{44}|^{2}\right) \left( |\rho_{23}|^{2} +|\rho_{33}|^{2}\right)}}\right).  
\end{equation}
\section{La discorde quantique géométrique}
Comme l'évaluation de la discorde quantique entropique implique une procédure d'optimisation, les résultats analytiques ne sont connus que pour quelques familles d'états à deux qubits. Pour surmonter ce problème, le fait que la discorde quantique entropique s'annulle pour les états classiquement corrélés a été utilisé pour définir une version géométrique de la discorde. Ceci en analogie complète avec les mesures géométriques de l'intrication qui sont définies en termes de distances par rapport à l'ensemble des états séparables. Dans cette direction, une condition nécessaire et suffisante pour l'existence d'une discorde quantique non nulle a été obtenue et une méthode géométrique de quantification de la discorde quantique a été proposée \cite{Dakic2010}. Pour commencer, nous énumérons quelques notations que nous avons utilisées. Nous notons $\rho$ l'opérateur densité d'un système bipartite $AB$ et ${{\Omega _0}}=\sum_{k}p_{k}\Pi_{k}^{A}\otimes\rho_{k}^{B}$ est l'ensemble des états de discorde nulle, avec $\Pi_{k}^{A}$ est le projecteur orthogonal dans l'espace de Hilbert $H_{A}$, et $\rho_{k}^{B}$ est un opérateur densité arbitraire dans l'espace de Hilbert $H_{B}$, $0\leq p_{k}\leq1$ et $\sum_{k}p_{k}=1$. De plus, $\|Z\|_{p}=\left[{\rm Tr}\left(Z^{\dagger}Z\right)^{\frac{p}{2}} \right]^{\frac{1}{p}}$ est la norme $p$ de Schatten, qui se réduit à la norme de Hilbert Schmidt pour $p=2$ et à la norme de trace pour $p=1$.\par

La discorde quantique géométrique quantifie la quantité de corrélation quantique présente dans un état quantique à l'aide de la distance de Schatten entre l'état considéré $\rho$ et son plus proche état classiquement corrélé $\chi$,
\begin{equation}
{{\cal Q}_g}_{p}\left( {{\rho _{AB}}} \right) = \mathop {\min }\limits_{\chi  \in {\Omega _0}} \left(\|{\rho  - \chi } \|_{p}\right)^{p},\label{DQGP}
\end{equation} 
De nombreuses mesures géométriques via différentes normes de Schatten ont été proposées. Nous citons ici deux mesures de la discorde quantique géométrique la plus utilisées. Ils sont définis respectivement par la norme de Hilbert-Schmidt et par la norme de trace. \par
Nous commençons par la determination analytique de la discorde quantique géométrique basée sur la distance de Hilbert-Schmidt. Elle est définie par \cite{Dakic2010}
\begin{equation}
{{\cal Q}_g}_{2}\left( {{\rho _{AB}}} \right) = \mathop {\min }\limits_{\chi  \in {\Omega _0}} {\left\| {\left. {\rho  - \chi } \right\|}_{2} \right.^2},\label{DG2}
\end{equation}
où le minimum est effectué sur touts les états classiques possibles $\chi $, et ${\left\| {\left.{\rho  - \chi } \right\|}_{2} \right.} =\sqrt{{\rm Tr}{\left( {\rho  - \chi } \right)^{2}}}$ est la norme de Hilbert-Schmidt. Le calcul de cette mesure nécessite un processus de minimisation plus simple. Nous montrerons ici comment évaluer cette quantité pour un état arbitraire à deux qubits. En effet, un état général à deux qubits peut être écrit dans la représentation de Bloch avec un triple $\left\{ {\vec{x}, \vec{y} ,T} \right\}$ comme
\begin{equation}
\rho  = \frac{1}{4}\left( {\openone \otimes \openone + \sum\limits_{i = 1}^3 {{x_i}{\sigma _i} \otimes \openone + } \sum\limits_{i = 1}^3 {{y_i}\openone \otimes {\sigma _i} + } \sum\limits_{i,j = 1}^3 {{T_{ij}}{\sigma _i} \otimes {\sigma _j}} } \right),\label{MG}
\end{equation}
où ${x_i} = tr\left( {\rho \left( {{\sigma _i} \otimes \openone} \right)} \right)$,  ${y_i} = tr\left( {\rho \left( {\openone \otimes {\sigma _i}} \right)} \right)$ sont les composantes des vecteurs de Bloch locaux, ${T_{ij}} = tr\left( {\rho \left( {{\sigma _i} \otimes {\sigma _j}} \right)} \right)$ sont les composantes du tenseur de corrélation et ${{\sigma _i}}$ ($i \in \left\{ {1,2,3} \right\}$) sont les matrices de spin de Pauli. Comme nous l'avons vu précédemment, les états de discorde nulle sont de la forme
\begin{equation}
\chi  = \sum\limits_{k = 1}^m {{p_k}\left| {{\psi _k}} \right\rangle \left\langle {{\psi _k}} \right|}  \otimes {\rho _k}.
\end{equation}
Pour les systèmes à deux qubits, l'état de discorde nulle est de la forme
\begin{equation}
\chi  = {p_1}\left| {{\psi _1}} \right\rangle \left\langle {{\psi _1}} \right| \otimes {\rho _1} + {p_2}\left| {{\psi _2}} \right\rangle \left\langle {{\psi _2}} \right| \otimes {\rho _2},
\end{equation}
où $\{\left| {{\psi _1}} \right\rangle,\left| {{\psi _2}} \right\rangle\}$ est une base orthonormée pour un seul qubit. Pour simplifier les calculs, nous introduisons la quantité $t=p_{1}-p_{2}$ et les trois vecteurs $\vec{e}=\left\langle{{\psi_1}}\right| \vec{\sigma}\left|{{\psi_1}}\right\rangle$ et $\vec{s_{\pm}}={\rm Tr}\left(p_{1}\rho_{1}\pm p_{2}\rho_{2}\right)\vec{\sigma}$. Ainsi, les vecteurs de Bloch locaux du premier et du deuxième qubit sont réduits à $\vec{x}=t\vec{e}$ et $\vec{y}=\vec{s}_{+}$, respectivement. Il est également facile de démontrer que le tenseur de corrélation devient $T=\vec{e}.\vec{s_{-}}^{t}$. Par conséquent, l'état de discorde nulle $\chi$ a une représentation de Bloch qui est associée au triple $\{t\vec{e},\vec{s}_{+},\vec{e}.\vec{s_{-}}^{t}\}$ où $\|\vec{e}\|^{2}=1$, $\|\vec{s}_{\pm}\|^{2}\leq1$ et $t\in \left[-1 ,1\right]$. En se basant sur l'expression de la distance de Hilbert-Schmidt entre les états $\rho$ et $\chi$, nous obtenons
\begin{align}
{\left\| {\left. {\rho  - \chi } \right\|}_{2} \right.^2} &= {\left\| \left.\rho  \right\|_{2}\right.^2} - 2{\rm Tr}\rho \chi  + {\left\|\left. \chi  \right\|_{2}\right.^2}\notag\\&=\frac{1}{4}\left(1+\|\left.\vec{x}\|_{2}\right.^2+\|\left.\vec{y}\|_{2}\right.^2+\|\left.T\|_{2}\right.^2 \right)\notag\\&-\frac{1}{2}\left( 1+t\vec{x}\vec{e}+\vec{y}\vec{s_{+}}+\vec{e}T\vec{s_{-}}\right)\notag\\&+\frac{1}{4}\left( 1+t^{2}+\|\left.\vec{s_{+}}\|_{2}\right.^2+\|\left.\vec{s_{-}}\|_{2}\right.^2\right).   \label{munDG}
\end{align}
Pour trouver une distance de Hilbert-Schmidt minimale entre l'état considéré $\rho$ et l'état de discorde nulle $\chi$, nous optimisons la distance en fonction des trois paramètres $\vec{s}_{\pm}$ et $t$. En ce sens, le minimum est obtenu lorsque la dérivée de la distance par rapport à ces paramètres est nulle. Ceci s'écrit alors
\begin{equation}
	\frac{\partial {\left\| {\left. {\rho  - \chi } \right\|}_{2} \right.^2}}{\partial t}=\frac{1}{2}\left(t-\vec{x} \vec{e}\right)=0,
\end{equation}
\begin{equation}
\frac{\partial {\left\| {\left. {\rho  - \chi } \right\|}_{2} \right.^2}}{\partial {\vec{s_{+}}}}=\frac{1}{2}\left(\vec{s_{+}}-\vec{y} \right)=0, 
\end{equation}
et
\begin{equation}
\frac{\partial {\left\| {\left. {\rho  - \chi } \right\|}_{2} \right.^2}}{\partial {\vec{s_{-}}}}=\frac{1}{2}\left(\vec{s_{-}}-T^{t}\vec{e}\right)=0. 
\end{equation}
La solution de ces trois équations ci-dessous donnent $t=\vec{x}\vec{e}$, $\vec{s_{+}}=\vec{y}$ et $\vec{s_{-}}=T^{t}\vec{e}$. Nous remplaçons ces solutions dans l'équation (\ref{munDG}) pour obtenu
\begin{equation}
{\left\| {\left. {\rho  - \chi } \right\|}_{2} \right.^2}=\frac{1}{4}\left[\|\left.\vec{x}\|_{2}\right.^2+\|\left.T\|_{2}\right.^2 -\vec{e}\left(\vec{x}\vec{x}^{t} +TT^{t}\right)\vec{e} \right]. \label{dis} 
\end{equation}
L'équation (\ref{dis}) donne une valeur minimale lorsque $\vec{e}$ est un vecteur propre de la matrice $K=\vec{x}\vec{x}^{t} +TT^{t}$ avec la plus grande valeur propre. Par conséquent, l'expression explicite de la discorde quantique géométrique pour une matrice densité générale (\ref{MG}) peut être obtenue comme
\begin{equation}
{{\cal Q}_g}_{2}\left( \rho  \right) = \frac{1}{4}\left( {{{\left\| x \right\|}_2}^{2} + {{\left\| T \right\|}_2}^{2} - {K_{\max }}} \right),
\end{equation}
où $K_{\max }$ est la valeur propre maximale de la matrice $K$. Pour les matrices densités arbitraires d'un système quantique à deux qubits,
\begin{equation}
\rho  = \left( {\begin{array}{*{20}{c}}
	{{\rho _{11}}}&{{\rho _{12}}}&{{\rho _{13}}}&{{\rho _{14}}}\\
	{{\rho _{21}}}&{{\rho _{22}}}&{{\rho _{23}}}&{{\rho _{24}}}\\
	{{\rho _{31}}}&{{\rho _{32}}}&{{\rho _{33}}}&{{\rho _{34}}}\\
	{{\rho _{41}}}&{{\rho _{42}}}&{{\rho _{43}}}&{{\rho _{44}}}
	\end{array}} \right),\label{matrice2G}
\end{equation}
il est facile de voir que la matrice de corrélation $T$ s'écrit en fonctions des éléments de la matrice densité $\rho$ comme
\begin{equation}
T = \left( {\begin{array}{*{20}{c}}
	{{\rho _{14}} + {\rho _{23}} + {\rho _{32}} + {\rho _{41}}}&{i\left( {{\rho _{14}} - {\rho _{23}} + {\rho _{32}} - {\rho _{41}}} \right)}&{{\rho _{13}} - {\rho _{24}} + {\rho _{31}} - {\rho _{42}}}\\
	{i\left( {{\rho _{14}} + {\rho _{23}} - {\rho _{32}} - {\rho _{41}}} \right)}&{ - {\rho _{14}} + {\rho _{23}} + {\rho _{32}} - {\rho _{41}}}&{i\left( {{\rho _{13}} - {\rho _{24}} - {\rho _{31}} + {\rho _{42}}} \right)}\\
	{{\rho _{12}} + {\rho _{21}} - {\rho _{34}} - {\rho _{43}}}&{i\left( {{\rho _{12}} - {\rho _{21}} - {\rho _{34}} + {\rho _{43}}} \right)}&{{\rho _{11}} - {\rho_{22}} - {\rho _{33}} + {\rho _{44}}}
	\end{array}} \right),
\end{equation}
et $x = \left\{ {{\rho _{13}} + {\rho _{24}} + {\rho _{31}} + {\rho _{42}},i\left( {{\rho _{13}} + {\rho _{24}} - {\rho _{31}} + {\rho _{42}}} \right),{\rho _{11}} + {\rho _{22}} - {\rho _{33}} - {\rho _{44}}} \right\}$. En utilisant ces résultats, nous trouvons
\begin{equation}
	{\left\| x \right\|}_2 + {\left\| T \right\|}_2={\lambda _1}+{\lambda _2}+{\lambda _3},
\end{equation}
avec ${\lambda _1}$, ${\lambda _2}$ et ${\lambda _3}$ sont les valeurs propres de la matrice $K$. Cela conduit à une expression beaucoup plus simple de la discorde quantique géométrique et il devient:
\begin{equation}
{{\cal Q}_g}_{2}\left( \rho  \right) = \frac{1}{4}\min \left\{ {{\lambda _1} + {\lambda _2},{\lambda _2} + {\lambda _3},{\lambda _1} + {\lambda _3}} \right\} \label{disgeo}.
\end{equation}
Dans la théorie de l'informations quantique, l'ensemble des états $X$ intervient dans de nombreux contextes. Il inclue des états de Bell purs ainsi que des états mixtes de Werner. Il est simple de calculer l'expression analytique de la discorde quantique géométrique pour des matrice densité de type (\ref{matrixX}). Après une simplification algébrique, nous obtenons
\begin{equation}
	{{\cal Q}_g}_{2}\left( \rho_{X}\right) =\frac{1}{4}\left[ 8\left(|\rho_{14}|^{2}+|\rho_{23}|^{2} \right) +2\left(\left(\rho_{11}-\rho_{33} \right)^{2} +\left(\rho_{22}-\rho_{44} \right)^{2} \right)-\max\{k_{1}+k_{2}+k_{3}\} \right],
\end{equation}
avec
\begin{equation}
	k_{1}=4\left(|\rho_{23}|^{2}-|\rho_{14}|^{2}\right), \hspace{1cm}k_{2}=4\left(|\rho_{23}|^{2}+|\rho_{14}|^{2}\right),\hspace{1cm}k_{3}=2\left(\left(\rho_{11}-\rho_{33}\right)^{2} +\left(\rho_{22}-\rho_{44} \right)^{2} \right).
\end{equation}
La relation de la discorde quantique géométrique (\ref{disgeo}) s'écrit dans ce cas comme
\begin{equation}
{{\cal Q}_g}_{2}\left( \rho  \right) = \frac{1}{4}\min \left\{ {{k_1} + {k_2},{k_2} + {k_3},{k_1} + {k_3}} \right\}.
\end{equation}
Pour les systèmes bipartites de dimension $2\times2$, la discorde quantique géométrique est liée à la discorde quantique entropique (\ref{QDD}) et à la négativité (\ref{neg}) par les inégalités suivantes \cite{Girolami20111,Girolami20112}:

\begin{equation}
	2{{\cal Q}_g}_{2}\left( \rho  \right)\geq {\cal Q}^{2}\left( {{\rho}} \right),\hspace{1cm}{\rm et}\hspace{1cm}2{{\cal Q}_g}_{2}\left( \rho  \right)\geq {\cal N}^{2}\left(\rho \right).
\end{equation}
Malgré sa simplicité de calcul, il a été souligné récemment que la discorde quantique géométrique basée sur la norme de Hilbert-Schmidt ne peut pas être considérée comme une bonne mesure des corrélations quantiques, car elle peut augmenter sous des opérations locales agissant sur le sous-système non mesuré, et elle est également non contractive sous les applications qui préservant les traces \cite{Piani2012}. Ce fait a conduit à une redéfinition et à des études plus approfondies de la discorde quantique géométrique. L'une de ces redéfinitions est la discorde quantique géométrique basée sur la distance de trace, qui utilise la norme-1 de Schatten qui obéit à la propriété de contractivité et qui est invariante sous des transformations unitaires. En utilisant l'équation (\ref{DQGP}) pour $p=1$, la discorde quantique géométrique, également appelée la discorde quantique de trace, se réduit à \cite{Paula2013,Nakano2013,Ciccarello2014}
\begin{equation}
{{\cal Q}_g}_{T}\left( {{\rho _{AB}}} \right) = \mathop {\min }\limits_{\chi  \in {\Omega _0}} {\left\| {\left. {\rho  - \chi } \right\|}_{1} \right.}, \label{DGT}
\end{equation}
où $\|Z\|_{1}={\rm Tr}\left[ \sqrt{Z^{\dagger}Z}\right]$ désigne la norme de trace (ou norme-1 de Schatten). L'expression de la discorde quantique basée sur la distance trace à été dérivée analytiquement pour les états $X$ arbitraires à deux qubits, et jusqu'à présent, aucun résultat analytique n'a été trouvé pour des états quantiques généraux. D'après les résultats rapportés dans la référence \cite{Ciccarello2014}, l'expression explicite de la discorde quantique géométrique basée sur la distance trace, pour les états $\rho_{X}$ de type (\ref{matrixX}), s'écrit sous la forme
\begin{equation}
{{\cal Q}_g}_{T}\left( {{\rho _{X}}} \right) =\sqrt{\frac{\gamma_{1}^{2}\max\{\gamma_{3}^{2},\gamma_{2}^{2}+x_{3}^{2}\}-\gamma_{2}^{2}\min\{\gamma_{3}^{2},\gamma_{1}^{2}\} }{\max\{\gamma_{3}^{2},\gamma_{2}^{2}+x_{3}^{2}\}-\min\{\gamma_{3}^{2},\gamma_{1}^{2}+\gamma_{1}^{2}-\gamma_{2}^{2}\}}}, \label{DGTX}
\end{equation}
avec
\begin{equation}
\gamma_{1}=2\left(\rho_{32}+\rho_{41} \right), \hspace{1cm}\gamma_{2}=2\left(\rho_{32}-\rho_{41} \right),\hspace{1cm}\gamma_{3}=1-2\left(\rho_{22}+\rho_{33} \right),\hspace{1cm}x_{3}=2\left(\rho_{11}+\rho_{22} \right)-1.
\end{equation}

A titre d'exemple de ces différentes formes de discorde quantique: entropique (\ref{QDD}), géométrique via la norme de Hilbert-Schmidt (\ref{DG2}) et via la norme de trace (\ref{DGT}), on considère la famille d'états $X$ qui sont diagonaux dans la base de Bell et qui sont paramétrisés par trois paramètres $c_{i}$ ($i=1,2,3$). Les matrices densité correspondantes sont de la forme 
\begin{equation}
\rho_{BD}=\frac{1}{4}\left(\openone_{4\times 4}+\sum_{i=1}^{3}c_{i}\sigma_{i}\otimes\sigma_{i} \right),\label{BD} 
\end{equation}
où $\openone_{4\times4}$ est la matrice d'identité et $0\leq c_{i}\leq1$. La forme matricielle de ces états est donnée par
\begin{equation}
\rho_{BD}=\frac{1}{4}\left( {\begin{array}{*{20}{c}}
	1+c_{3}&0&0&c_{1}-c_{2}\\
	0&1-c_{3}&c_{1}+c_{2}&0\\
	0&c_{1}+c_{2}&1-c_{3}&0\\
	c_{1}-c_{2}&0&0&1+c_{3}\end{array}} \right),
\end{equation}
qui sont des cas particuliers des états $X$. L'état $\chi$ le plus proche est toujours un état de Bell-diagonale avec un seul paramètre non nul $c_{k}$ correspondant à $c_{k}=\max\{|c_{1}|,|c_{2}|,|c_{3}|\}$. Alors en utilisant la relation (\ref{DGTX}), la discorde géométrique basée sur la distance trace des états de Bell-diagonaux peut être obtenue explicitement comme
\begin{equation}
{{\cal Q}_g}_{T}\left( {{\rho _{BD}}} \right)={\rm int}\{|c_{1}|,|c_{2}|,|c_{3}|\},
\end{equation}
où ${\rm int}\{|c_{1}|,|c_{2}|,|c_{3}|\}$ désigne la valeur intermédiaire parmi les valeurs absolues de $c_{1}$, $c_{2}$ et $c_{3}$. Dans ce type d'états, il est également facile de déterminer l'expression analytique de la discorde quantique entropique et de la discorde géométrique de Hilbert-Schmidt à l'aide des résultats obtenus dans les sections précédentes. Un calcul simple conduit à
\begin{equation}
{\cal Q}\left( {{\rho_{BD}}} \right)=\log_{2}\left[ \frac{4\lambda_{00}^{\lambda_{00}}\lambda_{01}^{\lambda_{01}}\lambda_{10}^{\lambda_{10}}\lambda_{11}^{\lambda_{11}}}{\left(1-c_{+} \right)^\frac{\left(1-c_{+} \right)}{2}\left(1+c_{+} \right)^\frac{\left(1+c_{+} \right)}{2} }\right],
\end{equation}
avec 
\begin{equation}
	\lambda_{ij}=\frac{1}{4}\left[ 1+\left( -1\right)^{i}c_{1} -\left( -1\right)^{i+j}c_{2}+\left(-1 \right)^{j} c_{3} \right],
\end{equation}
et
\begin{equation}
	{{\cal Q}_g}_{2}\left(\rho_{BD} \right) =\frac{1}{4}\left[c_{-}^{2}+{{\cal Q}_g}_{T}^{2}\left( {{\rho _{BD}}} \right) \right], 
\end{equation}
où $c_{+}=\max\{|c_{1}|,|c_{2}|,|c_{3}|\}$ et $c_{-}=\min\{|c_{1}|,|c_{2}|,|c_{3}|\}$ représente respectivement le maximum et le minimum parmi les valeurs absolues des paramétres $c_{1}$, $c_{2}$ et $c_{3}$. 
\section{La discorde quantique linéaire}
Malgré les nombreux efforts de la communauté scientifique, une expression analytique de la discorde quantique basée sur l'entropie de von Neumann fait encore défaut, même pour les systèmes à deux qubits. Très récemment, il a été montré que la procédure d'optimisation pour obtenir la discorde quantique est possible par l'approximation de l'entropie linéaire. Ceci est fait en remplaçant l'entropie de von Neumann par l'entropie linéaire dans l'expression des corrélations classiques. Cela fournit un bon outil pour trouver l'expression analytique de la discorde quantique des états bipartites de type qudit-qubit \cite{Ma2015}. Il a été trouvé que la discorde quantique basée sur l'entropie linéaire a un lien profond avec la discorde d'origine définie par l'entropie de von Neumann dans laquelle l'écart moyen est de l'ordre de $10^{-4}$. Dans cette section, nous proposons deux méthodes analytiques fiables pour déterminer les corrélations non-classiques basées sur l'entropie linéaire. La première méthode concerne des états bipartites de type qubit-qubit et la deuxième méthode concerne des états bipartites arbitraires de type qudit-qubit. Tout d'abord, nous rappelons que l'entropie linéaire d'un état $\rho$ est donnée par 
\begin{equation}
S_{2}\left(\rho \right)=2\left[ 1-{\rm tr}\left( \rho^{2}\right) \right]. \label{EL}
\end{equation}
Pour un système à $d$ niveaux, un état qudit dans la représentation de Fano-Bloch est donné par $\rho=(\openone_{d}+\vec{r}\cdot \vec{\gamma} )/d$, où $\openone_{d}$ désigne la matrice d'identité de dimension $d\times d$, $\vec{r}$ est un vecteur réel de dimension $d^{2}-1$, et $
\vec{\gamma} =\left(\gamma _{1},\gamma _{2},\ldots ,\gamma _{d^{2}-1}\right)^{T}$ est le vecteur des générateurs de l'algèbre de Lie $SU(d)$. Alors, l'entropie linéaire d'un état qudit devient
\begin{equation}
S_{2}\left(\frac{\openone_{d}+\vec{r}\vec{\gamma}}{d} \right)=\frac{2d^{2}-2d-4|\vec{r}|^{2}}{d^{2}}.\label{mm}
\end{equation}
De même, un état qubit peut être écrit comme $\rho=(\openone_{2}+\vec{r}_{B}\cdot \vec{\sigma} )/2$, où $\vec{r}_{B}$ est un vecteur tridimensionnel et $\vec{\sigma}=\left(\sigma_{1},\sigma_{2},\sigma_{3}\right)^{T}$ désigne les opérateurs de Pauli. En termes d'entropie linéaire (\ref{EL}), nous pouvons définir la corrélation classique (\ref{QCD}) comme suit
\begin{equation}
{\cal J_{\rm 2}}_{B}\left( {{\rho _{AB}}} \right)=\mathop {\max }\limits_{{\pi _B}^j} \left( {S_{2}\left( {{\rho _A}} \right) - \sum\limits_j {{p_{B,j}}S_{2}\left( {{\rho _{^{B,j}}}} \right)} } \right), \label{QCD2}
\end{equation}
Bien que la corrélation classique (\ref{QCD}) et, par conséquent, la discorde quantique (\ref{QDD}) soit extrêmement difficile à calculer en termes d'entropie de von Neumann, la corrélation classique (\ref{QCD2}) exprimée en termes d'entropie linéaire peut être calculée analytiquement. Considérons un système quantique bipartite $\rho_{AB}$ formé de deux sous-système; le sous-système $\rho_{A}$ de dimension $d$ et le sous-système $\rho_{B}$ de dimension $2$. Pour trouver l'expression explicite de la corrélation classique en termes d'entropie linéaire, nous écrivons d'abord un état quantique bipartite ${\rho _{AB}}$ comme \cite{Osborne2006} 
\begin{equation}
{\rho _{AB}} = \Lambda_{\rho}  \otimes \openone_{B}\left( {\left| {{v_{B'B}}} \right\rangle \left\langle {{v_{B'B}}} \right|} \right), \label{rhowithchannel}
\end{equation}
où $\left| v_{B'B} \right\rangle$ est la purification symétrique à deux qubits de l'opérateur de densité réduite $B$ sur un système auxiliaire $B'$, et $\Lambda_{\rho} $ est un canal qudit (une application complètement positive et préservant la trace), qui convertit un état qubit $B'$ en l'état qudit $A$. Nous notons que la purification symétrique à deux qubits signifie que les deux matrices densités réduites sont égales, c'est-à-dire  $v_{B'}=v_{B}=\rho_{B}$. L'action d'un canal qudit $\Lambda$ sur un état qubit $\rho$ peut être écrite comme
\begin{equation}
\Lambda_{\rho} =\frac{1}{d}\left[\openone_{d}+ (\mathbf{L} \vec{r} + \vec{l})\cdot\vec{\gamma}\right],
\end{equation}
où $\mathbf{L}$ désigne une matrice réelle de dimension $3\times\left(d^{2}-1\right)$ et leur éléments sont définis par $L_{ij}=\frac{1}{d}{\rm Tr}\left[\Lambda\left( {\sigma^{i}}\right).\gamma^{j} \right]$ (avec $i=1,2,3$ et $j=1,...,d^{2}-1$), et $\vec{l}$ est un vecteur tridimensionnel. En utilisant la décomposition spectrale de l'opérateur de densité réduite $\rho_{B}$, c-à-d $\rho_{B}=\sum_{i=0,1}\lambda_{i}\left| \varphi_{i}\right\rangle\left\langle\varphi_{i} \right|$, il est simple de vérifier que $\left| v_{B'B} \right\rangle=\sum_{i=0,1}\sqrt{\lambda_{i}}\left| \varphi_{i}\right\rangle\left| \varphi_{i}\right\rangle$. Dans cette base de Pauli, les décompositions possibles de $\rho_{B}$ en états purs sont représentées par tous les ensembles possibles de probabilités $\{p_{i}\}$ et de vecteurs unitaires $\{r_{i}\}$ pour lesquels $r_{B}=\sum_{i}p_{i}r_{i}$. L'action d'un canal qudit $\Lambda$ sur l'état de sous-système $B$ devient
\begin{equation}
\rho_{B}=\frac{\openone_{2}+\vec{r}_{B}\vec{\sigma}}{2},\hspace{1cm}   \xrightarrow[]{\Lambda:{\rm qubit}\longmapsto{\rm qudit}}\hspace{1cm} \Lambda\left(\rho_{B} \right)=\frac{\openone_{d}+\left(\mathbf{L} \vec{r}_{B}+\vec{l} \right)\gamma }{d}, 
\end{equation}
et en utilisant (\ref{mm}), l'entropie linéaire correspondante s'écrit
\begin{equation}
S_{2}\left( \Lambda\left(\rho_{B} \right)\right) =\frac{1}{d^{2}}\left(2d^{2}-2d-4\left(\mathbf{L} \vec{r}_{B}+\vec{l} \right)^{T}\left(\mathbf{L} \vec{r}_{B}+\vec{l} \right) \right).
\end{equation}
Les entropies linéaires de l'action d'un canal qudit sur les états purs  $\left| \varphi_{i}\right\rangle$ ($i=1,2$) s'obtiennent à partir de (\ref{mm}) et sont
\begin{equation}
S_{2}\left(\Lambda\left(\left| \varphi_{i}\right\rangle\left\langle \varphi_{i}\right|  \right)  \right)=\frac{1}{d^{2}}\left( 2d^{2}-2d-4\left(\mathbf{L} \vec{r}_{i}+\vec{l} \right)^{T}\left(\mathbf{L} \vec{r}_{i}+\vec{l} \right)\right).
\end{equation}
En utilisant ces résultats, il est simple de voir que la matrice densité réduite de l'état du qudit $\rho_{A}$ est exactement égale à l'action d'un canal qudit sur la matrice densité réduite $\rho_{B}$, cela signifie que $\rho_{A}={\Lambda \left( \rho _B \right)}$;
\begin{align}
\rho_{A}&={\rm Tr}_{B}\left[\sum_{i,j}\sqrt{\lambda_{i}\lambda_{j}}\Lambda\left(\left| \varphi_{i}\right\rangle\left\langle \varphi_{j}\right| \right)\otimes\left(\left| \varphi_{i}\right\rangle\left\langle \varphi_{j}\right| \right)  \right]\notag\\&=\left\langle\varphi_{k}\right|\left[\sum_{i,j}\sqrt{\lambda_{i}\lambda_{j}}\Lambda\left(\left| \varphi_{i}\right\rangle\left\langle \varphi_{j}\right| \right)\otimes\left(\left| \varphi_{i}\right\rangle\left\langle \varphi_{j}\right| \right)  \right]\left| \varphi_{k}\right\rangle \notag\\&= \sum_{i,j}\sqrt{\lambda_{i}\lambda_{j}}\Lambda\left(\left| \varphi_{i}\right\rangle\left\langle \varphi_{j}\right| \right)
\underbrace {\left\langle \varphi_{k}\right| \left| \varphi_{i}\right\rangle}_{\delta_{ki}}\underbrace {\left\langle \varphi_{j}\right| \left| \varphi_{k}\right\rangle}_{\delta_{jk}} \notag\\&=\sum_{i}\lambda_{i} \Lambda\left(\left| \varphi_{i}\right\rangle\left\langle \varphi_{i}\right| \right)=\Lambda \left( \rho _B \right). 
\end{align} 
Ainsi, la corrélation classique quantifiée par l'entropie linéaire coïncide avec la capacité linéaire Holevo pour le canal qubit $\chi_{2}\left(\rho_{B}, \Lambda\right)$ (voir l'équation ($4$) de la référence \cite{Osborne2006}). Nous obtenons donc
\begin{equation}
{\cal J_{\rm 2}}_{B}\left( {{\rho _{AB}}} \right) = \mathop {\max }\limits_{\left\{ {{p_i},{\varphi_i}} \right\}} \left[ {{S_2}\left( {\Lambda \left( {{\rho _B}} \right)} \right) - \sum\limits_i {{p_i}} {S_2}\left( {\Lambda \left( {\left| {{\varphi _i}} \right\rangle \left\langle {{\varphi _i}} \right|} \right)} \right)} \right],
\end{equation}
où le maximum est pris sur toutes les décompositions spectrales possibles de la matrice densité réduite $\rho_{B}$. En remplaçant les entropies linéaires par leurs expressions, on obtient alors
\begin{equation}
{\cal J_{\rm 2}}_{B}\left( {{\rho _{AB}}} \right) =\frac{4}{d^{2}}\mathop {\max }\limits_{\left\{ {{p_i},{\varphi _i}} \right\}}\left[p_{1}\Arrowvert\mathbf{L} \vec{r}_{1}+\vec{l}\Arrowvert^{2}+p_{2}\Arrowvert\mathbf{L} \vec{r}_{2}+\vec{l}\Arrowvert^{2}-\Arrowvert\mathbf{L} \vec{r}_{B}+\vec{l}\Arrowvert^{2} \right].\label{278} 
\end{equation}
En substituant $r_{i}=r_{B}+X_{i}$, on peut facilement vérifier que le calcul de ${\cal J_{\rm 2}}_{B}\left( {{\rho _{AB}}} \right)$ se réduit à déterminer le couple $\{p_{i},X_{i}\}$ sous les conditions $\sum_{i}p_{i}X_{i}=0$ et $\Arrowvert r_{B}+X_{i}\Arrowvert=1$. Dans cette image, l'équation (\ref{278}) devient
\begin{equation}
{\cal J_{\rm 2}}_{B}\left( {{\rho _{AB}}} \right) =\frac{4}{d^{2}}\mathop {\max }\limits_{\left\{ {{p_i},{X_i}} \right\}}\left[ \sum_{i}p_{i}X_{i}^{T}\mathbf{L}^{T}\mathbf{L}X_{i}\right]. \label{279}
\end{equation}
Supposons, sans perte de généralité, que la matrice $\mathbf{L}^{T}\mathbf{L}$ est diagonale avec des éléments diagonaux $\lambda^{x}\geqslant\lambda^{y}\geqslant\lambda^{z}$, alors $\mathbf{L}^{T}\mathbf{L}={\rm diag}\{\lambda^{x},\lambda^{y},\lambda^{z}\}$ et les contraintes $\Arrowvert r_{B}+X_{i}\Arrowvert=1$ conduisent à
\begin{equation}
\left( X_{i}^{x}\right)^{2} =1-\Arrowvert r_{B}\Arrowvert^{2}-2r_{B}^{T}X_{i}-\left( X_{i}^{y}\right)^{2}-\left( X_{i}^{z}\right)^{2}.
\end{equation}
En remplaçant ceci dans l'équation (\ref{279}), nous obtenons
\begin{equation}
{\cal J_{\rm 2}}_{B}\left( {{\rho _{AB}}} \right) =\frac{4}{d^{2}}\left[\lambda^{x}\left(1-\Arrowvert r_{B}\Arrowvert^{2} \right)+\mathop{\max }\limits_{\left\{ {{p_i},{X_i}} \right\}}\sum_{i}p_{i} \left(\left(\lambda^{y}-\lambda^{x} \right)\left(X_{i}^{y} \right)^{2}+\left(\lambda^{z}-\lambda^{x} \right)\left(X_{i}^{z} \right)^{2}\right)\right]. 
\end{equation}
La seconde partie de cette expression est inférieure ou égale à zaro, car $\lambda^{x}\geqslant\lambda^{y}\geqslant\lambda^{z}$. Elle est évidemment maximale si l'on choisit $X_{i}^{y}=X_{i}^{z}=0$ pour chaque $i$. Sachant que $S_{2}\left(\rho_{B} \right)=1-\Arrowvert r_{B}\Arrowvert^{2}$, l'expression analytique de la corrélation classique pour des états quantiques arbitraires de type qudit-qubit se récrit \cite{Ma2015}
\begin{equation}
{\cal J_{\rm 2}}_{B}\left( {{\rho _{AB}}} \right)=\frac{4}{d^{2}}\lambda_{\max
}\left( \mathbf{L}^{T}\mathbf{L}\right) S_{2}(\rho_{B}),\label{CC2}
\end{equation}
où $\lambda_{\max}\left( \mathbf{L}^{T}\mathbf{L}\right)$ représente la plus grande valeur propre de la matrice $\mathbf{L}^{T}\mathbf{L}$. Il est clair que le calcul de la corrélation classique en termes de l'entropie linéaire nécessite le calcul des éléments de la matrice $\mathbf{L}^{T}\mathbf{L}$. La question qui se pose alors est de savoir comment calculer les éléments de la matrice $\mathbf{L}$?

\subsection{Expression analytique de la discorde quantique linéaire pour les états qubit-qubit arbitraire}
Dans cette partie, nous présenterons une méthode efficace pour calculer les éléments de la matrice $\mathbf{L}$ pour des systèmes quantiques arbitraires de type qubit-qubit \cite{ShaukatSlaoui2020}. Dans la base de calcul $\left\{ {\left| {00} \right\rangle ,\left| {01} \right\rangle ,\left| {10} \right\rangle ,\left| {11} \right\rangle } \right\}$, la matrice densité des états à deux qubit (\ref{matrice2G}) peut être écrite comme
\begin{equation}
	{\rho _{AB}} = {\beta _{00}} \otimes \left| 0 \right\rangle \left\langle 0 \right| + {\beta _{01}} \otimes \left| 0 \right\rangle \left\langle 1 \right| + {\beta _{10}} \otimes \left| 1 \right\rangle \left\langle 0 \right| + {\beta _{11}} \otimes \left| 1 \right\rangle \left\langle 1 \right|,   \label{beta11}
\end{equation}
avec
\begin{eqnarray}
	{\beta _{00}} = \left( {\begin{array}{*{20}{c}}
			{{\rho _{11}}}&{{\rho _{13}}}\\
			{{\rho _{31}}}&{{\rho _{33}}}
	\end{array}} \right), \hspace{1cm}  {\beta _{01}} = \left( {\begin{array}{*{20}{c}}
			{{\rho _{21}}}&{{\rho _{23}}}\\
			{{\rho _{41}}}&{{\rho _{43}}}
	\end{array}} \right), \nonumber \\
	{\beta _{10}} = \left( {\begin{array}{*{20}{c}}
			{{\rho _{12}}}&{{\rho _{14}}}\\
			{{\rho _{32}}}&{{\rho _{34}}}
	\end{array}} \right), \hspace{1cm} {\beta _{11}} = \left( {\begin{array}{*{20}{c}}
			{{\rho _{22}}}&{{\rho _{24}}}\\
			{{\rho _{42}}}&{{\rho _{44}}}
	\end{array}} \right).
\end{eqnarray}
Les états purs de l'état $\rho_{B}$ peuvent être recréés dans cette base comme $\left| {{\varphi _0}} \right\rangle  = {\left( {{a_0},{a_1}} \right)^T}$ et $\left| {{\varphi _1}} \right\rangle  = {\left( {{b_0},{b_1}} \right)^T}$, où les $a_{i}$ et $b_{i}$ sont des paramètres réels et $T$ désigne la transposition matricielle usuelle. La purification symétrique à deux qubits de l'opérateur de densité réduite $\rho _B$ sur un système de qubit auxiliaire $B'$ conduit à l'état  $\left| {{r_{B'B}}} \right\rangle$ et nous avons
\begin{equation}
	\left| {{r_{B'B}}} \right\rangle \left\langle {{r_{B'B}}} \right| = \left( {\begin{array}{*{20}{c}}
			{A\bar A}&{AB}&{A\bar B}&{AD}\\
			{B\bar A}&{B\bar B}&{B\bar B}&{B\bar D}\\
			{B\bar A}&{B\bar B}&{B\bar B}&{B\bar D}\\
			{D\bar A}&{D\bar B}&{D\bar B}&{D\bar D}
	\end{array}} \right),
	\label{vector}
\end{equation}
où
\begin{equation}
	\left\{ \begin{array}{l}
		A = \sqrt {{\lambda _1}} {a_0}^2 + \sqrt {{\lambda _2}} {b_0}^2,\\
		B = \sqrt {{\lambda _1}} {a_0}{a_1} + \sqrt {{\lambda _2}} {b_0}{b_1},\\
		D = \sqrt {{\lambda _1}} {a_1}^2 + \sqrt {{\lambda _2}} {b_1}^2.
	\end{array} \right.
\end{equation}
Par conséquent, les éléments de l'équation (\ref{beta11})  peuvent être exprimés à l'aide des éléments de l'équation (\ref{vector}) comme suit
\begin{eqnarray}
	{\beta _{00}} &=& A\bar A\Lambda \left( {\left| 0 \right\rangle \left\langle 0 \right|} \right) + A\bar B\Lambda \left( {\left| 0 \right\rangle \left\langle 1 \right|} \right)
	+B\bar A\Lambda \left( {\left| 1 \right\rangle \left\langle 0 \right|} \right) + B\bar B\Lambda \left( {\left| 1 \right\rangle \left\langle 1 \right|} \right), \nonumber \\
	{\beta _{01}} &=& B\bar A \Lambda\left( {\left| 0 \right\rangle \left\langle 0 \right|} \right) + B\bar B \Lambda \left( {\left| 0 \right\rangle \left\langle 1 \right|} \right) 
	+ D\bar A \Lambda \left( {\left| 1 \right\rangle \left\langle 0 \right|} \right) + D\bar B \Lambda \left( {\left| 1 \right\rangle \left\langle 1 \right|} \right), \nonumber \\
	{\beta _{10}} &= &A\bar B\Lambda \left( {\left| 0 \right\rangle \left\langle 0 \right|} \right) + A\bar D\Lambda \left( {\left| 0 \right\rangle \left\langle 1 \right|} \right) 
	+ B\bar B\Lambda \left( {\left| 1 \right\rangle \left\langle 0 \right|} \right) + B\bar D\Lambda \left( {\left| 1 \right\rangle \left\langle 1 \right|} \right),\nonumber \\
	{\beta _{11}} &= &B\bar B\Lambda \left( {\left| 0 \right\rangle \left\langle 0 \right|} \right) + B\bar D\Lambda \left( {\left| 0 \right\rangle \left\langle 1 \right|} \right) + D\bar B\Lambda \left( {\left| 1 \right\rangle \left\langle 0 \right|} \right) + D\bar D\Lambda \left( {\left| 1 \right\rangle \left\langle 1 \right|} \right).\label{beta}
\end{eqnarray}
Les éléments de la matrice $\mathbf{L}$ peuvent être déterminés en résolvant l'équation (\ref{beta}) pour obtenir $\Lambda \left( {\left| i \right\rangle \left\langle j \right|} \right)$ à partir de laquelle on obtient les éléments ${L_{ij}} = {{{\rm tr}\left[ {\Lambda \left( {{\sigma _j}} \right){\sigma _i}} \right]} \mathord{\left/
		{\vphantom {{tr\left[ {\Lambda \left( {{\sigma _j}} \right){\sigma _i}} \right]} 2}} \right.
		\kern-\nulldelimiterspace} 2}$ en termes des éléments de la matrice densité $\rho$. A titre d'exemple, on prend le cas de l'état de Bell diagonale (\ref{BD}) comme dans les mesures précédentes. Dans ce cas précis, nous trouvons
\begin{equation}
{\beta _{00}} =\frac{1}{4} \left( {\begin{array}{*{20}{c}}
		{1+c_{3}}&{0}\\
		{0}&{1-c_{3}}
\end{array}} \right),\hspace{1cm}{\beta _{01}} =\frac{1}{4} \left( {\begin{array}{*{20}{c}}
{0}&{c_{1}+c_{2}}\\
{c_{1}-c_{2}}&{0}
\end{array}} \right),
\end{equation}
\begin{equation}
	{\beta _{10}} =\frac{1}{4} \left( {\begin{array}{*{20}{c}}
			{0}&{c_{1}-c_{2}}\\
			{c_{1}+c_{2}}&{0}
	\end{array}} \right),\hspace{1cm} {\beta _{11}} =\frac{1}{4} \left( {\begin{array}{*{20}{c}}
	{1-c_{3}}&{0}\\
	{0}&{1+c_{3}}
\end{array}} \right),
\end{equation}
et les quatre solutions de système différentiel (\ref{beta}) nous donnent
\begin{equation}
\Lambda \left( {\left| 0 \right\rangle \left\langle 0 \right|} \right)=2{\beta _{00}}, \hspace{0.7cm}\Lambda \left( {\left| 0 \right\rangle \left\langle 1 \right|} \right)=2{\beta _{01}}, \hspace{0.7cm}\Lambda \left( {\left| 1 \right\rangle \left\langle 0 \right|} \right)=2{\beta _{10}}, \hspace{0.7cm}\Lambda \left( {\left| 1 \right\rangle \left\langle 1\right|} \right)=2{\beta _{11}}.
\end{equation}
Cela signifie que $\mathbf{L}^{T}\mathbf{L}=4{\rm Diag}\{c_{1}^{2},c_{2}^{2},c_{3}^{2}\}$. Nous avons aussi $S_{2}\left( \rho_{B}\right)=1$, alors la formule analytique de corrélation classique en termes de l'entropie linéaire pour un état de Bell diagonale (\ref{BD}) est
\begin{equation}
{\cal J_{\rm 2}}_{B}\left( {{\rho _{BD}}} \right)=\max\{c_{1}^{2},c_{2}^{2},c_{3}^{2}\}.
\end{equation}
Comme deuxième exemple, considérons maintenant les états à deux qubits suivants
\begin{align}
\rho_{x}&=\frac{2-x}{6}\left| 00\right\rangle \left\langle 00\right|+\frac{1+x}{6}\left| 01\right\rangle \left\langle 01\right| +\frac{1}{6}\left| 01\right\rangle \left\langle 10\right|\notag\\&++\frac{1}{6}\left| 10\right\rangle \left\langle 01\right|+\frac{1+x}{6}\left| 10\right\rangle \left\langle 10\right|+\frac{2-x}{6}\left| 11\right\rangle \left\langle 11\right|,
\end{align}
qui sont paramétrés par le paramètre $x$ où $x\in\left[ 0,2\right]$. Après un simple calcul, nous avons $S_{2}\left(\rho_{B} \right)=1$ et l'action de canal qubit $\Lambda\left( \left| i\right\rangle \left\langle j\right|\right)$ est donné par
\begin{equation}
\Lambda\left( \left| 0\right\rangle \left\langle 0\right|\right)=\frac{2-x}{3}\left| 0\right\rangle \left\langle 0\right|+\frac{1+x}{3}\left| 1\right\rangle \left\langle 1\right|,
\end{equation}
\begin{equation}
\Lambda\left( \left| 0\right\rangle \left\langle 1\right|\right)=\frac{1}{3}\left| 1\right\rangle \left\langle 0\right|,\hspace{1cm}\Lambda\left( \left| 1\right\rangle \left\langle 0\right|\right)=\frac{1}{3}\left| 0\right\rangle \left\langle 1\right|,
\end{equation}
\begin{equation}
	\Lambda\left( \left| 1\right\rangle \left\langle 1\right|\right)=\frac{1+x}{3}\left| 0\right\rangle \left\langle 0\right|+\frac{2-x}{3}\left| 1\right\rangle \left\langle 1\right|.
\end{equation}
Nous obtenons alors
\begin{equation}
\mathbf{L}=\left( {\begin{array}{*{20}{c}}
		{\frac{1}{3}}&{0}&{0}\\
		{0}&{-\frac{1}{3}}&{0}\\
		{0}&{0}&{\frac{1-2x}{3}}
\end{array}} \right).
\end{equation}
Par conséquent, la corrélation classique sous l'entropie linéaire de l'état $\rho _{x}$ devient
\begin{equation}
{\cal J_{\rm 2}}_{B}\left( {{\rho _{x}}} \right)=\max_{x\in\left[ 0,2\right]}\left\lbrace\frac{1}{9},\frac{\left( 1-2x\right)^{2}}{9}\right\rbrace.
\end{equation}
D'autre part, en utilisant ces résultats, il est possible de dériver une autre formule analytique de la discorde quantique basée sur l'entropie de von Neumann pour les états à deux qubits de rang-$2$ autre que ce que nous discutons dans la section précédente. Pour obtenir cette formule, nous devons étudier les relations entre l'intrication de formation $E_{f}\left(\rho\right)$ (\ref{EF}), la concurrence ${\cal C}\left(\rho\right)$ (\ref{Concu}), le tangle $\tau_{g}\left(\rho\right)$ (\ref{tangel3}), la corrélation classique basée sur l'entropie linéaire ${\cal J_{\rm 2}}_{B}\left( \rho \right)$ (\ref{QCD2}), et la corrélation classique basée sur l'entropie de von Neumann ${\cal J}_{B}\left( \rho \right)$ (\ref{QCD}). Tout d'abord, le tangle d'un état bipartite qubit-qudit, $\rho_{AB}=\sum_{i}p_{i}\left| \psi_{i}\right\rangle\left\langle\psi_{i}\right|$, est défini comme le minimum sur les décompositions spectrales \cite{Osborne2005}
\begin{align}
\tau_{g}\left(\rho_{AB}\right)&=\min_{\{p_{i},\left| \psi_{i}\right\rangle\}}\sum_{i}p_{i} \left| {\left\langle \psi_{i}  \right|\left| {\tilde \psi_{i} } \right\rangle } \right|^{2}=\min_{\{p_{i},\left| \psi_{i}\right\rangle\}}\sum_{i}p_{i}{\cal C}^{2}\left( {\left| \psi_{i}  \right\rangle } \right),
\end{align}
qui est étroitement liée à la concurrence ${\cal C}\left( {\left| \psi_{i}  \right\rangle } \right)$. De plus, en raison de la convexité de ${\cal C}^{2}\left( {\left| \psi_{i}  \right\rangle } \right)=\left| {\left\langle \psi_{i}  \right|\left| {\tilde \psi_{i} } \right\rangle } \right|^{2}$, la concurrence au carré de l'état $\rho_{AB}$ satisfait l'inégalité  ${\cal C}^{2}\left(\rho_{AB}\right) \leqslant \tau_{g}\left(\rho_{AB}\right)$. Notez que le tangle n'est pas toujours égal au carré de la concurrence sauf que pour les états à deux qubit \cite{Osborne2005}. L'intrication de formation $E_{f}\left(\left| \psi\right\rangle_{AB} \right)$ et la concurrence ${\cal C}\left(\left| \psi\right\rangle_{AB}\right)$ d'un état pur $\left| \psi\right\rangle_{AB}$ sont définies par $E_{f}\left(\left| \psi\right\rangle_{AB}\right)=S\left(\rho_{B} \right)$  et ${\cal C}\left(\left| \psi\right\rangle_{AB}\right)=\sqrt{2\left[1-{\rm Tr} \left(\rho_{B}^{2} \right) \right] }$ , respectivement. Il est donc clair que
\begin{align}
	\tau_{g}\left(\rho_{AB}\right)&=\min_{\{p_{i},\left| \psi_{i}\right\rangle\}}\sum_{i}p_{i}S_{2}\left(\rho_{B}^{i} \right), 
\end{align}
où $\rho_{B}^{i}={\rm Tr}_{A}\left(\left| \psi_{i}\right\rangle\left\langle \psi_{i}\right|\right)$. Nous aurons également besoin de la relation entre l'intrication de formation et la concurrence, et comme nous l'avons vu dans le premier chapitre, l'intrication de formation est liée à la concurrence pour les états mixtes à deux qubit par cette relation 
\begin{equation}
	E_{f}\left(\rho_{AB}\right)=g\left({\cal C}^{2}\left(\rho_{AB}\right)\right), \hspace{1cm}{\rm avec}\hspace{1cm}g\left( x\right) =h\left(\frac{1+\sqrt{1-x}}{2} \right), 
\end{equation}
où $h\left(x \right)$ est l'entropie binaire. Considérons maintenant les états quantiques à deux qubits $\rho_{AB}$ de rang-$2$ et qui ont des décompositions spectrales $\rho_{AB}=\varsigma_{0}\left| \psi_{0}\right\rangle \left\langle\psi_{0}\right|+\varsigma_{1}\left| \psi_{1}\right\rangle \left\langle\psi_{1}\right|$, avec $\varsigma_{j}$ et $\left| \psi_{j}\right\rangle$, $j=0,1$, $\sum_{j}\varsigma_{j}=1$, sont respectivement les valeurs propres et les vecteurs propres de la matrice densité $\rho_{AB}$. En attachant un troisième qubit $E$, l'état $\rho_{AB}$ est purifié pour être $\left| \psi\right\rangle_{ABE}=\sqrt{\varsigma_{0}}\left| \psi_{0}\right\rangle\left| 0\right\rangle+\sqrt{\varsigma_{1}}\left| \psi_{1}\right\rangle\left| 1\right\rangle$ de sorte que $\rho_{AB}={\rm Tr}_{E}\left(\left| \psi\right\rangle_{ABE}\left\langle\psi\right|\right)$. Dans ce cas, nous avons $E_{f}\left(\rho_{BE}\right)=g\left({\cal C}^{2}\left(\rho_{BE}\right)\right)$ et à partir de la formule de Koashi-Winter, on obtient la discorde quantique basée sur l'entropie de von Neumann donnée par l'équation (\ref{QDrang2}). La relation de Koashi-Winter peut également être réécrite en terme de l'entropie linéaire comme \cite{Osborne2006}
\begin{equation}
	\tau_{g}\left(\rho_{AE}\right)+{\cal J_{\rm 2}}_{B}\left( \rho_{AB} \right)=S_{2}\left( \rho_{A}\right).
\end{equation}
Comme $\rho_{AE}$ est un état à deux qubits, nous avons $	\tau_{g}\left(\rho_{AE}\right)={\cal C}^{2}\left(\rho_{AE}\right)$. De plus, il est facile de voir que
\begin{equation}
S\left(\rho_{A} \right)=E_{f} \left(\left| \psi\right\rangle_{ABE} \right)=g\left( {\cal C}^{2}\left( \left| \psi\right\rangle_{ABE}\right) \right)=g\left( S_{2}\left( \rho_{A}\right)\right),  
\end{equation}
et
\begin{equation}
E_{f}\left(\rho_{AE}\right)=g\left( {\cal C}^{2}\left(\rho_{AE}\right)\right)=g\left(S_{2}\left( \rho_{A}\right)- {\cal J_{\rm 2}}_{B}\left( \rho_{AB} \right)\right).  
\end{equation}
En remplaçant ces deux expressions dans l'équation (\ref{RKW2}), on trouve
\begin{equation}
{\cal J}_{B}\left( \rho_{AB} \right)=S\left( \rho_{A}\right)-g\left(S_{2}\left( \rho_{A}\right)- {\cal J_{\rm 2}}_{B}\left( \rho_{AB} \right)\right).
\end{equation}
Par conséquent, la discorde quantique basée sur l'entropie de von Neumann est donnée par
\begin{equation}
{\cal Q}\left( {{\rho _{AB}}} \right) = S\left( {{\rho_{B}}} \right)- S\left( {{\rho_{AB}}} \right)+g\left(S_{2}\left( \rho_{A}\right)- {\cal J_{\rm 2}}_{B}\left( \rho_{AB} \right)\right).\label{QDDD}
\end{equation}
Cette dernière expression (\ref{QDDD}) fournit une formule analytique de la discorde quantique en termes d'entropie de von Neumann originale pour les états quantiques arbitraires de rang-$2$ à deux qubits. Il est très important de souligner ici que, contrairement à l'expression (\ref{QDrang2}), cette formule s'applique directement sur l'état de système $\rho_{AB}$. Nous n'avons donc pas besoin de purifier le système étudié. Prenons par exemple l'état de Horodecki \cite{Horst2013}, qui est un état de rang-$2$ et qui s'écrit comme un mélange d'un état de Bell, disons $\left|\varphi^{+} \right\rangle=\left(\left|01\right\rangle+\left|10\right\rangle \right)/\sqrt{2} $, et d'un état de vide $\left|00\right\rangle$, c'est-à-dire
\begin{equation}
\rho^{H}\left(p\right)=p\left|\varphi^{+}\right\rangle\left\langle\varphi^{+}\right|+\left(1+p \right)\left|00\right\rangle\left\langle00\right|,
\end{equation}
où $p\in\left[0,1\right]$. En utilisant la méthode ci-dessus, le canal qubit $\Lambda$ peut être explicitement calculé et nous obtenons
\begin{equation*}
\Lambda\left(\left|0\right\rangle\left\langle0\right|\right)=\frac{2\left(1-p \right) }{2-p}\left|0\right\rangle\left\langle0\right|+ \frac{p }{2-p}\left|1\right\rangle\left\langle1\right|,
\end{equation*}
\begin{equation}
\Lambda\left(\left|1\right\rangle\left\langle0\right|\right)=\sqrt{\frac{p}{2-p}}\left|1\right\rangle\left\langle0\right|,\hspace{0.5cm}\Lambda\left(\left|0\right\rangle\left\langle1\right|\right)=\sqrt{\frac{p}{2-p}}\left|0\right\rangle\left\langle1\right|,\hspace{0.5cm}\Lambda\left(\left|1\right\rangle\left\langle1\right|\right)=\left|0\right\rangle\left\langle0\right|.
\end{equation}
Alors la matrice $\mathbf{L}$ devient
\begin{equation}
	\mathbf{L}=\left( {\begin{array}{*{20}{c}}
			{\sqrt{\frac{p}{2-p}}}&{0}&{0}\\
			{0}&{-\sqrt{\frac{p}{2-p}}}&{0}\\
			{0}&{0}&{-\frac{p}{2-p}}
	\end{array}} \right).
\end{equation}
Il est simple de vérifier que $S_{2}\left(\rho^{H}\left(p\right)_{B}\right)=S_{2}\left(\rho^{H}\left(p\right)_{A}\right)=p\left(2-p \right)$ et $S\left(\rho^{H}\left(p\right)\right)=h\left( p\right)$. Ainsi, la discorde quantique basée sur l'entropie de von Neumann de $\rho^{H}\left(p\right)$ est donnée par
\begin{equation}
{\cal Q}\left( \rho^{H}\left(p\right) \right)=h\left(\frac{p}{2} \right)-h\left(p\right)+g\left(2p\left(1-p\right)\right).   
\end{equation}

\subsection{Expression analytique de la discorde quantique linéaire pour les états qubit-qudit arbitraire}
Nous proposons maintenant une méthode analytique fiable pour évaluer la corrélation classique basée sur l'entropie linéaire pour des états quantiques arbitraires de type qudit-qubit. Cette méthode se concentre principalement sur l'écriture à la fois de la matrice densité $\rho$ et de la matrice $\mathbf{L}$ dans la représentation de Fano-Bloch et sur le calcul de toutes les composantes de ces tenseurs de corrélation totale \cite{FadwaSlaoui2020}. Pour ce faire, nous écrivons $\rho_{AB}$ dans la représentation de Fano-Bloch comme
\begin{equation}
	{\rho _{AB}} = \frac{1}{{2d}}\sum\limits_{\alpha  = 0}^{{d^2} - 1} {\sum\limits_{\beta  = 0}^3 {{R_{\alpha \beta }}{\gamma ^\alpha } \otimes {\sigma ^\beta }} }, \label{rhoAB}
\end{equation}
où les coefficients ${R_{\alpha \beta }} = \frac{d}{2}{\rm Tr}\left( {{\rho _{AB}}{\gamma ^\alpha } \otimes {\sigma ^\beta }} \right)$ sont les composantes du tenseur de corrélation total d'un état qudit-qubit $\rho_{AB}$, avec $\gamma^{0}$ et $\sigma^{0}$ sont les matrices d'identités de dimensions respectives $d$ et $2$. De même, l'état ${\rho _{B'B}}$ peut être écrit comme
\begin{equation}
	{\rho _{B'B}} = \left| {{v_{B'B}}} \right\rangle \left\langle {{v_{B'B}}} \right| = \frac{1}{{4}}\sum\limits_{\alpha  = 0}^{3} {\sum\limits_{\beta  = 0}^3 {{{\cal R}_{\alpha \beta }}{\sigma ^{\alpha} } \otimes {\sigma ^\beta }} }, \label{ghobb}
\end{equation}
avec ${{\cal R}_{\alpha \beta }} = {\rm Tr}\left( {{\rho _{B'B}}{\sigma ^{\alpha} } \otimes {\sigma ^\beta }} \right)$. En utilisant l'équation (\ref{rhoAB}), l'opérateur densité réduite ${\rho _B}$ peut être réécrit, en termes de paramètres de Fano-Bloch, comme
\begin{align}
	{\rho _B} ={\rm Tr_{A}}\left( {{\rho _{AB}}} \right) =\frac{1}{{2d}}\sum\limits_{\alpha  = 0}^{{d^2} - 1} {\sum\limits_{\beta  = 0}^3 {{R_{\alpha \beta }}{\rm Tr}_{A}\left( {\gamma^\alpha}\right) \otimes {\sigma ^\beta }} }= \frac{1}{2}\sum\limits_{\beta  = 0}^3 {{R_{0\beta }}{\sigma ^\beta }}.\label{ghoB}
\end{align}
En outre, le canal qudit $\Lambda \left({\sigma ^{\alpha} }\right)$ peut également être reformulé par une fonction très utile qui s'écrit comme
\begin{equation}
	\Lambda \left({\sigma ^{\alpha}}\right) = \sum\limits_{\alpha  = 0}^3 {\sum\limits_{\delta  = 0}^{{d^2} - 1} {{{\cal L}_{\alpha \delta }}{\gamma ^\delta }}},
\end{equation}
avec ${\cal L}_{\alpha \delta}={Tr}\left(\Lambda\left({\sigma^{\alpha}}\right)\gamma^{\beta}\right)$. Nous obtenons alors
\begin{align}
	\left( {\Lambda  \otimes \openone} \right){\rho _{B'B}} &=\frac{1}{{4}}\sum\limits_{\alpha  = 0}^{3} {\sum\limits_{\beta  = 0}^3} {\cal R}_{\alpha \beta }\Lambda \left({\sigma ^{\alpha}}\right)\otimes\sigma^{\beta}\notag\\&=\frac{1}{{4}}\sum\limits_{\alpha  = 0}^{3} {\sum\limits_{\beta  = 0}^3} {\cal R}_{\alpha \beta }\sum_{\delta=0}^{d^{2}-1}{\cal L}_{\alpha \delta}\gamma^{\delta}\otimes\sigma^{\beta}\notag\\&=\frac{1}{{4}} {\sum\limits_{\beta  = 0}^3}\sum_{\delta=0}^{d^{2}-1}\left(\sum\limits_{\alpha  = 0}^{3}{\cal R}_{\alpha \beta } {\cal L}_{\alpha \delta}\right)\gamma^{\delta}\otimes\sigma^{\beta}\notag\\&=\frac{1}{{4}} {\sum\limits_{\beta  = 0}^3}\sum_{\delta=0}^{d^{2}-1}\left(\sum\limits_{\alpha  = 0}^{3}\left( {\cal R}^{T}\right)_{\beta\alpha} {\cal L}_{\alpha \delta}\right)\gamma^{\delta}\otimes\sigma^{\beta}\notag\\&= \frac{1}{4}\sum\limits_{\beta  = 0}^3 {\sum\limits_{\delta  = 0}^{{d^2} - 1} {{{\left( {{{\cal R}^T}{\cal L}} \right)}_{\beta \delta }}{\gamma ^\delta }}  \otimes {\sigma ^\beta }}.
\end{align}
En utilisant la définition (\ref{rhowithchannel}), il est facile de vérifier que
\begin{equation}
	\frac{1}{{2d}}\sum\limits_{\alpha  = 0}^{{d^2} - 1} {\sum\limits_{\beta  = 0}^3 {{R_{\alpha \beta }}} }  = \frac{1}{{2d}}\sum\limits_{\alpha  = 0}^{{d^2} - 1} {\sum\limits_{\beta  = 0}^3 {{{\left( {{{\cal R}^T}{\cal L}} \right)}_{\beta \alpha }}} }  = \frac{1}{4}\sum\limits_{\alpha  = 0}^{{d^2} - 1} {\sum\limits_{\beta  = 0}^3 {{{\left( {{{\cal L}^T}{\cal R}} \right)}_{\alpha \beta }}} }.
\end{equation}
Cela mène à l'écriture de la matrice ${\cal L}$ comme
\begin{equation}
	{\cal L}= \frac{2}{d}{\left( {{{\cal R}^T}} \right)^{ - 1}}{R^T}, \label{matrixL}
\end{equation}
où $R$, ${\cal R}$ et ${\cal L}$ sont les matrices dont les éléments sont $R_{\alpha \beta }$, ${\cal R}_{\alpha \beta }$ et ${\cal L}_{\alpha \beta }$, respectivement. Alors, la matrice $\mathbf{L}$ utilisée pour calculer la corrélation classique (\ref{CC2}) de tous les états qudit-qubit devient
\begin{equation}
\mathbf{L}=\sum\limits_{i= 0}^{3}\sum\limits_{j= 0}^{d^{2}-1}{\cal L}_{ij}.
\end{equation}
Ensuite, on peut facilement évaluer analytiquement la discorde quantique basée sur l'entropie linéaire pour tout système bipartite de type qudit–qubit comme
\begin{equation}
	{\cal Q}_{2}\left( {{\rho _{AB}}} \right) = {\cal I}\left( {{\rho _{AB}}} \right) -{\cal J_{\rm 2}}_{B}\left( \rho_{AB} \right).
\end{equation}

\section{Mesure des corrélations quantiques via l'incertitude quantique}
\subsection{L'information d'interchange (Skew Information)}
Les physiciens étudient la nature en effectuant des mesures et en prédisant leurs résultats. Dans un monde classique, les barres d'erreur sont exclusivement dues à des limitations technologiques, alors il est possible de mesurer deux observables quelconques avec une précision arbitraire. Cependant, un tel type de mesure n'est pas toujours possible dans les systèmes quantiques, car la mécanique quantique stipule que deux observables non commutables ne peuvent pas être mesurés conjointement avec une précision arbitraire, même si l'on pouvait accéder à un appareil de mesure sans défaut \cite{Heisenberg43,Oppenheim2010}. En outre, la mesure opérée sur un système quantique est complètement différente que celle d'un système classique, car non seulement le résultat obtenu est probabiliste, mais aussi l'état du système est modifié au cours de ce processus, et l'objectif est de déterminer, de façon plus précise, le résultat de cette mesure. L'idée d'exactitude est directement liée à l'incertitude associée au résultat de cette mesure et la relation d'incertitude donne la nature statistique des erreurs dans ce genre de mesure.\par

Dans la pratique standard et dans le cadre du formalisme de la mécanique quantique, l'incertitude d'une observable $K$ dans un état quantique $\rho$ est généralement quantifiée par la variance qui est donnée par la relation suivante \cite{Luo2005}
\begin{equation}
{\rm Var}\left(\rho,K\right):={\rm Tr}\left(\rho K_{0}^{2} \right)= {\rm Tr}\left( \rho K^{2}\right)-{\rm Tr}\left(\rho K\right)^{2},\label{Vr}
\end{equation}
où ${\rm Tr}$ désigne la trace et $K_{0}=K-{\rm Tr}\rho K$. En particulier, pour les états purs $\rho=\left|\psi \right\rangle\left\langle\psi \right|$, la variance se réduit à
\begin{equation}
	{\rm Var}\left(\left|\psi \right\rangle\left\langle\psi \right|,K\right)=\left\langle\psi \right| K^{2}\left| \psi\right\rangle -\left\langle\psi \right| K\left| \psi\right\rangle^{2}.
\end{equation}
La quantification de l'incertitude en terme de la variance est bien adaptée lorsque les états sont purs. Mais les choses se compliquent lorsque les états sont mixtes,  car l'incertitude inclut l'ignorance classique, due au mélange d'états, et une partie de la nature quantique résultant de la non commutativité entre l'état étudié et l'observable mesurée. Ainsi, la variance peut présenter des contributions d'origines quantiques et classiques. Il existe encore plusieurs méthodes permettant de quantifier l’incertitude, comme par exemple celles basées sur des mesures entropiques et qui sont largement utilisées comme indicateurs d'incertitude en théorie de l'information quantique. Cependant, ce type de mesure reste toujours insuffisant et limité, car ces quantificateurs sont toujours influencés par la mixité de l'état étudié.\par

Pour un état quantique $\rho$, une observable est dite certaine quantiquement, si l'erreur de mesure de l'observable est due uniquement à l'ignorance du mélange classique dans l'état $\rho$. Afin de déterminer la partie quantique de la variance (\ref{Vr}), Wigner et Yanase ont introduit la notion de l'information d'interchange (Skew Information en anglais) comme \cite{Wigner1963}
\begin{equation}
	I\left( {\rho ,K} \right): = {I_\rho }\left( K \right): = \frac{1}{2}{\rm Tr}\left\{ {{{\left( {i\left[ {\sqrt \rho  ,K} \right]} \right)}^2}} \right\},\label{IS}
\end{equation}
où $\left[.,.\right]$ désigne le commutateur. Il est important de noter que contrairement à la variance, l'information d'interchange de Wigner-Yanase n'est pas affectée par le mélange classique. Elle quantifie le degré de non commutativité entre un état quantique $\rho$ et une observable $K$ qui peut être considérée comme un Hamiltonien ou toute autre quantité conservée. De plus, l'information d'interchange fournit une mesure alternative du contenu d'information de l'état $\rho$ par rapport aux observables non commutatifs avec la quantité conservée $K$. En d'autres termes, elle représente la quantité d'information stockée dans l'état $\rho$ qui ne sont pas accessibles par la mesure de l'observable $K$. L'équation (\ref{IS}) peut être simplifiée comme
\begin{align}
	I\left( {\rho ,K} \right)& =-\frac{1}{2}{\rm Tr}\{\left(\sqrt{\rho}K-K\sqrt{\rho}\right)^{2}\}\notag \\&=- \frac{1}{2}\left\{ {{\rm Tr}\left( {\sqrt \rho  K\sqrt \rho  K} \right) - {\rm Tr}\left( {K\rho K} \right) - {\rm Tr}\left( {\sqrt \rho  {K^2}\sqrt \rho  } \right) + {\rm Tr}\left( {K\sqrt \rho  K\sqrt \rho  } \right)} \right\} \notag \\
	& = {\rm Tr}\left( {\rho {K^2}} \right) - {\rm Tr}\left( {\sqrt \rho  K\sqrt \rho  K} \right). \label{SII} 
\end{align}
Cette quantité a été généralisée par Dyson par la formule suivante \cite{Luo2003}
\begin{align}
	{I_{\rho ,\alpha }}\left( K \right) &= \frac{1}{2}{\rm Tr}\left\{ {\left( {i\left[ {{\rho ^\alpha },K} \right]} \right)\left( {i\left[ {{\rho ^{1 - \alpha }},K} \right]} \right)} \right\} \notag \\
	& = {\rm Tr}\left( {\rho {K^2}} \right) - {\rm Tr}\left( {{\rho ^\alpha }K{\rho ^{1 - \alpha }}K} \right).  \hspace{2cm} \alpha  \in \left[ {0,1} \right]
\end{align}
D'un autre côté, tous les états quantiques mixtes peuvent être écrits comme $\rho=\sum_{n}\lambda_{n}\left|\psi_{n}\right\rangle\left\langle\psi_{n}\right|$, où $\lambda_{n}$ sont les valeurs propres de la matrice densité $\rho$, et les vecteurs propres $\{\left|\psi_{n}\right\rangle\}$ constituent une base orthonormée telle que $\sum_{n}\left|\psi_{n}\right\rangle\left\langle\psi_{n}\right|=\openone$. On peut facilement montrer que
\begin{equation}
{\rm Tr}\left(\rho K^{2} \right)=\sum_{n,m}\lambda_{n}\mid\left\langle\psi_{n} \right|K\left|\psi_{m} \right\rangle\mid^{2},
\end{equation}
que l'on peut aussi écrire sous la forme symétrique suivante
\begin{equation}
{\rm Tr}\left(\rho K^{2} \right)=\sum_{n,m}\frac{\lambda_{n}+\lambda_{m}}{2}\mid\left\langle\psi_{n} \right|K\left|\psi_{m} \right\rangle\mid^{2}.
\end{equation}
Alors le deuxième terme de l'équation (\ref{SII}) devient
\begin{equation}
{\rm Tr}\left(K\sqrt{\rho}K\sqrt{\rho}\right)=\sum_{n,m}\sqrt{\lambda_{n}\lambda_{m}}\mid\left\langle\psi_{n} \right|K\left|\psi_{m} \right\rangle\mid^{2},
\end{equation}
et l'expression explicite de l'information d'interchange est donnée par
\begin{equation}
I\left(\rho,K \right)=\sum_{n,m}\frac{1}{2}\left(\sqrt{\lambda_{n}}-\sqrt{\lambda_{m}} \right)^{2}\mid\left\langle\psi_{n} \right|K\left|\psi_{m} \right\rangle\mid^{2}.
\end{equation}
L'information d'interchange $I\left( {\rho ,K} \right)$ a des propriétés intéressantes \cite{Wigner1963}, nous rappelons ici les plus pertinentes:
\begin{itemize}
	\item L'information d'interchange est toujours positive, $I\left( {\rho ,K} \right)\geq0$ et se réduit à la variance si l'état $\rho$ est pur ($\rho^{2}\equiv\rho$);
	\begin{equation}
I\left(\left|\psi \right\rangle \left\langle \psi\right|, K \right)={\rm Var}\left(\left|\psi \right\rangle\left\langle\psi \right|,K\right)=\left\langle\psi \right| K^{2}\left| \psi\right\rangle -\left\langle\psi \right| K\left| \psi\right\rangle^{2}.
	\end{equation}
\item L'information d'interchange est toujours plus petite que la variance de l'observable $K$. En effet
\begin{align}
I\left( {\rho ,K} \right)&={\rm Tr}\left(\rho K^{2}\right)-{\rm Tr}\left(\sqrt{\rho}K\sqrt{\rho}K\right)\notag\\&\leq {\rm Tr}\left(\rho K^{2}\right)-{\rm Tr}\left(\rho K\right)^{2}\notag\\&\equiv {\rm Var}\left( {\rho ,K} \right).
\end{align}
\item $I\left({\rho ,K}\right)$ est convexe car l'information d'interchange diminue lorsque plusieurs états sont mélangés; 
\begin{equation}
I\left(\sum_{i}\lambda_{i}\rho_{i},K \right)\leqslant\sum_{i} \lambda_{i}I\left( \rho_{i},K\right),
\end{equation}
pour tous les états quantiques $\rho_{i}$, et les constantes $\lambda_{i}$ satisfaisant $\sum_{i}\lambda_{i}=1$ et $0\leqslant\lambda_{i}\leqslant1$. Dans le cas contraire, la variance ${\rm Var}\left( {\rho ,K} \right)$ est concave dans l'état $\rho$.
\item L'information d'interchange est invariante sous une transformation unitaire $U$, si et seulement si $U$ et l'observable mesurée $K$ commutent, c-à-d; $\left[U,K\right]=0$. Dans ce cas, nous avons
\begin{equation}
UKU^{-1}=K,\hspace{2cm}\sqrt{U\rho U^{-1}}=U\sqrt{\rho}U^{-1},
\end{equation}
et donc
\begin{align}
I\left(U\rho U^{-1},K\right)&=-\frac{1}{2}{\rm Tr}\{\left[\sqrt{U\rho U^{-1}},K \right]^{2} \}\notag\\&= -\frac{1}{2}{\rm Tr}\{\left[\sqrt{U\rho U^{-1}},UKU^{-1}\right]^{2} \}\notag\\&=-\frac{1}{2}{\rm Tr}\{U\left[\sqrt{\rho},K\right]^{2}U^{-1}\}\notag\\&=-\frac{1}{2}{\rm Tr}\{\left[\rho,K \right]^{2}\}=I\left( {\rho ,K} \right).
\end{align}
\item Dans l'espace de Hilbert ${\cal H}_{A}\otimes{\cal H}_{B}$ d'un système composite $AB$, l'information d'interchange de l'état quantique $\rho_{AB}$ et la matrice de densité réduite $\rho_{A}$ associé au sous système $A$ satisfait la relation suivante
\begin{equation}
I\left(\rho_{AB}, K_{A}\otimes\openone_{B} \right)\geq I\left(\rho_{A}, K_{A}\right), 
\end{equation}
pour toutes les observables locales $K_{A}$ agissant sur l'espace de Hilbert ${\cal H}_{A}$. Ici $\openone_{B} $ est l'opérateur identité dans ${\cal H}_{B}$.
\item Afin d'obtenir une quantité intrinsèque fournissant le contenu informationnel de l'état $\rho$, Luo a introduit la moyenne \cite{Luo2006}
\begin{equation}
{\cal Q}_{I}\left(\rho \right)=\sum_{i=0}^{n^{2}-1} I\left(\rho,K^{i} \right), 
\end{equation}
où $\{K^{i}\}$ est un ensemble des $n^{2}$ observables sur un système quantique à $n$-dimensions. Ainsi, le contenu de l'information globale de l'état bipartite $\rho_{AB}$ en termes d'observables locales d'un système quantique à $n$-dimensions $A$ est
\begin{equation}
{\cal Q_{I}}_{A}\left(\rho_{AB} \right)=\sum_{i=0}^{n^{2}-1} I\left(\rho_{AB},K_{A}^{i}\otimes\openone_{B}\right), 
\end{equation}
\item La différence entre le contenu de l'information de l'état $\rho_{AB}$ et l'état $\rho_{A}\otimes\rho_{B}$ par rapport aux observables locales $K_{A}^{i}$, peut être interprétée comme une mesure des corrélations quantiques dans l'état $\rho_{AB}$. Nous pouvons donc définir une autre mesure des corrélations quantiques en terme de l'information d'interchange comme \cite{Luo2006}
\begin{equation}
{\cal F}_{A}\left(\rho_{AB} \right)={\cal Q_{I}}_{A}\left(\rho_{AB} \right)-{\cal Q_{I}}_{A}\left(\rho_{A}\otimes\rho_{B} \right)
\end{equation}
\end{itemize}
\subsection{L'incertitude quantique locale: Définition}
\begin{figure}[H]
	\centerline{\includegraphics[width=14cm]{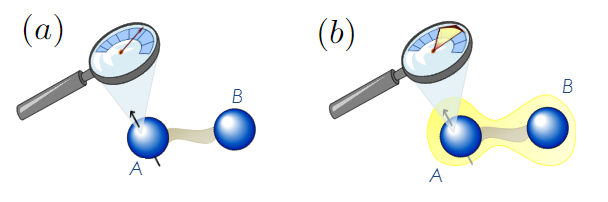}}
	\caption[Les corrélations quantiques conduisant à une incertitude quantique locale]{Pour l'état (a); Si $\rho$ n'est pas corrélé ou bien ne contient que des corrélations classiques, c'est-à- dire $\rho$ est de la forme $\rho=\sum_{i}p_{i}\left|i\right\rangle_{A}\left\langle i\right|\otimes\rho_{i,B}$ (avec $\{\left| i\right\rangle_{A}\}$ une base orthonormée du sous système $A$), il existe au moins une observable locale agissant sur le sous-système $A$ sans aucune incertitude quantique intrinsèque. Pour l'état (b); Si $\rho$ contient des corrélations quantiques non nulles quantifiées par l'intrication ou par la discorde quantique, toute mesure locale sur le sous-système $A$ est affectée par l'incertitude quantique. Cela signifie que les corrélations quantiques conduisent à une incertitude quantique locale.}
	\label{FLQU}
\end{figure}

En fait, l'incertitude quantique est intrinsèquement liée au concept de corrélation quantique. Par exemple, considérons un état de Bell à deux qubits $\left|\varphi^{+} \right\rangle=\left(\left|00 \right\rangle+ \left|11 \right\rangle\right)/\sqrt{2}$, où $\{\left|0 \right\rangle,\left|1\right\rangle\}$ désigne la base de calcul. Cet état est l'état propre de l'observable global $\sigma_{3}\otimes\sigma_{3}$, il n'y a donc pas d'incertitude sur le résultat de la mesure d'une telle observable (${\rm Var}\left(\left|\varphi^{+} \right\rangle\left\langle\varphi^{+}\right|,\sigma_{3}\otimes\sigma_{3} \right)=0$). Par contre, la mesure des observables de spin locales $\vec{\alpha}\vec{\sigma}\otimes\openone$ (où $\vec{\alpha}\neq0$ est un vecteur réel) est intrinsèquement incertaine pour l'opérateur densité $\left|\varphi^{+} \right\rangle\left\langle\varphi^{+}\right|$, puisqu'un état intriqué ne peut pas être un état propre d'une observable locale \cite{Girolami2013}. En particulier, la variance ${\rm Var}\left(\rho,K_{i} \right)$ pour une observable locale $K_{i}$ s'annule si et seulement si l'état n'est pas corrélé. Pour les états mixtes, l'information d'interchange $I\left(\rho,K\right)$ s'annule si et seulement si l'état $\rho$ n'est pas perturbé par la mesure de l'observable $K$. Si $K$ est une observable locale, alors les états qui sont invariants par la mesure locale sont les états qui ont une discorde quantique nulle par rapport à ce sous-système local. Par conséquent, l'incertitude quantique sur les observables locales est étroitement liée au concept de la discorde quantique; elle peut alors être utilisé comme un quantificateur des corrélations quantiques de type discorde \cite{Girolami2013} (voir la Fig.(\ref{FLQU})).

Très récemment, Girolami et ses collaborateurs ont défini l'incertitude quantique locale comme l'information d'interchange minimale pouvant être obtenue par une seule mesure locale \cite{Girolami2013}. Plus précisément, pour un état quantique bipartite $\rho_{AB}$, on considère l'observable locale $K^{\Lambda}=K_{A}^{\Lambda}\otimes\openone_{B}$ telle que $K_{A}^{\Lambda}$ est un opérateur hermitien, agissant sur les états du sous-système $A$, ayant un spectre non dégénéré $\Lambda$. L'incertitude quantique locale par rapport au sous-système $A$ est donnée par
\begin{equation}
{\cal U}_{A}^{\Lambda}\left( \rho_{AB}\right):=\min_{K_{A}^{\Lambda}}I\left(\rho_{AB}, K_{A}^{\Lambda}\otimes\openone_{B}\right), \label{DLQU}
\end{equation}
où la minimisation porte sur toutes les observables locales qui agissent sur le sous-système $A$. En effet, nous exigeons que $\Lambda$ soit non dégénéré, car il correspond aux observables informatives maximales du sous-système $A$. Ce quantificateur a de fortes raisons d'être considéré comme une mesure fidèle des corrélations quantiques des états quantiques bipartites. En effect, l'incertitude quantique locale est invariante sous les transformations unitaires locales, n'augmente pas sous les opérations locales sur le sous-système $B$, s'annule si et seulement si l'état quantique est classiquement corrélé \cite{Girolami2013}. Pour les états purs, l'incertitude quantique locale est une intrication monotone. Il s'en suit que l'incertitude quantique locale satisfait toutes les exigences physiques d'une mesure des corrélations quantiques.\par
Pour évaluer le minimum dans l'équation (\ref{DLQU}), il existe une exigence importante concernant les observables d'optimisation qui minimise l'information d'interchange, à savoir qu'ils doivent avoir un spectre fixe non dégénérés. En pratique, les observables avec ce spectre agissant sur le sous-système $A$ peuvent être paramétrés par \cite{Girolami2013}
\begin{equation}
K_{A}^{\Lambda}=V_{A}\Lambda V_{A}^{\dagger},
\end{equation}
où $V_{A}$ définit la base de mesure qui peut être modifiée arbitrairement sur le groupe unitaire spécial du sous-système $A$. Si nous limitons notre discussion au cas des systèmes quantiques bipartites de type qubit-qudit où le sous-système $A$ a une dimension $2$ et le sous-système $B$ a une dimension $d$, l'observable qubit à spectre fixe non dégénérer $\sigma_{3}$ peut être paramétrée par $K_{A}=V_{A}\sigma_{3} V_{A}^{\dagger}=\vec{s}.\vec{\sigma}$, où  $\Arrowvert\vec{s}\Arrowvert=1$ et $\sigma_{i}$ sont les matrices de Pauli, représentant aussi les générateurs de l'algèbre $su\left(2\right)$. Ce choix correspond à une incertitude quantique locale normalisée à l'unité pour les états purs maximalement intriqués. Dans ce cas, le spectre $\Lambda$ peut être supprimé dans l'équation (\ref{DLQU}) et l'incertitude quantique locale est équivalente à
\begin{align}
{\cal U}_{A}\left(\rho_{AB}\right)&=\min_{\vec{s}}I\left(\rho_{AB},\vec{s}.\vec{\sigma}\otimes\openone_{d} \right)\notag\\&=\min_{\vec{s}}\sum_{ij}s_{i}s_{j}\left[{\rm Tr}\{\rho_{AB}\sigma_{i}\sigma_{j}-\sqrt{\rho_{AB}}\left(\sigma_{i}\otimes\openone_{d} \right)\sqrt{\rho_{AB}}\left(\sigma_{j}\otimes\openone_{d}\right)\} \right]\notag\\&=1-\max\sum_{ij}s_{i}s_{j}\left[{\rm Tr}\{\sqrt{\rho_{AB}}\left(\sigma_{i}\otimes\openone_{d} \right)\sqrt{\rho_{AB}}\left(\sigma_{j}\otimes\openone_{d}\right)\} \right]\notag\\&=1-\max\left[ \vec{s}^{\dagger}{\cal W}\vec{s}\right],\label{LQUANLY}
\end{align}
où ${\cal W}$ est une matrice symétrique de dimension $3\times3$ dont les éléments sont
\begin{equation}
	w_{ij}={\rm Tr}\{\sqrt{\rho_{AB}}\left(\sigma_{i}\otimes\openone_{B} \right)\sqrt{\rho_{AB}}\left(\sigma_{j}\otimes\openone_{B}\right)\},\label{matriceW}
\end{equation}
avec $i,j=1,2,3$. Pour minimiser l'information d'inerchange, il est nécessaire de maximiser la quantité ${\cal W}$ sur tous les vecteurs unitaires $\vec{s}$. La valeur maximale coïncide avec la valeur propre maximale de la matrice ${\cal W}$. Par conséquent, la formule compacte de l'incertitude quantique locale est donnée par
\begin{equation}
{\cal U}_{A}\left(\rho_{AB}\right)=1-\max\{\xi_{1},\xi_{2},\xi_{3}\},\label{lqusimple}
\end{equation}
où les $\xi_{i}$ sont les valeurs propres de la matrice ${\cal W}=\left(w_{ij} \right)_{3\times3}$. Il est donc clair que le calcul de l'incertitude quantique locale requiert le calcul des éléments $w_{ij}$ de la matrice ${\cal W}$. Ainsi, cette mesure a l'avantage d'être simple lorsqu'il s'agit de calculer le maximum sur les paramètres liés aux mesures comme il faut le faire pour calculer la discorde quantique. En outre, l'incertitude quantique locale est réduite à l'intrication monotone pour les états purs. En fait, elle se réduit à l'entropie linéaire du sous-système réduit mesuré (c'est-à-dire la concurrence). Cette coïncidence intéressante ne se produit que pour les systèmes quantiques de dimension $2\otimes d$. Cependant, pour les systèmes quantiques à plusieurs niveaux (dimension de l'espace de Hilbert grande), il est difficile d'obtenir une forme compacte de l'incertitude quantique locale sans optimisation complexe. Pour clarifier ce point, nous prenons un état pur $\left| \psi \right\rangle=\sum_{mn}C_{mn}\left|mn \right\rangle $, les éléments de la matrice $\cal W$ deviennent
\begin{equation}
w_{ij}=\left\langle \psi\right|\sigma_{i}\otimes\openone_{d}\left|\psi\right\rangle\left\langle \psi\right|\sigma_{j}\otimes\openone_{d}\left|\psi\right\rangle.
\end{equation}
Le premier terme de l'équation ci-dessus devient comme suit
\begin{align}
\left\langle \psi\right|\sigma_{i}\otimes\openone_{d}\left|\psi\right\rangle=\sum_{mk}\left(C{C}^{\dagger} \right)_{km} \left\langle m\right|\sigma_i\left| k\right\rangle\label{yy}
\end{align}
La matrice densité réduite du sous-système $A$ peut être reécrite en fonction des coefficients $C_{mn}$ comme
\begin{align}
\rho_{A}=\sum_{mkl} C_{ml}C_{lk}^{\dagger}\left| m\right\rangle\left\langle k\right|=\sum_{mkl}\left(CC^{\dagger} \right)_{mk}\left|m\right\rangle\left\langle k\right|. \label{rhoaa}
\end{align}
Nous obtenons alors
\begin{equation}
\left\langle m\right|\rho_{A}\left| k\right\rangle =\left(CC^{\dagger} \right)_{mk}.\label{rr}
\end{equation}
Nous retournons à l'expression (\ref{yy}), en s'aidant de l'équation (\ref{rr}), on voit aisément qu'elle se réduit à la valeur moyenne de $\sigma_{i}$ par rapport à la base du sous-système $A$. Nous avons donc
\begin{align}
	\left\langle\psi\right|\sigma_{i}\otimes\openone_{d}\left|\psi\right\rangle=\sum_{k}\left\langle k\right|\rho_{A}\sigma_{i}\left| k\right\rangle={\rm Tr}_{A}\left(\rho_{A}\sigma_{i} \right)=\left\langle \sigma_{i}\right\rangle_{A}.
\end{align}
Cela mène à $w_{ij}=\left\langle \sigma_{i}\right\rangle_{A}\left\langle \sigma_{j}\right\rangle_{A}$, et la matrice ${\cal W}$ s'écrit sous la forme
\begin{equation}
	{\cal W}=\left( {\begin{array}{*{20}{c}}
			{\left\langle \sigma_{1}\right\rangle_{A}^{2}}&{\left\langle \sigma_{1}\right\rangle_{A}\left\langle \sigma_{2}\right\rangle_{A}}&{\left\langle \sigma_{1}\right\rangle_{A}\left\langle \sigma_{3}\right\rangle_{A}}\\
			{\left\langle \sigma_{2}\right\rangle_{A}\left\langle \sigma_{1}\right\rangle_{A}}&{\left\langle \sigma_{2}\right\rangle_{A}^{2}}&{\left\langle \sigma_{2}\right\rangle_{A}\left\langle \sigma_{3}\right\rangle_{A}}\\
			{\left\langle \sigma_{3}\right\rangle_{A}\left\langle \sigma_{1}\right\rangle_{A}}&{\left\langle \sigma_{2}\right\rangle_{A}\left\langle \sigma_{3}\right\rangle_{A}}&{\left\langle \sigma_{3}\right\rangle_{A}^{2}}
	\end{array}} \right).
\end{equation}
Après la diagonalisation de la matrice ${\cal W}$, on trouve une seule valeur propre non nulle qui s'écrit 
\begin{equation}
\xi_{\max}=\left\langle \sigma_{1}\right\rangle_{A}^{2}+\left\langle \sigma_{2}\right\rangle_{A}^{2}+\left\langle \sigma_{3}\right\rangle_{A}^{2},
\end{equation}
et l'incertitude quantique locale des états purs devient
\begin{equation}
{\cal U}_{A}\left( \left|\psi \right\rangle \left\langle \psi\right| \right)=1-\left[\left\langle \sigma_{1}\right\rangle_{A}^{2}+\left\langle \sigma_{2}\right\rangle_{A}^{2}+\left\langle \sigma_{3}\right\rangle_{A}^{2} \right]. 
\end{equation}
D'autre part, la représentation de Fano-Bloch de la matrice (\ref{rhoaa}) s'écrit $\rho_{A}=\frac{1}{2}\left(1+\vec{r}\vec{\sigma_{i}} \right)$, où $\vec{r}$ représente le vecteur de Bloch avec
\begin{equation}
\left\langle \sigma_{i}\right\rangle_{A}={\rm Tr}\left(\rho_{A}\sigma_{i} \right)=\frac{1}{2}\left({\rm Tr}\sigma_{i}+r_{j}{\rm Tr}\left(\sigma_{i}\sigma_{j} \right)\right)=r_{i}.  
\end{equation}
De même, nous avons
\begin{equation}
{\rm Tr}\left(\rho_{A}^{2} \right)=\frac{1}{2}\left(1+r_{i}^{2} \right)= \frac{1}{2}\left(1+\xi_{\max} \right).
\end{equation}
Par conséquent, pour les états purs $\rho_{AB}=\left|\psi \right\rangle \left\langle \psi\right|$, l'incertitude quantique locale coïncide avec l'entropie linéaire de l'intrication. Nous avons alors 
\begin{equation}
{\cal U}_{A}\left( \left|\psi \right\rangle \left\langle \psi\right| \right)=2\left[1-{\rm Tr} \left(\rho_{A}^{2} \right) \right]=S_{2}\left(\rho_{A} \right).
\end{equation}

Il est intéressant de noter que l'expression analytique de l'incertitude quantique locale (\ref{LQUANLY}) peut être appliquée à des systèmes multi-qubit sans aucune difficulté majeure. En effet, nous pouvons toujours considérer le système multi-qubit (à $N$ qubites) comme un système de dimension $2\otimes d$ où $d=2\otimes...\otimes2$ peut représenter les $N-1$ qubits restants sous la forme d'un système quantique avec un espace de Hilbert de dimension $d$. Comme il existe plusieurs bipartitions pour un système multi-qubit, il est possible que certaines bipartitions soient corrélées quantiquement, d'autres classiquement corrélées et d'autres peuvent être complètement décorrélées avec d'autres qubits. Par conséquent, nous devons définir l'incertitude quantique locale pour chaque bipartition. Pour chaque cas, nous obtenons une matrice symétrique ${\cal W}_{k}$, donc au total nous obtenons $N$ matrices symétriques. Après avoir calculé toutes ces incertitudes quantiques locales, nous pouvons calculer l'incertitude quantique locale moyenne pour un état multi-qubit donné. Dans ce but, nous considérons une matrice densité arbitraire $\rho_{N}$ associé à un système décrivant $N$ qubits et nous appliquons des mesures locales sur chaque qubit. Puis, les éléments de la matrice ${\cal W}_{k}$ de chaque bipartition $\rho_{k}$ (avec $k=1,2,...,N$) sont données par les expressions suivantes 
\begin{align}
&{\hat{w}}_{ij}^{1}={\rm Tr}\{\sqrt{\rho_{N}}\left(\sigma_{i}\otimes\openone_{2}...\otimes\openone_{2} \right)\sqrt{\rho_{N}}\left(\sigma_{j}\otimes\openone_{2}...\otimes\openone_{2}\right)\},\notag\\&{\hat{w}}_{ij}^{2}={\rm Tr}\{\sqrt{\rho_{N}}\left(\openone_{2}\otimes\sigma_{i}...\otimes\openone_{2} \right)\sqrt{\rho_{N}}\left(\openone_{2}\otimes\sigma_{j}...\otimes\openone_{2}\right)\},\notag\\&\hspace{0.2cm}:\hspace{1cm}:\hspace{1cm}:\hspace{1cm}:\hspace{1cm}:\hspace{1cm}:\hspace{1cm}:\hspace{1cm}:\notag\\&\hspace{0.2cm}:\hspace{1cm}:\hspace{1cm}:\hspace{1cm}:\hspace{1cm}:\hspace{1cm}:\hspace{1cm}:\hspace{1cm}:\notag\\&{\hat{w}}_{ij}^{N}={\rm Tr}\{\sqrt{\rho_{N}}\left(\openone_{2}\otimes\openone_{2}...\otimes\sigma_{i} \right)\sqrt{\rho_{N}}\left(\openone_{2}\otimes\openone_{2}...\otimes\sigma_{j}\right)\}.
\end{align}
Les incertitudes quantiques locales liées à chaque bipartition sont définies comme suit
\begin{align}
&{\cal U}_{1/23...N}\left( \rho_{N}\right)=1-\max\{\xi_{1}^{1},\xi_{2}^{1},\xi_{3}^{1}\},\notag\\&{\cal U}_{2/13...N}\left( \rho_{N}\right)=1-\max\{\xi_{1}^{2},\xi_{2}^{2},\xi_{3}^{2}\},\notag\\&\hspace{0.2cm}:\hspace{1cm}:\hspace{1cm}:\hspace{1cm}:\hspace{1cm}:\hspace{1cm}:\notag\\&\hspace{0.2cm}:\hspace{1cm}:\hspace{1cm}:\hspace{1cm}:\hspace{1cm}:\hspace{1cm}:\notag\\&{\cal U}_{N/12...N-1}\left( \rho_{N}\right)=1-\max\{\xi_{1}^{N},\xi_{2}^{N},\xi_{3}^{N}\},
\end{align}
où $\xi_{i}^{k}$ (avec $i=1,2,3$) sont les valeurs propres des matrices symétriques $\left({\cal W}_{k}\right)_{3\times3}$ qui peuvent être déterminées facilement. L'incertitude quantique locale, qui nous donne la quantité des corrélations quantiques globales dans l'état $\rho_{N}$, peut être considérée comme la valeur moyenne des incertitudes quantiques locales de chaque bipartition
\begin{equation}
{\cal U}\left( \rho_{N}\right)=\frac{1}{N}\left(\sum_{k=A}^{N}{\cal U}_{k/N_{k}} \right)\equiv\frac{1}{N}\left(\sum_{k=A}^{N}{\cal U}_{k/12...\bar{k}...N} \right),
\end{equation}
où $N_{k}=12...\bar{k}...N$ sont les ($N-1$)-qubits restants sauf le qubit $k$.\par
Il faut reconnaître que la quantification des corrélations dans des systèmes qudit-qudit ($d_{1}\otimes d_{2}$) reste un problème ouvert. Dans ce sens, une lueur d'espoir vient de la référence \cite{Wang2019}, dans laquelle les auteurs soulignent que la forme fermée de l'incertitude quantique locale peut être obtenue pour les états quantiques de dimension $d_{1}\otimes d_{2}$, en utilisant les générateurs de l'algèbre $su\left(d_{1}\right)$ dans l'optimisation en tant que groupe unitaire spécial de degré $d_{1}$. Dans ce cas, l'expression explicite de l'incertitude quantique locale s'écrit
\begin{equation}
{\cal U}_{A}\left(\rho \right)=\frac{2}{d_{1}}-\xi_{\max}\left(\hat{\cal W}\right) 
\end{equation}
où $\xi_{\max}\left(\hat{\cal W}\right)$ représente la valeur propre maximale de la matrice $\hat{\cal W}$, qui est une matrice d'ordre $\left(d_{1}^{2}-1 \right)\left(d_{1}^{2}-1 \right)$, dont les éléments sont donnés par

\begin{equation}
{\cal W}_{ij}={\rm Tr}\{\sqrt{\rho}\left( \lambda_{i}\otimes\openone_{d_{2}}\right)\sqrt{\rho}\left(\lambda_{j}\otimes\openone_{d_{2}} \right) \}-G_{ij}P,
\end{equation}
avec les $\lambda_{i}$ sont les générateurs de $su\left(d_{1}\right)$. Le vecteur ligne $G_{ij}$ et le vecteur colonne $P$, sont donnés respectivement par
\begin{equation}
G_{ij}=\left(g_{ij1},...,g_{ijk},...,g_{ij{d_{1}^{2}-1}} \right), \hspace{1cm}{\rm avec}\hspace{1cm}g_{ijk}=\frac{1}{4}{\rm Tr}\left(\lambda_{i}\lambda_{j}\lambda_{k}+\lambda_{j}\lambda_{i}\lambda_{k} \right),  
\end{equation}
et
\begin{equation}
P=\left({\rm Tr}\left(\rho\lambda_{1}\otimes\openone_{d_{2}} \right),...,{\rm Tr}\left(\rho\lambda_{k}\otimes\openone_{d_{2}} \right),...,{\rm Tr}\left(\rho\lambda_{d_{1}^{2}-1}\otimes\openone_{d_{2}} \right) \right)^{T}.
\end{equation}

\subsection{L'incertitude quantique locale maximale pour les états $X$ séparables à deux qubits}
Puisque l'incertitude quantique locale indique une valeur non nulle pour un état séparable, il sera intéressant d'étudier l'incertitude maximale que l'on peut obtenir avec ce type d'état \cite{Sen2016}. Cette valeur maximale fournit la limite de précision qui peut être atteinte en utilisant des états séparables en métrologie quantique (nous revenons sur ce sujet dans le chapitre $4$). Pour ce faire, nous devons maximiser l'incertitude quantique locale (\ref{lqusimple}) sur tous les états séparables ($S$) comme
\begin{align}
		\max_{\rho\in S}{\cal U}_{A}\left(\rho_{AB}\right)&=\max_{\rho\in S}\left(1-\max\{\xi_{1},\xi_{2},\xi_{3}\}\right),
\end{align}
cela signifie qu'il faut trouver l'état séparable parmi plusieurs états d'un ensemble S qui donnera la valeur propre minimale de la matrice $\cal W$. On considère ici un état $X$ à deux qubits comme dans l'équation (\ref{matrixX}). Ses éléments satisfont les conditions suivantes:
\begin{itemize}
	\item La condition de normalisation qui donne $\sum_{i=1}^{4}\rho_{ii}=1$.
	\item La matrice densité $\rho_{X}$ est positive, alors $\rho_{11}\rho_{44}\geqslant\rho_{14}^{2}$ et $\rho_{22}\rho_{33}\geqslant\rho_{23}^{2}$.
	\item $\rho_{X}$ est un opérateur hermitien ($\rho_{X}=\rho_{X}^{\dagger}$), donc $\rho_{14}=\rho_{41}^{*}$ et $\rho_{23}=\rho_{32}^{*}$. 
\end{itemize}
L'état $\rho_{X}$ a quatre valeurs propres réelles que nous notons par $\{\lambda_{1},\lambda_{2},\lambda_{3},\lambda_{4}\}$, et les vecteurs propres correspondants sont notés $\{\left|\vartheta_{1} \right\rangle,\left|\vartheta_{2} \right\rangle,\left|\vartheta_{3} \right\rangle,\left|\vartheta_{4} \right\rangle\}$. Ainsi, la matrice densité $\rho_{X}$ devient $\rho_{X}=\sum_{i=1}^{4}\lambda_{i}\left|\vartheta_{i} \right\rangle\left\langle \vartheta_{i}\right|$. Nous pouvons donc facilement écrire la racine carrée de cette matrice comme $\sqrt{\rho_{X}}=\sum_{i=1}^{4}\sqrt{\lambda_{i}}\left|\vartheta_{i} \right\rangle\left\langle \vartheta_{i}\right|$. Dans la même base de calcul de la matrice $\rho_{X}$, un calcul long conduit à
\begin{equation}
\sqrt{\rho_{X}}=\left( {\begin{array}{*{20}{c}}
			\Gamma_{1}&0&0&\Gamma_{5}\\
			0&\Gamma_{2}&\Gamma_{6}&0\\
			0&\Gamma_{6}&\Gamma_{3}&0\\
			\Gamma_{5}&0&0&\Gamma_{4}\end{array}} \right),
\end{equation}
avec
\begin{align}
&\Gamma_{1}=\frac{\sqrt{\lambda_{1}}\varpi_{1}^{2}}{\varpi_{1}^{2}+1}+\frac{\sqrt{\lambda_{2}}\varpi_{2}^{2}}{\varpi_{2}^{2}+1},\hspace{1cm}\Gamma_{2}=\frac{\sqrt{\lambda_{3}}\varpi_{3}^{2}}{\varpi_{3}^{2}+1}+\frac{\sqrt{\lambda_{4}}\varpi_{4}^{2}}{\varpi_{4}^{2}+1},\hspace{1cm}\Gamma_{3}=\frac{\sqrt{\lambda_{3}}}{\varpi_{3}^{2}+1}+\frac{\sqrt{\lambda_{4}}}{\varpi_{4}^{2}+1}\notag\\&\Gamma_{4}=\frac{\sqrt{\lambda_{1}}}{\varpi_{1}^{2}+1}+\frac{\sqrt{\lambda_{2}}}{\varpi_{2}^{2}+1},\hspace{1cm}\Gamma_{5}=\frac{\sqrt{\lambda_{1}}\varpi_{1}}{\varpi_{1}^{2}+1}+\frac{\sqrt{\lambda_{2}}\varpi_{2}}{\varpi_{2}^{2}+1},\hspace{1cm}\Gamma_{6}=\frac{\sqrt{\lambda_{3}}\varpi_{3}}{\varpi_{3}^{2}+1}+\frac{\sqrt{\lambda_{4}}\varpi_{4}}{\varpi_{4}^{2}+1},
\end{align}
où
\begin{align}
&\varpi_{1}=\frac{\rho_{11}-\rho_{44}+\lambda_{1}-\lambda_{2}}{2\rho_{41}}, \hspace{1cm}\varpi_{2}=\frac{\rho_{11}-\rho_{44}-\lambda_{1}+\lambda_{2}}{2\rho_{41}},\notag\\&\varpi_{3}=\frac{\rho_{22}-\rho_{33}+\lambda_{3}-\lambda_{4}}{2\rho_{32}},\hspace{1cm}\varpi_{4}=\frac{\rho_{22}-\rho_{33}-\lambda_{3}+\lambda_{4}}{2\rho_{32}}.
\end{align}
De plus, il est facile de montrer que $\left\langle \vartheta_{i}\lvert\vartheta_{i} \right\rangle=\varpi_{i}^{2}+1$, puis la matrice $\cal W$ est diagonale et ces éléments deviennent
\begin{align}
&\xi_{1}=2\left(\Gamma_{1}\Gamma_{3}+\Gamma_{2}\Gamma_{4}+2\Gamma_{5}\Gamma_{6} \right),\notag\\&\xi_{2}=2\left(\Gamma_{1}\Gamma_{3}+\Gamma_{2}\Gamma_{4}-2\Gamma_{5}\Gamma_{6} \right),\notag\\&\xi_{3}=\sum_{i=1}^{6}\Gamma_{i}^{2}-3\left(\Gamma_{5}^{2}+\Gamma_{6}^{2} \right). 
\end{align}
Jusqu'à présent, nous n'avons pas pris en compte la condition de séparabilité de l'état étudié. En utilisant les deux conditions de séparabilité du critère PPT (\ref{pptresults}), on trouve
\begin{equation}
\rho_{11}\rho_{44}\geqslant\rho_{23}^{2}, \hspace{1cm}{\rm et}\hspace{1cm}\rho_{22}\rho_{33}\geqslant\rho_{14}^{2}.
\end{equation}
Après avoir ajouté ces deux conditions à la contrainte de positivité, nous pouvons écrire le problème d'optimisation de l'incertitude quantique locale dans tous les états $X$ séparables comme une minimisation de $\xi_{\max}=\max\{\xi_{1},\xi_{2},\xi_{3}\}$ avec ces contraintes:
\begin{align}
&\rho_{14}\leqslant\min\left(\sqrt{\rho_{11}\rho_{44}},\sqrt{\rho_{22}\rho_{33}} \right), \hspace{1cm}\sum_{i=1}^{4}\rho_{ii}=1,\notag\\& \rho_{23}\leqslant\min\left(\sqrt{\rho_{11}\rho_{44}},\sqrt{\rho_{22}\rho_{33}} \right),\hspace{1cm}\rho_{ij}\geqslant0 \hspace{1cm}\forall i,j.
\end{align}
Nous observons que, pour des états $X$ à deux qubits avec des entrées positives, le produit $\Gamma_{5}\Gamma_{6}$ est toujours positif ($\Gamma_{5}\Gamma_{6}\geqslant0$). Cela implique que la valeur propre $\xi_{1}$ est toujours plus grande que $\xi_{2}$ de sorte que $\xi_{\max}=\max\{\xi_{1},\xi_{3}\}$. Par conséquent, pour déterminer l'incertitude quantique locale maximale dans les états $X$ séparables, nous devons envisager les deux situations distinctes suivantes:
\begin{itemize}
	\item[(i)] Si $\xi_{\max}=\xi_{1}$ alors l'incertitude quantique locale est donnée par
	\begin{equation}
		{\cal U}_{A}\left(\rho_{X}\right)=\max\{\left(\Gamma_{1}-\Gamma_{3}\right)^{2}+\left(\Gamma_{2}-\Gamma_{4}\right)^{2}+2\left(\Gamma_{5}-\Gamma_{6}\right)^{2}\}.
	\end{equation}
\item[(ii)] Si $\xi_{\max}=\xi_{3}$ alors l'incertitude quantique locale se réduit à
\begin{equation}
	{\cal U}_{A}\left(\rho_{X}\right)=4\max\left(\Gamma_{5}^{2}+\Gamma_{6}^{2}\right).
\end{equation}
\end{itemize} 
\section{La relation d'incertitude quantique entropique}
Le principe d'incertitude, formulé à l'origine par Heisenberg \cite{Heisenberg43}, illustre clairement la différence entre la mécanique quantique et classique. Ce principe limite les incertitudes sur les résultats de deux observables incompatibles, telles que la position et l'impulsion, sur une particule. En général, les relations d'incertitude limitent les connaissances potentielles que l'on peut avoir sur les propriétés physiques d'un système. Même avec une description complète de son état, il est impossible de prédire les résultats de toutes les mesures possibles sur la particule. Ce manque de connaissance, ou incertitude, a été quantifié par Heisenberg en utilisant l'écart type des observables de position et d'impulsion (que nous notons $\Delta X$ pour une observable $X$). Plus tard, Robertson et Schrödinger l'ont généralisé à des observables arbitraires $P$ et $Q$ non commutatives \cite{Robertson1929}. Cette généralisation prend la forme suivante
\begin{equation}
\left(\Delta P\right)\left(\Delta Q\right)\geqslant\frac{1}{2}|\left\langle\left[P,Q\right] \right\rangle|,
\end{equation}
où les écarts types des observables sont quantifiés en termes de valeurs moyennes $\left(\Delta P\right)\equiv\sqrt{\left\langle P^{2}\right\rangle-\left\langle P\right\rangle^{2}}$, $\left(\Delta Q\right)\equiv\sqrt{\left\langle Q^{2}\right\rangle-\left\langle Q\right\rangle^{2}}$ et le commutateur $\left[P,Q\right]\equiv PQ-QP$.\par 
En relation avec la théorie de l'information, il existe différentes formulations mathématiques du contenu physique de la relation d'incertitude. Outre la formulation standard en termes de la variance (\ref{Vr}), il existe une autre formulation en termes d'entropie, appelée la relation d'incertitude entropique. La principale différence entre ces formulations réside dans le fait que la relation d'incertitude entropique ne prend en compte que les probabilités des différents résultats de mesure. Cette relation, initialement proposée par Deutsch \cite{Deutsch1983}, a été reformulée en termes d'entropie de Shannon comme suit 
\begin{equation}
	{\cal H}\left(P\right) +{\cal H}\left(Q\right)\geqslant-2\log_{2}c\left(P,Q \right),
\end{equation}
avec ${\cal H}\left(X\right)$ désigne l'entropie de Shannon, et le terme $c\left(P,Q \right)\equiv\max_{i,j}\mid\left\langle p_{i}\mid q_{j}\right\rangle \mid^{2}$ quantifie la complémentarité des observables $P$ et $Q$, où $\{p_{i}\}$ et $\{q_{j}\}$ sont respectivement les valeurs propres correspondantes. Pour quantifier l'information de manière quantique, nous étendons cette étude vers le côté quantique et nous remplaçons l'entropie de Shannon par celle de von Neumann. En présence de la mémoire quantique (où l'environnement $B$), la relation d’incertitude quantique entropique devient \cite{Berta2010}
\begin{equation}
	S\left(P|B\right) +S\left(Q|B\right)\geqslant-2\log_{2}c\left(P,Q \right)+S\left(A| B\right). \label{yyi}
\end{equation}
Cette relation d'incertitude (\ref{yyi}) fournit une limite sur les incertitudes des résultats de mesure qui dépend de la quantité d'intrication entre la particule mesurée $A$ et la mémoire quantique $B$. $S\left(P|B\right)$, $S\left(Q|B\right)$ et $S\left(A|B\right)$ représentent les entropies conditionnelles quantiques, avec 
\begin{equation}
S\left(P|B\right)=S\left( \rho_{PB}\right)-S\left( \rho_{B}\right), \hspace{1cm}{\rm et}\hspace{1cm}S\left(Q|B\right)=S\left( \rho_{QB}\right)-S\left( \rho_{B}\right),
\end{equation}
où
\begin{align}
&\rho_{PB}=\sum_{i}\left( \left|p_{i} \right\rangle\left\langle p_{i}\right|\otimes\openone\right)\rho_{AB}\left( \left|p_{i} \right\rangle\left\langle p_{i}\right|\otimes\openone\right),\notag\\&\rho_{QB}=\sum_{i}\left( \left|q_{i} \right\rangle\left\langle q_{i}\right|\otimes\openone\right)\rho_{AB}\left( \left|q_{i} \right\rangle\left\langle q_{i}\right|\otimes\openone\right). 
\end{align}
La relation (\ref{yyi}) a été reformulée par Pati et ses collègues sous la forme suivante \cite{Pati2012}
\begin{equation}
	S\left(P|B\right) +S\left(Q|B\right)\geqslant-2\log_{2}c\left(P,Q \right)+S\left(A| B\right)+\max\left\lbrace 0,{\cal Q}_{A}\left( {{\rho _{AB}}} \right)- {\cal J}_{A}\left( {{\rho _{AB}}} \right)\right\rbrace,  
\end{equation}
où ${\cal Q}_{A}\left( {{\rho _{AB}}} \right)$ et ${\cal J}_{A}\left( {{\rho _{AB}}} \right)$ désignent respectivement la discorde quantique (\ref{QDD}) et la corrélation classique (\ref{QCD}) de l'état $\rho _{AB}$ après la mesure des observables $P$ et $Q$ sur le sous-système $A$. On prend les quantités $U_{B}^{P,Q}$ et $L_{B}^{P,Q}$ comme
\begin{align}
&U_{B}^{P,Q}=S\left(P|B\right) +S\left(Q|B\right),\notag\\&L_{B}^{P,Q}=-2\log_{2}c\left(P,Q \right)+S\left(A| B\right)+\max\left\lbrace 0,{\cal Q}_{A}\left( {{\rho _{AB}}} \right)- {\cal J}_{A}\left( {{\rho _{AB}}} \right)\right\rbrace,
\end{align}
on peut donc définir le gap d'incertitude dans un état quantique $\rho _{AB}$ correspondant au paire d'observables $\{P,Q\}$ comme $\varDelta^{P,Q}={U}_{B}^{P,Q}-L_{B}^{P,Q}$, qui est une grandeur non négative et peut caractériser la différence entre l'incertitude des résultats de mesures de $P$ et $Q$.

\section{Distributions des corrélations quantiques dans les systèmes quantiques multipartites}
\subsection{Monogamie des corrélations quantiques}
Au cours des dernières décennies, nous avons vu de nombreuses avancées pour la compréhension de l'intrication, et plus généralement, pour le développement de la théorie des corrélations quantiques. Néanmoins, le cadre avec plus de deux sous-systèmes reste un défi même d'un point de vue conceptuel, sans parler de leurs quantifications. Dans les systèmes multipartites, il est bien connu qu'une particule ne peut pas partager librement l'intrication avec deux ou plusieurs particules. Cette restriction, qui limite la fragilité des corrélations quantiques dans les états quantiques multipartites, est généralement appelée monogamie \cite{Coffman2000}. La propriété de la monogamie est importante dans de nombreuses tâches d'information quantique comme la cryptographie quantique par exemple. Dans ce contexte, le manque de monogamie est considéré comme un énorme obstacle à la mise en œuvre de la sécurité de l'information repose sur le fait que l'espion n'a pas les compétences pour corréler avec les parties de confiance \cite{Renes2006}. En outre, la propriété de monogamie quantifie la quantité d'informations qu'un espion pourrait obtenir sur la clé de secrète \cite{Masanes2009}.\par

La monogamie de l'intrication est l'un des phénomènes de la physique quantique qui la distingue de la physique classique. De plus, il donne lieu à la quantification et à la caractérisation de la distribution de l'intrication entre les sous-systèmes d'un système multipartite. Classiquement, trois bits aléatoires peuvent être corrélés au maximum. Cependant, il n'est pas possible de préparer trois qubits de manière à ce que deux qubits soient intriqués au maximum \cite{Coffman2000}, c'est-à-dire qu'un système quantique intriqué avec l'un des autres sous-systèmes limite son intrication avec les autres. Par exemple, pour un système quantique à trois qubits désignés par $A$, $B$ et $C$, si $A$ et $B$ sont dans un état maximalement intriqué, alors $A$ ne peut pas du tout être intriqué avec $C$. Ceci indique qu'il doit obéir à une relation de monogamie sur la quantité d'intrication entre les paires $AB$ et $AC$. La monogamie usuelle d'une mesure de l'intrication quantique $E$ implique que l'intrication $E\left(\rho_{A|BC} \right)$ entre le sous-système $A$ et l'autre partie $BC$ satisfait l'inégalité suivante
\begin{equation}
E\left(\rho_{A|BC} \right)\geqslant E\left(\rho_{AB}\right) +E\left(\rho_{AC}\right),\label{monogameE}
\end{equation}
où $E\left(\rho_{AB}\right)$ (resp.$E\left(\rho_{AC}\right)$) représente l'intrication entre les systèmes $A$ et $B$ (resp.$C$). L'inégalité (\ref{monogameE}) a été prouvée à l'origine pour des états arbitraires à trois qubits, en adoptant la concurrence (\ref{Concu}) comme mesure de l'intrication. Une question naturelle intéressante à cet égard est de savoir si la monogamie des corrélations est vraie pour des corrélations autres que l'intrication. La réponse à cette question a été abordée pour la discorde quantique dans la référence \cite{Bai2013}. Dans l'une de nos contributions, nous montrons également que l'incertitude quantique locale satisfait la relation de monogamie dans des états quantiques à trois qubits \cite{Slaoui2019} (voir la partie des contributions). La monogamie des corrélations quantiques est une propriété satisfaite par certaines mesures d'intrication dans un scénario multipartite (par exemple, la concurrence). Ici, il peut être généralisé pour une mesure des corrélations quantiques ${\cal Q}$ plus générale. Soit ${\cal Q}$ une mesure des corrélations quantiques arbitraire et $\rho_{AB_{1}B_{2}...B_{N-1}}$ est un état multipartite partagé entre $N$ parties, la condition de monogamie pour une mesure de corrélation quantique ${\cal Q}$ que les corrélations quantiques par paires dans l'état multipartite sont distribuées de telle manière que la relation suivante est satisfaite
\begin{equation}
{\cal Q}\left(\rho_{AB_{1}} \right)+{\cal Q}\left(\rho_{AB_{2}} \right) +...+{\cal Q}\left(\rho_{AB_{N-1}} \right)\leqslant {\cal Q}\left(\rho_{A|B_{1}B_{2}...B_{N-1}} \right),
\end{equation}
où $\rho_{AB_{i}}$ ($i=1,...,N-1$) sont les matrices densité réduites avec $\rho_{AB_{1}}={\rm Tr}_{B_{2}...B_{N-1}}\left[ \rho_{AB_{1}B_{2}...B_{N-1}}\right]$, etc. La violation de l'inégalité ci-dessus implique que la mesure de corrélation quantique ${\cal Q}$ est polygame pour l'état correspondant. Par souci de simplicité, nous désignons ${\cal Q}\left(\rho_{AB_{i}} \right)$ par ${\cal Q}_{AB_{i}}$, et ${\cal Q}\left(\rho_{A|B_{1}B_{2}...B_{N-1}} \right)$ par ${\cal Q}_{A|B_{1}B_{2}...B_{N-1}}$. On peut alors définir une quantité de monogamie $\delta_{{\cal Q}}$ pour l'état à $N$-partite $\rho_{AB_{1}B_{2}...B_{N-1}}$ comme
\begin{equation}
\delta_{{\cal Q}}={\cal Q}_{A|B_{1}B_{2}...B_{N-1}}-\sum_{i=1}^{N-1}{\cal Q}_{AB_{i}},
\end{equation}
et la positivité de $\delta_{{\cal Q}}$ pour tous les états quantiques implique la monogamie de la mesure ${\cal Q}$. Sinon, ${\cal Q}$ est polygame.

\subsection{Loi de conservation entre l'intrication et la discorde quantique}
\subsubsection{Pour les systèmes à trois qubits}
\'Elucider la façon dont la corrélation quantique est distribuée dans les systèmes multipartites est certainement important pour les technologies de traitement de l'information et de la communication dans des scénarios multi-utilisateurs. À cet effet, une série d'efforts majeurs ont été consacrés par la communauté scientifique pour comprendre la distribution des corrélations quantiques dans les systèmes multipartites. Il est bien connu que, pour un état pur bipartite, la définition de la discorde quantique (\ref{QDD}) coïncide avec celle de l'intrication de formation (\ref{EF}). Mais la question de savoir comment ces deux quantités sont liées pour les états mixtes est toujours restée ouverte. Par une extension du système bipartite mixte à sa version tripartite purifiée, Fanchini et ses collaborateurs \cite{Fanchini2011} ont montré comment l'intrication et la discorde quantique se répartissent dans un système pur tripartite arbitraire. Considérons d'abord un système arbitraire représenté par une matrice densité $\rho_{ABC}$ avec $A$ et $B$ représentant deux sous-systèmes et $C$ représentant le systéme de purification ou l'environnement, puisque, dans ce cas, $ABC$ est pure. Comme nous l'avons déjà vu, l'intrication de formation et la corrélation classique sont directement liées dans un système tripartite pur par la relation de Koashi-Winter (\ref{RKW2}),
\begin{equation}
E_{AB}+{\cal J}_{A|C}=S_{A}, \hspace{1cm}{\rm et} \hspace{1cm} E_{AC}+{\cal J}_{A|B}=S_{A},\label{abc}
\end{equation}
qui exprime une sorte de monogamie entre ces mesures distinctes, avec $E_{AB}\equiv E_{f}\left(\rho_{AB}\right)$ est l'intrication de formation entre les sous-systèmes $A$ et $B$, ${\cal J}_{A|C}\equiv{\cal J}_{A}\left(\rho_{AC}\right)$ est la corrélation classique entre $A$ et $C$, $S_{A}\equiv S\left( \rho_{A}\right)$ est l'entropie de von Neumann du sous-système $A$. La même chose s'applique à $E_{AC}$, $E_{BC}$, ${\cal J}_{A|B}$, ${\cal J}_{B|C}$, $S_{B}$ et $S_{C}$. En ajoutant aux deux côtés les informations mutuelles; ${\cal I}_{AC}$ entre $A$ et $C$, et ${\cal I}_{AB}$ entre $A$ et $B$, alors l'équation (\ref{abc}) peut facilement être réécrite pour relier l'intrication et la discorde comme
\begin{equation}
E_{AB}={\cal Q}_{A\mid C}+S_{A\mid C},\hspace{1cm}E_{AC}={\cal Q}_{A\mid B}+S_{A\mid B},
\end{equation}
où $S_{A\mid C}=S_{AC}-S_{C}$ et $S_{A\mid B}=S_{AB}-S_{B}$ sont les entropies conditionnelles. ${\cal Q}_{A\mid C}$ représente la discorde quantique de l'état $\rho_{AC}$ si la mesure est effectuée sur le sous-système $A$, et ${\cal Q}_{A\mid B}$ désigne la discorde quantique de l'état $\rho_{AB}$ si la mesure est effectuée sur le même sous-système $A$. On peut alors considérer le sous-système $A$ comme une particule centrale du système tripartite dans lequel il est toujours mesuré. D'autre part, toujours que nous avons un état pur, si nous le divisons en deux parties, alors les entropies de von Neumann correspondantes seront équivalentes. Par exemple, pour un état pur quadripartite $ABCD$, nous trouverons que $S_{AB}\equiv S_{CD}$, $S_{ABC}\equiv S_{D}$, et $S_{A}\equiv S_{BCD}$, etc. C'est un résultat bien connu de l'entropie de von Neumann (voir le livre de Nielsen et Chuang \cite{Nielsen2000} pour plus de détails). En utilisant ces résultats, nous constatons que
\begin{equation}
S_{A\mid B}+S_{A\mid C}=S_{AB}-S_{A}+S_{AC}-S_{C}=0.\label{123}
\end{equation}
Ainsi, la somme des intrications $E_{AB}$ et $E_{AC}$ conduit à la loi de conservation suivante
\begin{equation}
E_{AB}+E_{AC}={\cal Q}_{A\mid B}+{\cal Q}_{A\mid C}.\label{2183}
\end{equation}
Cela signifie que la somme de toutes les intrications bipartites partagés avec un sous-système particulier (ici le sous-système $A$), mesurées par l'intrication de formation, ne peut pas être augmentée sans augmenter, de la même quantité, la somme de toutes les discordes quantiques partagées avec ce même sous-système.\par

En outre, si l'on effectue la mesure sur toutes les parties du système tripartite, on peut trouver un autre type de la loi de conservation entre l'intrication de formation et la discorde quantique. Cette loi a une forme cyclique de sorte que nous n'avons pas de particule centrale comme dans le cas précédent. Pour le trouver, on effectue d'abord les mesures locales sur les sous-systèmes $B$ et $C$. On trouve ensuite
\begin{equation}
E_{BC}={\cal Q}_{B\mid A}+S_{B\mid A},\hspace{1cm}E_{CA}={\cal Q}_{C\mid B}+S_{C\mid B}.
\end{equation}
De la même manière que l'équation (\ref{123}), on voit facilement que
\begin{equation}
	S_{A\mid C}+S_{B\mid A}+S_{C\mid B}=S_{AC}-S_{C}+S_{AB}-S_{A}+S_{BC}-S_{B}=0,
\end{equation}
et la somme des intrications bipartites devient
\begin{equation}
E_{AB}+E_{BC}+E_{CA}={\cal Q}_{B\mid A}+{\cal Q}_{C\mid B}+{\cal Q}_{A\mid C}.\label{CLy3}
\end{equation}
La relation ci-dessus (\ref{CLy3}) montre que la somme des communications quantiques nécessaires pour simuler la corrélation présente dans chaque bipartitions, quantifiée par l'intrication de formation $E$, est égale à la somme des informations non accessibles par des mesures locales, quantifie par la discorde quantique ${\cal Q}$.
\subsubsection{Pour les systèmes à quatre particules}
Nous allons maintenant traiter le cas d'un système quantique pur à quatre particules $ABCD$, cela équivaut à un système tripartite mixte $ABC$ purifié par un système auxiliaire $D$. Si nous divisons ce système en trois parties, où deux parties ne contiennent qu'un seul sous-système et l'autre partie est composée de deux sous-systèmes, alors les relations de Koashi-Winter deviennent	
\begin{equation}
E_{A\mid BC}={\cal Q}_{A\mid D}+S_{A\mid D}, \hspace{1cm}E_{A\mid CD}={\cal Q}_{A\mid B}+S_{A\mid B}.
\end{equation}
Dans ce cas, la somme des entropies conditionnelles $S_{A\mid D}$ et $S_{A\mid B}$ ne s'annule pas. En utilisant la forte sous-additivité de l'entropie de von Neumann dans un état pur \cite{Lieb1973}, on obtient
\begin{equation}
	S_{A\mid D}+S_{A\mid B}=S_{AB}-S_{B}+S_{BC}-S_{ABC}\geqslant0.
\end{equation}
Par conséquent, l'inégalité de la distribution de l'intrication et de la discorde quantique dans les systèmes à quatre particules, si l'on considère le sous-système $A$ comme une particule centrale, est donnée par
\begin{equation}
E_{A\mid BC}+E_{A\mid CD}\geqslant{\cal Q}_{A\mid B}+{\cal Q}_{A\mid D}.
\end{equation}
De même, les relations de Koashi-Winter des intrications bipartites des sous-systèmes $B$ et $D$ avec la particule centrale $A$ sont
\begin{equation}
E_{AB}={\cal Q}_{A\mid CD}+S_{A\mid CD}, \hspace{1cm}E_{AD}={\cal Q}_{A\mid BC}+S_{A\mid BC},
\end{equation}
et avec un raisonnement similaire, nous obtenons la relation suivante
\begin{equation}
E_{AB}+E_{AD}\leqslant{\cal Q}_{A\mid BC}+{\cal Q}_{A\mid CD}.
\end{equation}
En outre, il est possible de trouver d'autres inégalités entre l'intrication de formation et la discorde quantique en calculant toutes les intrications possibles avec la particule centrale $A$. Pour ce faire, nous devons utiliser les relations fondamentales suivantes
\begin{equation}
E_{A|BC}={\cal Q}_{A|D}+S_{A|D},\hspace{1cm}E_{A|BD}={\cal Q}_{A|C}+S_{A|C},\hspace{1cm}E_{A|CD}={\cal Q}_{A|B}+S_{A|B}.
\end{equation}
Dans ce cas, la somme des entropies conditionnelles est positive car elle peut être écrite comme une combinaison de trois fortes inégalités de sous-additivité \cite{Lieb1973}. Nous avons alors
\begin{equation}
E_{A|BC}+E_{A|CD}+E_{A|DB}\geqslant{\cal Q}_{A|B}+{\cal Q}_{A|C}+{\cal Q}_{A|D}.
\end{equation}
Maintenant, si nous calculons tous les intrications qubit-qubit avec la particule centrale $A$, nous avons les inégalités suivantes
\begin{equation}
E_{AB}={\cal Q}_{A|CD}+S_{A|CD},\hspace{1cm}E_{AC}={\cal Q}_{A|DB}+S_{A|DB},\hspace{1cm}E_{AD}={\cal Q}_{A|CB}+S_{A|CB}.
\end{equation}
Cela conduit à une nouvelle relation de distribution de l'intrication et de la discorde qui s'écrit
\begin{equation}
E_{AB}+E_{AC}+E_{AD}\leqslant{\cal Q}_{A|CD}+{\cal Q}_{A|DB}+{\cal Q}_{A|CB}.
\end{equation}
Les inégalités mentionnées ci-dessus expliquent comment l'intrication et la discorde sont distribués lorsque l'une des particules est traitée comme une particule centrale. Ces résultats montrent qu'il est impossible de trouver une loi de conservation entre l'intrication de formation et la discorde quantique avec une particule centrale dans des systèmes à quatre particules. Mais si l'on effectue des mesures locales sur tous les sous-systèmes, cela donne-t-il une loi de conservation de type cyclique? pour répondre à cette question, il faut d'abord trouver les relations de Koashi-Winter liées à toutes les mesures locales. Dans ce cas, on trouve
\begin{align}
&E_{B|CD}={\cal Q}_{B|A}+S_{B|A},\hspace{2cm}E_{C|DA}={\cal Q}_{C|B}+S_{C|B},\notag\\&E_{D|AB}={\cal Q}_{D|C}+S_{D|C},\hspace{2cm}E_{A|BC}={\cal Q}_{A|D}+S_{A|D}.
\end{align}
Ici, la somme des entropies conditionnelles peut être écrite comme une combinaison de deux fortes inégalités de sous-additivité \cite{Lieb1973}
\begin{equation}
S_{B|A}+S_{A|D}+S_{C|B}+S_{D|C}=\left(S_{AB}+S_{BC}-S_{B}-S_{ABD} \right)+\left(S_{BC}+S_{CD}-S_{C}-S_{BCD} \right)\geqslant0,
\end{equation}
qui est toujours positive, on obtient alors l'inégalité suivante
\begin{equation}
E_{A|BC}+E_{B|CD}+E_{C|DA}+E_{D|AB}\geqslant{\cal Q}_{A|D}+{\cal Q}_{D|C}+{\cal Q}_{C|B}+{\cal Q}_{B|A}.
\end{equation}
Le dernier cas restant est de calculer toutes les intrications bipartites entre les deux sous-systèmes. Cela nous donnera les relations fondamentales suivantes
\begin{align}
&E_{CD}={\cal Q}_{C|AB}+S_{C|AB},\hspace{2cm}E_{AB}={\cal Q}_{A|CD}+S_{A|CD},\notag\\&E_{DA}={\cal Q}_{D|BC}+S_{D|BC},\hspace{2cm}E_{BC}={\cal Q}_{B|AD}+S_{B|AD}.
\end{align}
Avec le même raisonnement, la relation cyclique entre l'intrication de formation et la discorde quantique dans les systèmes à quatre particules s'écrit
\begin{equation}
E_{AB}+E_{BC}+E_{CD}+E_{DA}\leqslant{\cal Q}_{A|CD}+{\cal Q}_{B|DA}+{\cal Q}_{C|AB}+{\cal Q}_{D|BC}.
\end{equation}
\subsubsection{Pour les systèmes à cinq particules}
Pour les systèmes quantiques à cinq particules, il faut trouver les égalités de la distribution des corrélations d'une manière très similaire au cas d'un état en trois parties. Dans ce cas, nous pouvons trouver trois lois de conservation à partir de l'équation (\ref{2183}). Elles sont donnés par
\begin{align}
&E_{A|BC}+E_{A|DE}={\cal Q}_{A|BC}+{\cal Q}_{A|DE},\notag\\&E_{A|BD}+E_{A|CE}={\cal Q}_{A|BD}+{\cal Q}_{A|CE},\notag\\&E_{A|BE}+E_{A|CD}={\cal Q}_{A|BE}+{\cal Q}_{A|CD}.
\end{align}
Il est donc clair que la loi de conservation globale, qui contient tous les sous-systèmes d'un système à cinq particules, s'écrit comme suit
\begin{equation}
E_{A|BC}+E_{A|DE}+E_{A|BD}+E_{A|CE}+E_{A|BE}+E_{A|CD}={\cal Q}_{A|DE}+{\cal Q}_{A|BC}+{\cal Q}_{A|CE}+{\cal Q}_{A|BD}+{\cal Q}_{A|CD}+{\cal Q}_{A|BE}.
\end{equation}
Nous pouvons alors conclure que cette loi de conservation contient toutes les combinaisons d'intrication et de discorde de la particule centrale $A$ avec toutes les autres combinaisons possibles de deux particules. Cela signifie que le cas d'un système à cinq particules est très similaire au cas à trois particules, où la somme des intrications $E$ qu'une particule centrale $A$ partage avec toutes les autres combinaisons possibles de deux particules est égale à la somme de toutes les discordes quantiques ${\cal Q}$ entre les mêmes bipartitions. D'un autre côté, il est également très intéressant de voir si l'on peut retrouver la même loi de conservation cyclique (\ref{CLy3}) si les mesures locales s'appliquent à toutes les parties du système. À partir de l'équation fondamentale de Koashi-Winter, nous pouvons écrire les équations suivantes
\begin{align}
&E_{A|BC}={\cal Q}_{A|ED}+S_{A|ED}={\cal Q}_{A|ED}+S_{BC}-S_{ED}\notag\\&E_{C|DE}={\cal Q}_{C|BA}+S_{C|BA}={\cal Q}_{C|BA}+S_{DE}-S_{BA}\notag\\&E_{E|AB}={\cal Q}_{E|DC}+S_{E|DC}={\cal Q}_{E|DC}+S_{AB}-S_{DE}\notag\\&E_{B|CD}={\cal Q}_{B|AE}+S_{B|AE}={\cal Q}_{B|AE}+S_{CD}-S_{AE}\notag\\&E_{D|AE}={\cal Q}_{D|CB}+S_{D|CB}={\cal Q}_{D|CB}+S_{EA}-S_{CB}.
\end{align}
De la même façon que dans le cas à trois qubits, les entropies négatives annulent les entropies positives. Donc le cycle se ferme et la loi de conservation devient
\begin{equation}
E_{A|BC}+E_{B|CD}+E_{C|DE}+E_{D|EA}+E_{E|AB}={\cal Q}_{A|DE}+{\cal Q}_{B|EA}+{\cal Q}_{C|AB}+{\cal Q}_{D|BC}+{\cal Q}_{E|CD}.\label{lcimpaire}
\end{equation}
Ces résultats montrent que l'on peut généraliser la loi de conservation des corrélations quantiques aux systèmes multipartites purs si le nombre de parties $N$ est impair.
\subsubsection{Pour les systèmes à six particules}
Considérons maintenant le cas d'un système quantique pur à six particules $\rho_{123456}$, qui équivaut également à un système quantique mélange à cinq particules et purifié par un sixième système auxiliaire. La difficulté de trouver la loi de conservation consiste à diviser correctement les parties du système étudié, car il y a plus de façons de le faire en six parties qu'en cinq. Le meilleur choix dans les systèmes à six partis est d'écrire les équations fondamentales comme
\begin{align}
	&E_{1|23}={\cal Q}_{1|654}+S_{1654}-S_{654},\hspace{2cm}E_{4|56}={\cal Q}_{4|321}+S_{4321}-S_{321},\notag\\&E_{3|456}={\cal Q}_{3|21}+S_{321}-S_{21},\hspace{2cm}E_{6|123}={\cal Q}_{6|54}+S_{654}-S_{54},\notag\\&E_{6|12}={\cal Q}_{6|543}+S_{6543}-S_{543},\hspace{2cm}E_{3|45}={\cal Q}_{3|216}+S_{3216}-S_{216},\notag\\&E_{2|345}={\cal Q}_{2|16}+S_{216}-S_{16},\hspace{2cm}E_{5|612}={\cal Q}_{5|43}+S_{543}-S_{43},\notag\\&E_{5|61}={\cal Q}_{5|432}+S_{5432}-S_{432},\hspace{2cm}E_{2|34}={\cal Q}_{2|165}+S_{2165}-S_{165},\notag\\&E_{1|234}={\cal Q}_{1|65}+S_{165}-S_{65},\hspace{2cm}E_{4|561}={\cal Q}_{4|32}+S_{432}-S_{32}.
\end{align}
Dans ce cas, la somme des entropies s'annulle en raison de la propriété de l'entropie de von Neumann dans les états purs (c-à-d, $S_{321}=S_{654}$, $S_{65}=S_{4321}$, ect). Il en résulte la loi de conservation suivante
\begin{align}
	E_{1|23}+&E_{1|234}+E_{2|34}+E_{2|345}+E_{3|45}+E_{3|456}+E_{4|56}+E_{4|561}+E_{5|61}+\notag\\&E_{5|612}+E_{6|12}+E_{6|123}={\cal Q}_{1|654}+{\cal Q}_{1|65}+{\cal Q}_{2|165}+{\cal Q}_{2|16}+\notag\\&{\cal Q}_{3|216}+{\cal Q}_{3|21}+{\cal Q}_{4|321}+{\cal Q}_{4|32}+{\cal Q}_{5|432}+{\cal Q}_{5|43}+{\cal Q}_{6|543}+{\cal Q}_{6|54}.\label{lcpaire}
\end{align}

\subsubsection{Généralisation de la loi de conservation pour les systèmes multipartites à $N$-particules}
Pour expliquer comment la discorde et l'intrication quantique sont distribués dans les systèmes multipartites, il est intéressant de généraliser les équations (\ref{lcimpaire}) et (\ref{lcpaire}) aux systèmes multipartites avec $N$ particules. Pour ce faire, nous devons distinguer deux cas, la première loi de conservation s'applique aux systèmes quantiques multipartites comportant un nombre de parties $N$ paires, et le second au nombre de parties impaires. Dans les états multipartites purs, où $N$ est impair et $n=\left(N-2\right)/2$, nous pouvons écrire les relations de Koashi-Winter comme
\begin{align}
&E_{1|2,3,...,n+1}={\cal Q}_{1|N,N-1,...,n+2}+S_{2,3,...,n+1}-S_{N,N-1,...,n+2},\notag\\&E_{n+1|n+2,n+3,...,N}={\cal Q}_{n+1|n,n-1,...,1}+S_{n+2,n+3,...,N}-S_{n,n-1,...,1},\notag\\&E_{N|1,2,...,n}={\cal Q}_{N|N-1,N-2,...,n+1}+S_{1,2,..,n}-S_{N-1,N-2,...,n+1},\notag\\&E_{n|n+1,n+2,...,N-1}={\cal Q}_{n|n-1,n-2,...,N}+S_{n+1,n+2,...,N-1}-S_{n-1,n-2,...,N},\notag\\&E_{N-1|N,1,2,...,n-1}={\cal Q}_{N-1|N-2,N-3,...,n}+S_{N,1,2,...,n-1}-S_{N-2,N-3,...,n},\notag\\&\hspace{1cm}:\hspace{1.9cm}:\hspace{2cm}:\hspace{3cm}:\notag\\&E_{n+2|n+3,n+4,...,1}={\cal Q}_{n+2|n+1,n,...,2}+S_{n+3,n+4,...,1}-S_{n+1,n,...,2}.
\end{align}
La somme de toutes les équations ci-dessus conduit à l'annulation de toutes les entropies. Ensuite, la loi de conservation générale entre l'intrication et la discorde, pour les états multipartites purs avec un nombre impair de parties $N$, est donnée par
\begin{align}
E_{1|2,3,...,n+1}+&E_{2|3,4,...,n+2}+...+E_{N|1,2,...,n}={\cal Q}_{1|N,N-1,...,n+2}\notag\\&+{\cal Q}_{2|1,N,N-1,...,n+3}+...+{\cal Q}_{N|N-1,N-2,...,n+1}.
\end{align}
Cette loi de conservation généralisée nous donne une relation entre l'intrication de formation et la discorde quantique lorsqu'un seul sous-système est mesuré. Elle montre également quand la somme des informations quantiques nécessaires pour former l'intrication dans les parties bipartites est égale à la somme des informations localement inaccessibles par la mesure.\par
De la même manière, on peut généraliser l'équation (\ref{lcpaire}) aux états multipartites purs avec un nombre de particules $N$ pair. On prend dans ce cas $n=N/2$, alors la loi de conservation entre l'intrication de formation et la discorde entropique est
\begin{align}
E_{1|23..n}+&E_{1|23..n+1}+E_{2|34..n+1}+E_{2|34..n+2}+...+E_{N|12...n-1}+E_{N|12...n}={\cal Q}_{1|N-1,N-2,...,n+1}\notag\\&+{\cal Q}_{1|N-1,N-2,...,n}+{\cal Q}_{2|1,N-1,N-2,...,n+2}+{\cal Q}_{2|1,N-1,N-2,...,n+1}\notag\\&+...+{\cal Q}_{N|N-1,N-2,...,n}+{\cal Q}_{N|N-1,N-2,...,n-1}.
\end{align}
Ces lois de conservation et le principe de monogamie régissent la manière dont l'intrication et la discorde quantique sont distribuées dans les systèmes multipartites. Ces égalités relient les contraintes de l'intrication distribuée à la discorde distribuée et vice versa. Cela signifie que les propriétés de ces deux mesures sont profondément liées.


\chapter{La dynamique des systèmes quantiques ouverts: La cohérence et decohérence quantique}
Dans la plupart des situations réalistes, un système quantique doit être considéré comme un système quantique ouvert, couplé à un environnement externe qui induit à un mélange d'états du système, à savoir la décohérence \cite{Zurek2003}, et à un échange d'énergie entre le système et l'environnement, c'est-à-dire la dissipation \cite{Weiss2012}. Les systèmes quantiques réalistes ne sont jamais complètement isolés de leur environnement, et que lorsqu'un système quantique interagit avec son environnement, il se trouve en général rapidement et fortement intriqué avec un grand nombre de degrés de liberté environnementaux. Cette intrication influence considérablement ce que nous pouvons observer localement en mesurant le système, même lorsque d'un point de vue classique, l'influence de l'environnement sur le système (en termes de dissipation, de perturbations, de bruit, etc) est négligeable \cite{Zurek1991}. Parfois, les effets de ces systèmes ouverts sont faibles, mais ils ne peuvent jamais être ignorés. Cela est particulièrement important dans le domaine du traitement de l'information quantique, où l'existence d'un avantage quantique par rapport au traitement de l'information classique est souvent dérivée d'abord de la perspective idéalisée du système fermé, puis doit être étendue dans le cadre réaliste du système ouvert.\par
L'objectif principal de la théorie des systèmes quantiques ouverts est d'éviter d'avoir à interagir avec le système complet, comprenant à la fois le système quantique ouvert lui-même et son environnement, en décrivant la dynamique du système ouvert dans son espace de Hilbert réduit. La structure de l'état initial système-environnement est fondamentale pour déterminer l'évolution de la matrice densité réduite du système quantique ouvert $\rho_{S}\left(t\right)$, définie en traçant les degrés de liberté de l'environnement par rapport à la matrice densité du système complet $\rho_{T}\left(t\right)$. D'autre part, tout processus physique qui opère ou transforme l'état d'un système quantique est appelé processus quantique, et il y a toujours un hamiltonien qui donne lieu à un tel processus. En principe, l'évolution d'un système quantique fermé $S$ est toujours donnée par une transformation unitaire et l'hamiltonien n'agit que sur l'état du système $\rho_{S}$. Cependant, il est difficile d'isoler un système de son environnement, ce qui conduit à une interaction inévitable avec l'environnement. De plus, leur dynamique peut être classée en deux grandes classes, les processus quantiques Markoviens \cite{Norris1997} et les processus quantiques non-Markoviens \cite{Breuer2016}, selon que le processus d'évolution présente un comportement sans mémoire ou qu'il a des effets de mémoire.\par

En revanche, il est maintenant reconnu que l'origine de l'avantage quantique dans les tâches d'information quantique ne provient pas seulement de la corrélation quantique. En tant qu'autre concept fondamental de la science de l'information quantique, la cohérence quantique a fait l'objet de nombreuses discussions. En effet, la cohérence quantique découlant du principe de superposition quantique est un aspect fondamental de la physique quantique \cite{Nielsen2000}. Le laser \cite{Yang1962} et la superfluidité \cite{Penrose1956} sont des exemples de cohérence quantique, dont les effets sont évidents à l'échelle macroscopique. Cependant, le cadre de quantification de la cohérence n'a été étudié de façon méthodique que récemment. La première tentative d'aborder la classification de la cohérence quantique comme une ressource physique est donnée par Baumgratz et ses collaborateurs \cite{Baumgratz2014} qui ont établi un cadre rigoureux pour la quantification de la cohérence basée sur des mesures de distance dans un espace des états de dimension finie. D'autre part, l'un des principaux objectifs de la théorie de l'information quantique est de trouver des moyens efficaces de préserver la cohérence au sein d'un système quantique. La raison est double; Premièrement, la cohérence représente une caractéristique fondamentale des états quantiques et est liée à toutes les formes de corrélations quantiques \cite{Ficek2005}. Deuxièmement, la cohérence elle-même est une ressource précieuse pour de nombreuses nouvelles technologies quantiques, mais l'interaction inévitable des dispositifs quantiques avec l'environnement décohèrent souvent les états d'entrée et induit une perte de la cohérence, ce qui affaiblit la supériorité de ces technologies quantiques \cite{Nielsen2000}.\par

La première partie de ce chapitre fournit une description de base des systèmes quantiques ouverts à l'interaction avec leur environnement. Nous commencerons par fournir les axiomes de base nécessaires pour étudier et comprendre les systèmes fermés ainsi que les systèmes ouverts. Dans le fil de notre développement, on s'attèle en particulier sur le cas correspondant à une situation très simple, où un système principal interagit avec un réservoir où l'interaction est décrite par le modèle de Jaynes-Cummings \cite{Wolfgang2001}. Dans cette partie, nous dérivons l'équation maîtresse de Born-Markov à partir de l'équation de Liouville-von Neumann. Ensuite, nous dérivons la forme de Lindblad de l'équation maîtresse markovienne qui est un outil important pour le traitement des évolutions non-unitaires. Vers la fin de ce chapitre, nous présentons les outils et le formalisme mathématique de la théorie de la cohérence et ses méthodes de quantification. Puis, nous donnons les liens entre la cohérence quantique et les mesures des corrélations quantiques, et comment on peut quantifier la cohérence en termes d'intrication et de la discorde quantique.
\section{Dynamique d'un système quantique ouvert}
En général, les systèmes physiques sont dynamiques et subissent des processus d'évolution avec le temps. Un processus d'évolution pour un système isolé et fermé est unitaire. Cependant, aucun système quantique ne peut rester isolé de son environnement. Il y a toujours une interaction entre un système et son environnement. L'évolution conjointe du système et de l'environnement est considérée comme une opération unitaire alors que l'évolution locale du système principal peut être non-unitaire. Ce processus non-unitaire provoque un flux d'informations entre le système et l'environnement, qui peut modifier l'entropie du système, et donc modifier la corrélation quantique présente dans le système.
\subsection{Représentation de Kraus}
Une bonne compréhension de la dynamique des systèmes quantiques ouverts est très importante dans de nombreux domaines de la physique, allant de l'optique quantique au traitement de l'information quantique \cite{Davies1976}. En général, on peut supposer une interaction entre le système ouvert, désigné par $S$, et l'environnement, désigné par $E$. Cet environnement est un système quantique auquel est associé un espace de Hilbert de dimension arbitraire. L'ensemble du système $S+E$ évolue de façon unitaire, décrivant une évolution unitaire conjointe donnée par $U\left(t\right)=e^{-iHt}$. Si l'état total initial est $\rho\left(0\right)$, l'équation de Schrödinger conduit à l'état
\begin{equation}
\rho\left(t \right)=U\left(t\right)\rho\left(0 \right)U^{\dagger} \left(t\right).
\end{equation}
Puisque l'opérateur densité de l'environnement $\rho_{E}$ est positif et normalisé, alors il présente une décomposition spectrale sur une base orthonormée avec des valeurs propres positives. Donc nous avons
\begin{equation}
\rho_{E}\left( 0\right)=\sum_{\nu}\vartheta_{\nu}\left| \nu\right\rangle_{E} \left\langle \nu\right|,
\end{equation}
où $\vartheta_{\nu}$ sont les valeurs propres (c'est-à-dire les probabilités) et $\{\left| \nu\right\rangle\}$ sont les vecteurs propres orthonormés correspondants. Il faut mentionner ici que l'état initial du système ouvert S peut varier, mais l'état initial de l'environnement E est considéré comme déterminé par des conditions externes. Dans ce contexte, il est naturel de se demander quelle est l'évolution temporelle du système ouvert $S$? Pour répondre à cette question, nous pourrions suivre les arguments présentés dans le livre de Breuer et Petruccione \cite{Breuer2002} et trouver l'expression explicite de l'opérateur densité $\rho_{S}\left(t \right)$. Premièrement, l'état du système est trouvé en effectuant une trace partielle sur tout le degré de liberté de l'environnement, c'est-à-dire $\rho_{S}\left(t\right)={\rm Tr}_{E}\left[\rho\left(t\right)\right]$. On peut effectuer la trace partielle dans la base orthonormée des états propres de l'environnement comme
\begin{align}
\rho_{S}\left(t\right)&={\rm Tr}_{E}\left[U\left(t\right)\rho\left(0 \right)U^{\dagger}\left(t\right)\right]\notag\\&=\sum_{\nu}\left\langle \nu\right|U\left(t\right)\rho\left(0 \right)U^{\dagger}\left(t\right)\left| \nu\right\rangle. \label{3.3}
\end{align}
Dans la plupart des études sur la dynamique des systèmes ouverts, on suppose que l'état de système ouvert et son environnement est factorisé (séparé) à $t=0$, c'est-à-dire qu'il est complètement découplé et qu'ils sont décrits par l'opérateur densité de la forme
\begin{equation}
\rho_{T}\left(0\right)=\rho_{S}\left(0\right)\otimes\rho_{E}\left(0\right).
\end{equation}
En utilisant l'équation (\ref{3.3}), on obtient alors
\begin{align}
\rho_{S}\left(t\right)&=\sum_{\mu}\left\langle \mu\right|\left[U\left(t\right)\rho_{S}\left(0 \right)\otimes\sum_{\nu}\vartheta_{\nu}\left| \nu\right\rangle \left\langle \nu\right|U^{\dagger}\left(t\right)\right]\left|\mu\right\rangle\notag\\&=\sum_{\mu\nu}\sqrt{\vartheta_{\nu}}\left\langle\mu\right|U\left(t\right)\left|\nu\right\rangle_{E}\rho_{S}\left(0\right)\sqrt{\vartheta_{\nu}}\left\langle \nu\right|U^{\dagger}\left(t\right)\left| \mu\right\rangle_{E}\notag\\&=\sum_{\mu\nu}K_{\mu\nu}\left(t\right)\rho_{S}\left(0\right)K_{\mu\nu}^{\dagger}\left(t\right),  \label{repKraus}   
\end{align}
avec les opérateurs $\{K_{\mu\nu}\}$ sont appelés les opérateurs de Kraus et sont donnés par
\begin{equation}
K_{\mu\nu}\left(t\right)=\sqrt{\vartheta_{\nu}}\left\langle\mu\right|U\left(t\right)\left|\nu\right\rangle.\label{kraus}
\end{equation}
Cette équation (\ref{kraus}) définissant l'évolution du système en termes d'opérateurs de Kraus. D'autre part, l'état du système doit être normalisé à tout moment, donc cela nécessite
\begin{align}
{\rm Tr}\left[\rho_{S}\left(t\right)\right]&= {\rm Tr}\left[\sum_{\mu\nu}K_{\mu\nu}\left(t\right)\rho_{S}\left(0\right)K_{\mu\nu}^{\dagger}\left(t\right) \right]={\rm Tr}\left[\sum_{\mu\nu}K_{\mu\nu}^{\dagger}\left(t\right)K_{\mu\nu}\left(t\right)\rho_{S}\left(0\right) \right]=1.   
\end{align}
Il est facile de vérifier que l'état du système est normalisé si les opérateurs de Kraus satisfont l'identité suivante
\begin{equation}
\sum_{\mu\nu}K_{\mu\nu}^{\dagger}\left(t\right)K_{\mu\nu}\left(t\right)=\openone.
\end{equation}
Cette condition peut être vérifiée en utilisant directement la définition des opérateurs de Kraus, donnée par l'équation (\ref{kraus}). Ceci est obtenu comme
\begin{align}
\sum_{\mu\nu}K_{\mu\nu}^{\dagger}\left(t\right)K_{\mu\nu}\left(t\right)&=\sum_{\mu\nu}\vartheta_{\nu}\left\langle\mu\right|U\left(t\right)\left|\nu\right\rangle\left\langle\nu\right|U^{\dagger}\left(t\right)\left|\mu\right\rangle\notag\\&=\sum_{\nu}\vartheta_{\nu}\left\langle \nu\right| U^{\dagger}\left(t\right) \left(\sum_{\mu}\left|\mu\right\rangle \left\langle\mu\right|\right)U\left(t\right)\left| \nu\right\rangle\notag\\&= \sum_{\nu}\vartheta_{\nu}\left\langle \nu|\nu\right\rangle=\sum_{\nu}\vartheta_{\nu}=1. 
\end{align}
Notez que si nous avons $U\left(t\right)=U_{S}\left(t\right)\otimes U_{E}\left(t\right)$, les opérateurs de Kraus dans ce cas particulier deviennent
\begin{equation}
K_{\mu\nu}\left(t\right)=U_{S}\left(t\right)\sqrt{\vartheta_{\nu}}\left\langle \mu\right|U_{E}^{\dagger}\left(t\right)\left|\nu\right\rangle\equiv \kappa_{\mu\nu}U_{S}\left(t\right),
\end{equation}
avec $\kappa_{\mu\nu}=\sqrt{\vartheta_{\nu}}\left\langle \mu\right|U_{E}^{\dagger}\left(t\right)\left|\nu\right\rangle$, alors la condition de normalisation implique
\begin{equation}
\sum_{\mu\nu}K_{\mu\nu}^{\dagger}\left(t\right)\sum_{\mu\nu}K_{\mu\nu}\left(t\right)=\sum_{\mu\nu}\kappa_{\mu\nu}\kappa_{\mu\nu}^{*}U_{S}^{\dagger}\left(t\right)U_{S}\left(t\right)=\openone,
\end{equation}
cela signifie que $\sum_{\mu\nu}\kappa_{\mu\nu}\kappa_{\mu\nu}^{*}=1$. En conséquence, l'état qui décrit l'évolution d'un système $S$ s'écrit comme suit
\begin{align}
\rho_{S}\left(t\right)=\sum_{\mu\nu}\kappa_{\mu\nu}U_{S}\left(t\right)\rho_{S}\left(0\right)\kappa_{\mu\nu}^{*}U_{S}^{\dagger}\left(t\right)=U_{S}\left(t\right)\rho_{S}\left(0\right)U_{S}^{\dagger}\left(t\right).\label{kr1}
\end{align}
Cette équation (\ref{kr1}) représente le cas d'une évolution dans un système fermé. Ainsi, la représentation des opérateurs de Kraus (\ref{repKraus}) est plus générale que l'équation de Schrödinger, car elle contient cette dernière comme un cas particulier.

\subsection{Principales propriétés des opérateurs de Kraus}
Comme nous l'avons vu dans l'équation (\ref{repKraus}), l'évolution de l'état d'un système quantique ouvert $\rho_{S}$ peut être exprimée comme une évolution unitaire du système tolat composite $\{S+E\}$ et en effectuant une trace partielle sur tout les degrés de liberté de l'environnement $E$. Cela conduit à la représentation de Kraus suivante
\begin{equation}
\rho_{S}\left(t\right)={\rm Tr}_{E}\left[U\left(t\right)\left(\rho_{S}\left(0\right)\otimes\rho_{E}\left(0\right)\right)U^{\dagger}\left(t\right)\right]=\sum_{\beta}K_{\beta}\left(t\right)\rho_{S}\left(0\right)K_{\beta}^{\dagger}\left(t\right),\label{kkraus}
\end{equation}
où nous avons combiné les indices $\mu\nu$ précédents en un seul indice $\beta\equiv\left(\mu\nu\right)$. À partir de l'équation (\ref{kkraus}), nous pouvons considérer la représentation de Kraus comme une application ou un canal où l'état initial passe à l'état final du système ouvert $S$. Il est donc logique d'écrire
\begin{equation}
\rho_{S}\left(t\right)=\varPi\left[\rho_{S}\left(0\right) \right]\hspace{1cm} \longleftrightarrow \hspace{1cm} \varPi:\rho_{S}\left(0\right)\mapsto\rho_{S}\left(t\right),
\end{equation}
avec $\varPi\left[Y\right]\equiv\sum_{\beta}K_{\beta}YK_{\beta}^{\dagger}$. Nous pouvons facilement identifier les trois propriétés clés de l'application du Kraus $\varPi$ comme suit
\begin{enumerate}
	\item[$\divideontimes$] L'application $\varPi$ est préserve la trace, c'est-à-dire
	\begin{align}
{\rm Tr}\left[\varPi\left(\rho \right)\right]=\sum_{\beta}{\rm Tr}\left(K_{\beta}\rho K_{\beta}^{\dagger}\right)=\sum_{\beta}{\rm Tr}\left(K_{\beta}^{\dagger}K_{\beta}\rho\right)={\rm Tr}\left[\underbrace {\sum_{\beta}K_{\beta}^{\dagger}K_{\beta}}_{\openone}\rho\right]={\rm Tr} \left[\rho \right].
	\end{align}
\item[$\divideontimes$] L'application $\varPi$ est linéaire. Nous avons alors
\begin{align}
\varPi\left[a\rho_{a}+b\rho_{b}\right]&=\sum_{\beta}{\rm Tr}\left(K_{\beta}a\rho_{a}K_{\beta}^{\dagger}\right)+\sum_{\beta}{\rm Tr}\left(K_{\beta}b\rho_{b}K_{\beta}^{\dagger}\right)\notag\\&=a\sum_{\beta}{\rm Tr}\left(K_{\beta}\rho_{a}K_{\beta}^{\dagger}\right)+b\sum_{\beta}{\rm Tr}\left(K_{\beta}\rho_{b}K_{\beta}^{\dagger}\right)\notag\\&=a\varPi\left[\rho_{a}\right]+b\varPi\left[\rho_{b}\right]. 
\end{align}
\item[$\divideontimes$] $\varPi$ est positive. Cela signifie que l'application $\varPi$ transforme les opérateurs positifs en opérateurs positifs. Supposons que l'opérateur ${\varTheta}>0$, c'est-à-dire qu'il n'a que des valeurs propres positives et que nous puissions écrire ${\varTheta}=\sum_{i}\eta_{i}\left|i\right\rangle \left\langle i\right|$ avec $\eta_{i}\geqslant0$. Pour prouver que $\varPi\left[{\varTheta}\right] >0$, il suffit de montrer que $\left\langle\nu\right|\varPi\left[{\varTheta}\right]\left|\nu\right\rangle\geqslant0$ pour tout $\left|\nu\right\rangle\in{\cal H}_{S}$. Pour vérifier cela, nous prenons $\left|\chi_{\beta}\right\rangle=K_{\beta}^{\dagger}\left|\nu\right\rangle$, donc
\begin{equation}
\left\langle \nu\right|\varPi\left[{\varTheta}\right]\left|\nu\right\rangle=\sum_{\beta}\left\langle \nu\right| K_{\beta}{\varTheta}K_{\beta}^{\dagger}\left|\nu\right\rangle=\sum_{\beta}\left\langle\chi_{\beta} \right|{\varTheta}\left|\chi_{\beta} \right\rangle=\sum_{i\beta}\eta_{i}|\left\langle\chi_{\beta}|i\right\rangle|^{2}\geqslant0.
\end{equation}	
\end{enumerate}
Par conséquent $\varPi\left[{\varTheta}\right] >0$, et $\varPi$ lui-même est une application positive. De plus, nous pouvons dessiner un diagramme commutatif qui montre l'action de cette application dynamique $\varPi\left(t\right)$ comme suit: 
\begin{equation}
\begin{CD}\rho\left(0\right)=\rho_{S}\left(0\right)\otimes\rho_{E}     @>{\rm Evolution \hspace{0.1cm}unitaire}>>  \rho\left(t\right)= U\left(t\right)\left[\rho_{S}\left(0\right)\otimes\rho_{E}\right]U^{\dagger}\left(t\right)\\@VV{\rm Tr_{E}}V        @VV{\rm Tr_{E}}V\\\rho_{S}\left(0 \right)     @>{\rm Application\hspace{0.1cm}dynamique}>>  \rho_{S}\left(t\right)=\varPi\left[\rho_{S}\left(0\right)\right]\end{CD}
\end{equation}

\subsection{Représentation de Kraus pour une condition initiale générale}

\subsubsection{Pour les états séparables}
Dans cette partie, nous considérons le cas où l'état initial est un état séparable et voyons comment écrire l'état final dans la représentation de Kraus. l'état séparable s'écrit
\begin{equation}
\rho\left(0\right)=\sum_{i}p_{i}\rho_{S}^{i}\otimes\rho_{E}^{i},
\end{equation}
où $\rho_{S}^{i}$ et $\rho_{E}^{i}$ sont respectivement les états du système et de l'environnement et l'état initial du système est $\rho_{S}\left( 0\right)={\rm Tr}_{E}\left[\rho\left(0\right)\right]=\sum_{i}p_{i}\rho_{S}^{i}$. La décomposition spectrale de la matrice densité de l'environnement est
\begin{equation}
\rho_{E}^{i}=\sum_{\nu_{i}}\xi_{\nu_{i}}\left| \nu_{i}\right\rangle\left\langle \nu_{i}\right|.
\end{equation}
Une trace sur les degrés de liberté de l'environnement conduit à
\begin{align}
\rho_{S}\left(t\right)&=\sum_{\mu,i,\nu_{i}}p_{i}\xi_{\nu_{i}}\left\langle \mu\right| U\left(t\right)\rho_{S}^{i}\otimes\left| \nu_{i}\right\rangle\left\langle \nu_{i}\right| U^{\dagger}\left(t\right)\left|\mu\right\rangle\notag\\&=\sum_{\mu,i,\nu_{i}}\xi_{\nu_{i}}\left\langle \mu\right| U\left(t\right) \left|\nu_{i}\right\rangle p_{i}\rho_{S}^{i}\left\langle\nu_{i}\right| U^{\dagger}\left(t\right)\left| \mu\right\rangle.  \label{320}
\end{align}
Si l'on suppose que tous les états $\rho_{E}^{i}$ commut, c'est-à-dire qu'ils sont en diagonale sur la même base de sorte que $\nu_{i}\equiv\nu$, $\xi_{\nu}^{i}\equiv\xi_{\nu}$ et $\rho_{E}^{i}\left( 0\right)=\sum_{\nu}\xi_{\nu}\left| \nu\right\rangle\left\langle \nu\right|$, l'équation (\ref{320}) se réduit à
\begin{equation}
\rho_{S}\left(t\right)=\sum_{\mu\nu}\xi_{\nu}\left\langle \mu\right|U\left(t\right) \left| \nu\right\rangle\sum_{i}p_{i}\rho_{S}^{i}\left\langle \nu\right| U^{\dagger}\left(t\right)\left| \mu\right\rangle,
\end{equation}
et cela peut être représenté comme une application agissant sur l'état $\rho_{S}\left( 0\right)=\sum_{i}p_{i}\rho_{S}^{i}$, c'est-à-dire que nous revenons à nouveau à l'équation (\ref{kraus}). Cela n'est vrai que pour le cas où $\rho_{E}^{i}=\rho_{E}$, mais dans le cas géneral où $\rho_{E}^{i}\neq\rho_{E}$, nous ne pouvons pas écrire l'état d'évolution du système $S$ dans la représentation de Kraus.
\subsubsection{Pour les états initiaux généraux}
En utilisant une base orthonormée générale pour l'espace de Hilbert conjoint ${\cal H}_{S+E}$, nous pouvons toujours écrire l'état bipartite comme
\begin{equation}
\rho\left(0\right)=\sum_{ijkl}\xi_{ijkl}\left|i\right\rangle \left\langle j\right|\otimes\left|k\right\rangle \left\langle l\right|.
\end{equation}
Ensuite, l'état initial du système $S$ correspondant devient
\begin{equation}
\rho_{S}\left(0\right)={\rm Tr}_{E}\left[\rho\left(0\right) \right]=\sum_{ijk}\xi_{ijkk}\left|i\right\rangle \left\langle j\right|.
\end{equation}
Si nous suivons les mêmes étapes que dans la dérivation des opérateurs de Kraus, nous obtenons
\begin{align}
\rho_{S}\left(t\right)&=\sum_{\mu}\left\langle \mu\right| U\left(t\right) \rho\left(0\right) U^{\dagger}\left(t\right)\left| \mu\right\rangle\notag\\&=\sum_{\mu ijkl} \xi_{ijkl}\left\langle \mu\right| U\left(t\right) \left|i\right\rangle \left\langle j\right|\otimes\left| k\right\rangle \left\langle l\right| U^{\dagger}\left(t\right)\left| \mu\right\rangle\notag\\&=\sum_{\mu k}\left\langle \mu\right| U\left(t\right)\left|k\right\rangle \left(\sum_{ij}\xi_{ijkk}\left|i\right\rangle\left\langle j\right| \right)\left\langle k\right| U^{\dagger}\left(t\right)\left| \mu\right\rangle\notag\\&+\sum_{\mu,i,j,k\neq l}\left\langle \mu\right| U\left(t\right)\left|k\right\rangle \left(\sum_{ij}\xi_{ijkl}\left|i\right\rangle\left\langle j\right| \right)\left\langle l\right| U^{\dagger}\left(t\right)\left| \mu\right\rangle. \label{rhosig}
\end{align}
Le premier terme de l'équation (\ref{rhosig}) comporte des opérateurs de Kraus $\left\langle \mu\right| U\left(t\right)\left|k\right\rangle$. Cependant, le second terme ne peut pas être écrit sous la forme de Kraus, de sorte que nous n'obtenons pas d'application qui transforme l'état $\rho_{S}\left( 0\right)$ en état $\rho_{S}\left( t\right)$.

\section{Equation Maîtresse de Born-Markov} 
Les systèmes quantiques ouverts sont généralement difficiles à étudier car cette étude implique certaines corrélations entre le système et l'environnement, de sorte que l'état du système $S$ ne peut pas être décrit par une transformation unitaire. Par conséquent, il faut procéder à des séries d'approximations afin de dériver l'équation d'évolution de l'état du système, de la résoudre d'une certaine manière, et donc d'appréhender la dynamique du système étudié.\par
Considérons une situation physique où un système principal $S$ est couplé à un autre système quantique $R$, appelé réservoir. Ici, ${\cal H}_{S}$ et ${\cal H}_{R}$ sont, respectivement, les espaces de Hilbert du système principal $S$ et du réservoir $R$, de sorte que l'état global du système total sera représenté par un vecteur appartient à l'espace du produit tensoriel ${\cal H}_{S}\otimes{\cal H}_{R}$. L'Hamiltonien du système total est
\begin{equation}
{\hat H }\left( t\right)={\hat H}_{S}\otimes{\hat\openone}_{R}+{\hat\openone}_{S}\otimes{\hat H}_{R}+{\hat H}_{SR},
\end{equation}
où ${\hat H}_{S}$ est l'Hamiltonien du système principal $S$, ${\hat H}_{R}$ est celui de l'environnement (réservoir $R$), ${\hat H}_{SR}$ décrit l'opérateur qui correspond à l'interaction entre le système $S$ et son environnement $R$, ${\hat\openone}_{S}$ et ${\hat\openone}_{R}$ sont les opérateurs identités de l'espace de Hilbert ${\cal H}_{S}$ et de ${\cal H}_{R}$ respectivement. Par souci de simplicité, nous ignorons le symbole du produit tensoriel $\otimes$ et nous écrivons
\begin{equation}
	{\hat H }\left( t\right)={\hat H}_{S}+{\hat H}_{R}+{\hat H}_{SR}.
\end{equation}
L'état du système total est décrit par la matrice densité $\rho_{SR}$, puis la matrice densité réduite $\rho_{S}$ du système principal $S$ est obtenue en prenant la trace partielle sur tous les degrés de liberté du réservoir $R$. Puisque le système total est supposé être fermé, son évolution est donc décrite par l'équation de Liouville-von Neumann suivante
\begin{equation}
\frac{d}{dt}\rho_{SR}\left(t\right)=-\frac{i}{\hbar}\left[{\hat H}_{S}+{\hat H}_{R}+{\hat H}_{SR},\rho_{SR}\left(t\right) \right]. 
\end{equation}
Dans la représentation d'interaction, la matrice densité $\rho_{SR}$ et l'Hamiltonien ${\hat H}_{SR}$ deviennent
\begin{equation}
	\tilde{\rho}_{SR}\left( t\right)=e^{\frac{i}{\hbar}\left( {\hat H}_{S}+{\hat H}_{R}\right)t}\rho_{SR}\left(t\right)e^{-\frac{i}{\hbar}\left( {\hat H}_{S}+{\hat H}_{R}\right)t},
\end{equation}
et
\begin{equation}
{\hat H }_{I}\left( t\right)=e^{\frac{i}{\hbar}\left( {\hat H}_{S}+{\hat H}_{R}\right)t}{\hat H}_{SR}e^{-\frac{i}{\hbar}\left( {\hat H}_{S}+{\hat H}_{R}\right)t}.\label{HI}
\end{equation}
Il convient de noter ici que la représentation d'interaction est une façon de traiter les problèmes d'évolution des Hamiltoniens qui dépendant du temps, et qu'elle est plus adéquate à utiliser dans certains cas que la représentation de Schrödinger. En outre, la représentation d'interaction ne modifie pas la physique, c'est-à-dire que la valeur moyenne d'un opérateur arbitraire dans la représentation d'interaction coïncide avec sa valeur moyenne dans la représentation de Schrödinger. Ainsi, l'équation de Liouville-von Neumann pour $\tilde{\rho}_{SR}$ sera
\begin{equation}
	\frac{d}{dt}\tilde{\rho}_{SR}\left(t\right)=-\frac{i}{\hbar}\left[{\hat H }_{I}\left( t\right),\tilde{\rho}_{SR}\left(t\right)\right]. \label{66}
\end{equation}
Après intégration, nous obtenons
\begin{equation}
\tilde{\rho}_{SR}\left(t\right)=\tilde{\rho}_{SR}\left(0\right)-\frac{i}{\hbar}\int_{0}^{t}\left[{\hat H }_{I}\left( t'\right),\tilde{\rho}_{SR}\left(t'\right)\right]dt'. \label{88}
\end{equation}
Nous insérons le résultat (\ref{88}) dans l'équation (\ref{66}), on trouve alors
\begin{equation}
\frac{d}{dt}\tilde{\rho}_{SR}\left(t\right)=-\frac{i}{\hbar}\left[{\hat H }_{I}\left( t\right),\tilde{\rho}_{SR}\left(0\right)\right]-\frac{1}{\hbar^{2}}\left[{\hat H }_{I}\left(t\right),\int_{0}^{t}\left[{\hat H }_{I}\left(t'\right),\tilde{\rho}_{SR}\left(t'\right) \right] dt'\right]. 
\end{equation}
Pour connaître l'évolution du système principal $\tilde{\rho}_{S}\left(t\right)$, nous prenons la trace partielle des degrés de liberté du réservoir comme
\begin{equation}
	\frac{d}{dt}\tilde{\rho}_{S}\left(t\right)=-\frac{i}{\hbar}{Tr}_{R}\left\lbrace\left[{\hat H }_{I}\left( t\right),\tilde{\rho}_{SR}\left(0\right)\right]\right\rbrace-\frac{1}{\hbar^{2}}{Tr}_{R}\left\lbrace \left[{\hat H }_{I}\left(t\right),\int_{0}^{t}\left[{\hat H }_{I}\left(t'\right),\tilde{\rho}_{SR}\left(t'\right) \right] dt'\right]\right\rbrace.\label{oo}
\end{equation}
Dans la représentation d'interaction de l'Hamiltonien du système total (\ref{HI}), ${\hat H }_{I}$ dépend de ${\hat H}_{SR}$, alors l'Hamiltonien ${\hat H}_{SR}$ peut toujours être défini de sorte que le premier terme de l'équation (\ref{oo}) sera égal à zéro, c'est-à-dire $\left[{\hat H }_{I}\left( t\right),\tilde{\rho}_{SR}\left(0\right)\right]=0$. Par conséquent, nous obtenons
\begin{equation}
	\frac{d}{dt}\tilde{\rho}_{S}\left(t\right)=-\frac{1}{\hbar^{2}}{Tr}_{R}\left\lbrace \left[{\hat H }_{I}\left(t\right),\int_{0}^{t}\left[{\hat H }_{I}\left(t'\right),\tilde{\rho}_{SR}\left(t'\right) \right] dt'\right]\right\rbrace.\label{eqmatp}
\end{equation}
Dans le cas général, il est difficile de résoudre exactement l'équation (\ref{eqmatp}), le prototype le plus simple à résoudre est fourni par un processus markovien pour lequel la dynamique est stationnaire dans le temps et tous les effets de mémoire sont ignorés. Ceci signifie que l'environnement ne se souvient plus de ses interactions passées avec le système en raison de la dispersion des corrélations dans les nombreux degrés de liberté environnementaux. Ce processus est basé sur deux approximations qui restent valables dans plusieurs cas:\par
\textbf{Approximation de Born:} La première approximation est l'approximation de Born qui suppose que le couplage système-environnement est suffisamment faible pour que l'influence du système considéré sur le réservoir soit absente pendant l'évolution, et donc le système principal $S$ n'aura que peu d'infuence sur le réservoir $R$. Cela signifie que l'état du système total puisse être factorisé pour $t>0$ comme $\rho\left(t\right)\equiv\rho_{S}\left(t\right)\otimes\rho_{R}\left(t\right)$, et l'état du réservoir $\rho_{R}$ restant constant au cours du temps.\par
\textbf{Approximation de Markov:} La seconde approximation appelée l'approximation de Markov repose sur le fait que les effets de mémoire du réservoir sont négligeables à long terme. Cela qui signifie que la dynamique du système principal S à l'instant $t$ ne dépend pas de son état passé $\rho_{S}\left(t'\right)$ (avec $t'<t$), de sorte que $\rho_{S}\left(t'\right)$ est remplacé dans l'intégrale (\ref{eqmatp}) par $\rho_{S}\left(t\right)$ à l'instant présent.\par
Il s'en suit que l'équation d'évolution dans le cadre de l'approximation de
Born-Markov est donnée par
\begin{equation}
	\frac{d}{dt}\tilde{\rho}_{S}\left(t\right)=-\frac{1}{\hbar^{2}}\int_{0}^{\infty}{Tr}_{R}\left\lbrace \left[{\hat H }_{I}\left(t\right),\left[{\hat H }_{I}\left(t'\right),\tilde{\rho}_{S}\left(t\right)\tilde{\rho}_{R}\right] \right]\right\rbrace dt',\label{EBM}
\end{equation}
où nous supposons que l'intégration peut être étendue à l'infini sans changer son résultat. L'équation (\ref{EBM}) est l'équation Maîtresse de Born-Markov \cite{Breuer2002}. Il s'agit d'une approche familiére dans l'étude des systèmes ouverts dans de nombreux contextes, de son développement original dans l'étude de la relaxation des spins dans les champs magnétiques aux applications modernes en imagerie médicale par résonance magnétique, en science des matériaux ou en biophysique \cite{Pollard1996,Jean1992,Ishizaki2010}.

\begin{figure}[H]
	\centerline{\includegraphics[width=14cm]{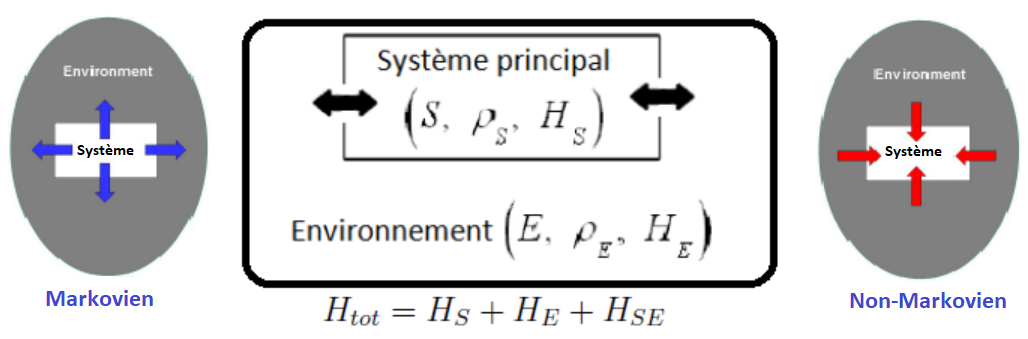}}
	\caption{Schéma d'interaction entre un système quantique ouvert $S$ et son environnement $E$ dans le cas markovien et non-markovien.}
	\label{decoherence}
\end{figure}
\section{La forme de Lindblad de l'équation Maîtresse Markovienne}
\subsection{L'équation générale de Lindblad}
En réalité, on est loin d'obtenir une équation générale d'évolution non unitaire de la matrice densité $\tilde{\rho}_{S}\left(t\right)$, car on s'attend à des effets de mémoire. En outre, non seulement l'information s'échappe du système vers l'environnement, mais elle passe également de l'environnement au système. Dans de nombreux modèles de décohérence markovienne, l'équation maîtresse markovienne est largement utilisée dans la description de la dynamique de la décohérence. Elle permet un calcul relativement simple de la dynamique réduite tout en fournissant, dans de nombreux cas d'intérêt pratique, une bonne approximation de la dynamique exacte et des données observées expérimentalement. Dans ce sens, Lindblad \cite{Lindblad1976} a donné une structure mathématique bien définie à l'équation d'évolution principale (\ref{EBM}). Pour rendre cette équation plus compréhensible, nous présentons ici sa déduction dans le cas spécifique de deux systèmes; le système principal $S$ et le réservoir $R$ avec une interaction entre eux qui ressemble à celle du modèle de Jaynes Cummings. Nous écrivons d'abord l'hamiltonien de l'interaction système-réservoir sous la forme la plus générale suivante
\begin{equation}
{\hat H}_{SR}=\hbar\left({\hat S}{\hat R}^{\dagger}+{\hat S}^{\dagger}{\hat R}\right),\label{Hint}
\end{equation}
où ${\hat S}$ est un opérateur général qui agit uniquement sur le système principal $S$, et ${\hat R}$ est un opérateur qui agit uniquement sur le réservoir $R$. Maintenant, nous considérons que ${\hat S}$ commute avec ${\hat H}_{S}$, c'est-à-dire $\left[{\hat S},{\hat H}_{S}\right]=0$. Cela signifie que l'opérateur ${\hat S}$ n'est pas affecté par la transformation de l'image d'interaction, nous pouvons alors écrire ${\hat S}\left(t\right)\equiv{\hat S}$.\par
Généralement, les systèmes physiques peuvent être représentés soit par un qubit (c-à-d une particule de spin-$1/2$) si l'espace d'état du système est discret, soit par une particule décrite par des coordonnées continues de l'espace de phase. De même, un large éventail d'environnements peut être modélisé comme une collection d'oscillateurs harmoniques quantiques (un environnement d'oscillateur qui représente un quasi-continuum de modes bosoniques délocalisés) ou par des qubits (un environnement de spin qui représente une collection de modes localisés avec les modes discrets). Considérons ici l'Hamiltonien de réservoir qui est défini par une collection de bosons comme
\begin{equation}
{\hat H}_{R}=\hbar\sum_{k}w_{k}{\hat a}_{k}^{\dagger}{\hat a}_{k},
\end{equation}
où ${\hat a}_{k}^{\dagger}$ et ${\hat a}_{k}$ sont les opérateurs de création et d'annihilation de réservoir, les quantités $w_{k}$ sont les fréquences caractéristiques de chaque mode $k$. L'opérateur ${\hat R}$ qui apparaît dans l'équation (\ref{Hint}) est défini par
\begin{equation}
{\hat R}=\sum_{k}g_{k}^{*}{\hat a}_{k},\label{RINT}
\end{equation}
où $g_{k}$ sont des coefficients complexes représentant des constantes de couplage. Ensuite, dans l'image d'interaction, nous avons
\begin{equation}
\tilde{R}\left( t\right)=e^{\frac{i}{\hbar}{\hat H}_{R}t} {\hat R} e^{\frac{-i}{\hbar}{\hat H}_{R}t}.\label{Rint}
\end{equation}
En développant chaque exponentielle et en utilisant les relations de commutation bosonique, l'équation (\ref{Rint}) conduit à
\begin{equation}
\tilde{R}\left( t\right)=\sum_{k}g_{k}^{*} {\hat  a}_{k} e^{-iw_{k}t},\hspace{1cm}{\rm et}\hspace{1cm}\tilde{R}^{\dagger}\left(t\right)=\sum_{k}g_{k} {\hat  a}_{k}^{\dagger} e^{iw_{k}t}.\label{RR}
\end{equation}
Il est intéressant de noter que l'interaction (\ref{Hint}) avec la définition (\ref{RINT}) ressemble à celle de Jaynes-Cummings \cite{Wolfgang2001}, qui représente un seul atome à deux niveaux interagissant avec un seul mode du champ de rayonnement, et dans l'une de nos contributions \cite{SlaouiShaukat2018}, nous avons utilisé le même modèle pour un système de deux atomes à deux niveaux en présence de modes de champ électromagnétique (voir la partie des contributions). En adaptant ces définitions, le commutateur dans l'équation maîtresse de Born-Markov (\ref{EBM}) sera évalué analytiquement. Nous commençons d'abord par
\begin{align}
\left[{\hat H}\left(t\right),\left[{\hat H}\left(t'\right),\rho_{R} \tilde{\rho}_{S}\left( t\right)\right] \right]=\hbar \left[{\hat S}{\tilde{R}}^{\dagger}\left(t\right),\left[{\hat H}\left(t'\right),\rho_{R} \tilde{\rho}_{S}\left( t\right)\right] \right]+\hbar \left[{\hat S}^{\dagger}{\tilde{R}}\left(t\right),\left[{\hat H}\left(t'\right),\rho_{R} \tilde{\rho}_{S}\left( t\right)\right] \right].\label{42}
\end{align}
Le développement de chaque terme de l'équation (\ref{42}) conduit à
\begin{align}
\left[{\hat S}{\tilde{R}}^{\dagger}\left(t\right),\left[{\hat H}\left(t'\right),\rho_{R} \tilde{\rho}_{S}\left( t\right)\right] \right]&=\hbar{\hat S}{\hat S}\tilde{\rho}_{S}\left( t\right){\tilde{R}}^{\dagger}\left(t\right){\tilde{R}}\left(t'\right)\rho_{R}+\hbar{\hat S}{\hat S}^{\dagger}\tilde{\rho}_{S}\left( t\right){\tilde{R}}^{\dagger}\left(t\right){\tilde{R}}\left(t'\right)\rho_{R}\notag\\&-\hbar{\hat S}{\tilde{\rho}_{S}\left( t\right)}{\hat S}{\tilde{R}}^{\dagger}\left(t\right)\rho_{R}{\tilde{R}}^{\dagger}\left(t'\right)-\hbar{\hat S}{\tilde{\rho}_{S}\left( t\right)}{\hat S}^{\dagger}{\tilde{R}}^{\dagger}\left(t\right)\rho_{R}{\tilde{R}}\left(t'\right)\notag\\&-\hbar{\hat S}{\tilde{\rho}_{S}\left( t\right)}{\hat S}{\tilde{R}}^{\dagger}\left(t'\right)\rho_{R}{\tilde{R}}^{\dagger}\left(t\right)-\hbar{\hat S}^{\dagger}{\tilde{\rho}_{S}\left( t\right)}{\hat S}{\tilde{R}}\left(t'\right)\rho_{R}{\tilde{R}}^{\dagger}\left(t\right)\notag\\&+\hbar{\tilde{\rho}_{S}\left( t\right)}{\hat S}{\hat S}\rho_{R}{\tilde{R}}^{\dagger}\left(t'\right){\tilde{R}}^{\dagger}\left(t\right)+\hbar{\tilde{\rho}_{S}\left( t\right)}{\hat S}^{\dagger}{\hat S}\rho_{R}{\tilde{R}}\left(t'\right){\tilde{R}}^{\dagger}\left(t\right). \label{43}
\end{align}
De la même façon, nous pouvons voir que
\begin{align}
	\left[{\hat S}^{\dagger}{\tilde{R}}\left(t\right),\left[{\hat H}\left(t'\right),\rho_{R} \tilde{\rho}_{S}\left( t\right)\right] \right]&=\hbar{\hat S}^{\dagger}{\hat S}\tilde{\rho}_{S}\left( t\right){\tilde{R}}\left(t\right){\tilde{R}}^{\dagger}\left(t'\right)\rho_{R}+\hbar{\hat S}^{\dagger}{\hat S}^{\dagger}\tilde{\rho}_{S}\left( t\right){\tilde{R}}\left(t\right){\tilde{R}}\left(t'\right)\rho_{R}\notag\\&-\hbar{\hat S}^{\dagger}{\tilde{\rho}_{S}\left( t\right)}{\hat S}{\tilde{R}}\left(t\right)\rho_{R}{\tilde{R}}^{\dagger}\left(t'\right)-\hbar{\hat S}^{\dagger}{\tilde{\rho}_{S}\left( t\right)}{\hat S}^{\dagger}{\tilde{R}}\left(t\right)\rho_{R}{\tilde{R}}\left(t'\right)\notag\\&-\hbar{\hat S}{\tilde{\rho}_{S}\left( t\right)}{\hat S}^{\dagger}{\tilde{R}}^{\dagger}\left(t'\right)\rho_{R}{\tilde{R}}\left(t\right)-\hbar{\hat S}^{\dagger}{\tilde{\rho}_{S}\left( t\right)}{\hat S}^{\dagger}{\tilde{R}}\left(t'\right)\rho_{R}{\tilde{R}}\left(t\right)\notag\\&+\hbar{\tilde{\rho}_{S}\left( t\right)}{\hat S}{\hat S}^{\dagger}\rho_{R}{\tilde{R}}^{\dagger}\left(t'\right){\tilde{R}}\left(t\right)+\hbar{\tilde{\rho}_{S}\left( t\right)}{\hat S}^{\dagger}{\hat S}^{\dagger}\rho_{R}{\tilde{R}}\left(t'\right){\tilde{R}}\left(t\right).\label{44}
\end{align}
A l'aide de l'équation (\ref{RR}), il est facile de voir que
\begin{equation}
{\rm Tr}_{R}\left\lbrace{\tilde{R}}\left(t\right){\tilde{R}}\left(t'\right)\rho_{R} \right\rbrace ={\rm Tr}_{R}\left\lbrace{\tilde{R}}^{\dagger}\left(t\right){\tilde{R}}^{\dagger}\left(t'\right)\rho_{R} \right\rbrace =0.
\end{equation}
En substituant ces résultats dans les équations (\ref{43}) et (\ref{44}), et en utilisant les propriétés cycliques de la trace, nous obtenons 
\begin{align}
{\rm Tr}_{R}\left\lbrace \left[{\hat S}{\tilde{R}}^{\dagger}\left(t\right),\left[{\hat H}\left(t'\right),\rho_{R} \tilde{\rho}_{S}\left( t\right)\right] \right]\right\rbrace&=\hbar\left[{\hat S}{\hat S}^{\dagger}{\tilde{\rho}_{S}\left( t\right)}-{\hat S}^{\dagger}{\tilde{\rho}_{S}\left( t\right)}{\hat S} \right]{\rm Tr}_{R}\left\lbrace{\tilde{R}}^{\dagger}\left(t\right){\tilde{R}}\left(t'\right)\rho_{R}\right\rbrace \notag\\&+ \hbar\left[{\tilde{\rho}_{S}\left( t\right)}{\hat S}^{\dagger}{\hat S}-{\hat S}{\tilde{\rho}_{S}\left( t\right)}{\hat S}^{\dagger}\right]{\rm Tr}_{R}\left\lbrace{\tilde{R}}\left(t'\right){\tilde{R}}^{\dagger}\left(t\right)\rho_{R}\right\rbrace,
\end{align}
et
\begin{align}
	{\rm Tr}_{R}\left\lbrace \left[{\hat S}^{\dagger}{\tilde{R}}\left(t\right),\left[{\hat H}\left(t'\right),\rho_{R} \tilde{\rho}_{S}\left( t\right)\right] \right]\right\rbrace&=\hbar\left[{\hat S}^{\dagger}{\hat S}{\tilde{\rho}_{S}\left( t\right)}-{\hat S}{\tilde{\rho}_{S}\left( t\right)}{\hat S}^{\dagger} \right]{\rm Tr}_{R}\left\lbrace{\tilde{R}}\left(t\right){\tilde{R}}^{\dagger}\left(t'\right)\rho_{R}\right\rbrace \notag\\&+ \hbar\left[{\tilde{\rho}_{S}\left( t\right)}{\hat S}{\hat S}^{\dagger}-{\hat S}^{\dagger}{\tilde{\rho}_{S}\left( t\right)}{\hat S}\right]{\rm Tr}_{R}\left\lbrace{\tilde{R}}^{\dagger}\left(t'\right){\tilde{R}}\left(t\right)\rho_{R}\right\rbrace.
\end{align}
Les résultats donnés par ces équations nous permettent de revenir à l'équation (\ref{42}) et de l'écrire comme
\begin{align}
{\rm Tr}_{R}\left\lbrace \left[{\hat H}\left(t\right),\left[{\hat H}\left(t'\right),\rho_{R} \tilde{\rho}_{S}\left( t\right)\right] \right]\right\rbrace&=\hbar^{2}\left[{\hat S}{\hat S}^{\dagger}{\tilde{\rho}_{S}\left( t\right)}-{\hat S}^{\dagger}{\tilde{\rho}_{S}\left( t\right)}{\hat S} \right]{\rm Tr}_{R}\left\lbrace{\tilde{R}}^{\dagger}\left(t\right){\tilde{R}}\left(t'\right)\rho_{R}\right\rbrace \notag\\&+ \hbar^{2}\left[{\tilde{\rho}_{S}\left( t\right)}{\hat S}^{\dagger}{\hat S}-{\hat S}{\tilde{\rho}_{S}\left( t\right)}{\hat S}^{\dagger}\right]{\rm Tr}_{R}\left\lbrace{\tilde{R}}\left(t'\right){\tilde{R}}^{\dagger}\left(t\right)\rho_{R}\right\rbrace\notag\\&+\hbar^{2}\left[{\hat S}^{\dagger}{\hat S}{\tilde{\rho}_{S}\left( t\right)}-{\hat S}{\tilde{\rho}_{S}\left( t\right)}{\hat S}^{\dagger} \right]{\rm Tr}_{R}\left\lbrace{\tilde{R}}\left(t\right){\tilde{R}}^{\dagger}\left(t'\right)\rho_{R}\right\rbrace \notag\\&+ \hbar^{2}\left[{\tilde{\rho}_{S}\left( t\right)}{\hat S}{\hat S}^{\dagger}-{\hat S}^{\dagger}{\tilde{\rho}_{S}\left( t\right)}{\hat S}\right]{\rm Tr}_{R}\left\lbrace{\tilde{R}}^{\dagger}\left(t'\right){\tilde{R}}\left(t\right)\rho_{R}\right\rbrace.\label{48}
\end{align}
Pour plus de simplicité, nous allons définir les fonctions suivantes
\begin{equation}
F\left( t\right)=\int_{0}^{t}{\rm Tr}_{R}\left\lbrace{\tilde{R}}\left(t\right){\tilde{R}}^{\dagger}\left(t'\right)\rho_{R} \right\rbrace dt', \hspace{0.75cm}{\rm et}\hspace{0.75cm}G\left(t\right)=\int_{0}^{t}{\rm Tr}_{R}\left\lbrace{\tilde{R}}^{\dagger}\left(t'\right){\tilde{R}}\left(t\right)\rho_{R} \right\rbrace dt'.\label{FG}
\end{equation}
Nous avons alors
\begin{equation}
F^{*}\left( t\right)=\int_{0}^{t}{\rm Tr}_{R}\left\lbrace{\tilde{R}}\left(t'\right){\tilde{R}}^{\dagger}\left(t\right)\rho_{R} \right\rbrace dt',\hspace{0.75cm}{\rm et}\hspace{0.75cm}G^{*}\left( t\right)=\int_{0}^{t}{\rm Tr}_{R}\left\lbrace{\tilde{R}}^{\dagger}\left(t\right){\tilde{R}}\left(t'\right)\rho_{R} \right\rbrace dt',
\end{equation}
où $*$ désigne la conjugaison complexe. Ensuite, en remplaçant l'équation (\ref{48}) dans l'équation d'évolution de Born-Markov (\ref{EBM}), nous trouvons facilement l'expression suivante
\begin{align}
\frac{d}{dt}\tilde{\rho}_{S}\left(t\right)=&-\left[{\hat S}{\hat S}^{\dagger}{\tilde{\rho}_{S}\left( t\right)}-{\hat S}^{\dagger}{\tilde{\rho}_{S}\left( t\right)}{\hat S} \right]G^{*}\left(t\right)- \left[{\tilde{\rho}_{S}\left( t\right)}{\hat S}^{\dagger}{\hat S}-{\hat S}{\tilde{\rho}_{S}\left( t\right)}{\hat S}^{\dagger}\right]F^{*}\left(t\right) \notag\\&-\left[{\hat S}^{\dagger}{\hat S}{\tilde{\rho}_{S}\left( t\right)}-{\hat S}{\tilde{\rho}_{S}\left( t\right)}{\hat S}^{\dagger} \right]F\left( t\right)-\left[{\tilde{\rho}_{S}\left( t\right)}{\hat S}{\hat S}^{\dagger}-{\hat S}^{\dagger}{\tilde{\rho}_{S}\left( t\right)}{\hat S}\right]G\left(t\right).\label{FGGF}
\end{align}
Dans ce qui suit, nous donnons quelques caractéristiques spécifiques de l'environnement pour examiner ces approximations en détail.
\subsection{La spécification du réservoir}
Premièrement, nous considérons l'état initial du réservoir thermique comme un état de vide
\begin{equation}
\rho_{R}=\left(\left|0\right\rangle\left|0\right\rangle...\left|0\right\rangle  \right)\otimes\left(\left\langle 0\right|\left\langle 0\right|...\left\langle 0\right|\right). \label{evide} 
\end{equation}
Ensuite, l'évaluation des fonctions $F\left(t\right)$ et $G\left(t\right)$ définies dans l'équation (\ref{FG}) est faite en considérant les définitions de $\tilde{R}\left(t\right)$, $\tilde{R}^{\dagger}\left(t\right)$ et $\rho_{R}$ dans les équations (\ref{RR}) et (\ref{evide}). Si nous utilisons la base des états du réservoir dénoteé par $\left\lbrace\left| r\right\rangle \right\rbrace$, la trace partielle apparaissant dans les expressions de $F\left(t\right)$ et $G\left(t\right)$ peut être évaluée comme
\begin{align}
{\rm Tr}_{R}\left\lbrace {\hat R}\left(t \right){\hat R}^{\dagger}\left(t' \right)\rho_{R} \right\rbrace&=\sum_{r}\left\langle r\right| {\hat R}\left(t \right){\hat R}^{\dagger}\left(t' \right)\left(\left|0\right\rangle\left|0\right\rangle...\left|0\right\rangle  \right)\otimes\left(\left\langle 0\right|\left\langle 0\right|...\left\langle 0\right|\right)\left| r\right\rangle\notag\\&= \left(\left\langle 0\right|\left\langle 0\right|...\left\langle 0\right|\right)\sum_{r}\left|r \right\rangle\left\langle r\right|{\hat R}\left(t \right){\hat R}^{\dagger}\left(t'\right)\left(\left|0\right\rangle\left|0\right\rangle...\left|0\right\rangle  \right)\notag\\&=\left(\left\langle 0\right|\left\langle 0\right|...\left\langle 0\right|\right){\hat R}\left(t \right){\hat R}^{\dagger}\left(t'\right)\left(\left|0\right\rangle\left|0\right\rangle...\left|0\right\rangle  \right), 
\end{align}
et
\begin{align}
{\rm Tr}_{R}\left\lbrace {\hat R}^{\dagger}\left(t'\right){\hat R}^{\dagger}\left(t\right)\rho_{R} \right\rbrace=\left(\left\langle 0\right|\left\langle 0\right|...\left\langle 0\right|\right){\hat R}^{\dagger}\left(t' \right){\hat R}\left(t\right)\left(\left|0\right\rangle\left|0\right\rangle...\left|0\right\rangle  \right).
\end{align}
En substituant les opérateurs ${\hat R}\left(t\right)$ et ${\hat R}^{\dagger}\left(t\right)$ par ces expressions, on trouve alors
\begin{equation}
{\rm Tr}_{R}\left\lbrace {\hat R}\left(t\right){\hat R}^{\dagger}\left(t'\right)\rho_{R} \right\rbrace=\sum_{k,k'}g_{k}^{*}g_{k'}e^{-i\left( w_{k}t-w_{k'}t'\right)}\left(\left\langle 0\right|\left\langle 0\right|...\left\langle 0\right|\right){\hat a}_{k}{\hat a}_{k}^{\dagger}\left(\left|0\right\rangle\left|0\right\rangle...\left|0\right\rangle\right),\label{F55}
\end{equation}
et
\begin{equation}
	{\rm Tr}_{R}\left\lbrace {\hat R}^{\dagger}\left(t'\right){\hat R}\left(t\right)\rho_{R} \right\rbrace=\sum_{k,k'}g_{k}^{*}g_{k'}e^{-i\left( w_{k}t-w_{k'}t'\right)}\left(\left\langle 0\right|\left\langle 0\right|...\left\langle 0\right|\right){\hat a}_{k}^{\dagger}{\hat a}_{k}\left(\left|0\right\rangle\left|0\right\rangle...\left|0\right\rangle\right)\equiv0.
\end{equation}
Cela signifie que la fonction $G\left(t\right)\equiv{G^{*}}\left(t\right)\equiv0$, et donc les fonctions $F\left(t\right)$ et $F^{*}\left(t\right)$ qui apparaîtront uniquement dans l'équation d'évolution de Born-Markov (\ref{FGGF}). Nous rappelons que les opérateurs bosoniques satisfont la relation de commutation
\begin{equation}
{\hat a}_{k}{\hat a}_{k}^{\dagger}=\delta_{k,k'}+{\hat a}_{k}^{\dagger}{\hat a}_{k}.
\end{equation}
Nous pouvons reécrire alors l'équation (\ref{F55}) comme
\begin{align}
	{\rm Tr}_{R}\left\lbrace {\hat R}\left(t\right){\hat R}^{\dagger}\left(t'\right)\rho_{R} \right\rbrace=&\sum_{k,k'}g_{k}^{*}g_{k'}e^{-i\left( w_{k}t-w_{k'}t'\right)}\left(\left\langle 0\right|\left\langle 0\right|...\left\langle 0\right|\right){\hat a}_{k}^{\dagger}{\hat a}_{k}\left(\left|0\right\rangle\left|0\right\rangle...\left|0\right\rangle  \right)\notag\\&+\sum_{k,k'}g_{k}^{*}g_{k'}e^{-i\left( w_{k}t-w_{k'}t'\right)}\delta_{k,k'}.
\end{align}
Par conséquent, il en résulte que
\begin{equation}
{\rm Tr}_{R}\left\lbrace {\hat R}\left(t\right){\hat R}^{\dagger}\left(t'\right)\rho_{R} \right\rbrace=\sum_{k}|g_{k}|^{2} e^{-iw_{k}\left(t-t'\right)} \hspace{0.7cm}{\rm et\hspace{0.1cm} donc}\hspace{0.7cm} F\left( t\right)=\sum_{k}|g_{k}|^{2}\int_{0}^{t}e^{-iw_{k}\left(t-t'\right)}dt'.\label{359}
\end{equation}
Dans l'expression de $F\left( t\right)$ donnée par l'équation (\ref{359}), si nous adoptons la définition générale de la densité d'états qui codent toutes les propriétés physiques de l'environnement comme
\begin{equation}
J\left( w\right)=\sum_{k}|g_{k}|^{2}\delta\left(w-w_{k}\right),  
\end{equation}
où $\delta$ est la distribution de Dirac, alors la somme sur les indices $k$ peut être remplacée par une intégrale sur un continuum de fréquences comme
\begin{equation}
F\left(t\right)=\int_{0}^{\infty}J\left( w\right) dw\int_{0}^{t}e^{-iw\left(t-t'\right)}dt'.
\end{equation}
Pour simplifier, nous introduisons la nouvelle variable $\tau=t-t'$, alors $d\tau=-dt'$ et  $\int_{0}^{t}dt'=-\int_{t}^{0}d\tau=\int_{0}^{t}d\tau$.
Cela conduit à
\begin{equation}
F\left(t\right)=\int_{0}^{\infty}J\left(w\right) \int_{0}^{t}e^{-iw\tau}d\tau.\label{362}
\end{equation}
Dans l'approximation de Markov, la limite $t\rightarrow\infty$ est prise dans l'intégrale de temps, nous prenons donc les bornes du deuxième intégrale de l'équation (\ref{362}) entre zéro et l'infini. En outre, nous utiliserons la relation
\begin{align}
\int_{0}^{\infty} e^{-iw\tau}d\tau=&\lim_{\eta\mapsto0^{+}}\int_{0}^{\infty}e^{-iw\tau-\eta\tau}d\tau\notag\\&=\lim_{\eta\mapsto0^{+}}\frac{\eta-iw}{\eta^{2}+w^{2}}\notag\\&=\lim_{\eta\mapsto0^{+}}\frac{\eta}{\eta^{2}+w^{2}}-\lim_{\eta\mapsto0^{+}}\frac{iw}{\eta^{2}+w^{2}}\notag\\&=\pi\delta\left( w\right) -iP\frac{1}{w},
\end{align}
avec $P$ représente la partie principale de Cauchy, la fonction $F\left(t\right)$ peut s'écrire comme suit
\begin{equation}
F\left(t\right)=\pi\int_{0}^{\infty}J\left( w\right) \delta\left( w\right) dw-iP\int_{0}^{\infty}\frac{J\left( w\right)}{w}dw.
\end{equation}
Il vient alors que la fonction $F$ d'une densité d'états générale s'écrit comme suit
\begin{equation}
F=\frac{\gamma+i\varepsilon}{2},
\end{equation}
avec 
\begin{equation}
\gamma=2\pi\int_{0}^{\infty}J\left( w\right) \delta\left( w\right) dw, \hspace{1cm}{\rm et}\hspace{1cm}\varepsilon=-2P\int_{0}^{\infty}\frac{J\left( w\right)}{w}dw.
\end{equation}
Par conséquent, l'équation maîtresse de Born-Markov (\ref{EBM}) devient
\begin{align}
\frac{d}{dt}\tilde{\rho}_{S}\left(t\right)=&-\frac{\gamma}{2}\left[\tilde{\rho}_{S}\left(t\right){\hat S}^{\dagger}{\hat S}-{\hat S}\tilde{\rho}_{S}\left(t\right){\hat S}^{\dagger}+{\hat S}^{\dagger}{\hat S}\tilde{\rho}_{S}\left(t\right)-{\hat S}\tilde{\rho}_{S}\left(t\right){\hat S}^{\dagger}\right]\notag\\&+i\frac{\varepsilon}{2} \left[\tilde{\rho}_{S}\left(t\right){\hat S}^{\dagger}{\hat S}-{\hat S}\tilde{\rho}_{S}\left(t\right){\hat S}^{\dagger}-{\hat S}^{\dagger}{\hat S}\tilde{\rho}_{S}\left(t\right)-{\hat S}\tilde{\rho}_{S}\left(t\right){\hat S}^{\dagger} \right].\label{367}
\end{align}
Le terme $\varepsilon$ dans l'équation (\ref{367}) dépend de l'importance de la relation entre la forme de la densité spectrale du réservoir et sa fréquence. En fait, nous pouvons choisir la forme de densité d'état $J\left(w\right)$ pour donner $\varepsilon=0$, où nous pouvons étendre la limite inférieure d'intégration à $-\infty$. Cette approximation est appliquée dans de nombreux cas pour dériver une équation maîtresse markovienne sous la forme de Lindblad \cite{Lindblad1976} qui décrit l'évolution temporelle du système étendu (par exemple, pour un système représenté par un atome à deux niveaux interagissant avec un environnement bosonique structuré à zéro température, la densité d'état $J\left(w\right)$ nous donne $\varepsilon=0$). Dans ce cas précis, le résultat final est

\begin{equation}
\frac{d}{dt}\tilde{\rho}_{S}\left(t\right)=\gamma\left[ {\hat S}\tilde{\rho}_{S}\left(t\right){\hat S}^{\dagger}-\frac{1}{2}\left\lbrace {\hat S}^{\dagger}{\hat S},\tilde{\rho}_{S}\left(t\right) \right\rbrace\right]. \label{368}
\end{equation}
Revenons donc à l'image d'origine via
\begin{align}
\frac{d}{dt}\tilde{\rho}_{S}\left(t\right)&=\frac{d}{dt}\left[e^{\frac{i}{\hbar}{\hat H}_{S}t} \rho_{S}\left( t\right) e^{\frac{-i}{\hbar}{\hat H}_{S}t}\right]\notag\\&=e^{\frac{i}{\hbar}{\hat H}_{S}t} \frac{d\rho_{S}\left( t\right)}{dt}e^{\frac{-i}{\hbar}{\hat H}_{S}t}+\frac{i}{\hbar}e^{\frac{i}{\hbar}{\hat H}_{S}t}\left[{\hat H}_{S},\rho_{S}\left( t\right) \right] e^{\frac{-i}{\hbar}{\hat H}_{S}t}.
\end{align}
Avec la même procédure, l'équation (\ref{368}) devient
\begin{align}
\left[ {\hat S}\tilde{\rho}_{S}\left(t\right){\hat S}^{\dagger}-\frac{1}{2}\left\lbrace {\hat S}^{\dagger}{\hat S},\tilde{\rho}_{S}\left(t\right) \right\rbrace\right] =e^{\frac{i}{\hbar}{\hat H}_{S}t}\left[ {\hat S}\rho_{S}\left(t\right){\hat S}^{\dagger}-\frac{1}{2}\left\lbrace {\hat S}^{\dagger}{\hat S},\rho_{S}\left(t\right) \right\rbrace\right]e^{\frac{-i}{\hbar}{\hat H}_{S}t}. 
\end{align}
Il s'ensuit que l'équation d'évolution de Lindblad dans le cadre de l'approximation de Born-Markov est donnée par
\begin{align}
\frac{d\rho_{S}\left( t\right)}{dt}=-\frac{i}{\hbar}\left[{\hat H}_{S}, \rho_{S}\left( t\right)\right]+\gamma\left[ {\hat S}\rho_{S}\left(t\right){\hat S}^{\dagger}-\frac{1}{2}\left\lbrace {\hat S}^{\dagger}{\hat S},\rho_{S}\left(t\right) \right\rbrace\right].\label{371}
\end{align}
En revanche, l'équation (\ref{371}) est généralement présentée avec plusieurs opérateurs ${\hat S}$, sous la forme d'une combinaison linéaire d'opérateurs Lindbladiens, souvent désignés par des opérateurs ${\hat L}$. Les opérateurs ${\hat L}$ sont appelés les opérateurs de Lindblad et, dans le cas général, l'équation de Lindblad prend la forme suivante
\begin{align}
	\frac{d\rho_{S}\left( t\right)}{dt}=\underbrace {-\frac{i}{\hbar}\left[{\hat H}_{S}, \rho_{S}\left( t\right)\right]}_{\rm Partie\hspace{0.1cm} unitaire}+\underbrace {\gamma\sum_{j}\left[ {\hat L}_{j}\rho_{S}\left(t\right){\hat L}_{j}^{\dagger}-\frac{1}{2}\left\lbrace {\hat L}_{j}^{\dagger}{\hat L}_{j},\rho_{S}\left(t\right) \right\rbrace\right]}_{\rm Partie\hspace{0.1cm} non \hspace{0.1cm}unitaire\hspace{0.1cm}(dissipateur)}.\label{LinL}
\end{align}
En raison de nos hypothèses simplificatrices, il est important de souligner que toutes ces simplifications limitent la validité de cette dérivation à des cas plus généraux, mais elle fournit une illustration détaillée de la signification physique de chaque terme apparaissant dans l'équation de Lindblad (\ref{LinL}).
\section{Cohérence quantique comme une ressource en théorie de l'information quantique}
L'une des caractéristiques fondamentales qui séparent la physique quantique de la physique classique est l'idée de la superposition quantique, également connue sous le nom de la cohérence quantique. Habituellement, les caractéristiques de la cohérence et de la décohérence semblent s'opposer l'une à l'autre. Ainsi, divers quantificateurs de la cohérence pourraient être examinés du point de vue de leur diminution au cours de processus de décohérence \cite{Yao2017}. Formellement, un état quantique présente une cohérence quantique lorsqu'il est décrit par une matrice densité qui n'est pas diagonale par rapport à la base orthogonale des bits classiques. Cette partie se concentre sur la compréhension de la cohérence quantique dans le cadre mathématique des théories des ressources. Il s'agit d'une ressource à exploiter et qui permet de distinguer de façon quantitative les états quantiques par rapport aux états classiques. Toutefois, malgré son importance cruciale dans le développement de la science de l'information quantique, ce n'est que très récemment qu'un cadre théorique rigoureux a été établi par Baumgratz et ses collègues \cite{Baumgratz2014} pour quantifier la quantité de la cohérence quantique contenue dans les états quantiques. Des idées similaires ont déjà été présentées par Aberg \cite{Aberg2006}, mais sans fournir une formulation complète.
\subsection{Théorie de ressource de la cohérence quantique}
\subsubsection{Les contraintes}
Par analogie avec la théorie de l'intrication quantique qui établit les ensembles des états séparables ("classiques") et des opérations locales et de communication classique (LOCC), la quantification de la cohérence quantique est basée sur les concepts des états incohérents et des opérations incohérentes. Pour un état quantique associé à un espace de Hilbert ${\cal H}$ de dimension $d$, nous fixons la base orthogonale $\left\lbrace \left|j \right\rangle \right\rbrace_{j=1}^{d}$ comme une base de référence. Si un état quantique est diagonal lorsqu'il est exprimé sur cette base de référence locale, il est appelé état incohérent et prend la forme
\begin{equation}
\delta=\sum_{j=1}^{d}\delta_{j}\left|j\right\rangle\left\langle j\right|,
\end{equation}
avec $\delta\in {\cal I}$, où ${\cal I}$ est l'ensemble des états incohérents. D'autre part, une opération incohérente est une application linéaire complètement positive et préservant les traces qui transforme un état incohérent à un état incohérent et aucune création de cohérence n'a pu être observée. En outre, les opérations incohérentes peuvent être caractérisées par des opérateurs de Kraus $\left\lbrace K_{j} \right\rbrace$ satisfaisant $\sum_{j=1}^{d}K_{j}^{\dagger}K_{j}\equiv\openone$. En fait, Baumgratz a identifié deux classes différentes d'opérations incohérentes:
\begin{itemize}
	\item Les opérations incohérentes complètement positives et préservant les traces qui agissent comme $\Lambda\left( \rho\right)=\sum_{j}K_{j}\rho K_{j}^{\dagger}$. Ici, tous les $K_{j}$ sont de même dimension, et doivent obéir à la propriété $K_{j}\delta K_{j}^{\dagger}/p_{j}\in {\cal I}$ pour chaque état incohérent $\delta\in {\cal I}$, où $p_{j}={\rm Tr}\left(K_{j}\rho K_{j}^{\dagger}\right)$ désigne la probabilité d'obtenir le résultat $j$. 
	\item Les opérations incohérentes pour lesquelles les résultats de mesure de sortie sont conservés. Ils exigent également que $K_{j}\delta K_{j}^{\dagger}/p_{j}\in {\cal I}$ soit satisfait pour tous $\delta\in {\cal I}$. Mais la dimension de $K_{j}$ peut être différente.
\end{itemize}
Baumgratz et al ont établis dans la référence \cite{Baumgratz2014} que toute mesure appropriée de la cohérence $C\left(\rho\right)$, qui transforme les états quantiques en un nombre réel positif, doit satisfaire les conditions suivantes:
\begin{enumerate}
	\item[(C1)]\underline{\textit{Non-négativité}:} $C\left(\rho\right)\geq 0$, et $C\left(\rho\right)=0$ si et seulement si $\rho\in{\cal I}$.
	\item[(C2)]\underline{\textit{Monotonicité}:} $C\left(\rho\right)$ n'augmente pas sous l'action d'une opération incohérente, i.e.,
	\begin{equation}
C\left(\Lambda_{I}\left( \rho\right) \right)\leq C\left(\rho\right),
	\end{equation}
 pour toutes les applications incohérentes $\Lambda_{I}$.
	\item[(C3)]\underline{\textit{Convexité dans le mélange des états}:} $C\left(\rho\right)$ est une fonction convexe de l'état $\rho$, i.e.,
	\begin{equation}
\sum_{j}p_{j}C\left( \rho_{j}\right)\geq C\left(\sum_{j}p_{j}\rho_{j} \right), 
	\end{equation}
avec $p_{j}\geq1$ et $\sum_{j}p_{j}=1$.	
	\item[(C4)]\underline{\textit{Forte monotonie}:} $C\left(\rho\right)$ n'augmente pas en moyenne lors d'opérations sélectives incohérentes, c'est-à-dire
	\begin{equation}
C\left( \rho\right)\geq\sum_{j}p_{j}C\left( \rho_{j}\right),
	\end{equation}
avec des probabilités $p_{j}={\rm Tr}\left(K_{j}\rho K_{j}^{\dagger}\right)$, les états après la mesure $\rho_{j}=\left(K_{j}\rho K_{j}^{\dagger}\right)/p_{j}$, et les opérateurs incohérents $K_{j}$.
\end{enumerate}
En général, les conditions (C3) et (C4) peuvent être remplacées par une seule condition, à savoir l'exigence d'une additivité de la cohérence pour les états indépendants du sous-espace. Ce cadre alternatif pour quantifier la cohérence quantique a été proposé par Ya et ses collègues dans la référence \cite{Yu2016}. Considérons deux états $\rho_{1}$ et $\rho_{2}$ dans les deux sous-espaces différents, la quantité de la cohérence contenue dans l'état $\rho=p_{1}\rho_{1}\oplus p_{2}\rho_{2}$ (avec $p_{1}$ et $p_{2}$ étant des probabilités) ne doit être ni plus ni moins que la cohérence moyenne de l'état $\rho_{1}$ et l'état $\rho_{2}$. Par conséquent, une mesure raisonnable de la cohérence devrait satisfaire la condition
\begin{equation}
	C\left(p_{1}\rho_{1}\oplus p_{2}\rho_{2}\right)=p_{1}C\left( \rho_{1}\right)+p_{2}C\left(\rho_{2}\right).
\end{equation}
Il convient de noter que la condition (C4) est plus forte que la condition (C2), car sa combinaison avec (C3) implique automatiquement la condition (C2). En général, une fonction à valeur réelle $C\left(\rho\right)$ est appelée mesure de la cohérence si elle satisfait aux quatre conditions ci-dessus. Si seules les conditions (C1), (C2) et (C4) sont satisfaites, la fonction $C\left(\rho\right)$ est généralement appelée une cohérence monotone. Cette analyse de Baumgratz et ses collaborateurs a attiré l'attention de nombreux chercheurs, et diverses mesures de la cohérence quantique, qui satisfont aux exigences physiques de non contractivité et de monotonicité mentionnées ci-dessus, ont été proposées depuis lors. Toutes ces mesures peuvent être classées en deux catégories; La première concerne celles basées sur les fonctions d'entropie. La seconde catégorie concerne les mesures géométriques qui quantifient la cohérence comme sa distance par rapport à l'état incohérent le plus proche.
\subsubsection{Cohérence bipartite}
La cohérence quantique est habituellement associée à la capacité d'un état quantique à présenter des phénomènes d'interférence quantique \cite{Walls1995}. Les effets de la cohérence sont généralement attribués aux éléments hors diagonale d'une matrice de densité par rapport à une base de référence particulière définie par le scénario physique considéré. Ici, nous étendons le cadre de la théorie de ressource de la cohérence quantique au scénario bipartite. En particulier, pour un système à deux qubits associé à un espace de Hilbert ${\cal H}_{A}\otimes{\cal H}_{B}$, on fixe la base de calcul $\left\lbrace\left| i\right\rangle_{A}\otimes\left|j\right\rangle_{B} \right\rbrace$ comme base de référence, et on définit les états incohérents comme ceux dont la matrice de densité $\sigma_{AB}$ est diagonale, c'est-à-dire
\begin{equation}
\sigma_{AB}=\sum_{k}p_{k}\delta_{k}^{A}\otimes\tau_{k}^{B},\label{einhbip}
\end{equation}
où $p_{k}$ sont des probabilités et les états $\delta_{k}^{A}$ et $\tau_{k}^{B}$ sont des états incohérents sur le sous-système $A$ et $B$, respectivement. Cela veut dire que
\begin{equation}
	\delta_{k}^{A}=\sum_{i}p_{ik}^{'}\left|i\right\rangle_{A}\left\langle i\right|, \hspace{1cm}	\tau_{k}^{B}=\sum_{j}p_{jk}^{''}\left|j\right\rangle_{B}\left\langle j\right|,
\end{equation}
avec des probabilités $p_{ik}^{'}$ et $p_{jk}^{''}$. Il convient de noter que les états bipartites incohérents indiqués par l'équation (\ref{einhbip}) sont toujours séparables. Pour un système quantique à deux qubits, les opérations incohérentes bipartites sont de la forme
\begin{equation}
\Lambda_{AB}\left[ \sigma_{AB}\right]=\sum_{l}K_{l}\sigma_{AB}K_{l}^{\dagger},
\end{equation}
avec $K_{l}$ représentent des opérateurs de Kraus incohérents, tels que $K_{l}{\cal I}K_{l}^{\dagger}\in{\cal I}$ où ${\cal I}$ est maintenant l'ensemble des états incohérents bipartites. En général, il n'est pas possible de créer une cohérence à partir d'un état incohérent à deux qubits en utilisant la porte CNOT. On peut donc considérer qu'il s'agit d'un exemple important de l'opération bipartite incohérente la plus pratique et la plus simple. Ce type d'opération incohérente transforme tout état incohérent pur $\left| i\right\rangle \otimes\left| j\right\rangle$ en un autre état incohérent pur, et nous pouvons l'écrire comme
\begin{equation}
\Lambda_{\rm CNOT}\left( \left| i\right\rangle \otimes\left| j\right\rangle\right)=\left| i\right\rangle\otimes\left|{\rm mod}\left(i+j,2\right)\right\rangle. \label{cnot}
\end{equation}
\subsection{Quantifier la cohérence quantique}
\subsubsection{Mesures de la cohérence basées sur la distance}
Dans le domaine de la science de l'information quantique, les approches géométriques ont été utilisées pour traiter une énorme classe de problèmes tels que la caractérisation et la quantification de diverses caractéristiques quantiques. De manière analogue à la théorie des ressources de l'intrication pour laquelle les opérations libres sont décrites par les opérateurs LOCC, les états libres correspondent aux états séparables, et l'intrication peut être définie par une distance entre l'état considéré et l'ensemble des états séparables, il est naturel de quantifier la cohérence d'un état en utilisant une mesure de distance car la cohérence est également placée dans un cadre théorique de ressources. Une façon générale de construire des mesures de cohérence consiste à minimiser une fonction de distance sur tous les états incohérents. On peut alors quantifier le degré de la cohérence contenu dans un état $\rho$ en utilisant sa distance minimale par rapport à l'état incohérent le plus proche \cite{Baumgratz2014},
\begin{equation}
C_{D}\left(\rho\right)=\min_{\delta\in{\cal I}}D\left(\rho,\delta \right),\label{cohd}
\end{equation}
où $D\left(\rho,\delta \right)$ désigne une mesure de distance entre les états quantiques et la minimisation porte sur l'ensemble des états incohérents ${\cal I}$. Selon la définition de l'équation (\ref{cohd}), la condition (C1) est satisfaite pour la mesure de distance qui donne $D\left(\rho,\delta\right)=0$ si et seulement si $\rho\equiv\delta$, tandis que la condition (C2) peut être satisfaite lorsque $D$ est monotone sous l'action des opérations incoherénts, c'est-à-dire $D\left(\rho,\delta\right)\geq D\left(\Lambda\left(\rho\right),\Lambda\left(\delta\right)\right)$. De plus, la condition (C3) est également satisfaite si $D$ est conjointement convexe, c'est-à-dire $D\left(\sum_{i}p_{i}\rho_{i},\sum_{i}p_{i}\delta_{i}\right)\leq\sum_{i}p_{i}D\left(\rho_{i}, \delta_{i}\right)$. Nous présenterons ici les définitions de certains quantificateurs de cohérence basés sur les distances couramment utilisées dans la littérature.
\paragraph{L'entropie relative de la cohérence:} En fait, l'entropie relative a été adoptée pour quantifier l'intrication et la discorde quantique. Dans cette direction, Baumgratz et ses collaborateurs ont montré qu'elle peut également servir d'outil valable pour quantifier la cohérence quantique. Pour tout état quantique $\rho$ sur l'espace de Hilbert ${\cal H}$, l'entropie relative de la cohérence est définie comme
\begin{equation}
C_{r}\left(\rho\right):=\min_{\delta\in{\cal I}}S\left( \rho\parallel\delta\right), \label{Cr}
\end{equation}
où $S\left( \rho\parallel\delta\right)={\rm Tr}\left(\rho\log_{2}\rho-\rho\log_{2}\delta\right)$ est l'entropie relative. En ce qui concerne les propriétés de l'entropie relative, il est assez facile de vérifier que cette mesure satisfait les conditions des mesures de cohérence. En particulier, il existe une solution qui permet d'évaluer facilement les expressions analytiques. Prenons un espace de Hilbert ${\cal H}$ avec une base fixe $\left\lbrace\left|i \right\rangle\right\rbrace_{i=1}^{d}$, la matrice densité $\rho=\sum_{i,j}\rho_{i,j}\left|i\right\rangle\left\langle j\right|$ et $\rho_{\rm diag}=\sum_{i}\rho_{ii}\left|i\right\rangle\left\langle i\right|$. En utilisant les propriétés d'entropie relative, il est facile d'obtenir
\begin{equation}
C_{r}\left(\rho\right)=S\left(\rho_{\rm diag}\right)-S\left(\rho \right).\label{384}
\end{equation}
On remarque que l'état incohérent $\rho_{\rm diag}$ est généré en supprimant tous les éléments hors diagonale et en laissant les éléments diagonaux dans la matrice densité $\rho$. Cette opération est appelée un canal complètement décohérant et on note alors
\begin{equation}
\rho_{\rm diag}=\Pi\left[\rho \right]=\sum_{i=1}^{d}\Pi_{i}\rho\Pi_{i},
\end{equation}
où $\Pi_{i}=\left|i\right\rangle \left\langle i\right|$ sont des projecteurs unidimensionnels, et $\sum_{i}\Pi_{i}=\openone_{\cal H}$ avec $\openone_{\cal H}$ est opérateur d'identité sur l'espace de Hilbert ${\cal H}$. En outre, certaines propriétés de base ont été données. Par exemple, nous pouvons obtenir
\begin{equation}
C_{r}\left(\rho\right)\leq S\left(\rho_{\rm diag}\right)\leq \log_{2}d,
\end{equation}
et $C_{r}\left(\rho\right)=S\left(\rho_{\rm diag}\right)$ si et seulement si l'état quantique $\rho$ est un état pur. En particulier, s'il existe des états purs tels que $C_{r}\left(\rho\right)=\log_{2}d$, ces états purs sont appelés les états maximalement cohérents. Les auteurs de la référence \cite{Baumgratz2014} ont défini un état maximalement cohérent comme un état qui permet la génération déterministe de tous les autres états quantiques par des opérations incohérentes, qui prend la forme suivante
\begin{equation}
\left| \psi_{d}\right\rangle:=\frac{1}{\sqrt{d}}\sum_{i=1}^{d}\left|i \right\rangle.
\end{equation}
\paragraph{La norme $l_{1}$ de la cohérence:} De manière intuitive, la superposition correspond aux éléments non diagonaux de la description de l'opérateur densité d'un état quantique par rapport à la base de référence sélectionnée. En partant de cette considération, Baumgratz et ses collègues ont montré que la norme $l_{1}$ peut être considérée comme un quantificateur prometteur de la cohérence quantique dans les systèmes multipartites. Explicitement, la cohérence est définie comme
\begin{equation}
C_{l_{1}}\left( \rho\right):=\min_{\delta\in{\cal I}}\parallel\rho-\delta\parallel_{l_{1}}=\sum_{i\neq j}|\left\langle i\right|\rho\left| j\right\rangle|.\label{Cl1}
\end{equation}
Elle est égale à la somme des valeurs absolues des éléments hors diagonale de la matrice densité $\rho$, et est favorisée pour sa facilité d'évaluation. Outre la condition de convexité, elle satisfait également l'inégalité $C_{l_{1}}\left( p_{1}\rho_{1}+p_{2}\rho_{2}\right)\geq|p_{1}C_{l_{1}}\left(\rho_{1}\right)-p_{2}C_{l_{1}}\left(\rho_{2}\right)|$ \cite{Dai2017}. De plus, pour tout état pur bipartite, la norme $l_{1}$ de la cohérence est égale au double de sa négativité \cite{Vidal2002}, qui est une mesure de l'intrication quantique. D'autre part, une relation entre la norme $l_{1}$ de la cohérence (\ref{Cl1}) et l'entropie relative de la cohérence (\ref{Cr}) a également été établie pour les états purs $\rho=\left|\psi\right\rangle\left\langle \psi\right|$. Cette relation est donnée par \cite{Rana2016}
\begin{equation}
C_{l_{1}}\left(\left|\psi\right\rangle \right)\geq\max\left\lbrace C_{r}\left( \left|\psi\right\rangle\right), 2^{C_{r}\left( \left|\psi\right\rangle\right)}-1\right\rbrace.
\end{equation}
Il a également été prouvé que \cite{Rana2017}
\begin{equation}
C_{l_{1}}^{2}\left(\left| \psi\right\rangle\right) \leq \frac{d\left(d-1\right)C_{r}\left(\left| \psi\right\rangle\right) }{\sqrt{2}},
\end{equation}
avec $d={\rm Rang}\left(\left| \psi\right\rangle\right)$ est le rang de l'état $\left|\psi\right\rangle$. Si nous avons $d>2$, on peut aussi obtenir $C_{l_{1}}\left(\left| \psi\right\rangle\right)-C_{r}\left(\left| \psi\right\rangle\right)\leq d-1-\log_{2}d$.\par
En fait, grâce à la forme paramétrique d'une matrice densité arbitraire, tout état quantique $\rho$ à $d$ dimension peut être décomposé comme \cite{Byrd2003}
\begin{equation}
\rho=\frac{1}{d}\openone_{d}+\frac{1}{2}\sum_{i=1}^{d^{2}-1}x_{i}{\hat X}_{i},\label{repgho}
\end{equation}
avec $x_{i}={\rm Tr}\left(\rho{\hat X}_{i}\right)$ et les $\left\lbrace {\hat X}_{i}\right\rbrace$ sont les générateurs de l'algèbre $su\left(d\right)$ (ils coïncident avec les matrices de Pauli lorsque $d=2$ et avec les matrices de Gell-Mann lorsque $d=3$). Ensuite, si l'on classe les éléments de ces générateurs ${\hat X}_{i}$ ($i=1,...,d^{2}-1$) comme
\begin{equation}
\left\lbrace {\hat X}_{i}\right\rbrace_{i=1}^{d^{2}-1}=\left\lbrace {\hat U}_{ij},{\hat V}_{jk},{\hat W}_{l} \right\rbrace, 
\end{equation}
où $j,k\in \left\lbrace1,2,...,d\right\rbrace$ et $l\in \left\lbrace1,...,d-1\right\rbrace$, avec
\begin{equation}
{\hat U}_{jk}=\left|j\right\rangle\left\langle k\right|+\left| k\right\rangle\left\langle j\right|,\hspace{1cm}{\hat V}_{jk}=-i\left(\left|j\right\rangle\left\langle k\right|-\left| k\right\rangle\left\langle j\right|\right), \label{393}
\end{equation}
et
\begin{equation}
{\hat W}_{l}=\sqrt{\frac{2}{l\left(l+1\right)}}\sum_{j=1}^{l}\left( \left|j\right\rangle\left\langle j\right|-l\left| l+1\right\rangle\left\langle l+1\right|\right). \label{394}
\end{equation}
Alors, le vecteur des générateurs $\vec{X}$ peut être étiqueté comme suit
\begin{align}
	\vec{X}=&\left\lbrace {\hat X}_{1},...,{\hat X}_{\frac{d\left(d-1\right)}{2}},{\hat X}_{\frac{d\left(d-1\right)}{2}+1},...,{\hat X}_{d\left(d-1\right)},{\hat X}_{d\left(d-1\right)+1},...,{\hat X}_{d^{2}-1} \right\rbrace\notag\\& =\left\lbrace {\hat U}_{12},...,{\hat U}_{d\left(d-1\right)},{\hat V}_{12},...,{\hat V}_{d\left(d-1\right)},{\hat W}_{1},...,{\hat W}_{\left(d-1 \right)}\right\rbrace.
\end{align}
Dans cette représentation (\ref{repgho}), l'expression explicite de la norme $l_{1}$ de la cohérence d'un système à $d$ dimensions, donnée par l'équation (\ref{Cl1}), s'écrit
\begin{equation}
C_{l_{1}}\left( \rho\right):=\sum_{i\neq j}|\rho_{ij}|=\sum_{r=1}^{d\left(d-1\right)/2}\sqrt{x_{r}^{2}+x_{r+d\left(d-1\right)/2}^{2}}.
\end{equation}
D'autre part, il est bien connu que le bruit est responsable de l'augmentation du mélange d'un système quantique. Il apparaît donc comme un paramètre intuitif pour comprendre la décohérence. Pour chaque état quantique, l'interaction avec le monde extérieur ou la décohérence affecte sa pureté, et le bruit introduit une mixité dans le système quantique. Cette interaction entraîne une perte d'informations et sa caractérisation est donc une tâche importante dans les protocoles d'information quantique. Une question naturelle qui se pose est de savoir comment les quantités physiques importantes dans la théorie de l'information quantique, telles que la cohérence et l'intrication quantique, peuvent s'opposer à la mixité des systèmes quantiques? Dans ce sens, Singh et ses collaborateurs \cite{Singh2015} ont obtenu une relation de complémentarité entre la cohérence quantique qui est quantifiée par la norme $l_{1}$ et la mixité qui est quantifié en termes d'entropie linéaire normalisée. Cette relation nous donne les restrictions imposées par la mixité d’un système sur la quantité maximale de la cohérence quantique. Si nous utilisons la mixité basée sur l'entropie linéaire normalisée \cite{Peters2004} d'un état arbitraire à $d$ dimension, qui est donnée par
\begin{equation}
M_{l}\left( \rho\right):=\frac{d}{d-1}\left( 1-{\rm Tr}\left(\rho^{2} \right)\right)=1-\frac{d}{2\left(d-1 \right)}\sum_{i=1}^{d^{2}-1}x_{i}^{2},
\end{equation}
nous pouvons donc facilement montrer que pour un système quantique général à $d$ dimension, la somme de la cohérence au carré et de la mixité est toujours inférieure ou égale à l'unité \cite{Singh2015}, à savoir
\begin{equation}
\frac{C_{l_{1}}^{2}\left(\rho\right)}{\left(d-1\right)^{2} }+M_{l}\left( \rho\right)\leq 1.\label{397}
\end{equation}
Dans ce sens, lorsque les systèmes quantiques perdent leur cohérence en présence d'un environnement et perdent leur pureté, la relation (\ref{397}) nous donne la cohérence maximale que nous puissions avoir. Ceci est en principe similaire à l'intrication maximale qu'un état mixte peut avoir. C'est physiquement pertinent car cela nous indique la quantité de ressources (la cohérence ou bien l'intrication) que nous pouvons obtenir d'un état mixte pour des applications dans un protocole quantique pertinent.
\paragraph{Cohérence basée sur la distance de trace:} En dehors de la norme $l_{1}$, on peut se demander si la norme générale des distances $l_{p}$ et de Schatten peut être adoptée pour définir les mesures de la cohérence. En général, cette question n'a pas encore reçu de réponse. L'une des mesures de distance largement utilisées, la distance de trace (c'est-à-dire la norme-$1$ de Schatten), a été proposée comme un candidat possible pour la mesure de cohérence dans la référence \cite{Baumgratz2014}. Elle est définie comme
\begin{equation}
C_{\rm Tr}\left(\rho\right):=\min_{\delta\in{\cal I}}\parallel\rho-\delta\parallel_{1},
\end{equation}
avec $\parallel A\parallel_{1}={\rm Tr}\sqrt{A^{\dagger}A}$ désigne la norme trace de la matrice $A$. Pour une famille d'états $X$ à plusieurs qubits, il a déjà été prouvé que $C_{\rm Tr}\left(\rho\right)$ est une cohérence monotone, et l'état incohérent optimal correspondant est donné par $\rho_{\rm diag}$ \cite{Bromley2015}. De plus, pour l'état $\rho$ avec tous ses éléments non diagonaux égaux les uns aux autres, c'est-à-dire $\rho_{ij}=\alpha$ ($\forall i\neq j$), la cohérence basée sur la distance de trace peut être dérivée analytiquement comme

\begin{equation}
C_{\rm Tr}\left(\rho_{X}\right)=2\left(d-1\right)|\alpha|,
\end{equation}
où $d$ est la dimension de l'état $\rho_{X}$ et l'état incohérent le plus proche est $\delta=\rho_{X_{\rm diag}}$.
\subsubsection{Cohérence géométrique basée sur la fidélité}
De façon analogue à l'intrication géométrique \cite{Wei2003} qui est une mesure d'intrication, les auteurs de la référence \cite{Streltsov2015} définissent la cohérence géométrique comme
\begin{equation}
C_{g}\left( \rho\right)=1-\sqrt{\max_{\delta\in{\cal I}}F\left(\rho,\delta\right)}, \label{Cg}
\end{equation}
où $F\left(\rho,\delta\right)=\parallel\sqrt{\rho}\sqrt{\delta}\parallel_{1}^{2}$ est la fidélité entre l'état $\rho$ et l'état $\delta$ \cite{Uhlmann1976}. En outre, il a été démontré que $C_{g}\left( \rho\right)$ remplit les conditions (C1), (C2), (C3) et (C4) pour l'ensemble des états incohérentes et des opérations incohérentes. Cela signifie que la cohérence géométrique (\ref{Cg}) est une cohérence monotone. Si $\rho$ est un état d'un système quantique à un seul qubit, $C_{g}\left( \rho\right)$ a une expression analytique fermée. Tout d'abord, a partir de la représentation sphérique de Bloch d'un état quantique, les états $\rho$ et $\delta$ peuvent être exprimés comme
\begin{equation}
\rho=\frac{\openone_{2\times2}+\vec{r}.\vec{\sigma}}{2},\hspace{2cm}\delta=\frac{\openone_{2\times2}+\vec{s}.\vec{\sigma}}{2},
\end{equation}
où $\openone_{2\times2}$ est l'opérateur d'identité, $\vec{r}=\left(r_{x},r_{y},r_{z}\right)$ et $\vec{s}=\left(s_{x},s_{y},s_{z}\right)$ sont les vecteurs de Bloch et $\vec{\sigma}=\left(\sigma_{x},\sigma_{y},\sigma_{z}\right)$ est un vecteur de matrices de Pauli. Il est donc facile de montrer que la fidélité pour les qubits a la forme explicite suivante
\begin{equation}
F\left(\rho,\delta\right)=\frac{1}{2}\left[1+\vec{r}.\vec{s}+\sqrt{\left(1-|r|^{2}\right)\left(1-|s|^{2} \right)}\right].\label{Fidelite}
\end{equation}
Puisque $\delta$ est l'état incohérent, alors le vecteur de Bloch $\vec{s}$ peut être exprimé comme $\vec{s}=\left(0,0,s_{z}\right)$, alors l'équation (\ref{Fidelite}) se réduit à
\begin{equation}
	F\left(\rho,\delta\right)=\frac{1}{2}\left[1+r_{z}.s_{z}+\sqrt{\left(1-r_{x}^{2}-r_{y}^{2}-r_{z}^{2}\right)\left(1-s_{z}^{2}\right)}\right].
\end{equation}
Pour maximiser la fidélité $F\left(\rho,\delta\right)$, nous devons prendre une dérivée par rapport à $s_{z}$, nous avons donc
\begin{equation}
\frac{\partial F\left(\rho,\delta\right)}{\partial s_{z}}=\frac{1}{2}\left[r_{z}-\sqrt{\frac{1-r_{x}^{2}-r_{y}^{2}-r_{z}^{2}}{1-s_{z}^{2}}}s_{z}\right].
\end{equation}
Après quelques opérations algébriques simples, nous pouvons obtenir
\begin{equation}
\max_{\delta\in{\cal I}}F\left( \rho,\delta\right)=\frac{1}{2}\left[1+\sqrt{1-r_{x}^{2}-r_{y}^{2}} \right].
\end{equation}
Par conséquent, la formule compacte de la cohérence géométrique (\ref{Cg}) est donnée par
\begin{equation}
C_{g}\left(\rho \right)=1-\frac{\sqrt{2}}{2}\sqrt{1+\sqrt{1-r_{x}^{2}-r_{y}^{2}}}.
\end{equation}
\subsubsection{Relation entre la cohérence quantique et la discorde quantique}
Soit $\left\lbrace \left| i\right\rangle_{A}\right\rbrace$ et $\left\lbrace \left| j\right\rangle_{B}\right\rbrace$ les bases de référence locales des sous-systèmes $A$ et $B$, respectivement. Nous utilisons généralement leur produit tensoriel $\left\lbrace \left|ij\right\rangle_{AB}\right\rbrace$ comme une base de référence du système composite $AB$.  Pour un état quantique $\rho_{AB}$, sa cohérence totale est $C_{r}\left(\rho_{AB}\right)$, tandis que $C_{r}\left(\rho_{A}\right)$ et $C_{r}\left(\rho_{B}\right)$ sont appelées les cohérences locales. Chaque fois que l'état $\rho_{AB}$ est un état qui s'écrit sous forme d'un produit, la somme des cohérences locales est égale à la cohérence totale. La cohérence mesurée à partir de l'entropie relative admet la propriété de super-additivité suivante \cite{Xi2015}
\begin{equation}
C_{r}\left(\rho_{AB}\right)\geq C_{r}\left(\rho_{A}\right)+C_{r}\left(\rho_{B}\right).
\end{equation}
Ainsi, la définition de la cohérence quantique corrélée par rapport à l'entropie relative de la cohérence est définie comme \cite{Tan2016}
\begin{equation}
C_{r}^{cc}\left(\rho_{AB}\right)\equiv C_{r}\left(\rho_{AB}\right)- C_{r}\left(\rho_{A}\right)-C_{r}\left(\rho_{B}\right).
\end{equation}
Evidemment, la cohérence quantique corrélée est la cohérence totale entre les sous-systèmes. Pour un état $\rho_{AB}$, quelles que soient les bases de référence des sous-systèmes, sa cohérence quantique corrélée est toujours nulle, si et seulement si l'état $\rho_{AB}$ n'a pas de corrélations. En ce sens, la cohérence quantique corrélée peut être vue comme la mesure de la corrélation quantique comme la discorde quantique. En fait, les mesures locales de von Neumann des sous-systèmes $A$ et $B$ sont désignées par $\Pi_{A}=\left\lbrace\left| i\right\rangle_{A} \left\langle i\right|\right\rbrace$ et $\Pi_{B}=\left\lbrace\left| j\right\rangle_{B} \left\langle j\right|\right\rbrace$, respectivement. Par un calcul direct, nous obtenons que la consommation de cohérence quantique corrélée pour tout état $\rho_{AB}$ sous la mesure locale $\Pi_{B}$ coïncide avec la discorde quantique ${\cal Q}_{A|B}\left( \rho_{AB}\right)$ \cite{Yadin2016}, cela signifie que
\begin{equation}
C_{r}^{cc}\left( \rho_{AB}\right)-C_{r}^{cc}\left( \Pi_{B}\left( \rho_{AB}\right)\right)={\cal Q}_{A|B}\left( \rho_{AB}\right).
\end{equation}
D'ailleurs, la cohérence quantique corrélée reste inchangée sous sa mesure locale $\Pi_{B}$, c'est-à-dire $C_{r}^{cc}\left( \rho_{AB}\right)=C_{r}^{cc}\left( \Pi_{B}\left( \rho_{AB}\right)\right)$, si et seulement s'il existe une décomposition $\rho_{AB}=\sum_{k}p_{k}\rho_{A}^{k}\otimes\rho_{B}^{k}$ telle que $p_{k}$ sont des probabilités et que tous les états $\rho_{B}^{k}$ sont parfaitement discernables par la mesure de von Neumann dans la base de référence du sous-système mesuré $\left\lbrace\left|j\right\rangle_{B}\right\rbrace$. En d'autres termes, si la quantité de corrélations quantiques totales dans le système quantique bipartite $\rho_{AB}$ est égale à la quantité de corrélations classiques sous la mesure de von Neumann, la quantité de la cohérence quantique n'est pas affectée par la mesure locale $\Pi_{B}$.
\subsection{Concurrence de la cohérence quantique}
Une nouvelle mesure de la cohérence quantique pour un système quantique de dimension arbitraire, appelée la concurrence de la cohérence, a été proposée par Xianfei et ses collègues \cite{Qi2017}. Cette mesure peut être compris comme l'analogue de la concurrence d'intrication qui est basée essentiellement sur l'utilisation de matrices de Gell-Mann généralisées. En outre, elle satisfait à toutes les exigences imposées par la théorie des mesures de la cohérence, ce qui prouve qu'il s'agit d'une bonne mesure de la cohérence quantique.\par
Comme nous l'avons déjà vu, les matrices de Gell-Mann généralisées sont les générateurs de l'algèbre $su\left(d\right)$ définies comme les trois types de matrices différents; les matrices symétriques ${\hat U}_{jk}$ et les matrices antisymétriques ${\hat V}_{jk}$ (\ref{393}) pour $\left\lbrace1\leq j<k\leq d\right\rbrace$, et les matrices diagonales ${\hat W}_{l}$ (\ref{394}) pour $\left\lbrace 1\leq l\leq d-1\right\rbrace$. En utilisant la forme paramétrique de la matrice densité de dimension arbitraire (\ref{repgho}), nous pouvons donner une nouvelle expression de la norme $l_{1}$ de la cohérence $C_{l_{1}}\left(\rho\right)$ basée sur des matrices de Gell-Mann symétriques généralisées ${\hat U}_{jk}$. Cela permet d'obtenir
\begin{align}
C_{l_{1}}\left(\rho\right)&=2\sum_{1\leq j<k\leq d}|\rho_{jk}|\notag\\&=\sum_{1\leq j<k\leq d}|\sqrt{\eta_{1}^{jk}}-\sqrt{\eta_{2}^{jk}}|,\label{3111}
\end{align}
où $\eta_{1}^{jk}$ et $\eta_{2}^{jk}$ sont les valeurs propres non nulles de la matrice $\rho{\hat U}_{jk}\rho^{*}{\hat U}_{jk}$, et $\rho^{*}$ désigne le conjugué complexe dans la base standard. Pour montrer l'équation ci-dessus (\ref{3111}), il suffit juste de le prouver que
\begin{equation}
2|\rho_{jk}|=|\sqrt{\eta_{1}^{jk}}-\sqrt{\eta_{2}^{jk}}|.\label{3112}
\end{equation}
Après un calcul fastidieux mais simple, on peut facilement montrer que les valeurs propres non nulles de la matrice $\rho{\hat U}_{jk}\rho^{*}{\hat U}_{jk}$ sont $\left(|\rho_{jk}|+\sqrt{\rho_{jj}\rho_{kk}}\right)^{2}$ et $\left(|\rho_{jk}|-\sqrt{\rho_{jj}\rho_{kk}}\right)^{2}$. Par conséquent, les racines carrées des valeurs propres non nulles sont $|\rho_{jk}|+\sqrt{\rho_{jj}\rho_{kk}}$ et $\sqrt{\rho_{jj}\rho_{kk}}-|\rho_{jk}|$, ce qui implique l'équation (\ref{3112}) comme nécessaire. Par cette écriture, la concurrence de la cohérence, pour un état pur $\left|\psi\right\rangle$ de dimension arbitraire, est définie comme suit
\begin{equation}
C_{c}\left(\left|\psi\right\rangle \right):=\sum_{1\leq j<k\leq d}|\left\langle \psi\right|{\hat U}_{jk}\left| \psi^{*}\right\rangle|\equiv C_{l_{1}}\left(\left|\psi\right\rangle \left\langle \psi\right|\right).
\end{equation}
Cela signifie que la concurrence de la cohérence est égale à la norme $l_{1}$ de cohérence pour les états purs. En règle générale, si une mesure de la cohérence est définie pour tous les états purs, elle peut être étendue à tous les états mixtes en utilisant les mêmes méthodes de mesures d'intrication. A cet effet, la concurrence de la cohérence d'un état mixte $\rho$ prend la forme 
\begin{equation}
C_{c}\left(\rho\right)=\min_{\left\lbrace p_{i},\left|\psi_{i}\right\rangle\right\rbrace }\sum_{i}p_{i}C_{c}\left(\left|\psi_{i}\right\rangle\right),
\end{equation}
où la minimisation est prise sur toutes les décompositions à l'état pur de $\rho=\sum_{i=1}^{d}p_{i}\left|\psi_{i}\right\rangle\left\langle \psi_{i}\right|$, $\sum_{i}p_{i}=1$. $C_{c}\left(\left|\psi_{i}\right\rangle\right)$ est la norme $l_{1}$ de l'état pur $\left|\psi_{i}\right\rangle$. C'est très similaire à la concurrence d'intrication (\ref{Concu}). Bien que ce quantificateur de la cohérence soit une mesure de cohérence acceptable parce qu'il est valable pour tous les états purs, il n'est pas facile à calculer en général puisque ce calcul implique des minimisations. Pour les états mixtes de type $X$, la décomposition optimale à l'état pur a été trouvée, puis l'expression analytique de la concurrence de la cohérence a été donnée \cite{Zhao2020}. Mais la situation devient beaucoup plus compliquée pour les systèmes généraux à $d$ dimensions.
\section{Conversion de la cohérence locale en intrication collective}
Suite au travail fondamental de Baumgratz \cite{Baumgratz2014}, des recherches fructueuses ont été menées, dont certaines ont été principalement consacrées à la recherche de nouvelles mesures appropriées de la cohérence quantique \cite{Girolami20142,Yuan2015}, ou à l'étude des états à cohérence maximale \cite{Peng2016}, à la distribution de la cohérence quantique dans les systèmes multipartites \cite{Radhakrishnan2016}, et à la relation entre la cohérence et d'autres mesures de corrélation quantique \cite{Streltsov2015,Yao2015,Ma2016}. En ce qui concerne la théorie de la cohérence quantitative, des efforts considérables ont été consacrés à l'élaboration de nombreuses mesures de la cohérence, alors que l'on en sait beaucoup moins sur les relations entre ces mesures, et en particulier sur leur lien avec les autres ressources de la théorie de l'information quantique.\par

La cohérence quantique et l'intrication quantique sont deux manifestations fondamentales de la théorie quantique qui peuvent chacune être caractérisées dans le cadre d'une théorie des ressources opérationnelles. En fait, il n'est pas nécessaire de chercher très loin pour trouver un lien important entre les opérations incohérentes et l'intrication quantique qui est l'une des ressources les plus importantes du traitement de l'information quantique. Nous considérons par exemple la tâche de la génération d'intrication. Cette procédure est généralement modélisée en rassemblant deux ou plusieurs sous-systèmes quantiques initialement dans un état de produit, c'est-à-dire $\rho\otimes\sigma$, puis en appliquant une opération d'intrication conjointe. Toutefois, si seules des opérations incohérentes ont été utilisées, cela ne sera possible que si l'état $\rho$ ou bien l'état $\sigma$ possède déjà une cohérence initiale. La raison en est que lorsque l'état $\rho\otimes\sigma$ est un état bipartite incohérent, toute opération incohérente agissant sur les deux systèmes laissera l'état totale incohérent, et donc l'état final n'est pas intriqué. En revanche, si l'état total est $\left|\pm\right\rangle\left|0\right\rangle$, avec $\left|\pm\right\rangle=\left(\left|0\right\rangle\pm \left|1\right\rangle\right)/\sqrt{2}$, alors une application de la porte ${\rm CNOT}$ (\ref{cnot}) donne l'état intriqué $\left(\left|0\right\rangle+\left|1\right\rangle\right)/\sqrt{2}$. Cet exemple révèle que la cohérence, ou du moins les opérations génératrices de la cohérence, est une condition préalable à la production de l'intrication. Sur la base de cette idée, Streltsov et ses collègues ont montré que chaque état cohérent peut être utilisé pour générer une intrication d'une manière similaire à cet exemple \cite{Streltsov2015}.\par
\begin{figure}[H]
	\centerline{\includegraphics[width=14cm]{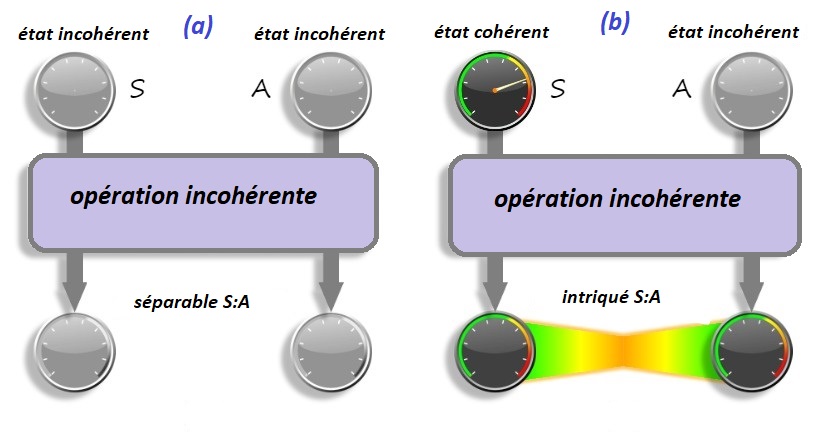}}
	\caption{(a) Si l'état d'entrée est incohérent, les opérations incohérentes ne peuvent pas générer l'intrication quantique, et donc l'état de sortie est un état séparable. (b) A l'inverse, tout état cohérent différent de zéro dans l'état d'un système d'entrée peut être converti en intrication par une opération incohérente sur l'état $S$ et un auxiliaire incohérent $A$.}
	\label{cohint}
\end{figure}

Il est bien connu que l'intrication découle du principe de superposition, qui est également le fondement de la cohérence, ce qui nous pousse à se demander comment une ressource peut émerger quantitativement de l'autre. En général, la cohérence quantique est traitée par l'idée que tous les objets ont des propriétés ondulatoires. Si l'onde d'un objet est divisée en deux parties qui décrivent chacune un état quantique, alors ces deux ondes peuvent interférer de manière cohérente l'une avec l'autre de façon à former un seul état qui est une superposition de ces deux états. Pour le second phénomène qui est l'intrication quantique, les états qui sont en superposition sont les états communs de deux particules intriquées, et non ceux des deux ondes divisées à partir d'une seule particule. À cet égard, le but du cette partie est de fournir une connexion quantitative et opérationnelle claire entre la cohérence et l'intrication. Dans la référence \cite{Streltsov2015}, les auteurs ont montré que tous les états quantiques affichant une cohérence dans une base de référence sont des ressources utiles pour la création d'intrication via des opérations incohérentes. Ce résultat nous permet de définir une classe générale de mesure de la cohérence pour un système quantique de dimension arbitraire, en termes de l'intrication bipartite maximale qui peut être généré via des opérations incohérentes appliquées au système $S$ et à un état auxiliaire incohérent $A$ (voir la Fig.(\ref{cohint})). Par ailleurs, ces résultats montrent clairement l'équivalence qualitative et quantitative entre la cohérence et l'intrication à un niveau fondamental, et fournissent un schéma opérationnel intuitif permettant d'échanger ces deux ressources non classiques pour des applications appropriées dans les technologies quantiques.
En se basant sur la figure (\ref{cohint}), nous disons que l'état final composite d'un système $S$ et un système auxiliaire $A$ initialisé dans un état incohérent de référence $\left|0\right\rangle\left\langle 0\right|^{A}$ pour toute opération incohérente $\Lambda^{SA}$ est séparable, c'est-à-dire l'état $\Lambda^{SA}\left[ \rho^{S}\otimes\left|0\right\rangle\left\langle 0\right|^{A}\right]$ est séparable, si l'état initial $\rho^{S}$ d'un système $S$ à $d$ dimensions est incohérent. En d'autres termes, l'intrication peut être générée par des opérations incohérentes si l'état initial $\rho^{S}$ est cohérent. Au contraire, les systèmes $S$ avec des états incohérents ne peuvent pas être utilisés pour la création d'intrication de cette manière, puisque l'état final $\Lambda^{SA}\left[ \rho^{S}\otimes\left|0\right\rangle\left\langle 0\right|^{A}\right]$ est séparable. Nous pouvons alors conclure que pour toute mesure de l'intrication $E$, l'intrication maximale générée entre $S$ et un auxiliaire incohérent $A$ par des opérations incohérentes définit une cohérence $C_{E}$ sur l'état initial de $S$. Ce phénomène a été démontré en utilisant la porte CNOT à deux qubits (\ref{cnot}) et en considérant les quantificateurs basés sur la distance entre l'état $S$ et l'état séparable le plus proche pour l'intrication $E_{D}$, et la distance entre l'état $S$ et l'état incohérent le plus proche pour la cohérence $C_{D}$ \cite{Streltsov2015}. Nous avons donc
\begin{equation}
E_{D}\left(\rho\right)=\min_{\sigma\in {\cal S}}D\left(\rho,\sigma \right),\hspace{1cm}C_{D}\left(\rho\right)=\min_{\sigma\in {\cal I}}D\left(\rho,\sigma \right),
\end{equation}
avec ${\cal S}$ est l'ensemble des états séparables et ${\cal I}$ est l'ensemble des états incohérents. En outre, pour que ces mesures soient valables, nous insistons pour que la distance $D$ soit contractive dans le cadre des opérations quantiques. Cela nous donne
\begin{equation}
D\left(\Lambda\left(\rho \right),\Lambda\left(\sigma \right) \right) \leq D\left(\rho,\sigma\right), 
\end{equation}
pour toute application $\Lambda$ complètement positive et préservant des traces. D'autre part, si l'état $\sigma^{S}$ est l'état incohérent le plus proche de $\rho^{S}$, c'est-à-dire $C_{D}\left(\rho^{S}\right)=D\left(\rho^{S}, \sigma^{S}\right)$, alors la contractivité de la distance $D$ implique l'égalité suivante
\begin{equation}
	D\left(\rho^{S},\sigma^{S}\right)=D\left(\rho^{S}\otimes \left|0\right\rangle\left\langle 0\right|^{A},\sigma^{S}\otimes \left|0\right\rangle\left\langle 0\right|^{A}\right).  
\end{equation}
Il convient de noter que l'application d'une opération incohérente $\Lambda^{SA}$ à l'état incohérent $\sigma^{S}\otimes \left|0\right\rangle\left\langle 0\right|^{A}$ conduit à un autre état incohérent et qui est séparable (voir la figure (\ref{cohint})). Nous trouvons alors
\begin{align}
	C_{D}\left(\rho^{S} \right)&=D\left(\rho^{S}\otimes \left|0\right\rangle\left\langle 0\right|^{A},\sigma^{S}\otimes \left|0\right\rangle\left\langle 0\right|^{A}\right)\notag\\&\geq D\left(\Lambda^{SA} \left[\rho^{S}\otimes \left|0\right\rangle\left\langle 0\right|^{A}\right],\Lambda^{SA} \left[\sigma^{S}\otimes \left|0\right\rangle\left\langle 0\right|^{A}\right]\right)\notag\\&\geq\min_{\delta\in S} D\left(\Lambda^{SA} \left[\rho^{S}\otimes \left|0\right\rangle\left\langle 0\right|^{A}\right],\delta\right)=E_{D}^{S:A}\left(\Lambda^{SA} \left[\rho^{S}\otimes \left|0\right\rangle\left\langle 0\right|^{A}\right] \right).   
\end{align}
Par conséquent, pour toute mesure de la distance $D$, la quantité d'intrication $E_{D}$ générée à partir d'un état $\rho^{S}$ par une opération incohérente $\Lambda^{SA}$ est limitée par la quantité de la cohérence $C_{D}$, de sorte que
\begin{equation}
E_{D}^{S:A}\left(\Lambda^{SA} \left[\rho^{S}\otimes \left|0\right\rangle\left\langle 0\right|^{A}\right]\right)\leq C_{D}\left(\rho^{S}\right).\label{3119}
\end{equation}
Cela nous permet d'unifier les théories des ressources de l'intrication et de la cohérence via leurs comportements combinés. D'autre part, il faut reconnaître que les quantificateurs de l'intrication et de la cohérence sont différentes les unes des autres, cela est lié à la mesure de la distance que nous choisissons, et chaque mesure a des caractéristiques majeures. Les auteurs de la référence \cite{Streltsov2015} soulignent que l'inégalité (\ref{3119}) peut être saturée lorsque $D$ est spécifié comme l'entropie relative et si la dimension du système auxiliaire $A$ n'est pas inférieure à celle du système $S$ (c-à-d $d_{A}\geq d_{S}$). Dans ce cas, il existe toujours une opération incohérente $\Lambda^{SA}$, tel que l'entropie relative de l'intrication $E_{r}$ (\ref{Er}) et l'entropie relative de la cohérence $C_{r}$ (\ref{Cr}) satisfont l'égalité suivante
\begin{equation}
	E_{r}^{S:A}\left(\Lambda^{SA} \left[\rho^{S}\otimes \left|0\right\rangle\left\langle 0\right|^{A}\right]\right)= C_{r}\left(\rho^{S}\right).
\end{equation}
L'équation ci-dessus montre que toute quantité non nulle de l'entropie relative de la cohérence $C_{r}$ dans un système $S$ peut être convertie en entropie relative d'intrication $E_{r}$ entre $S$ et un auxiliaire $A$ initialement incohérente, par des opérations incohérentes $\Lambda^{SA}$. Dans le cas où la distance $D$ est choisie comme étant l'entropie relative quantique, nous pouvons prouver la borne (\ref{3119}) si nous considérons l'opération unitaire incohérente comme
\begin{align}
U=\sum_{i=0}^{d_{S}-1}\sum_{j=0}^{d_{S}-1}\left|i \right\rangle\left\langle i\right|^{S}\otimes\left|{\rm mod}\left(i+j,d_{S} \right)\right\rangle \left\langle j\right|^{A}+\sum_{i=0}^{d_{S}-1}\sum_{j=d_{S}}^{d_{A}-1}\left|i \right\rangle\left\langle i\right|^{S}\otimes \left|j \right\rangle\left\langle j\right|^{A},\label{UCNOT}
\end{align}
avec $d_{S}$ et $d_{A}$ sont les dimensions du système $S$ et du système auxiliaire $A$, respectivement. Notez que pour deux qubits, l'équation (\ref{UCNOT}) est équivalent à la porte CNOT (\ref{cnot}). On peut facilement constater qu'elle fait évoluer l'état $\rho^{S}\otimes\left|0\right\rangle\left\langle 0\right|^{A}$ vers l'état
\begin{equation}
\Lambda^{SA}\left[\rho^{S}\otimes\left|0\right\rangle\left\langle 0\right|^{A}\right]=U\left[\rho^{S}\otimes\left|0\right\rangle\left\langle 0\right|^{A}\right]U^{\dagger}=\sum_{i,j}\rho_{ij}\left|i\right\rangle\left\langle j\right|^{S}\otimes\left|i\right\rangle\left\langle j\right|^{A},
\end{equation}
où $\rho_{ij}$ sont les éléments matriciels de $\rho^{S}=\sum_{ij}\rho_{ij}\left|i\right\rangle\left\langle j\right|^{S}$. Dans la prochaine étape, nous utilisons le fait que pour tout état quantique $\varsigma^{SA}$, l'entropie relative de l'intrication est bornée comme suit \cite{Plenio2000}
\begin{equation}
E_{r}^{S:A}\left( \varsigma^{SA}\right)\geq S\left(\varsigma^{S} \right)-S\left(\varsigma^{SA}\right),
\end{equation}
avec $S\left(\varsigma^{S} \right)$ représente l'entropie de von Neumann de l'état réduit $\varsigma^{S}$. Si on applique cette inégalité à l'état $\Lambda^{SA}\left[\rho^{S}\otimes\left|0\right\rangle\left\langle 0\right|^{A}\right]$, on constate
\begin{equation}
E_{r}^{S:A}\left(\Lambda^{SA}\left[\rho^{S}\otimes\left|0\right\rangle\left\langle 0\right|^{A}\right]\right)\geq S\left(\sum_{i}\rho_{ii}\left|i\right\rangle \left\langle i\right|^{S} \right)-S\left(\rho^{S}\right).
\end{equation}
Remarquant que le côté droit de cette inégalité est égal à l'entropie relative de la cohérence $C_{r}\left(\rho^{S} \right)$ (voir l'équation (\ref{384})). On obtient alors
\begin{equation}
E_{r}^{S:A}\left(\Lambda^{SA}\left[\rho^{S}\otimes\left|0\right\rangle\left\langle 0\right|^{A}\right]\right)\geq C_{r}\left(\rho^{S} \right).
\end{equation}
Ce résultat montre que la cohérence et l'intrication peuvent être convertis l'un à l'autre dans ce scénario particulièrement intéressant. Sur la base de ces résultats, l'intrication maximale $E^{S:A}$ générée par rapport à tout intrication monotone donnée via des opérations incohérentes, définit une cohérence monotone $C_{E}$ comme suit
\begin{equation}
C_{E}\left(\rho^{S}\right)=\lim_{d_{A}\longrightarrow \infty}\left\lbrace \max_{\Lambda^{SA}}E^{S:A}\left(\Lambda^{SA}\left[\rho^{S}\otimes\left|0\right\rangle \left\langle 0\right|^{A}\right]\right)\right\rbrace,
\end{equation}
où $d_{A}$ est la dimension de système auxiliaire, et la maximisation porte sur toutes les opérations incohérentes $\Lambda^{SA}$. A ce propos, il est intéressant de souligner que $C_{E}$ est une cohérence monotone chaque fois que $E^{S:A}$ est une intrication monotone. En particulier, $C_{E}$ remplit toutes les conditions de la théorie de la cohérence pour l'ensemble des opérations incohérentes $\Lambda^{SA}$ chaque fois que $E^{S:A}$ remplit les conditions correspondantes dans la théorie de l'intrication. En outre, cette procédure peut induire des mesures de la cohérence, y compris l'entropie relative de la cohérence et la cohérence géométrique, à partir des mesures de l'intrication. Fait intéressant, $C_{E}=C_{r},C_{g}$ quand $E=E_{r},E_{g}$, respectivement. Cependant, on sait peu de choses sur les mesures qui peuvent être induites de cette manière au-delà de ces exemples, et le lien entre la cohérence et l'intrication est loin d'être clair.

\chapter{Théorie de l'estimation des paramètres et sa relation avec les corrélations quantiques}
La métrologie quantique, ou la théorie de l'estimation quantique, est d'une importance capitale dans le développement de dispositifs de haute précision dans plusieurs domaines technologiques \cite{Giovannetti2006,Giovannetti2011}. Son objectif principal est de réaliser des mesures de haute précision en estimant les paramètres inconnus qui spécifient un système quantique donné, à l'aide d'effets quantiques. En ce sens, elle vise à développer de nouvelles méthodes pour améliorer la limite de précision des paramètres physiques au-delà des méthodes métrologiques classiques. En bref, elle s'intéresse à la plus grande précision possible dans diverses tâches d'estimation des paramètres, et à la recherche des schémas de mesure qui atteignent cette précision. À l'origine, la métrologie était axée sur les mesures effectuées à l'aide de systèmes classiques, tels que les systèmes mécaniques décrits par la physique classique ou les systèmes optiques modélisés par l'optique ondulatoire classique. Nous devons donc nous demander comment pouvons nous déduire les limites de précision de l'estimation des paramètres, ainsi que les méthodes qui peuvent améliorer la précision dans les systèmes quantiques? Y a-t-il une limite fondamentale? Dans le schéma classique, les premières réponses sont apparues vers les années 1940 avec les travaux de Rao \cite{Rao1945} et Cramer \cite{cramer1946}, qui ont trouvé indépendamment une limite inférieure à la variance d'un estimateur arbitraire. Ces résultats se sont étendus au cas multiparamétrique par Darmois \cite{Darmois1945}. Cette limite, généralement appelée la limite de Cramér-Rao, est intimement liée à l'information de Fisher, introduite par Fisher dans les années 1920 \cite{Fisher1923}. L'information de Fisher joue donc un rôle central dans la théorie de l'estimation. Sa maximisation sur toutes les mesures quantiques possibles définit l'information quantique de Fisher \cite{Braunstein1994,Braunstein1996} et fournit une limite quantique inférieure à la borne de Cramér-Rao.\par

En revanche, la théorie de l'estimation quantique est le cadre mathématique permettant d'aborder l'optimisation d'une mesure quantique. Elle s'applique aux situations où il s'agit de prédire la valeur d'un paramètre en effectuant une série répétée de mesures sur des préparations identiques du système, puis en traitant les données pour estimer la valeur du paramètre inconnu \cite{Paris2009}. Pour accomplir toutes ces tâches, nous aurions besoin de phénomènes quantiques comme ressources, tels que l'intrication et la cohérence quantique, afin d'améliorer la sensibilité d'un système. En même temps, les systèmes quantiques imposent des restrictions intrinsèques à la sensibilité qui peut être obtenue, par exemple grâce aux relations d'incertitude de Heisenberg, qui stipulent que les variables complémentaires ne peuvent pas être mesurées simultanément avec une précision illimitée. Pour déterminer comment les propriétés quantiques améliorent mais aussi restreignent la sensibilité du système, nous modélisons sa dynamique et nous nous concentrons sur une quantité clé qui est l'information quantique de Fisher. \par

L'information quantique de Fisher est un concept fondamental en théorie de l'estimation des paramètres quantiques, en raison de son lien important avec la borne de Cramér-Rao quantique. Selon cette borne, une plus grande précision est obtenue pour les petites variances, qui correspondent aux plus grandes valeurs de l'information quantique de Fisher. Ainsi, le but principal de tous les protocoles de la métrologie quantique est d'atteindre la plus petite valeur de la variance. Cependant, des études récentes ont révélé de larges liens entre l'information quantique de Fisher et d'autres aspects de la mécanique quantique, y compris la transition de phase quantique \cite{Ye2016}, la caractérisation des corrélations quantiques \cite{Kim2018}, la thermodynamique quantique \cite{Hasegawa2020}, le contrôle de l'intrication \cite{Blondeau2017} et la limite de vitesse quantique \cite{Taddei2013}. Ces liens indiquent que c'est plus qu'un concept en métrologie quantique, mais plutôt une quantité fondamentale en mécanique quantique.\par

Dans ce chapitre, nous commençons par introduire les concepts de base de la théorie de l'estimation avant d'aborder l'avantage quantique des schémas de détection. Nous procédons ensuite à la dérivation de l'information de Fisher classique, qui est une mesure d'information pouvant être liée à la variance d'un paramètre. Nous montrons ensuite comment l'information de Fisher classique peut être généralisée pour toutes les mesures possibles à l'information de Fisher quantique. La motivation pour étudier ces quantités devient claire lorsque nous introduisons et dérivons la borne de Cramér-Rao, qui relie l'information de Fisher quantique et classique à la variance d'un paramètre. Cette limite quantique de Cramér-Rao est toujours atteinte à saturation dans le cas où un seul paramètre est estimé, et dans ce cas, les corrélations quantiques nous aident à améliorer la précision et l'efficacité des protocoles de la métrologie quantique \cite{Zhang2014,Hyllus2010}. À l'inverse, il est difficile de saturer cette limite dans le cas d'une estimation multiparamétrique, en raison de l'incompatibilité entre les mesures optimales des différents paramètres estimés. L'objet pertinent dans le problème d'estimation multiparamétrique est donné par ce qu'on appelle la matrice d'information quantique de Fisher \cite{Braunstein1994}. À cette fin, dans la deuxième partie de ce chapitre, nous fournissons les techniques complètes sur le calcul de cette matrice, et nous montrons que la stratégie simultanée de plusieurs paramètres est toujours avantageuse et peut fournir une meilleure précision que la stratégie individuelle dans les procédures d'estimation multiparamétrique. La caractérisation des corrélations quantiques en termes d'information quantique de Fisher sera également discutée.

\section{Théorie de l'estimation classique}
La théorie de l'estimation est un concept mathématique important utilisé dans de nombreuses applications de communication et de traitement du signal. Cette théorie est utile pour estimer l'information désirée dans les données reçues elle est donc utilisée dans toute une série d'applications allant du radar au traitement vocal \cite{Steven1993}. Dans cette section, nous énumérerons les concepts de base des statistiques ainsi que le processus d'estimation. Comme nous l'avons dit, les principaux outils mathématiques utilisés par la métrologie appartiennent à la statistique. De plus, nous nous intéressons particulièrement à la théorie de l'estimation, qui montre comment estimer correctement une quantité à partir d'un échantillon de données. Les données peuvent être de n'importe quel type. Par exemple, l'échantillon de données peut être un ensemble des résultats de tirages au sort, ou même les longueurs d'onde des photons sortant d'un échantillon radioactif. Supposons qu'un processus non déterministe produise un ensemble de données $x=\left\lbrace x_{1},x_{2},...,x_{n}\right\rbrace$, qui est une occurrence particulière (ou réalisation) d'un ensemble de variables aléatoires indépendantes et identiquement distribuées désignées par $X=\left\lbrace X_{1},X_{2},...,X_{n}\right\rbrace$. La fonction de densité de probabilité correspondant à toute variable aléatoire de cet ensemble est $p\left(x;\theta\right)$, où $x\in R$ est une réalisation possible d'une variable aléatoire et $\theta$ est un nombre réel (par exemple, cela pourrait être un paramètre physique, comme la phase dans un interféromètre optique) qui est paramétrisé par $p\left(x;\theta\right)$. Ainsi, la probabilité qu'un résultat de mesure de notre ensemble de données se trouve dans l'intervalle $x_{0}\leq x\leq x_{f}$ est donnée par l'intégrale de $p\left(x;\theta\right)$ comme suit: $\int_{x_{0}}^{x_{f}}p\left(x;\theta\right)dx$.\par
En général, le problème central de l'estimation des paramètres est le suivant: Pour un ensemble de données $x$ de $n$ résultats d'essais, avec quelle précision pouvons-nous prédire la valeur de $\theta$? En effet, en physique, le paramètre $\theta$ est une propriété physique du système étudié et nous voulons déterminer sa valeur à partir des données expérimentales $x_{i}$. Pour cela, un estimateur est utilisé pour transformer les données $x$ en une prédiction de $\theta$. Nous désignerons un tel estimateur par $\boldsymbol{\hat{\theta}}\left(x \right)$, qui se trouve être une variable aléatoire puisqu'elle dépend des valeurs prises des variables aléatoires dans $X$. Nous pouvons maintenant reformuler le problème d'estimation comme suit; Pour tous ensemble de données $x$, quelle est la petite valeur de l'erreur de notre estimateur? Plus précisément, l'erreur minimale est donnée par la variance de cet estimateur ${\rm Var}_{\theta}\left(\hat{\theta}\right)$. Elle est définie par
\begin{align}
{\rm Var}_{\theta}\left(\hat{\theta}\right)&:=\textbf{E}\left[\left(\boldsymbol{\hat{\theta}}\left(x\right) -\theta\right)^{2}\right]\notag\\&=\int p\left(x_{1};\theta\right)p\left(x_{2};\theta\right)...p\left(x_{n};\theta\right)\left(\boldsymbol{\hat{\theta}}\left(x\right)-\theta\right)^{2}dx_{1}dx_{2}...dx_{n}\notag\\&=\int p\left(x;\theta\right)\left(\boldsymbol{\hat{\theta}}\left(x\right)-\theta\right)^{2}dx^{n},\label{Delta}
\end{align}
elle est utilisée pour quantifier la performance de notre tâche d'estimation, avec $\textbf{E}$ est une espérance mathématique. Dans l'équation (\ref{Delta}), nous avons utilisé le fait que $dx^{n}=dx_{1}dx_{2}...dx_{n}$ et $p\left(x;\theta\right)=p\left(x_{1};\theta\right)p\left(x_{2};\theta\right)...p\left(x_{n};\theta\right)$.\par 
Il est donc clair que la loi de $X$ sera complètement connue si la fonction de distribution de probabilité $p\left(x;\theta\right)$ est connue, cela implique nécessairement la connaissance de la valeur des paramètres inconnus $\theta$. Dans ce contexte, la théorie de l'estimation vise à approcher numériquement la valeur des paramètres inconnus sans connaître la densité de probabilité et la loi de probabilité qui régit notre problème. Dans ce formalisme, nous faisons l'hypothèse implicite que $\theta$ est connaissable avec une précision arbitraire à condition d'employer un estimateur approprié $\boldsymbol{\hat{\theta}}\left(x\right)$ et d'avoir accès à un ensemble de données $x$.

\subsection{Modèles statistiques classiques}

Considérons une expérience aléatoire dont les résultats sont décrits par une variable aléatoire $X$, avec un espace de probabilité $\left(X,x,p\left(x\right) \right)$ et une densité de probabilité $p\left(x\right)$. La tâche consiste à reconstruire $p\left(x\right)$, que l'on appelle la véritable densité de probabilité, à partir de $N$ points d'échantillonnage indépendants ou observations de $X$. Il existe de nombreuses façons d'aborder le problème de l'apprentissage de $p\left(x\right)$, mais si la forme fonctionnelle de $p\left(x\right)$ est déjà connue, ou peut être supposée avec une précision raisonnable, une approche paramétrique est tout à fait naturelle. La densité de probabilité réelle $p\left(x\right)$ est supposée appartenir à une famille paramétrique de densités de probabilité $\left\lbrace p\left(x;\theta\right)\right\rbrace_{\theta\in\Theta}$, où $\Theta\subset\mathbb{R}^{n}$ est l'espace des paramètres.\par 
Un modèle statistique classique $S_{C}$ est une famille de densités de probabilité sur $X$ paramétrée par $n$ paramètres réels $\theta\in\Theta\subset\mathbb{R}^{n}$ et noté par
\begin{equation}
S_{C}=\left\lbrace \forall x\in X; p\left(x;\theta\right),\hspace{0.25cm}\left(X,x,p\left(x\right) \right)\mapsto\Theta,\hspace{0.25cm}\theta\equiv\left(\theta^{1},\theta^{2},...,\theta^{n}\right)\in\Theta\right\rbrace,
\end{equation}
où l'application de paramétrisation $\left(X,x,p\left(x\right) \right)\mapsto\Theta$ est injective et peut être aussi dérivable $n^{ime}$ fois que si nécessaire par rapport aux paramètres estimés, c'est-à-dire que toutes les dérivées possibles $\partial_{\theta^{1}}...\partial_{\theta^{n}}p\left(x;\theta\right)$ existent. Il est très intéressant de souligner que la densité de probabilité P est toujours positive et normalisée. Nous avons donc deux cas; Si $X$ est dénombrable, alors $p\left(x;\theta\right)$ est décrit dans un espace discret et normalisé tel que 
\begin{equation}
\sum_{x\in X}p\left(x;\theta\right)=1, \hspace{1cm}\forall \theta\in\Theta
\end{equation}
Si $X$ est indénombrable, alors $p\left(x;\theta\right)$ est décrite dans un espace continu et normalisé de sorte que
\begin{equation}
\int_{x}p\left(x;\theta\right)dx=1, \hspace{1cm}\forall \theta\in\Theta.
\end{equation}
Pour simplifier notre discussion, nous utilisons cette dernière notation tout au long de ce chapitre.
\subsection{L'information de Fisher classique}
Les mesures sur les systèmes physiques produisent des résultats probabilistes, et l'estimation des paramètres physiques, décrivant un système ou régissant son évolution, est un problème d'inférence statistique. La source des erreurs statistiques peut être liée aux imperfections de l'expérience ou des mesures de lecture, ou être plus fondamentale, imposée par exemple, par les relations d'incertitude de Heisenberg dans une configuration quantique. Dans le cas classique, une quantité qui est la clé de cette discussion est le Score \cite{Cox1979}, que nous appelons ${\cal L}_{\theta}\left(x\right)$. Le Score est la dérivée partielle du logarithme naturel de la fonction de vraisemblance, et indique donc la sensibilité du système à un changement infinitésimal de la valeur du paramètre $\theta$. Le Score est défini comme
\begin{equation}
{\cal L}_{\theta}\left(x\right) =\frac{\partial}{\partial \theta}\log p\left(x;\theta\right),
\end{equation}
qui mesure la sensibilité de la fonction de densité de probabilité $p\left(x;\theta\right)$ aux changements du paramètre $\theta$. Cela découle du fait que
\begin{align}
\textbf{E}\left({\cal L}_{\theta}\left(x\right)\right)&=\int p\left(x;\theta \right)\frac{\partial}{\partial \theta}\log p\left(x;\theta\right)dx\notag\\&=\int p\left(x;\theta \right)\frac{1}{p\left(x;\theta \right)}\frac{\partial}{\partial \theta}p\left(x;\theta \right)dx\notag\\&=\int \frac{\partial}{\partial \theta}p\left(x;\theta \right)dx=0.\label{46}
\end{align}
L'information de Fisher classique ${\cal F}\left(\theta\right)$ dans une seule observation d'une fonction de densité de probabilité $p\left(x;\theta\right)$ est la variance du Score ${\cal L}_{\theta}\left(x\right)$, c'est-à-dire
\begin{align}
{\cal F}\left(\theta\right)&:= {\cal F}\left( p\left(x;\theta\right)\right):={\rm Var}\left( {\cal L}_{\theta}\left(x\right)\right)\notag\\&=\textbf{E}\left[\left(\frac{\partial}{\partial \theta}\log p\left(x;\theta\right) \right)^{2}\right]\notag\\&=\int p\left(x;\theta \right)\left(\frac{\partial}{\partial \theta}\log p\left(x;\theta\right) \right)^{2}dx.\label{FC}
\end{align}
De manière formelle, l'information de Fisher (\ref{FC}) quantifie l'information fournie par un échantillon sur le paramètre $\theta$. En complément, une information de Fisher proche de zéro indique un échantillon avec peu d'information sur la valeur de $\theta$. Nous notons ici une propriété importante et utile de l'information de Fisher classique; à savoir, la non-négativité et l'additivité pour les événements non corrélés. Autrement dit, si $p\left(\boldsymbol{x};\theta\right)=p\left(x_{1};\theta\right)p\left(x_{2};\theta\right)...p\left(x_{n};\theta\right)$, alors
\begin{align}
{\cal F}\left(p\left(\boldsymbol{x};\theta\right)\right)&={\rm Var}\left[\frac{\partial}{\partial \theta}\log p\left(x_{1};\theta\right)+...+\frac{\partial}{\partial \theta}\log p\left(x_{n};\theta\right)\right]\notag\\&=n{\rm Var}\left[\frac{\partial}{\partial \theta}\log p\left(x_{1};\theta\right) \right]\notag\\&=n{\cal F}\left(\theta\right).
\end{align}
Par conséquent, l'information de Fisher dans un échantillon aléatoire est juste $n$ fois l'information dans une seule observation. Cela devrait avoir un sens intuitif, car cela signifie que plus d'observations produisent plus d'informations.
\subsection{Borne de Cramér-Rao}
La borne de Cramér-Rao fournit une limite sur la façon dont l'estimation peut être effectuée. Elle fixe une limite fondamentale à la variance de l'estimateur $\boldsymbol{\hat{\theta}}$ quelle que soit la façon dont il est construit. Elle est exprimée par la formule suivante \cite{Rao1945,cramer1946}
\begin{equation}
{\rm Var}\left(\boldsymbol{\hat{\theta}}\right)\geq \frac{1}{n{\cal F}\left(\theta\right)},\label{BCR}
\end{equation}
où $n$ est le nombre d'essais (le nombre d'entrées dans l'ensemble de données $\boldsymbol{x}$). Cette borne (\ref{BCR}) déclare que l'erreur de tout estimateur non biaisé $\boldsymbol{\hat{\theta}}$ de $\theta$ est limitée par l'inverse de l'information de Fisher. En d'autres termes, cela indique que la variance d'un estimateur $\boldsymbol{\hat{\theta}}$ est au moins aussi élevée que l'inverse de l'information de Fisher, c'est-à-dire que plus l'information de Fisher est grande, plus on peut s'attendre à une précision d'estimation élevée. Ainsi, l'information de Fisher est la quantité qui produit une limite ultime sur la précision pouvant être atteinte dans un problème d'estimation des paramètres. Un estimateur pour lequel l'égalité est respectée est dit efficace, mais, avec un nombre fini de mesures, il n'y a aucune garantie qu'un tel estimateur existe.\par
Nous fournissons maintenant une preuve de la limite de Cramér-Rao en suivant une approche similaire à celle de la référence \cite{Kok2010}. Afin de la prouver (\ref{BCR}), nous devons faire l'hypothèse que l'estimateur $\boldsymbol{\hat{\theta}}\left(x\right)$ doit être non biaisé. Ce qui signifie qu'en moyenne, l'estimateur $\boldsymbol{\hat{\theta}}$ est égal à la vraie valeur de $\theta$;
\begin{equation}
\textbf{E}\left(\hat{\theta}\left(\boldsymbol{x} \right)\right)\equiv\theta.\label{biaise}
\end{equation}
En déplaçant $\theta$ vers le côté gauche et en écrivant explicitement la moyenne en termes de la fonction de densité de probabilité $p\left(\boldsymbol{x};\theta\right)$. L'équation (\ref{biaise}) devient
\begin{align}
\int p\left(\boldsymbol{x};\theta\right)\left(\textbf{E}\left(\hat{\theta}\left(\boldsymbol{x} \right)\right)-\theta \right)d\boldsymbol{x}^{n}=0. 
\end{align}
En dérivant cette expression par rapport à $\theta$, nous obtenons
\begin{align}
\frac{\partial}{\partial\theta}\int p\left(\boldsymbol{x};\theta\right)\left(\textbf{E}\left(\hat{\theta}\left(\boldsymbol{x} \right)\right)-\theta \right)d\boldsymbol{x}^{n}=\int \frac{\partial}{\partial\theta}\left(p\left(\boldsymbol{x};\theta\right) \right)\left(\textbf{E}\left(\hat{\theta}\left(\boldsymbol{x} \right)\right)-\theta \right)d\boldsymbol{x}^{n}-\int p\left(\boldsymbol{x};\theta\right)d\boldsymbol{x}^{n}=0,\label{412}
\end{align}
où nous avons utilisé le fait que l'estimateur ne dépend pas explicitement de $\theta$, donc $\frac{\partial}{\partial\theta}\left(\hat{\theta}\left(\boldsymbol{x} \right)-\theta \right)=-1$. Pour toute fonction de densité de probabilité, nous avons $\int p\left(\boldsymbol{x};\theta\right)d\boldsymbol{x}^{n}=1$. Par conséquent, l'équation (\ref{412}) nous donne
\begin{align}
\int \frac{\partial p\left(\boldsymbol{x};\theta\right)}{\partial\theta}\left(\hat{\theta}\left(\boldsymbol{x} \right)-\theta \right)d\boldsymbol{x}^{n}=1.\label{413}
\end{align}
Notons que, $\partial p\left(\boldsymbol{x};\theta\right)/\partial\theta$ peut être encore écrite de la façon suivante
\begin{align}
\frac{\partial}{\partial\theta}\left(\prod_{k=1}^{n}p\left(x_{k};\theta\right) \right)&=\frac{\partial p\left(x_{1};\theta\right)}{\partial\theta}\prod_{k\neq1}^{n}p\left(x_{k};\theta\right)+\frac{\partial p\left(x_{2};\theta\right)}{\partial\theta}\prod_{k\neq2}^{n}p\left(x_{k};\theta\right)+...+\frac{\partial p\left(x_{n};\theta\right)}{\partial\theta}\prod_{k\neq n}^{n}p\left(x_{k};\theta\right)\notag\\&=\sum_{l=1}^{n}\left[ \frac{\partial p\left(x_{l};\theta\right)}{\partial\theta}\prod_{k\neq l}^{n}p\left(x_{k};\theta\right)\right],
\end{align}
où $\prod_{k\neq l}^{n}p\left(x_{k};\theta\right)$ désigne la multiplication de tous les termes $p\left(x_{k};\theta\right)$ pour $k\in\left\lbrace 1,2,...,n\right\rbrace$ à l'exception du $l$-ème terme. En écrivant $\prod_{k\neq l}^{n}p\left(x_{k};\theta\right)=p\left(\boldsymbol{x};\theta\right)/p\left(x_{l};\theta\right)$, on obtient
\begin{align}
\frac{\partial p\left(\boldsymbol{x};\theta\right)}{\partial\theta}=\sum_{l=1}^{n}\frac{1}{p\left(x_{l};\theta\right)}\frac{\partial p\left(x_{l};\theta\right)}{\partial\theta}p\left(\boldsymbol{x};\theta\right)=p\left(\boldsymbol{x};\theta\right)\sum_{l=1}^{n}\frac{\partial\log p\left(x_{l};\theta\right)}{\partial\theta}.
\end{align}
L'équation précédente peut être utilisée pour réexprimer l'équation (\ref{413}) comme suit
\begin{align}
\int p\left(\boldsymbol{x};\theta\right)\sum_{l=1}^{n}\frac{\partial\log p\left(x_{l};\theta\right)}{\partial\theta}\left(\hat{\theta}\left(\boldsymbol{x} \right)-\theta \right)d\boldsymbol{x}^{n}&=\int\left[\sqrt{p\left(\boldsymbol{x};\theta\right)}\sum_{l=1}^{n}\frac{\partial\log p\left(x_{l};\theta\right)}{\partial\theta} \right]\notag\\&\times\left[\sqrt{p\left(\boldsymbol{x};\theta\right)}\left(\hat{\theta}\left(\boldsymbol{x} \right)-\theta \right) \right]d\boldsymbol{x}^{n}=1, \label{416}
\end{align}
où nous avons décidé d'écrire la fonction de densité de probabilité $p\left(\boldsymbol{x};\theta\right)$ comme un produit de ses racines carrées. L'équation (\ref{416}) a donc la forme suivante
\begin{equation}
\int f\left(\boldsymbol{x}\right) g\left(\boldsymbol{x} \right)d\boldsymbol{x}^{n}=1
\end{equation}
avec
\begin{equation}
f\left(\boldsymbol{x} \right)\equiv\sqrt{p\left(\boldsymbol{x};\theta\right)}\sum_{l=1}^{n}\frac{\partial\log p\left(x_{l};\theta\right)}{\partial\theta},\hspace{1cm}{\rm et}\hspace{1cm}g\left(\boldsymbol{x} \right)\equiv\sqrt{p\left(\boldsymbol{x};\theta\right)}\left(\hat{\theta}\left(\boldsymbol{x} \right)-\theta \right). 
\end{equation}
Nous allons maintenant utiliser l'inégalité de Cauchy-Schwarz \cite{Helstrom1976}:
\begin{equation}
\left(\int f\left(\boldsymbol{x} \right) g\left(\boldsymbol{x} \right)d\boldsymbol{x}^{n}\right)^{2}\leq\int f\left(\boldsymbol{x}\right)^{2}d\boldsymbol{x}^{n}\int g\left(\boldsymbol{x}^{\prime}\right)^{2}d{\boldsymbol{x}^{\prime}}^{n}.
\end{equation}
Nous élevons au carré les deux côtés de l'équation (\ref{416}) et nous appliquons l'inégalité de Cauchy-Schwarz pour obtenir
\begin{equation}
1\leq \left[\int p\left(\boldsymbol{x};\theta\right)\left(\sum_{l=1}^{n}\frac{\partial\log p\left(x_{l};\theta\right)}{\partial\theta}\right)^{2}d\boldsymbol{x}^{n}\right]\left[\underbrace {\int p\left(\boldsymbol{x}^{\prime};\theta\right) \left(\hat{\theta}\left(\boldsymbol{x}^{\prime} \right)-\theta \right)^{2}d{\boldsymbol{x}^{\prime}}^{n}}_{{\rm Var}\left(\boldsymbol{\hat{\theta}}\right)}\right].  
\end{equation}
Notons que le deuxième terme de la dernière équation est simplement la variance de l'estimateur $\boldsymbol{\hat{\theta}}$. Alors
\begin{equation}
{\rm Var}\left(\boldsymbol{\hat{\theta}}\right)\geq\frac{1}{\int p\left(\boldsymbol{x};\theta\right)\left(\sum_{l=1}^{n}\frac{\partial\log p\left(x_{l};\theta\right)}{\partial\theta}\right)^{2}d\boldsymbol{x}^{n}}.\label{421}
\end{equation}
Pour compléter la preuve, il suffit de montrer que le dénominateur de l'équation (\ref{421}) est égal à $n{\cal F}\left(\theta\right)$. En développant le terme de sommation dans le dénominateur, nous pouvons l'exprimer comme suit
\begin{align}
\int p\left(\boldsymbol{x};\theta\right)\left(\sum_{l=1}^{n}\frac{\partial\log p\left(x_{l};\theta\right)}{\partial\theta}\right)^{2}d\boldsymbol{x}^{n}&=\int p\left(\boldsymbol{x};\theta\right)\sum_{k=1}^{n}\sum_{l=1}^{n} \frac{\partial\log p\left(x_{l};\theta\right)}{\partial\theta}\frac{\partial\log p\left(x_{k};\theta\right)}{\partial\theta}d\boldsymbol{x}^{n}\notag\\&=\int p\left(\boldsymbol{x};\theta\right)\sum_{l=1}^{n}\left( \frac{\partial\log p\left(x_{l};\theta\right)}{\partial\theta}\right)^{2}d\boldsymbol{x}^{n}\notag\\&+\int p\left(\boldsymbol{x};\theta\right)\sum_{k,l=1,k\neq l}^{n} \frac{\partial\log p\left(x_{l};\theta\right)}{\partial\theta}\frac{\partial\log p\left(x_{k};\theta\right)}{\partial\theta}d\boldsymbol{x}^{n}.\label{422}
\end{align}
Le dernier terme de cette l'équation peut être encore réécrit comme
\begin{align}
\int p\left(\boldsymbol{x};\theta\right)\sum_{k,l=1,k\neq l}^{n} \frac{\partial\log p\left(x_{l};\theta\right)}{\partial\theta}&\frac{\partial\log p\left(x_{k};\theta\right)}{\partial\theta}d\boldsymbol{x}^{n}=\int dx_{1}dx_{2}...dx_{n}p\left(x_{1};\theta\right)p\left(x_{2};\theta\right)...p\left(x_{n};\theta\right)\notag\\&\times\sum_{k,l=1,k\neq l}^{n} \frac{\partial\log p\left(x_{l};\theta\right)}{\partial\theta}\frac{\partial\log p\left(x_{k};\theta\right)}{\partial\theta},
\end{align}
puis de retirer la sommation et d'intégrer certains des termes comme suit
\begin{align}
&\sum_{k\neq1}\left(\int p\left(x_{1};\theta\right)\frac{\partial\log p\left(x_{1};\theta\right)}{\partial\theta}dx_{1}\right)\left(\int p\left(x_{k};\theta\right) \frac{\partial\log p\left(x_{k};\theta\right)}{\partial\theta}dx_{k}\right)\left(\int \prod_{m\neq1}^{n}p\left(x_{m};\theta\right)dx_{m}\right)+\notag\\& \sum_{k\neq2}\left(\int p\left(x_{2};\theta\right)\frac{\partial\log p\left(x_{2};\theta\right)}{\partial\theta}dx_{2}\right)\left(\int p\left(x_{k};\theta\right) \frac{\partial\log p\left(x_{k};\theta\right)}{\partial\theta}dx_{k}\right)\left(\int \prod_{m\neq2}^{n}p\left(x_{m};\theta\right)dx_{m}\right)+... 
\end{align}
Chacun des termes ci-dessus est un produit des valeurs d'espérance des Scores. Par conséquent, nous pouvons réécrire cette expression comme
\begin{align}
&\sum_{k\neq1}\textbf{E}\left( \frac{\partial\log p\left(x_{1};\theta\right)}{\partial\theta}\right)\textbf{E}\left( \frac{\partial\log p\left(x_{k};\theta\right)}{\partial\theta}\right)+\sum_{k\neq2}\textbf{E}\left( \frac{\partial\log p\left(x_{2};\theta\right)}{\partial\theta}\right)\textbf{E}\left( \frac{\partial\log p\left(x_{k};\theta\right)}{\partial\theta}\right)\notag\\&\hspace{2cm}+...+\sum_{k\neq n}\textbf{E}\left( \frac{\partial\log p\left(x_{n};\theta\right)}{\partial\theta}\right)\textbf{E}\left( \frac{\partial\log p\left(x_{k};\theta\right)}{\partial\theta}\right)\notag\\&= \sum_{k\neq1}\textbf{E}\left({\cal L}_{\theta}\left(x_{1}\right) \right)\textbf{E}\left({\cal L}_{\theta}\left(x_{k}\right) \right)+\sum_{k\neq2}\textbf{E}\left({\cal L}_{\theta}\left(x_{2}\right) \right)\textbf{E}\left({\cal L}_{\theta}\left(x_{k}\right) \right)+...+\sum_{k\neq n}\textbf{E}\left({\cal L}_{\theta}\left(x_{n}\right) \right)\textbf{E}\left({\cal L}_{\theta}\left(x_{k}\right) \right)\notag\\&=0.  
\end{align}
Nous avons démontré précédemment que la valeur d'espérance d'un score est toujours égale à zéro (voir l'équation (\ref{46})). Par conséquent, l'ensemble de cette expression est égale à zéro et nous pouvons simplifier l'équation (\ref{422}) comme suit
\begin{align}
\int p\left(\boldsymbol{x};\theta\right)\left(\sum_{l=1}^{n} \frac{\partial\log p\left(x_{l};\theta\right)}{\partial\theta}\right)^{2}d\boldsymbol{x}^{n}&=\int p\left(\boldsymbol{x};\theta\right)\sum_{l=1}^{n} \left(\frac{\partial\log p\left(x_{l};\theta\right)}{\partial\theta} \right)^{2}d\boldsymbol{x}^{n}\notag\\&=\sum_{l=1}^{n}\int dx_{1}...dx_{n}p\left(x_{1};\theta\right)...p\left(x_{n};\theta\right)\sum_{l=1}^{n} \left(\frac{\partial\log p\left(x_{l};\theta\right)}{\partial\theta}\right)^{2}\notag\\&=\sum_{l=1}^{n}\left[\int p\left(x_{l};\theta\right)\left(\frac{\partial\log p\left(x_{l};\theta\right)}{\partial\theta}\right)^{2}dx_{l}\right]\left[\underbrace {\int\prod_{k\neq l}^{n}p\left(x_{k};\theta\right)dx_{k}}_{=1}\right]\notag\\&=\sum_{l=1}^{n}\textbf{E}\left[\left(\frac{\partial\log p\left(x_{l};\theta\right)}{\partial\theta}\right) \right]. \label{426} 
\end{align}
Chaque terme de l'équation (\ref{426}) est l'information de Fisher classique pour la variable aléatoire $X_{l}$ avec la fonction de densité de probabilité $p\left(x_{l};\theta\right)$. Puisque les variables aléatoires considérées sont indépendantes et distribuées de manière identique, toutes les $p\left(x_{l};\theta\right)$ sont identiques. Par conséquent, nous avons
\begin{equation}
\int p\left(\boldsymbol{x};\theta\right)\left(\sum_{l=1}^{n}\frac{\partial\log p\left(x_{l};\theta\right)}{\partial\theta}\right)^{2}d\boldsymbol{x}^{n}=\sum_{l=1}^{n}{\cal F}\left( p\left(x_{l};\theta\right)\right)=n{\cal F}\left(\theta\right).\label{427}
\end{equation}
Enfin, en combinant l'équation (\ref{427}) avec l'équation (\ref{421}), on obtient la fameuse limite de Cramér-Rao (\ref{BCR}). Nous allons énumérer ici quelques propriétés de l'information de Fisher classique ${\cal F}\left(\theta\right)$: 
\begin{itemize}
	\item Par rapport à d'autres mesures d'information, telles que l'entropie de Shannon ou l'information mutuelle, l'information de Fisher classique est dimensionnelle. Nous acquérons des unités parce que les densités de probabilité $p\left(x_{l};\theta\right)$ sont généralement dimensionnelles et parce que la dérivée $\partial_{\theta}$ a des unités de $\theta^{-1}$. Cela est nécessaire pour relier ${\cal F}\left(\theta\right)$ à la variance ${\rm Var}\left(\theta\right)$ à travers l'inégalité de Cramér-Rao (\ref{BCR}).
	\item L'information de Fisher classique est une quantité strictement positive. Ceci peut être démontré en considérant une forme plus générale de ${\cal F}\left(\theta\right)$ qui prend en compte de nombreuses variables $\left\lbrace \theta_{1},...,\theta_{d}\right\rbrace$ (c-à-d le cas de l'estimation multi-paramètres) et en examinant la forme des matrices résultantes, dont on peut montrer qu'elles sont positives.
	\item L'information de Fisher classique est liée à l'information mutuelle, mais il s'agit de deux quantités fondamentalement différentes. Pour une compréhension intuitive, l'information mutuelle ${\cal I}\left( X:Y\right)$ (\ref{IMC}) mesure la corrélation entre les variables aléatoires $X$ et $Y$, tandis que ${\cal F}\left(\theta\right)$ s'intéresse à la probabilité qu'un estimateur $\boldsymbol{\hat{\theta}}$ s'approche de près de la "vraie" valeur de $\theta$.
\end{itemize}
\subsection{Extension à la métrologie classique multiparamétrique}
La mission principale ici est d'estimer un ensemble de paramètres inconnus. Dans ce cas, les différents paramètres que nous souhaitons estimer sont construits dans le vecteur défini dans l'espace des paramètres $\Theta$, appelé le vecteur de paramètre $\boldsymbol{ \theta}\left(\boldsymbol{x} \right)=\left(\theta_{1}\left( \boldsymbol{x}\right),...,\theta_{d}\left(\boldsymbol{x}\right)\right)^{T}\in\mathbb{R}^{d}$. Alors, la réalisation du vecteur d'estimation est représentée par $\boldsymbol{\hat{\boldsymbol{ \theta}}}\left( \boldsymbol{x}\right)=\left(\hat{\theta}_{1}\left(\boldsymbol{x} \right),...,\hat{\theta}_{d}\left(\boldsymbol{x}\right)\right)^{T}$. Pour tout estimateur $\boldsymbol{\hat{\boldsymbol{ \theta}}}\left(\boldsymbol{x}\right)$ de $\boldsymbol{\boldsymbol{\theta}}\left(\boldsymbol{x}\right)$, l'espérance mathématique est donnée par $\textbf{E}\left[\left(\boldsymbol{\boldsymbol{\theta}}- \boldsymbol{\hat{\boldsymbol{\theta}}}\right) \left(\boldsymbol{\boldsymbol{\theta}}- \boldsymbol{\hat{\boldsymbol{\theta}}}\right)^{T}\right]$. Ensuite, l'erreur quadratique moyenne est égale à la matrice de covariance ${\rm Cov}\left( \boldsymbol{\hat{\boldsymbol{ \theta}}}\right)$. L'un des résultats centraux de la théorie des probabilités classique, à savoir l'inégalité de Cramér-Rao, imposant une limite inférieure à la matrice de covariance \cite{Paris2009} 
\begin{equation}
{\rm Cov}\left( \boldsymbol{\hat{\boldsymbol{ \theta}}}\right)\geq \mathbb{F}\left(\boldsymbol{\boldsymbol{ \theta}} \right)^{-1},
\end{equation}
où $\mathbb{F}\left(\boldsymbol{\boldsymbol{ \theta}}\right)$ est la matrice d'information de Fisher classique avec des éléments
\begin{align}
\left[\mathbb{F}\left(\boldsymbol{\boldsymbol{ \theta}} \right)\right]_{ij}&=\textbf{E}\left[\left(\frac{\partial\log p\left(\boldsymbol{x};\boldsymbol{\boldsymbol{ \theta}}\right)}{\partial\theta_{i}}\right)\left(\frac{\partial\log p\left(\boldsymbol{x};\boldsymbol{\boldsymbol{ \theta}}\right)}{\partial\theta_{j}}\right) \right]\notag\\&=\sum_{\boldsymbol{x}\in X}\left(\frac{\partial\log p\left(\boldsymbol{x};\boldsymbol{\boldsymbol{ \theta}}\right)}{\partial\theta_{i}}\frac{\partial\log p\left(\boldsymbol{x};\boldsymbol{\boldsymbol{ \theta}}\right)}{\partial\theta_{j}}\right)p\left(\boldsymbol{x};\boldsymbol{\boldsymbol{ \theta}}\right),
\end{align}
qui dépendent de la distribution de probabilité $p\left(\boldsymbol{x};\boldsymbol{\boldsymbol{ \theta}}\right)$ des résultats $\boldsymbol{x}$. L'inégalité de Cramér-Rao ne s'applique qu'aux distributions de probabilité qui satisfont la condition de régularité suivante
\begin{equation}
\textbf{E}\left[\frac{\partial\log p\left(\boldsymbol{x};\boldsymbol{\boldsymbol{ \theta}}\right)}{\partial\theta_{i}}\right]=0,\hspace{1cm}\forall \theta\in\Theta.\label{430}
\end{equation}
Si un tel estimateur localement non biaisé existe, la borne de Cramér-Rao peut toujours être saturée. Cependant, l'identification des conditions de saturation de la borne de Cramér-Rao est un sujet complexe qui implique une variété de détails techniques. En particulier, pour la distribution de probabilité satisfaisant l'équation (\ref{430}), il existe un estimateur local non biaisé $\boldsymbol{\hat{\boldsymbol{ \theta}}}$ saturant la borne de Cramér-Rao si \cite{Kay1993}
\begin{equation}
\frac{\partial\log p\left(\boldsymbol{x};\boldsymbol{\boldsymbol{ \theta}}\right)}{\partial\theta}=\mathbb{F}\left(\boldsymbol{\boldsymbol{ \theta}} \right)\left(\boldsymbol{\hat{\boldsymbol{ \theta}}}-\boldsymbol{\boldsymbol{ \theta}}\right).
\end{equation}
\section{Théorie de l'estimation quantique}
La nature probabiliste universelle de la mécanique quantique nous amène à travailler régulièrement avec des probabilités. De plus, si l'on étudie des domaines liés à des expériences ou à des réalisations physiques, cette nature probabiliste de la mécanique quantique devient encore plus visible. D'autre part, des caractéristiques exotiques telles que l'intrication découlent de la mécanique quantique, qui sont directement liées aux propriétés probabilistes du système quantique. La présente section vise à décrire les systèmes quantiques du point de vue de la métrologie.
\subsection{Modèles statistiques quantiques}
Par analogie avec le cas classique, un modèle statistique quantique $S_{Q}$ est défini par une famille d'opérateurs densité dans un espace de Hilbert $\cal{H}$ et est paramétré par $n$ paramètres réels $\theta\in \Theta\subset\mathbb{R}^{n}$, avec $\Theta$ est un espace de paramètres estimé. Cela signifie que
\begin{equation}
S_{Q}=\left\lbrace\rho_{\theta}=\Lambda_{\theta}\left(\rho\right): \hspace{0.5cm}\theta=\left(\theta^{1},\theta^{2},...,\theta^{n}\right)\in\Theta\right\rbrace, 
\end{equation}
où l'application de paramétrage $\Lambda_{\theta}$ est injective (c'est-à-dire qu'il existe un inverse de l'application $\Lambda_{\theta}$), et $\rho_{\theta}$ peut être différencié autant de fois que nécessaire par rapport aux paramètres.\par
Un modèle statistique quantique se présente typiquement de la manière suivante: Le système $S$ est préparé au temps $t=0$ dans un état initial $\rho\equiv\rho\left(t=0\right)$, et passe ensuite par un canal quantique $\Lambda_{\theta}$, qui dépend de la vraie valeur de $\theta$ d'un ou plusieurs paramètres. Alors le modèle associé est défini comme $\rho_{\theta}:=\Lambda_{\theta}\left(\rho\right)$. Un exemple caractéristique est le canal unitaire généré par l'hamiltonien du système, à savoir
\begin{equation}
\rho_{\theta}=U_{t}\rho U_{t}^{\dagger}, \hspace{1cm}{\rm avec}\hspace{1cm} U_{t}=\exp\left(-itH_{\theta}\right),
\end{equation}
où le paramètre $\theta$ est appelé le paramètre hamiltonien. De plus, il est intéressant de distinguer ici les paramètres hamiltoniens de phase (ou de décalage) et les paramètres généraux. Dans le premier cas, le paramètre apparaît linéairement, comme une constante multiplicative globale, c'est-à-dire que nous pouvons écrire comme $H_{\theta}=\theta K$. Dans le deuxième cas, le paramètre intervient de manière non linéaire, de sorte que les vecteurs propres du Hamiltonien $H_{\theta}$ dépendent en général de $\theta$. Cependant, le codage dynamique n'est pas la seule possibilité. Pour certains modèles, ce codage est statique. Un exemple typique est celui d'un modèle thermique qui décrit l'état d'équilibre d'un système quantique en interaction avec un réservoir thermique, dans lequel
\begin{equation}
\rho_{\beta}=\frac{\exp\left(-\beta H\right)}{Z}, \hspace{1cm}{\rm avec}\hspace{1cm} Z={\rm Tr}\left(\exp\left(-\beta H\right)\right),
\end{equation}
où le paramètre d'intérêt, conventionnellement désigné par $\beta$, est l'inverse de la température du réservoir et $H$ est Hamiltonien du système.\par 
Dans ces deux scénarios, n'importe quel modèle statistique quantique $S=\left\lbrace \rho_{\boldsymbol{\theta}}\right\rbrace_{\boldsymbol{\theta}\in\Theta}$, effectuant une mesure avec des opérateurs $\left\lbrace \Pi_{x}\right\rbrace_{x\in X}$, donne lieu à un modèle statistique classique via la relation $p_{\theta}\left(x \right)=p\left(x;\theta\right)={\rm Tr}\left(\rho_{\theta}\Pi_{x} \right)$, où l'espace d'échantillon $X$ est supposé être dénombrable. A titre d'exemple, nous prenons un état d'un système quantique à deux niveaux que nous pouvons décomposer sur la base des états propres $\left\lbrace\left|0 \right\rangle,\left|1\right\rangle \right\rbrace$ en fonction de deux angles $\left\lbrace\theta^{1}\equiv\theta,\theta^{2}\equiv\varphi\right\rbrace $ comme suit 
\begin{equation}
\left|\psi \right\rangle=\cos\left(\frac{\theta}{2} \right)\left|0 \right\rangle+e^{i\varphi}\sin\left(\frac{\theta}{2} \right)\left|1 \right\rangle,
\end{equation}
avec $0\leq\theta\leq\pi$, $0\leq\varphi\leq2\pi$ et évoluant dans un espace de Hilbert à deux dimensions comme le montre la figure (\ref{Bloch}). L'état du qubit est décrit par le vecteur de Bloch $\vec{r}$ comme suit:
\begin{equation}
\hat{\rho}=\frac{1}{2}\sum_{i=0}^{3}r_{i}\hat{\sigma_{i}}, \hspace{1cm}\hat{\sigma_{0}}=\hat{\openone}_{2\times2}, \hspace{1cm}r_{i}={\rm Tr}\left(\hat{\sigma_{i}}\hat{\rho}\right),\label{4.6}
\end{equation}
où $\sigma_{i}$ est l'opérateur de Pauli dans la direction $i$ ($i = x, y, z$). Alors l'état d'un système à deux niveaux est visualisé sur la sphère de Bloch par le vecteur de Bloch
\begin{equation}
	\vec{r}=\left( {\begin{array}{*{20}{c}}
			\left\langle {\hat{\sigma_{x}}}\right\rangle \\
			\left\langle {\hat{\sigma_{y}}}\right\rangle\\
			\left\langle {\hat{\sigma_{z}}}\right\rangle\end{array}} \right)=\left( {\begin{array}{*{20}{c}}
			\cos\theta\sin\varphi \\
			\sin\theta\sin\varphi\\
			\cos\varphi\end{array}} \right).
\end{equation}
Si l'état $\hat{\rho}$ est un état pur, la matrice densité (\ref{4.6}) se réduit à
\begin{equation}
\hat{\rho}=\frac{1}{2}\left( {\begin{array}{*{20}{c}}
		1+\cos\frac{\theta}{2}&1-i\cos\frac{\theta}{2}\sin\frac{\theta}{2}\\
		1+i\cos\frac{\theta}{2}\sin\frac{\theta}{2}&1-\cos\frac{\theta}{2}\end{array}} \right).\label{4.8}
\end{equation}
Supposons que nous effectuons la mesure de la valeur de projection (PVM) de l'opérateur $\sigma_{z}$ sur l'état (\ref{4.8}), qui est décrit par l'ensemble des opérateurs de projection
\begin{equation}
\hat{\Pi}_{+}\equiv\frac{1}{2}\left(\hat{\openone}+ \hat{\sigma_{z}}\right),\hspace{1cm} \hat{\Pi}_{-}\equiv\frac{1}{2}\left(\hat{\openone}- \hat{\sigma_{z}}\right),
\end{equation}
alors la distribution de probabilité du résultat de cette mesure est calculée comme suit
\begin{equation}
p\left(+,\theta \right)\equiv{\rm Tr} \left(\hat{\Pi}_{+}\hat{\rho}\right)=\frac{1}{2}\left(1+\cos\frac{\theta}{2} \right),\hspace{1cm} p\left(-,\theta\right)\equiv{\rm Tr} \left(\hat{\Pi}_{-}\hat{\rho}\right)=\frac{1}{2}\left(1-\cos\frac{\theta}{2}\right).
\end{equation}
Par conséquent, le modèle statistique de cette distribution de probabilité est paramétré par
\begin{equation}
x=\left\lbrace +,-\right\rbrace,\hspace{1cm} p\left(\pm,\theta \right)=\frac{1}{2}\left(1\pm\cos\frac{\theta}{2} \right), 
\end{equation}
et l'espace des paramètres $\Theta$ devient
\begin{equation}
	\Theta=\left\lbrace\boldsymbol{\theta}=\left(\theta,\varphi\right),\hspace{0.5cm}0\leq\theta\leq\pi, \hspace{0.5cm} 0\leq\varphi\leq2\pi, \hspace{0.5cm}n=2 \right\rbrace.
\end{equation}

\subsection{Avantage de la stratégie quantique dans la théorie de l'éstimation}

Grâce à des propriétés telles que la cohérence et l'intrication quantique, les systèmes quantiques sont extrêmement prometteurs en tant que capteurs quantiques. Plusieurs raisons expliquent pourquoi les systèmes quantiques sont capables de surpasser les dispositifs classiques. Tout d'abord, et c'est sans doute le plus important, la limite de mesure classique peut être dépassée en utilisant des systèmes quantiques. Pour les systèmes classiques, la fluctuation statistique est de l'ordre de $\Delta\theta\propto1/\sqrt{N}$ en raison du théorème de la limite centrale, où $N$ est le nombre de détecteurs utilisés pour la mesure \cite{Giovannetti2004}. Par capteurs, nous voulons dire le nombre de mesures effectuées, ou le nombre de systèmes de capteurs avec lesquels le système source interagit. Il peut s'agir du nombre de photons dans une cavité ou des atomes dans un réseau de capteurs. La fluctuation statistique est parfois écrite explicitement comme $\Delta\theta\propto1/\sqrt{NM}$, qui est connue sous le nom de la limite quantique standard, c'est le résultat du bruit de projection quantique avec $M$ est le nombre de répétitions de la même mesure \cite{Braginskii1975}.\par

En comparaison, les systèmes quantiques peuvent obtenir une meilleure précision par rapport à $1/\sqrt{N}$. Il a été démontré que le fait d'injecter du vide comprimé dans l'un des ports d'un interféromètre augmentait la précision à environ $1/N^{3/4}$ \cite{Caves1981,Barnett2003}. En outre, l'ajout de l'intrication dans le schéma, qui a été inclus en injectant des états intriqués dans les ports de l'interféromètre, donne une précision de $1/N$, ce qui est une amélioration du facteur $\sqrt{N}$ \cite{Dowling1998}. Cette limite est souvent appelée la limite de Heisenberg, et les résultats indiquent qu'il s'agit bien de la véritable limite quantique \cite{Ou1997}.\par

\begin{figure}[H]
	\centerline{\includegraphics[width=14cm]{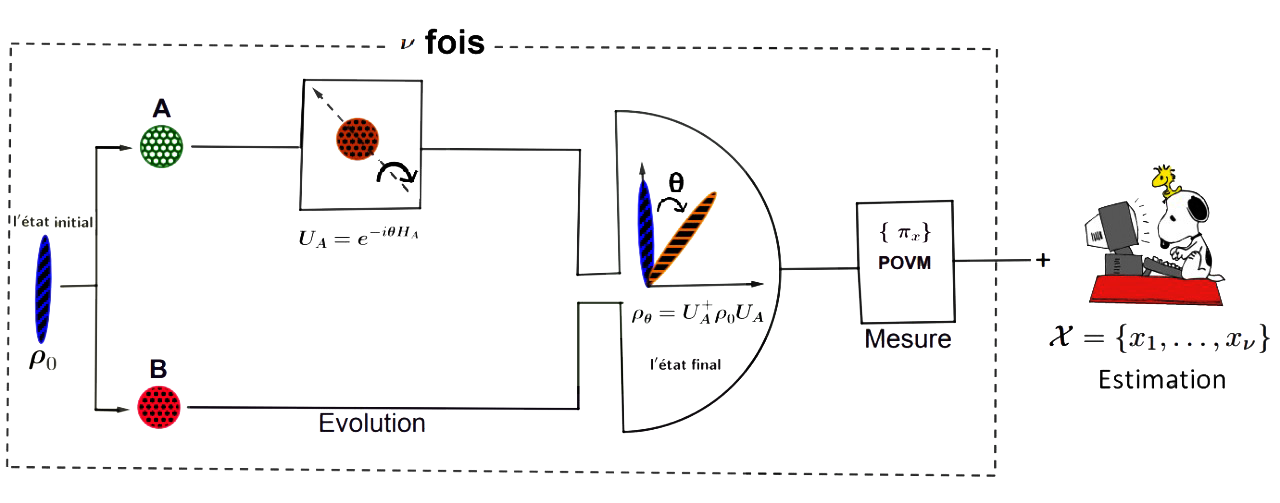}}
	\caption{Schéma d'un processus complet de la métrologie quantique, qui contient quatre étapes: (1) préparation de l'état initial; (2) paramétrisation; (3) mesure quantique; (4) l'estimation classique.}
	\label{fig4.1}
\end{figure}

Le schéma le plus élémentaire d'une configuration métrologique dans le présent contexte est le suivant. Tout d'abord, un état $\rho$ est préparé, suivi d'une évolution générale représentée par une application $\Lambda_{\theta}$ dans laquelle le paramètre inconnu $\theta$ est imprimé sur l'état. Enfin, l'état sortant est caractérisé par une certaine quantité mesurée $\left\lbrace\pi_{x}\right\rbrace$, qui nous permet de déduire la valeur du paramètre $\theta$. La figure (\ref{fig4.1}) illustre les principales étapes de la métrologie quantique.
\subsection{L'information quantique de Fisher}
\subsubsection{Définition générale}
L'information classique de Fisher définit une limite sur la précision asymptotique liée à une stratégie de mesure spécifique. Il est évident que différentes stratégies extraient différentes quantités d'informations. En général, en employant un schéma de mesure particulier qui obtient des informations sur une propriété donnée, on exclut la possibilité de mesurer des observables complémentaires dans la même expérience. Pour un système quantique, la précision est cependant finalement limitée par le codage du paramètre inconnu $\theta$ dans l'état $\rho_{\theta}$ du système. La variation de $\rho_{\theta}$ avec des changements infinitésimaux de la valeur de $\theta$ définit donc une limite quantique supérieure sur l'information de Fisher classique, ${\cal F}\left( p\left(x;\theta\right)\right)\leq{\cal F}_{Q}\left(\rho_{\theta}\right)$, avec l'information quantique de Fisher ${\cal F}_{Q}\left(\rho_{\theta}\right)$ est donnée par
\begin{equation}
{\cal F}_{Q}\left(\rho_{\theta}\right)={\rm Tr}\left(\hat{L}_{\theta}^{2}\rho_{\theta}\right), \label{QF}
\end{equation}
où la dérivée logarithmique symétrique $\hat{L}_{\theta}$ est un opérateur auto-adjoint qui se rapporte à la variation de l'état $\rho_{\theta}$ avec des changements infinitésimaux de la valeur du paramètre inconnu $\theta$. Ceci est déterminée en résolvant l'équation de Lyapunov suivante
\begin{equation}
\frac{\partial\rho_{\theta}}{\partial\theta}=\frac{1}{2}\left(\hat{L}_{\theta}\rho_{\theta}+\rho_{\theta}\hat{L}_{\theta} \right).\label{DLS}
\end{equation}
Autrement dit, l'information quantique de Fisher (\ref{QF}) peut être formellement dérivée en maximisant l'information classique de Fisher (\ref{FC}) sur toutes les mesures possibles qui peuvent être faites sur l'état $\rho$, et qualitativement, est une mesure de la quantité d'informations qu'un état contient sur le paramètre $\theta$. La solution générale pour la dérivée logarithmique symétrique $L_{\theta}$ est
\begin{equation}
\hat{L}_{\theta}=2\int_{0}^{\infty}\exp\left[-\rho_{\theta}t \right]\frac{\partial\rho_{\theta}}{\partial\theta}\exp\left[-\rho_{\theta}t \right]dt,
\end{equation}
qui n'est pas une équation facile à utiliser. Nous allons dériver ici une forme alternative explicitement en termes de valeurs propres $\lambda_{i}$ et de vecteurs propres $\left|\vartheta_{i} \right\rangle$ de la matrice densité $\rho_{\theta}$. En écrivant $\rho_{\theta}$ dans sa base diagonale; $\rho_{\theta}\left|\vartheta_{i} \right\rangle=\lambda_{i}\left|\vartheta_{i} \right\rangle$. Nous pouvons alors écrire
\begin{align}
\left(\frac{\partial\rho_{\theta}}{\partial\theta} \right)_{ij}&=\left\langle\vartheta_{i}\right|\frac{\partial\rho_{\theta}}{\partial\theta}\left|\vartheta_{j} \right\rangle\notag\\&=\frac{1}{2}\left[\left\langle\vartheta_{i}\right|\hat{L}_{\theta}\rho_{\theta}\left|\vartheta_{j} \right\rangle\notag+\left\langle\vartheta_{i}\right|\rho_{\theta}\hat{L}_{\theta}\left|\vartheta_{j} \right\rangle\notag \right]\notag\\&=\frac{1}{2}\left[\lambda_{j}\left(\hat{L}_{\theta}\right)_{ij}+\lambda_{i}\left(\hat{L}_{\theta}\right)_{ij}\right].  
\end{align}
Nous résolvons ensuite pour trouver $L_{\theta}$ comme
\begin{equation}
\left(\hat{L}_{\theta}\right)_{ij}=\frac{2}{\lambda_{i}+\lambda_{j}}\left\langle\vartheta_{i}\right|\frac{\partial\rho_{\theta}}{\partial\theta}\left| \vartheta_{j}\right\rangle,\label{dls}
\end{equation}
où le dénominateur comprend uniquement les termes qui satisfont $\lambda_{i}+\lambda_{j}\neq0$. 

\subsubsection{Information de Fisher quantique pour les états purs}
Pour l'instant, nous allons rester sur le scénario beaucoup plus simple dans lequel notre état reste un état pur, de sorte que $\rho_{\theta}^{2}=\rho_{\theta}=\left|\psi_{\theta} \right\rangle\left\langle\psi_{\theta}\right|$. Alors
\begin{equation}
\frac{\partial\rho_{\theta}}{\partial\theta}=\frac{\partial\rho_{\theta}^{2}}{\partial\theta}=\rho_{\theta}\frac{\partial\rho_{\theta}}{\partial\theta}+\frac{\partial\rho_{\theta}}{\partial\theta}\rho_{\theta}.
\end{equation}
En comparant cela avec l'équation (\ref{DLS}), nous obtenons
$\hat{L}_{\theta}=2\frac{\partial\rho_{\theta}}{\partial\theta}$. On trouve alors
\begin{align}
{\cal F}_{Q}\left(\rho_{\theta}\right)=4{\rm Tr}\left[\rho_{\theta}\left(\frac{\partial\rho_{\theta}}{\partial\theta}\right)^{2} \right]=4\left[\left\langle\hat{\psi_{\theta}}|\hat{\psi_{\theta}}\right\rangle - |\left\langle \hat{\psi_{\theta}}|\psi_{\theta}\right\rangle|^{2} \right],  
\end{align}
où $\left|\hat{\psi_{\theta}}\right\rangle=\partial_{\theta}\left|\psi_{\theta}\right\rangle$. La dernière expression, combinée à l'hypothèse des transformations unitaires, nous amène à la dernière forme simplifiée de l'information de Fisher quantique. Elles agissent sur les états purs comme
\begin{equation}
{\cal U}_{\theta}\left|\psi_{0}\right\rangle\equiv\left|\psi_{\theta}\right\rangle,
\end{equation}
où ${\cal U}_{\theta}=\exp\left[-i\theta \hat{H}_{\theta} \right]$ et avec $\hat{H}_{\theta}$ est un opérateur hermitien. Pour ces états, l'information de Fisher quantique prend la forme suivante:
\begin{equation}
{\cal F}_{Q}\left(\left|\psi_{\theta} \right\rangle\left\langle\psi_{\theta}\right|\right)\equiv{\cal F}_{Q}\left(\left|\psi_{0}\right\rangle,\hat{H}_{\theta}\right)=4{\rm Var}\left(\hat{H}_{\theta}\right),
\end{equation}
qui est proportionnelle à la variance de l'opérateur hermitien $\hat{H}_{\theta}$. En outre, une propriété importante de l'information quantique de Fisher est sa convexité. Cette caractéristique implique que ${\cal F}_{Q}\left(\rho\right)$ atteint sa valeur maximale sur des états purs et par conséquent, les états quantiques mixtes ne peuvent pas augmenter la sensibilité d'estimation réalisable. On peut donc généraliser la dernière formule comme
\begin{equation}
{\cal F}_{Q}\left(\rho\right)\leq4{\rm Var}\left(\hat{H}_{\theta}\right),
\end{equation}
avec égalité pour les états purs.
\subsubsection{Information de Fisher quantique pour les états mixtes}
Dans cette partie, nous fournissons une dérivation de l'information quantique de Fisher pour les états quantiques mixtes. En utilisant la même notation que dans les références \cite{Pang2014} et \cite{Jing2014}, nous réécrivons l'état $\rho_{\theta}\equiv\rho$ comme $\rho=\sum_{i}p_{i}\left|\psi_{i}\right\rangle\left\langle \psi_{i}\right|$ pour mieux distinguer les états propres et les vecteurs propres. Dans ce cas, l'information quantique de Fisher peut en général être exprimée sous la forme
\begin{align}
{\cal F}_{Q}\left(\rho\right)=4\sum_{i=0}^{s}\left(\left(\partial_{\theta}\sqrt{p_{i}} \right)^{2}+ p_{i}\left\langle\partial_{\theta}\psi_{i}|\partial_{\theta}\psi_{i}\right\rangle-p_{i}|\left\langle\psi_{i}|\partial_{\theta}\psi_{i} \right\rangle|^{2}\right) -8\sum_{i\neq j}\frac{p_{i}p_{j}}{p_{i}+p_{j}}|\left\langle\psi_{i}|\partial_{\theta}\psi_{j} \right\rangle|^{2},
\end{align}
où l'opérateur hermitien $\hat{H}_{\theta}$ est défini par $\hat{H}_{\theta}=-i{\cal U}_{\theta}^{\dagger}\partial_{\theta}{\cal U}_{\theta}$. Pour dériver cette expression, nous supposons que $\left| \psi_{i}\right\rangle$ est la base propre de l'état évolué avec des valeurs propres $p_{i}$. La dérivée logarithmique symétrique (\ref{dls}) est alors donnée par
\begin{equation}
\hat{L}_{\theta}=2\sum_{ij}\frac{\left\langle \psi_{i}\right| \partial_{\theta}\rho\left| \psi_{j}\right\rangle }{p_{i}+p_{j}}\left|\psi_{i} \right\rangle\left\langle\psi_{j}\right|. 
\end{equation}
Ensuite, en superposant la définition de la dérivée logarithmique symétrique (\ref{DLS}) avec deux états de la base propre, on obtient

\begin{align}
\left\langle \psi_{i}\right|\partial_{\theta}\rho\left| \psi_{j}\right\rangle&=\frac{1}{2}\left(\left\langle \psi_{i}\right|\hat{L}_{\theta}\rho\left| \psi_{j}\right\rangle+\left\langle \psi_{i}\right|\rho \hat{L}_{\theta}\left| \psi_{j}\right\rangle \right)\notag\\&=\frac{1}{2}\left(\left\langle \psi_{i}\right|\hat{L}_{\theta}\sum_{k}p_{k}\left|\psi_{k}\right\rangle\left\langle\psi_{k}\right| \left| \psi_{j}\right\rangle+\left\langle \psi_{i}\right|\sum_{k}p_{k}\left|\psi_{k}\right\rangle\left\langle\psi_{k}\right|\hat{L}_{\theta} \left| \psi_{j}\right\rangle \right)\notag\\&=\frac{1}{2}\left(\left\langle \psi_{i}\right|\hat{L}_{\theta}p_{j}\left|\psi_{j}\right\rangle+\left\langle \psi_{i}\right|p_{i}\hat{L}_{\theta}\left|\psi_{j}\right\rangle  \right)\notag\\&=\frac{1}{2}\left(p_{i}+p_{j} \right)\left(\hat{L}_{\theta}\right)_{ij},     
\end{align}
dans lequel nous avons défini $\left(\hat{L}_{\theta}\right)_{ij}=\left\langle \psi_{i}\right| \hat{L}_{\theta}\left|\psi_{j}\right\rangle$. Nous pouvons alors écrire l'information quantique de Fisher comme suit
\begin{align}
{\cal F}_{Q}\left(\rho\right)&=\sum_{i}\left\langle \psi_{i} \right| \hat{L}_{\theta}^{2}\rho\left|\psi_{i}\right\rangle\notag\\&=\sum_{ij}\left\langle \psi_{i} \right| \hat{L}_{\theta}\left|\psi_{j} \right\rangle\left\langle\psi_{j} \right| \hat{L}_{\theta}\rho\left|\psi_{i}\right\rangle\notag\\&=\sum_{ij}\left\langle \psi_{i} \right| \hat{L}_{\theta}\left|\psi_{j} \right\rangle\left\langle\psi_{j} \right| \hat{L}_{\theta}\left|\psi_{i}\right\rangle p_{i}\notag\\&=\sum_{ij}\left(\hat{L}_{\theta}\right)_{ij}\left(\hat{L}_{\theta}\right)_{ji}p_{i}.
\end{align}
Nous pouvons maintenant réécrire la dérivée logarithmique symétrique de la façon suivante
\begin{equation}
\left(\hat{L}_{\theta}\right)_{ij}=\frac{2\left\langle \psi_{i}\right|\partial_{\theta}\rho\left| \psi_{j}\right\rangle}{p_{i}+p_{j}}.
\end{equation}
Grâce à cette expression, l'information quantique de Fisher peut être écrite comme suit
\begin{equation}
{\cal F}_{Q}\left(\rho\right)=\sum_{i=0}^{s}\sum_{j=0}^{t}\frac{4p_{i}}{\left(p_{i}+p_{j}\right)^{2}}|\left\langle \psi_{i}\right|\partial_{\theta}\rho\left| \psi_{j}\right\rangle|^{2}.
\end{equation}
Maintenant, nous avons introduit les deux indices $s$ et $t$. L'état ne prend en charge que $s$, donc tous les indices $p_{t}=0$ pour $t\geq s$, car l'opérateur $\partial_{\theta}\rho$ est hermitien. Ensuite, nous pouvons montrer à partir de la décomposition spectrale de $\rho$ que
\begin{align}
\left\langle \psi_{i}\right|\partial_{\theta}\rho\left| \psi_{j}\right\rangle&=\sum_{k}\left(\left\langle \psi_{i}\right|\partial_{\theta}p_{k}\left| \psi_{k}\right\rangle\left\langle \psi_{k}| \psi_{j}\right\rangle+p_{k}\left\langle \psi_{i}| \partial_{\theta}\psi_{k}\right\rangle\left\langle \psi_{k}| \psi_{j}\right\rangle+p_{i} \left\langle \psi_{i}|\psi_{k}\right\rangle\left\langle \partial_{\theta}\psi_{k}| \psi_{j}\right\rangle\right)\notag\\&= \partial_{\theta}p_{i}\delta_{ij}+p_{j}\left\langle \psi_{i}| \partial_{\theta}\psi_{j}\right\rangle+p_{i}\left\langle \partial_{\theta}\psi_{i}| \psi_{j}\right\rangle.
\end{align}
Maintenant, on utilise le fait que la résolution de l'identité devient nulle lorsqu'elle est différenciée, c'est à dire
\begin{equation}
\partial_{\theta}\openone\equiv\sum_{i}\left| \partial_{\theta}\psi_{i}\right\rangle\left\langle\psi_{i}\right|+\left|\psi_{i}\right\rangle\left\langle\partial_{\theta}\psi_{i}\right|=0,
\end{equation}
ce qui implique $\left\langle\partial_{\theta}\psi_{i}|\psi_{j}\right\rangle=-\left\langle\psi_{i}|\partial_{\theta}\psi_{j}\right\rangle$. Nous trouvons alors que
\begin{equation}
\left\langle \psi_{i}\right|\partial_{\theta}\rho\left| \psi_{j}\right\rangle=\partial_{\theta}p_{i}\delta_{ij}+\left(p_{j}-p_{i} \right)\left\langle\psi_{i}|\partial_{\theta}\psi_{j} \right\rangle.
\end{equation}
Nous pouvons maintenant insérer ceci dans l'équation ci-dessus pour trouver
\begin{align}
{\cal F}_{Q}\left(\rho\right)&=\sum_{i=0}^{s}\sum_{j=0}^{t}\frac{4p_{i}}{\left(p_{i}+p_{j} \right)^{2}}|\partial_{\theta}p_{i}\delta_{ij}+\left(p_{j}-p_{i} \right)\left\langle\psi_{i}|\partial_{\theta}\psi_{j} \right\rangle|^{2}\notag\\&=\sum_{i=0}^{s}\frac{1}{p_{i}}\left(\partial_{\theta}p_{i} \right)^{2} +\sum_{i=0}^{s}\sum_{j=0}^{t}\frac{4p_{i}\left(p_{i}-p_{j} \right)^{2} }{\left(p_{i}+p_{j} \right)^{2}}|\left\langle\psi_{i}|\partial_{\theta}\psi_{j} \right\rangle|^{2}.
\end{align}
Mais maintenant, nous remarquons que la deuxième somme peut être divisée en deux sommes: une qui court sur le support jusqu'à $s$ et l'autre $j$ le support de $s+1$ à $t$. Alors
\begin{align}
\sum_{i=0}^{s}\sum_{j=0}^{t}\frac{4p_{i}\left(p_{i}-p_{j} \right)^{2} }{\left(p_{i}+p_{j} \right)^{2}}|\left\langle\psi_{i}|\partial_{\theta}\psi_{j} \right\rangle|^{2}=\sum_{i,j=0}^{s}\frac{4p_{i}\left(p_{i}-p_{j} \right)^{2} }{\left(p_{i}+p_{j} \right)^{2}}|\left\langle\psi_{i}|\partial_{\theta}\psi_{j} \right\rangle|^{2}+\sum_{i=0}^{s}\sum_{j=s+1}^{t}4p_{i}|\left\langle\psi_{i}|\partial_{\theta}\psi_{j} \right\rangle|^{2}.
\end{align}
Il faut noter que tous les $p_{s+1}=0$, donc la deuxième somme peut s'écrire comme suit
\begin{equation}
\sum_{j=s+1}^{t}\left|\psi_{j} \right\rangle \left\langle\psi_{j}\right|=\openone-\sum_{j=0}^{s}\left|\psi_{j} \right\rangle \left\langle\psi_{j}\right|.
\end{equation}
En insérant ceci dans l'expression ci-dessus, nous trouvons
\begin{align}
\sum_{i=0}^{s}\sum_{j=s+1}^{t}4p_{i}|\left\langle\psi_{i}|\partial_{\theta}\psi_{j} \right\rangle|^{2}&=\sum_{i=0}^{s}\sum_{j=s+1}^{t}4p_{i}\left\langle\partial_{\theta}\psi_{i}|\psi_{j}\right\rangle\left\langle\psi_{j}|\partial_{\theta}\psi_{i} \right\rangle\notag\\&=\sum_{i=0}^{s}4p_{i}\left\langle \partial_{\theta}\psi_{i}\right|\left[\openone-\sum_{j=0}^{s}\left|\psi_{j} \right\rangle \left\langle\psi_{j}\right| \right] \left|\partial_{\theta}\psi_{i}\right\rangle\notag\\&= \sum_{i=0}^{s}4p_{i}\left\langle\partial_{\theta}\psi_{i}|\partial_{\theta}\psi_{i}\right\rangle-\sum_{i,j=0}^{s}4p_{i}|\left\langle\psi_{j}|\partial_{\theta}\psi_{i} \right\rangle|^{2}.
\end{align}
En rassemblant tous ces éléments, on trouve
\begin{align}
{\cal F}_{Q}\left(\rho\right)&=\sum_{i=0}^{s}\frac{1}{p_{i}}\left(\partial_{\theta}p_{i} \right)^{2} +\sum_{i,j=0}^{s}\frac{4p_{i}\left(p_{i}-p_{j} \right)^{2} }{\left(p_{i}+p_{j} \right)^{2}}|\left\langle\psi_{i}|\partial_{\theta}\psi_{j} \right\rangle|^{2}+\sum_{i=0}^{s}4p_{i}\left\langle\partial_{\theta}\psi_{i}|\partial_{\theta}\psi_{i}\right\rangle-\sum_{i,j=0}^{s}4p_{i} |\left\langle\psi_{j}|\partial_{\theta}\psi_{i} \right\rangle|^{2}\notag\\&=\sum_{i=0}^{s}\frac{1}{p_{i}}\left(\partial_{\theta}p_{i} \right)^{2}+\sum_{i=0}^{s}4p_{i}\left\langle\partial_{\theta}\psi_{i}|\partial_{\theta}\psi_{i}\right\rangle-8\sum_{i,j=0}^{s}\frac{p_{i}p_{j}}{p_{i}+p_{j}} |\left\langle\psi_{i}|\partial_{\theta}\psi_{j} \right\rangle|^{2}.\label{466}
\end{align}
De plus, nous pouvons aussi écrire la première partie de cette expression comme suit
\begin{equation}
\sum_{i=0}^{s}\frac{1}{p_{i}}\left(\partial_{\theta}p_{i} \right)^{2}=4\sum_{i=0}^{s}\left(\partial_{\theta}\sqrt{p_{i}} \right)^{2}.
\end{equation}
Nous remarquons maintenant que la troisième somme de l'équation (\ref{466}) a à la fois des valeurs diagonales et non diagonales. Par conséquent, cette expression peut être encore simplifiée, et l'information quantique de Fisher est finalement exprimée par:
\begin{align}
{\cal F}_{Q}\left(\rho\right)&=4\sum_{i=0}^{s}\left(\partial_{\theta}\sqrt{p_{i}} \right)^{2}+\sum_{i=0}^{s}4p_{i}\left\langle\partial_{\theta}\psi_{i}|\partial_{\theta}\psi_{i}\right\rangle-8\left(\sum_{i}\delta_{ij}+\sum_{i\neq j} \right)\frac{p_{i}p_{j}}{p_{i}+p_{j}} |\left\langle\psi_{i}|\partial_{\theta}\psi_{j} \right\rangle|^{2}\notag\\&=4\sum_{i=0}^{s}\left(\left(\partial_{\theta}\sqrt{p_{i}} \right)^{2}+ p_{i}\left\langle\partial_{\theta}\psi_{i}|\partial_{\theta}\psi_{i}\right\rangle-p_{i}|\left\langle\psi_{i}|\partial_{\theta}\psi_{i} \right\rangle|^{2}\right) -8\sum_{i\neq j}\frac{p_{i}p_{j}}{p_{i}+p_{j}}|\left\langle\psi_{i}|\partial_{\theta}\psi_{j} \right\rangle|^{2}.
\end{align}
On remarque maintenant que lorsque les états $\left|\psi_{i}\right\rangle$ évoluent dans le temps, on obtient
$\left|\psi_{i}\left(t\right)\right\rangle={\cal U}_{\theta}\left(t \right)\left|\psi_{i}\left(0\right)\right\rangle$, ce qui implique $\partial_{\theta}\left|\psi_{i}\left(t\right)\right\rangle=\partial_{\theta}{\cal U}_{\theta}\left(t \right)\left|\psi_{i}\left(0\right)\right\rangle$. Nous définissons maintenant le générateur $\hat{H}_{\theta}=-i{\cal U}_{\theta}^{\dagger}\partial_{\theta}{\cal U}_{\theta}$, qui simplifiera la manipulation de cette expression. En utilisant la propriété d'identité ${\cal U}_{\theta}^{\dagger}{\cal U}_{\theta}=\openone$, nous voyons que
\begin{equation}
{\cal U}_{\theta}^{\dagger}\partial_{\theta}{\cal U}_{\theta}=-\left( \partial_{\theta}{\cal U}_{\theta}^{\dagger} \right){\cal U}_{\theta},
\end{equation}
où nous avons ajouté les parenthèses pour désigner la dérivée partielle. Avec cette identité, on note que
\begin{equation}
\hat{H}_{\theta}^{2}=-{\cal U}_{\theta}^{\dagger}\left(\partial_{\theta}{\cal U}_{\theta}\right){\cal U}_{\theta}^{\dagger}\partial_{\theta}{\cal U}_{\theta}=\left(\partial_{\theta}{\cal U}_{\theta}\right)^{\dagger}\partial_{\theta}{\cal U}_{\theta}.
\end{equation}
De plus, dans le cas d'un opérateur d'évolution unitaire ${\cal U}_{\theta}$, la contribution classique disparaît ($\partial_{\theta}\sqrt{p_{i}}=0$).  Ainsi nous procédons à écrire
\begin{align}
	{\cal F}_{Q}\left(\rho\right)=4\sum_{i=0}^{s}p_{i}&\left( \left\langle\psi_{i}\left(0 \right)\right|\hat{H}_{\theta}^{2}\left| \psi_{i}\left(0 \right)\right\rangle-\left\langle\psi_{i}\left(0 \right)\right|\hat{H}_{\theta}\left| \psi_{i}\left(0 \right)\right\rangle\right)\notag\\& -8\sum_{i\neq j}\frac{p_{i}p_{j}}{p_{i}+p_{j}}|\left\langle\psi_{i}\left(0 \right)\right|\hat{H}_{\theta}\left| \psi_{i}\left(0 \right)\right\rangle|^{2}.
\end{align}
Le tableau suivant résume les propriétés de l'estimation classique et quantique ainsi que la relation entre elles:
\begin{center}
\begin{tabular}{|l|c|r|}
	\hline \rowcolor{lightgray}\multicolumn{2}{c|}{Estimation des paramètres}  \\
	\hline\rowcolor{lightgray} Dans le cas classique & Dans le cas quantique  \\
	\hline  Densité de probabilité $p\left(x;\theta\right) $ & Matrice densité $\rho_{\theta}=\Lambda_{\theta}\left(\rho_{0}\right)$  \\
	\hline  Le Score ${\cal L}_{\theta}\left(x\right) =\frac{\partial}{\partial \theta}\log p\left(x;\theta\right)$ & La dérivée logarithmique symétrique $\hat{L}_{\theta}$  \\
	\hline  $\frac{\partial p\left(x;\theta\right)}{\partial\theta}=\frac{1}{2}\left( {\cal L}_{\theta}p\left(x;\theta\right)+p\left(x;\theta\right){\cal L}_{\theta} \right) $ & $\frac{\partial\rho_{\theta}}{\partial\theta}=\frac{1}{2}\left(\hat{L}_{\theta}\rho_{\theta}+\rho_{\theta}\hat{L}_{\theta} \right)$ \\
	\hline  ${\rm Var}\left({\cal L}_{\theta}\right)=\sum_{x}p\left(x;\theta\right){\cal L}_{\theta}^{2}={\cal F}\left(\theta\right)$ & ${\rm Var}\left(\hat{L}_{\theta}\right)={\rm Tr}\left(\rho\hat{L}_{\theta}^{2}\right)={\cal F}_{Q}\left(\rho\right)$ \\
	\hline  $\bar{{\cal L}_{\theta}}=\sum_{x}p\left(x;\theta\right){\cal L}_{\theta}=0$ &$\left\langle \hat{L}_{\theta}\right\rangle={\rm Tr}\left(\rho\hat{L}_{\theta}\right)=\frac{\partial}{\partial \theta}{\rm Tr}\left(\rho\right)=0$  \\
	\hline 
\end{tabular}
\end{center}
\begin{itemize}
	\item {\bf Remarque:}\\ Il existe toujours, pour tout $\rho$, une mesure optimale telle que ${\cal F}\left(\theta\right)={\cal F}_{Q}\left(\rho_{\theta} \right)$. Cette mesure optimale est donnée par la projection sur les états propres $\left|l\right\rangle$ de $\hat{L}_{\theta}$. Dans ce cas, nous avons $p\left( l;\theta\right)=\left\langle l\right|\rho_{\theta}\left|l\right\rangle$. Alors
	\begin{equation}
		\frac{\partial p\left( l;\theta\right)}{\partial\theta}=\left\langle l\right|\frac{\partial \rho_{\theta}}{\partial\theta}\left|l\right\rangle=l p\left( l;\theta\right).
	\end{equation}
Il est donc facile de vérifier que
\begin{equation}
{\cal F}\left(\theta\right)=\sum_{l}\frac{1}{p\left( l;\theta\right)}\left( \frac{\partial p\left( l;\theta\right)}{\partial \theta}\right)^{2}=\sum_{l}l^{2}p\left( l;\theta\right)={\rm Tr}\left( \rho_{\theta} \hat{L}_{\theta}^{2}\right)={\cal F}_{Q}\left(\rho_{\theta} \right).
\end{equation}
\end{itemize}
\section{Métrologie quantique multiparamétrique}
Jusqu'à présent, nous n'avons considéré qu'un seul paramètre à estimer. Mais il existe des tâches pour lesquelles il est important d'estimer plusieurs paramètres \cite{Humphreys2013,Baumgratz2016}. Ces scénarios incluent par exemple l'estimation simultanée de plusieurs phases et l'estimation de plusieurs spins pointant dans des directions différentes ou, en général, de paramètres correspondant à des générateurs unitaires non commutables. La théorie décrite dans les sections précédentes peut être naturellement généralisée à l'estimation multiparamètres \cite{Berry2015,Vaneph2013}. Pour déterminer la précision dans ce cas, il faut calculer la matrice d'information quantique de Fisher, notée $\boldsymbol{ {\cal F}}_{Q}\left(\boldsymbol{\theta}\right)$, qui décrit les limites de distinction des états quantiques infiniment proches $\rho_{\boldsymbol{\theta}}$ et $\rho_{\boldsymbol{\theta}+d\boldsymbol{\theta}}$ (où $\boldsymbol{\theta}=\left\lbrace\theta_{1},..\theta_{\mu},\theta_{\nu},..\theta_{n} \right\rbrace$). En général, des éléments plus grands de cette matrice prédisent une meilleure capacité de distinction, ce qui conduit à une meilleure précision dans l'estimation du vecteur de paramètres $\boldsymbol{\theta}$.\par 

D'autre part, la façon d'augmenter la matrice d'information quantique de Fisher est une question difficile en métrologie quantique. En fait, l'estimation d'un seul paramètre joue un rôle important à bien des égards, en raison de l'existence d'un état final optimal contenant une quantité maximale de l'information quantique de Fisher \cite{Yuan2017,Liu2013}. Les problèmes réalistes peuvent généralement impliquer plusieurs paramètres, car il n'existe pas d'état optimal dans lequel la matrice d'information quantique de Fisher est plus grande que les autres états \cite{Matsumoto2012}. Ceci est lié au fait que les mesures optimales pour l'estimation de différents paramètres ne sont pas nécessairement commuables. En outre, l'inégalité de Cramér-Rao n'est pas toujours saturable car les mesures de différents paramètres peuvent être incompatibles \cite{Rehacek2018,Ragy2016}. Toutes ces restrictions font de l'estimation simultanée de plusieurs paramètres une tâche importante en métrologie quantique. Dans le même contexte, les techniques de calcul de la matrice de l'information quantique de Fisher ont connu un développement rapide dans divers scénarios et modèles. Cependant, il n'existe pas de documents qui résument ces techniques de manière structurée. Par conséquent, cette partie fournit des techniques complètes pour calculer la matrice $\boldsymbol{ {\cal F}}_{Q}\left(\boldsymbol{\theta}\right)$ dans une variété de scénarios.

\subsection{Représentation matricielle de l'information quantique de Fisher}
Commençons par considérer le cas général, c'est-à-dire une famille d'états quantiques $\rho_{\boldsymbol{\theta}}$ qui dépendent d'un ensemble de $n$ paramètres différents $\boldsymbol{\theta}=\left\lbrace \theta_{\mu}\right\rbrace$, $\mu=1,..,n$. Nous pouvons déduire les opérateurs de la dérivée logarithmique symétrique pour chaque paramètre impliqué comme suit

\begin{equation}
\frac{\partial\rho_{\boldsymbol{\theta}}}{\partial\theta_{\mu}}=\frac{1}{2}\left(\hat{L}_{\theta_{\mu}}\rho_{\boldsymbol{\theta}}+\rho_{\boldsymbol{\theta}}\hat{L}_{\theta_{\mu}} \right).\label{DLSM}
\end{equation}
Les éléments de la matrice de l'information quantique de Fisher sont définis comme \cite{Paris2009}
\begin{equation}
\boldsymbol{ {\cal F}}_{\theta_{\mu}\theta_{\nu}}\left(\rho_{\boldsymbol{\theta}}\right):=\frac{1}{2}{\rm Tr}\left(\rho_{\boldsymbol{\theta}}\left\lbrace\hat{L}_{\theta_{\mu}},\hat{L}_{\theta_{\nu}} \right\rbrace\right),\label{QFIM}
\end{equation}
où $\left\lbrace,\right\rbrace$ représente l'anti-commutation et $\hat{L}_{\theta_{\mu}}\left(\hat{L}_{\theta_{\nu}} \right)$ est la dérivée logarithmique symétrique pour le paramètre $\theta_{\mu}\left(\theta_{\nu}\right)$, qui est déterminé par l'équation (\ref{DLSM}). On sait bien que ${\rm Tr}\left(\rho_{\boldsymbol{\theta}}\hat{L}_{\theta_{\mu}}\right)$, donc en utilisant l'équation ci-dessus (\ref{QFIM}), $\boldsymbol{ {\cal F}}_{\theta_{\mu}\theta_{\nu}}\left(\rho_{\boldsymbol{\theta}}\right)$ peut aussi être exprimée de la manière suivante
\begin{equation}
\boldsymbol{ {\cal F}}_{\theta_{\mu}\theta_{\nu}}\left(\rho_{\boldsymbol{\theta}}\right)={\rm Tr}\left(\hat{L}_{\theta_{\nu}}\partial_{\theta_{\mu}}\rho_{\boldsymbol{\theta}} \right)=-{\rm Tr}\left(\rho_{\boldsymbol{\theta}}\partial_{\theta_{\mu}} \hat{L}_{\theta_{\nu}}\right).
\end{equation}
D'après l'équation (\ref{QFIM}), l'entrée diagonale de la matrice $\boldsymbol{ {\cal F}}\left(\rho_{\boldsymbol{\theta}}\right)$ est
\begin{equation}
\boldsymbol{ {\cal F}}_{\theta_{\mu}\theta_{\mu}}\left(\rho_{\boldsymbol{\theta}}\right)={\rm Tr}\left(\rho_{\boldsymbol{\theta}}\hat{L}_{\theta_{\mu}}^{2} \right),
\end{equation}
qui est exactement l'information quantique de Fisher pour le paramètre $\theta_{\mu}$.\par
Les propriétés de l'information de Fisher quantique ont été bien présentées ci-dessus. De même, la matrice d'information quantique de Fisher possède également quelques propriétés puissantes qui ont été largement appliquées dans la pratique. Nous présentons ici ces propriétés de la façon suivante \cite{Petz2008}
\begin{description}
	\item[*] $\boldsymbol{ {\cal F}}\left(\rho_{\boldsymbol{\theta}}\right)$ est symétrique réel, c-à-d $\boldsymbol{ {\cal F}}_{\theta_{\mu}\theta_{\nu}}=\boldsymbol{ {\cal F}}_{\theta_{\nu}\theta_{\mu}}\in\mathbb{R}^{2}$.
	\item[*] $\boldsymbol{ {\cal F}}\left(\rho_{\boldsymbol{\theta}}\right)$ est semi-défini positif, c-à-d $\boldsymbol{ {\cal F}}\left(\rho_{\boldsymbol{\theta}}\right)\geq0$. Si $\boldsymbol{ {\cal F}}>0$, alors $\left[\boldsymbol{ {\cal F}}^{-1}\right]_{\theta_{\mu}\theta_{\mu}}=1/\boldsymbol{ {\cal F}}_{\theta_{\mu}\theta_{\mu}}$ pour tout $\theta_{\mu}$.
	\item[*] $\boldsymbol{ {\cal F}}\left(\rho_{\boldsymbol{\theta}}\right)=\boldsymbol{ {\cal F}}\left(U\rho_{\boldsymbol{\theta}}U^{\dagger}\right)$ pour une opération unitaire $U$.
	\item[*] $\boldsymbol{ {\cal F}}$  est monotone sous une carte complètement positive et préservant la trace $\Lambda$, à savoir, $\boldsymbol{ {\cal F}}\left(\Lambda\left( \rho\right) \right)\leq\boldsymbol{ {\cal F}}\left( \rho\right)$.
	\item[*] Convexité: $\boldsymbol{ {\cal F}}\left(p\rho_{1}+\left(1-p\right)\rho_{2} \right)\leq p\boldsymbol{ {\cal F}}\left(\rho_{1}\right)+\left(1-p \right)\boldsymbol{ {\cal F}}\left(\rho_{2}\right)$ pour $p\in\left[0,1\right]$.
\end{description}
\subsection{Expression analytique de la matrice d'information quantique de Fisher}
Nous examinons ici les techniques de calcul de la matrice d'information quantique de Fisher et quelques résultats analytiques pour des cas spécifiques. La dérivation traditionnelle de cette matrice suppose généralement que le rang de la matrice densité est complet, c'est-à-dire que toutes les valeurs propres de la matrice densité $\rho$ sont positives. Plus précisément, si nous écrivons $\rho=\sum_{i}\lambda_{i}\left|\vartheta_{i}\right\rangle\left\langle \vartheta_{i}\right|$, où $\lambda_{i}$ et $\left|\vartheta_{i}\right\rangle$ sont la valeur propre et l'état propre correspondant, on suppose généralement que $\lambda_{i}>0$, pour tout $0\leq i\leq {\rm dim}\left(\rho \right)-1$. Dans cette hypothèse, les éléments de la matrice d'information quantique de Fisher peuvent être écrits comme suit
\begin{equation}
	\boldsymbol{ {\cal F}}_{\theta_{\mu}\theta_{\nu}}=\sum_{i,j=0}^{d-1}\frac{2\textsf{Re}\left(\left\langle \vartheta_{i}\right|\partial_{\theta_{\mu}} \rho\left|\vartheta_{j} \right\rangle\left\langle \vartheta_{j}\right|\partial_{\theta_{\nu}} \rho\left|\vartheta_{i} \right\rangle\right) }{\lambda_{i}+\lambda_{j}},
\end{equation}
où $\textsf{Re}$ désigne la partie réelle et $d$ est la dimension de la matrice densité. On peut facilement voir que si la matrice densité n'est pas de rang complet, il peut y avoir des termes divergents dans l'équation ci-dessus. Pour l'étendre aux matrices de densité générales, nous pouvons supprimer les termes divergents comme suit
\begin{equation}
	\boldsymbol{ {\cal F}}_{\theta_{\mu}\theta_{\nu}}=\sum_{i,j=0,\lambda_{i}+\lambda_{j}\neq0}^{d-1}\frac{2\textsf{Re}\left(\left\langle \vartheta_{i}\right|\partial_{\theta_{\mu}} \rho\left|\vartheta_{j} \right\rangle\left\langle \vartheta_{j}\right|\partial_{\theta_{\nu}} \rho\left|\vartheta_{i} \right\rangle\right) }{\lambda_{i}+\lambda_{j}}.
\end{equation}
En substituant la décomposition spectrale de $\rho$ dans l'équation ci-dessus, on peut la réécrire sous forme de
\begin{align}
\boldsymbol{{\cal F}}_{\theta_{\mu}\theta_{\nu}}&=\sum_{i=0}^{d-1}\frac{\left(\partial_{\theta_{\mu}}\lambda_{i} \right) \left(\partial_{\theta_{\nu}}\lambda_{i} \right)}{\lambda_{i}}+\sum_{i\neq j,\lambda_{i}+\lambda_{j}\neq j}^{d-1}\frac{2\left(\lambda_{i}-\lambda_{j}\right)^{2}}{\left(\lambda_{i}+\lambda_{j}\right)}\textsf{Re}\left(\left\langle \vartheta_{i}|\partial_{\theta_{\mu}}\vartheta_{j} \right\rangle \left\langle\partial_{\theta_{\nu}}\vartheta_{j}|\vartheta_{i}\right\rangle \right)\notag\\&= \sum_{i}\frac{\left(\partial_{\theta_{\mu}}\lambda_{i} \right) \left(\partial_{\theta_{\nu}}\lambda_{i} \right)}{\lambda_{i}}+\sum_{i}4\lambda_{i}\textsf{Re}\left(\left\langle \partial_{\theta_{\mu}}\vartheta_{i}|\partial_{\theta_{\nu}}\vartheta_{i} \right\rangle\right)+\sum_{i\neq j}\frac{8\lambda_{i}\lambda_{j}}{\lambda_{i}+\lambda_{j}}\left(\left\langle \partial_{\theta_{\mu}}\vartheta_{i}|\vartheta_{j} \right\rangle \left\langle\vartheta_{j}|\partial_{\theta_{\nu}}\vartheta_{i}\right\rangle \right).
\end{align}
En raison de la relation entre la matrice d'information de Fisher quantique et l'information de Fisher quantique, on peut facilement obtenir
\begin{align}
\boldsymbol{{\cal F}}_{\theta_{\mu}\theta_{\mu}}=&{\cal F}_{Q}\left(\rho_{\theta_{\mu}} \right)=\sum_{i}\frac{\left(\partial_{\theta_{\mu}}\lambda_{i} \right)^{2}}{\lambda_{i}}+\sum_{i}4\lambda_{i}\left(\left\langle \partial_{\theta_{\mu}}\vartheta_{i}|\partial_{\theta_{\mu}}\vartheta_{i} \right\rangle\right)\notag\\&-\sum_{i\neq j}\frac{8\lambda_{i}\lambda_{j}}{\lambda_{i}+\lambda_{j}}|\left\langle \partial_{\theta_{\mu}}\vartheta_{i}|\vartheta_{j} \right\rangle|^{2}.
\end{align}
La représentation de Bloch est un autre outil bien utilisé également dans la théorie de l'estimation quantique. Pour une matrice densité à d-dimensions, elle peut être exprimée par
\begin{equation}
\rho=\frac{1}{d}\left(\openone+\sqrt{\frac{d\left(d-1\right)}{2}}\vec{r}.\vec{K} \right), 
\end{equation}
où $\vec{r}=\left(r_{1},r_{2},..,r_{m},..\right)^{T}$ est le vecteur de Bloch et $\vec{K}$ est un vecteur de dimension ($d^{2}-1$) du générateur de $su\left(d \right)$ qui satisfait ${\rm Tr}\left(K_{i}\right)=0$. La relation d'anti-commutation pour eux est $\left\lbrace K_{i},K_{j} \right\rbrace=\frac{4}{d}\delta_{ij}\openone+\sum_{m=1}^{d^{2}-1}v_{ijm}K_{m}$, et la relation de commutation est $\left[K_{i},K_{j} \right]=i\sum_{m=1}^{d^{2}-1}\epsilon_{ijm}K_{m}$, où $v_{ijm}$ et $\epsilon_{ijm}$ sont les constantes de structure symétriques et antisymétriques. Watanabe et ses collègues \cite{Watanabe2010,Watanabe2011} ont récemment fourni la formule de la matrice d'information quantique de Fisher pour un vecteur de Bloch général en considérant le vecteur de Bloch lui-même comme des paramètres à estimer. Dans la représentation de Bloch d'une matrice de densité à d-dimensions, la matrice d'information quantique de Fisher peut être exprimée comme suit \cite{Watanabe2014}  
\begin{equation}
\boldsymbol{{\cal F}}_{\theta_{\mu}\theta_{\nu}}=\left(\partial_{\theta_{\nu}}\vec{r} \right)^{T}\left(\frac{d}{2\left(d-1\right)}G-\vec{r}\vec{r}^{T} \right)\partial_{\theta_{\mu}}\vec{r}, 
\end{equation}
où $G$ est une matrice symétrique réelle dont les éléments sont
\begin{equation}
G_{ij}=\frac{1}{2}{\rm Tr}\left(\rho\left\lbrace K_{i},K_{j} \right\rbrace \right)=\frac{2}{d}\delta_{ij}+\sqrt{\frac{d-1}{2d}}\sum_{m}v_{ijm}r_{m}.
\end{equation}
Le scénario le plus largement utilisé de ce théorème est celui des systèmes à un seul qubit, dans lequel $\rho=\left(\openone+\vec{r}.\vec{\sigma}\right)$ avec $\vec{\sigma}=\left(\sigma_{x},\sigma_{y},\sigma_{z}\right)$ le vecteur des matrices de Pauli. Dans ce cas, la matrice d'information quantique de Fisher se réduit à
\begin{equation}
\boldsymbol{{\cal F}}_{\theta_{\mu}\theta_{\nu}}=\left(\partial_{\theta_{\mu}}\vec{r} \right)\left(\partial_{\theta_{\nu}}\vec{r} \right)+\frac{\left(\vec{r}.\partial_{\theta_{\mu}}\vec{r} \right) \left(\vec{r}.\partial_{\theta_{\nu}}\vec{r} \right) }{1-|\vec{r}|^{2}},
\end{equation}
où $|\vec{r}|$ est la norme de $\vec{r}$. Pour un état pur à un seul qubit, on trouve $\boldsymbol{{\cal F}}_{\theta_{\mu}\theta_{\nu}}=\left(\partial_{\theta_{\mu}}\vec{r} \right)\left(\partial_{\theta_{\nu}}\vec{r} \right)$.\par
Pour un état général à deux qubits, le calcul de la matrice d'information quantique de Fisher nécessite la diagonalisation d'une matrice densité $4\times4$, ce qui est difficile à résoudre analytiquement. Cependant, certains états spéciaux à deux qubits, tels que l'état $X$ (\ref{matrixX}), peuvent être diagonalisés analytiquement. En effet, l'état $\rho_{X}$ à deux qubits peut être réécrit sous la forme d'une diagonale de blocs sous la forme $\rho_{X}=\rho^{(0)}\oplus\rho^{(1)}$, où $\oplus$ représente la somme directe et
\begin{equation}
	\rho^{(0)}=\left( {\begin{array}{*{20}{c}}
			\rho_{11}&\rho_{14}\\
			\rho_{41}&\rho_{44}\end{array}} \right),\hspace{1cm}{\rm et}\hspace{1cm}\rho^{(1)}=\left( {\begin{array}{*{20}{c}}
			\rho_{22}&\rho_{23}\\
			\rho_{32}&\rho_{33}\end{array}} \right).
\end{equation}
Il faut noter que $\rho^{(0)}$ et $\rho^{(1)}$ ne sont pas des matrices densité car leur trace n'est pas normalisée. La matrice d'information quantique de Fisher pour cet état peut être écrite comme $\boldsymbol{{\cal F}}_{\theta_{\mu}\theta_{\nu}}=\boldsymbol{{\cal F}}_{\theta_{\mu}\theta_{\nu}}^{(0)}+\boldsymbol{{\cal F}}_{\theta_{\mu}\theta_{\nu}}^{(1)}$, où $\boldsymbol{{\cal F}}_{\theta_{\mu}\theta_{\nu}}^{(0)}(\boldsymbol{{\cal F}}_{\theta_{\mu}\theta_{\nu}}^{(1)})$ est la matrice d'information quantique de Fisher pour $\rho^{(0)}(\rho^{(1)})$ \cite{Liu2014}. Les valeurs propres de $\rho^{(i)}$ sont 
\begin{equation}
\lambda_{\pm}^{(i)}=\frac{1}{2}\left({\rm Tr}\rho^{(i)}\pm\sqrt{\left({\rm Tr}\rho^{(i)}\right)^{2}-4{\rm det}\rho^{(i)} } \right), 
\end{equation}
et les états propres correspondants sont
\begin{equation}
\left|\vartheta_{\pm}^{(i)} \right\rangle={\cal N}_{\pm}^{(i)}\left(\frac{1}{2{\rm Tr}\left(\rho^{(i)}\sigma_{+} \right) }\left[{\rm Tr}\left(\rho^{(i)}\sigma_{+} \right)\pm \sqrt{\left({\rm Tr}\rho^{(i)}\right)^{2} -4{\rm det}\rho^{i}} \right],1\right)^{T},
\end{equation}
avec ${\cal N}_{\pm}^{(i)}$ ($i=0,1$) est le coefficient de normalisation. Ici, la forme spécifique de $\sigma_{z}$ et $\sigma_{+}$ sont
\begin{equation}
\sigma_{z}=\left( {\begin{array}{*{20}{c}}
		1&0\\
		0&-1\end{array}} \right),\hspace{2cm}\sigma_{+}=\left( {\begin{array}{*{20}{c}}
		0&1\\
		0&0\end{array}}\right).
\end{equation}
Sur la base des informations ci-dessus, $\boldsymbol{{\cal F}}_{\theta_{\mu}\theta_{\nu}}^{(i)}$ peut être écrit de manière spécifique comme suit
\begin{align}
\boldsymbol{{\cal F}}_{\theta_{\mu}\theta_{\nu}}^{(i)}=&\sum_{k=\pm}\frac{\left(\partial_{\theta_{\mu}}\lambda_{k}^{(i)} \right)\left(\partial_{\theta_{\nu}}\lambda_{k}^{(i)} \right)}{\lambda_{k}^{(i)}}+\lambda_{k}^{(i)}\boldsymbol{{\cal F}}_{\theta_{\mu}\theta_{\nu}}\left(\left|\vartheta_{k}^{(i)} \right\rangle \right)\notag\\&-\frac{16{\det\rho^{(i)}}}{{\rm Tr}\rho^{(i)}}\textsf{Re}\left(\left\langle\partial_{\theta_{\mu}}\vartheta_{+}^{(i)}| \vartheta_{-}^{(i)}\right\rangle\left\langle\vartheta_{-}^{(i)}|\partial_{\theta_{\nu}} \vartheta_{+}^{(i)}\right\rangle\right),  
\end{align}
où $\boldsymbol{{\cal F}}_{\theta_{\mu}\theta_{\nu}}\left(\left|\vartheta_{k}^{(i)} \right\rangle \right)$ est la matrice d'information quantique de Fisher pour l'état $\left|\vartheta_{k}^{(i)} \right\rangle$.

\subsection{Matrice d'information quantique de Fisher par la méthode de vectorisation}
Trés récemment, Šafránek \cite{Safranek2018} a fourni une autre méthode pour calculer la matrice d'information quantique de Fisher en utilisant la matrice densité dans l'espace de Liouville, qui ne nécessite pas la diagonalisation de la matrice densité $\rho$ est prouvée et aussi bien efficace. Dans cet l'espace, la matrice densité est un vecteur contenant toutes les éléments de la matrice densité dans l'espace de Hilbert. Dans ce sens, nous considérons une application algébrique qui transforme une matrice en un vecteur colonne pour définir les éléments de la matrice d'information quantique de Fisher sans diagonaliser la matrice densité. Soit $\emph{M}^{n\times n}$ l'espace des matrices réelles (ou complexes) $n\times n$. Pour toute matrice $A\in \emph{M}^{n\times n}$, l'opérateur ${\rm vec}\left[A\right]$ est défini comme suit \cite{Gilchrist} 
\begin{equation}
{\rm vec}\left[A \right]=\left(a_{11},...,a_{n1},a_{12},...,a_{n2},a_{1n},...,a_{nn} \right)^{T}.
\end{equation}
De plus, en utilisant l'expression $A=\sum_{k,l=1}^{n}a_{kl}\left|k\right\rangle\left\langle l\right|$, l'opérateur ${\rm vec}$ est réécrit
\begin{equation}
{\rm vec}\left[A\right]=\left(\openone_{n\times n}\otimes A\right)\sum_{i=1}^{n}e_{i}\otimes e_{i}. 
\end{equation}
où $e_{i}$ désigne les éléments de la base de calcul de $\emph{M}^{n\times n}$. Cela signifie que l'opérateur ${\rm vec}$ crée un vecteur colonne à partir d'une matrice $A$ en superposant les vecteurs colonnes de $A$ les uns au-dessous des autres. En utilisant les propriétés du produit de Kronecker \cite{Schacke2004}, on peut trouver les résultats
\begin{align}
{\rm vec}\left[AB\right]&=\left(\openone_{n\times n}\otimes A \right){\rm vec}\left[B\right]=\left(B^{T}\otimes\openone_{n\times n} \right){\rm vec} \left[A\right],\notag\\& {\rm Tr}\left(A^{T}B \right)={\rm vec}\left[A\right]^{\dagger} {\rm vec}\left[B\right],\notag\\&{\rm vec}\left[AXB\right]=\left(B^{\dagger}\otimes A\right){\rm vec}\left[X\right], \label{vecAXB}
\end{align}
pour toutes les matrices $A$, $B$ et $X$. En utilisant les propriétés données par les équations ci-dessus (\ref{vecAXB}), il est facile de vérifier que la matrice d'information quantique de Fisher, donnée par l'équation (\ref{QFIM}), peut être exprimée par \cite{Safranek2018}
\begin{equation}
\boldsymbol{{\cal F}}_{\theta_{\mu}\theta_{\nu}}=2{\rm vec}\left[\partial_{\theta_{\mu}}\rho\right]^{\dagger}\left(\rho\otimes\openone+\openone\otimes\rho^{*} \right)^{-1}{\rm vec}\left[\partial_{\theta_{\nu}}\rho \right],
\end{equation}
où $\rho^{*}$ est le conjugué de $\rho$, et l'opérateur de dérivé logarithmique symétrique dans l'espace de Liouville, noté ${\rm vec}\left[\hat{L}_{\theta_{\mu}}\right]$, se réduit à
\begin{equation}
	{\rm vec}\left[\hat{L}_{\theta_{\mu}}\right]=2\left(\rho\otimes\openone+\openone\otimes\rho^{*} \right)^{-1}{\rm vec}\left[\partial_{\theta_{\mu}}\rho\right].
\end{equation}
Cette méthode a l'avantage d'être analytiquement calculable pour un système arbitraire. Il est basé uniquement sur le calcul de l'inverse de la matrice $\left(\rho\otimes\openone+\openone\otimes\rho^{*} \right)$. Dans l'une de nos contributions \cite{Bakmou2019}, nous avons utilisé cette méthode pour étudier la comparaison entre les stratégies d'estimation simultanée et individuelle dans un système à deux qubits. Cependant, pour des systèmes plus grands, il peut être nécessaire d'employer des méthodes efficaces telles que la décomposition de Cholesky \cite{Krishnamoorthy2013}. Si cette matrice n'est pas inversible, de nouvelles expressions, pour la matrice d'information quantique de Fisher ainsi que pour les dérivés logarithmiques symétriques, sont présentées dans la même référence \cite{Safranek2018}.

\section{Rôle des corrélations non classiques en métrologie quantique}
Des efforts importants ont été déployés récemment pour évaluer la dynamique de l'information quantique de Fisher afin d'établir la validité de l'intrication quantique en métrologie quantique \cite{Blondeau2017,Giovannetti2004}. Il a été démontré que, dans les processus unitaires, l'intrication conduit à une amélioration notable de la précision de l'estimation des paramètres \cite{Blondeau2016,Huelga1997}. L'intrication peut même être exploitée comme ressource quantique pour dépasser la limite quantique standard et atteindre la limite de Heisenberg \cite{Giovannetti2004}. Ce fait soulève une question importante; une telle augmentation de l'information quantique de Fisher peut-elle être utilisée comme signature de l'intrication quantique? En outre, il est naturel de se demander si les corrélations quantiques au-delà de l'intrication peuvent être liées à la précision dans les protocoles de métrologie quantique. Il serait possible de relier et de quantifier les correlations quantiques en termes de l'information quantique de Fisher. Récemment, plusieurs études ont été menées dans cette direction \cite{Pezze2009,Rivas2010}. Un résultat important est de trouver une nouvelle mesure des corrélations quantiques en termes de l'information quantique de Fisher \cite{Kim2018}. En fait, l'information quantique de Fisher locale a été introduite pour traiter les corrélations quantiques par paire de type discorde. Ce quantificateur nous permet de mieux comprendre comment les corrélations quantiques contribuent à établir la précision en métrologie quantique. Il possède les propriétés souhaitables que tout bon quantificateur de corrélation quantique devrait satisfaire. En effet, il est non négatif et disparaît pour les états bipartis à discorde nulle (les états classiquement corrélés). Elle est invariante sous toute opération unitaire locale et coïncide avec la discorde géométrique pour les états quantiques purs \cite{Kim2018}.\par

Considérons d'abord un état quantique bipartite $d_{A}\times 
d_{B}$ agissant sur un espace de Hilbert $H_{A}\otimes H_{B}$. Dans le cas où une seule partie, disons la partie $A$, est conduite avec l'observable fixe $K_{A}=K_{A}\otimes\openone_{B}$, cela signifie que $\rho_{\theta}^{AB}=e^{-i\theta K_{A}}\rho^{AB}e^{i\theta K_{A}}$, ${\cal F}\left(\rho^{AB},K_{A} \right)$ est appelée l'information quantique de Fisher locale sur la partie $A$. Si ${\cal F}\left(\rho^{AB},K_{A} \right)=0$, cela signifie qu'il n'y a pas de changement dans l'évolution en raison de l'observable $K_{A}$ et qu'aucune information ne peut être obtenue par une mesure. Sur cette base, l'information quantique de Fisher locale est une mesure permettant de quantifier les corrélations quantiques en termes d'information de Fisher quantique sur la partie $A$ et elle est donnée par \cite{Kim2018}
\begin{equation}
{\cal Q}_{F}\left(\rho^{AB}\right)=\min_{K_{A}}{\cal F}\left(\rho^{AB},K_{A}\right).\label{LQFI}
\end{equation}
Cette nouvelle mesure de type discorde est très similaire à l'incertitude quantique locale (\ref{DLQU}) qui quantifie également les corrélations quantiques. Puisque ces deux mesures sont tous basés sur la notion d'incertitude quantique, il est intéressant d'étudier la relation et l'interaction entre elles. C'est la question principale que nous développons dans l'une de nos contributions \cite{SlaouiB2019}. Ici, nous donnons la relation entre l'incertitude quantique locale (\ref{DLQU}) et l'information quantique de Fisher locale (\ref{LQFI})  pour les systèmes quantiques de type qubit-qudit. Tout d'abord, pour ce type d'état, l'information quantique de Fisher locale se réduit à
\begin{equation}
{\cal Q}_{F}\left(\rho^{AB}\right)=\min_{K_{A}}\left[{\rm Tr}\left(\rho K_{A}^{2} \right)-\sum_{i\neq j}\frac{2p_{i}p_{j}}{p_{i}+p_{j}}|\left\langle\psi_{i}\right| K_{A}\left|\psi_{j} \right\rangle|^{2}\right],\label{497}
\end{equation}
où nous avons utilisé la décomposition spectrale de $\rho$, c'est-à-dire $\rho=\sum_{i}p_{i}\left|p_{i} \right\rangle\left\langle p_{i}\right|$ avec $p_{i}\geq0$ et $\sum_{i}p_{i}=1$. Ensuite, la forme générale d'un hamiltonien local est $K_{A}=\vec{\sigma}\vec{r}$, avec $|\vec{r}|=1$ et $\vec{\sigma}=\left(\sigma_{x},\sigma_{y},\sigma_{z}\right)$ sont les matrices de Pauli usuelles. On peut observer que ${\rm Tr}\left(\rho K_{A}^{2} \right)=1$ et que le deuxième terme de l'équation (\ref{497}) peut être exprimé comme suit
\begin{align}
\sum_{i\neq j}\frac{2p_{i}p_{j}}{p_{i}+p_{j}}|\left\langle\psi_{i}\right| K_{A}\left|\psi_{j} \right\rangle|^{2}&=\sum_{i\neq j}\sum_{l,k=1}^{3}\frac{2p_{i}p_{j}}{p_{i}+p_{j}}\left\langle\psi_{i}\right| \sigma_{l}\otimes\openone_{B}\left|\psi_{j} \right\rangle\left\langle\psi_{j}\right| \sigma_{k}\otimes\openone_{B}\left|\psi_{i} \right\rangle\notag\\&=\vec{r}^{\dagger}.M.\vec{r},
\end{align}
où les éléments de la matrice symétrique $M$, de dimension $3\times3$, sont donnés par
\begin{equation}
M_{lk}=\sum_{i\neq j}\frac{2p_{i}p_{j}}{p_{i}+p_{j}}\left\langle\psi_{i}\right| \sigma_{l}\otimes\openone_{B}\left|\psi_{j} \right\rangle\left\langle\psi_{j}\right| \sigma_{k}\otimes\openone_{B}\left|\psi_{i} \right\rangle\notag. \label{MatrixM}
\end{equation}
De même que dans l'expression de l'incertitude quantique locale, pour minimiser ${\cal F}\left(\rho^{AB},K_{A}\right)$, il est nécessaire de maximiser la quantité $\vec{r}^{\dagger}.M.\vec{r}$ sur tous les vecteurs unitaires $\vec{r}$. La valeur maximale coïncide avec la valeur propre maximale de $M$. Par conséquent,  l'expression analytique de l'information quantique de Fisher locale ${\cal Q}_{F}\left(\rho^{AB}\right)$ est donnée par
\begin{equation}
{\cal Q}_{F}\left(\rho^{AB}\right)=1-\lambda_{\rm max}\left(M \right),
\end{equation}
où $\lambda_{\rm max}$ désigne la valeur propre maximale de la matrice symétrique $M$ définie par (\ref{MatrixM}). Il est extrêmement important de souligner ici que cette quantité coïncide avec la discorde géométrique pour tout état quantique pur \cite{Kim2018}. Cela signifie qu'elle peut être considérée comme une mesure raisonnable des corrélations quantiques et que nous pouvons quantifier les corrélations quantiques en mesurant uniquement l'information quantique de Fisher des observables locales. De plus, l'information quantique de Fisher locale fournit un outil pour comprendre le rôle des corrélations quantiques au-delà de l'intrication dans l'amélioration de la précision et de l'efficacité des protocoles de métrologie quantique. En fait, l'information de Fisher quantique est associée à l'incertitude quantique locale. Il a été démontré que dans l'évolution unitaire de la matrice densité, l'information de Fisher quantique et l'information  d'interchange (\ref{IS}) satisfont l'inégalité suivante \cite{SlaouiB2019}
\begin{equation}
I\left(\rho,K \right)\leq {\cal F}\left(\rho,K\right)\leq2I\left(\rho,K \right), 
\end{equation}
à partir duquel on obtient
\begin{equation}
{\cal U}\left( \rho\right)\leq {\cal Q}_{F}\left(\rho\right) \leq2{\cal U}\left( \rho\right).
\end{equation}
D'un autre côté, il est bien connu que l'incertitude quantique locale est majorisée par l'information quantique de Fisher, à savoir
\begin{equation}
{\cal U}\left( \rho\right)\leq I\left(\rho,K \right)\leq {\cal F}\left(\rho,K \right).
\end{equation}
Par conséquent, selon le théorème de Cramér-Rao (\ref{BCR}), la précision du paramètre peut être limitée par l'incertitude quantique locale et par l'information quantique de Fisher locale comme suit \cite{SlaouiB2019}
\begin{equation}
{\rm Var}\left(\theta \right)_{\rm min}\leq \frac{1}{{\cal U}\left( \rho\right)},\hspace{1cm}{\rm et}\hspace{1cm} {\rm Var}\left(\theta \right)_{\rm min}\leq \frac{1}{{\cal Q}_{F}\left(\rho\right)}.
\end{equation}

\chapter{Contributions}
\section{Contribution 1: The dynamics of local quantum uncertainty and trace distance discord for two-qubit $X$ states under decoherence: a comparative study}

\hspace{2.5cm} {\bf Quantum Information Processing, 17 (2018) 1-24.}

\subsection{Résumé:}
Une approche analytique pour évaluer l'incertitude quantique locale (\ref{DLQU}) est en général une tâche difficile en raison d'une procédure d'optimisation de l'information d'interchange sur des mesures généralisées locales, qui reste un problème ouvert important même dans le cas le plus simple d'un système à deux qubits. C'est exactement dans ce sens que nous développons notre travail. Le but de cette contribution est de développer une méthode analytique pour évaluer les corrélations quantiques dans des systèmes à deux qubits au moyen du concept d'incertitude quantique locale. Nous avons présenté la forme générale de la matrice $W$ (\ref{matriceW}) pour les états arbitraires de type $X$. Cette méthode constitue un outil alternatif pour évaluer les expressions analytiques de corrélations quantiques englobées dans des états $X$ à deux qubits.\par
En outre, nous étudions le comportement dynamique de cette mesure et la comparons à deux quantificateurs bien connus des corrélations quantiques dans les systèmes bipartites, à savoir la discorde géométrique basée sur la distance de trace (\ref{DGT}) et la concurrence (\ref{Concu}), sous deux modèles de décohérence différents. Nous concentrons davantage sur le comportement et les limites de chaque mesure. Dans le premier modèle, les deux qubits sont couplés individuellement à deux réservoirs bosoniques indépendants, en tenant compte de trois types de réservoirs; sub-ohmique, ohmique et super-ohmique. Dans le second modèle, nous étudions la dynamique des corrélations dans un système de deux atomes à deux niveaux, interagissant avec un champ de rayonnement quantifié, initialement préparé dans un état séparable.

\subsection{Contenu de la publication $1$:}
\section{Contribution 2: Universal evolution of non-classical correlations due to collective spontaneous emission}
\hspace{2.5cm} {\bf The European Physical Journal Plus, 133 (2018) 413.}
\subsection{Résumé:}

La dynamique des corrélations quantiques au-delà de l'intrication dans les systèmes quantiques ouverts a été largement étudiée dans la littérature, mais peu de ces études se concentrent sur la façon dont le système est corrélé avec son environnement. Ici, nous étudions ce problème dans une situation réaliste, en prenant deux atomes à deux niveaux comme qubits identiques couplés à un environnement Markovien commun en présence de modes de champ électromagnétique. Nous prenons en compte l'interaction dipôle-dipôle, l'amortissement collectif et, plus important encore, la distance interatomique entre eux. Nous étudions comment la distance interatomique influence la façon dont l'un de ces atomes se trouve en corrélation quantique avec l'environnement et, comme nous le montrerons, cette séparation apparaît comme une variable cruciale.\par

Afin d'étudier l'évolution dynamique des corrélations quantiques, l'équation Maîtresse (\ref{eqmatp}) dans les approximations de Born-Markov a été résolue analytiquement, où nous utilisons la représentation de l'état collectif de Dicke, qui suppose que le système à deux atomes se comporte comme un système unique à quatre niveaux. À partir de cette solution analytique, nous avons calculé les quantités pertinentes pour nos objectifs; la concurrence, la discorde quantique géométrique, et l'incertitude quantique locale pour plusieurs états initiaux. Lorsque les deux atomes sont initialement dans un état intriqué à excitation double ou nulle, les différentes mesures des corrélations quantiques se désintègrent de manière exponentielle en fonction de chaque paramètre d'émission spontanée. Pendant ce temps, l'intrication est générée spontanément avec un comportement oscillatoire, qui dépend à la fois de l'amortissement collectif et de l'interaction qubit-qubit. Contrairement à ce cas, toutes les corrélations montrent une augmentation rapide, suivie d'une décroissance très lente lorsque ces deux atomes sont initialement préparés dans la superposition des états symétrique et antisymétrique maximalement intriqués. Ces résultats montrent que l'amortissement collectif et l'interaction dipôle-dipôle jouent un rôle clé dans le renforcement des corrélations non classiques au cours du processus de décohérence intrinsèque.\par

De plus, nous avons discuté l'évolution de ces corrélations pour des atomes placés très près l'une de l'autre. De manière contre-intuitive, nous avons montré que cette distance critique peut être celle qui minimise la perte de corrélation quantique et que nous pouvons la préserver pour de petites distances.

\subsection{Contenu de la publication 2:}


\section{Contribution 3: The dynamic behaviors of local quantum uncertainty for three-qubit $X$ states under decoherence channels}

\hspace{2.5cm} {\bf Quantum Information Processing, 18 (2019) 250.}
\subsection{Résumé:}

Comme nous l'avons déjà évoqué, l'incertitude quantique locale a des applications intéressantes et significatives dans le traitement de l'information quantique. Jusqu'à présent, cette mesure des corrélations quantiques de type discorde n'a été calculée explicitement que pour un ensemble assez limité d'états quantiques à deux qubits, et les expressions pour des états quantiques plus généraux ne sont pas connues. Dans cette contribution, nous avons traité le problème de calcul de l'incertitude quantique locale pour les états $X$ tripartites, qui ont été d'intérêt dans divers contextes du domaine. A titre d'illustration, nous avons considéré les états à trois qubits de type GHZ et de type Bell. Nous avons calculé la discorde quantique entropique (\ref{QDD}), la négativité tripartite (\ref{neg}) et l'incertitude quantique locale (\ref{DLQU}) de ces états et nous avons comparé le comportement de chaque mesure. En outre, nous avons discuté la propriété de monogamie de l'incertitude quantique locale dans ces états.\par

Suite à ce résultat, nous avons traité les effets de la décohérence sur les corrélations quantiques générées par les canaux d'amortissement de phase, de dépolarisation et d'inversion de phase. Nous avons montré que la décohérence influence fortement les corrélations de ces états initiaux pour les canaux de déphasage et de dépolarisation. De plus, pour les canaux d'inversion de phase, nous remarquons une résistance de l'incertitude quantique locale, ce qui permet une comparaison directe de la fidélité de ces corrélations dans les systèmes quantiques ouverts.

\subsection{Contenu de la publication 3:}

\section{Contribution 4: A comparative study of local quantum Fisher information and local quantum uncertainty in Heisenberg $XY$ model}
\hspace{2.5cm} {\bf Physics Letters A, 383 (2019) 2241-2247}

\subsection{Résumé:}
Bien que la capacité de l'intrication à améliorer la métrologie quantique a été bien explorée dans des scénarios idéaux, les contraintes expérimentales, telles que le bruit, les états mixtes et la restriction aux mesures locales, rendent généralement impossible l'atteinte de la limite quantique ultime. Dans ce contexte, une étude plus générale du rôle des corrélations quantiques dans la métrologie quantique est essentielle, car elle peut conduire à des schémas de mesure plus généraux qui tirent parti des propriétés non classiques. La discorde quantique est une mesure des corrélations non classiques, dont l'intrication est un sous-ensemble. L'exploration de la signification opérationnelle de la discorde quantique en tant que ressource dans les tâches de la métrologie quantique est d'une importance essentielle pour notre compréhension des corrélations non classiques.\par
Dans ce travail, nous montrons que l'information quantique de Fisher locale joue un rôle essentiel dans l'évaluation des corrélations quantiques. Cela est dû à sa relation avec le concept de l'incertitude quantique locale. Nous montrons que l'incertitude quantique locale est majorée par l'information quantique de Fisher locale dans toute tâche métrologique de l'estimation de phase et que l'analogue quantique de l'inégalité de Cramér-Rao classique peut s'écrire en termes de ces deux quantificateurs de correlations quantiques de type discorde. De plus, l'incertitude quantique locale d'un état quantique mixte garantit une précision minimale quantifiée par l'information quantique de Fisher dans le protocole d'estimation de phase optimale.\par

D'un point de vue analytique, nous proposons un formalisme relativement simple pour dériver des expressions analytiques des corrélations quantiques. Grâce à ce formalisme, nous analysons le comportement de l'incertitude quantique locale et le comparons au comportement de l'information quantique de Fisher locale dans le modèle de Heisenberg $XY$. Nous considérons deux situations particulières. La première concerne le modèle $XY$ anisotrope et la seconde situation concerne le modèle $XY$ isotrope soumis à un champ magnétique externe. Dans ces deux situations, l'information quantique de Fisher locale et l'incertitude quantique locale présentent une variation similaire. Plus important encore, l'information de Fisher quantique locale est toujours supérieure à l'incertitude quantique locale. Ces résultats confirment que ces deux mesures offrent un outil prometteur pour comprendre le rôle des corrélations quantiques autres que l'intrication dans l'amélioration de la précision et de l'efficacité des protocoles de métrologie quantique.

\subsection{Contenu de la publication 4:}
\section{Contribution 5: Influence of Stark-shift on quantum coherence and non-classical correlations for two two-level atoms interacting with a single-mode cavity field}
\hspace{1cm} {\bf Physica A: Statistical Mechanics and its Applications, 558 (2020) 124946}

\subsection{Résumé:}
Comme nous le savons de la mécanique quantique, la présence de symétries conduit à des niveaux d'énergie dégénérés. Par exemple, considérons un électron atomique avec le potentiel $V\left(r \right)$, nous disons que ce système est invariant par rotation et nous nous attendons à une certaine forme de dégénérescence pour chaque niveau atomique. La disparition de la dégénérescence peut être réalisée en brisant la symétrie. La symétrie rotationnelle peut être manifestement brisée en considérant un champ électrique ou magnétique externe uniforme le long d'une direction fixe. Un champ magnétique se couple avec le moment magnétique de l'atome, tandis qu'un champ électrique se couple avec le moment dipolaire électrique de l'atome. Dans les deux cas, on vérifie le fractionnement des niveaux d'énergie atomique dû au couplage avec le champ externe. Dans le cas d'un champ magnétique uniforme, ce phénomène est connu sous le nom d'effet Zeeman. Dans le cas où le champ externe est un champ électrique uniforme, on l'appelle l'effet Stark. \par

Dans cet article, nous nous intéressons à la manière dont l'effet du décalage de Stark et le nombre de photons d'état cohérents affectent à la fois les comportements de corrélation quantique et de cohérence quantique. Le modèle considéré est constitué de deux atomes à deux niveaux couplés à un champ de cavité électromagnétique monomode multi-photons en présence du décalage de Stark. Pour atteindre cet objectif, nous supposons que le champ est initialement préparé dans un état cohérent et que les deux atomes sont initialement préparés dans un état excité. Ensuite, nous avons étudié l'évolution de notre modèle physique, et la solution exacte de l'évolution est dérivée en analysant l'équation de Schrödinger avec quelques simplifications. Nous avons observé que l'évolution temporelle de la cohérence quantique, quantifiée par la divergence quantique de Jensen-Shannon, est presque similaire à l'évolution temporelle des corrélations quantiques mesurée par la discorde quantique entropique. Mais nous avons remarqué que la quantité de cohérence quantique est toujours plus grande et va au-delà de la discorde quantique en raison du fait que la cohérence quantique totale dans les systèmes multipartites a des contributions de la cohérence locale sur les sous-systèmes et de la cohérence collective entre eux. En l'absence et en présence des photons d'état cohérent, l'effet du décalage de Stark sur le système est examiné. Nous avons montré que le système est sensible à la variation à la fois du paramètre de décalage de Stark et du nombre de photons d'état cohérent initial. De plus, ces deux quantificateurs diminuent avec des valeurs croissantes des paramètres de décalage de Stark lorsque le nombre de photons augmente. \par

Dans une vue globale, nous pensons que ce travail montre que les protocoles d'information quantique basés sur des ressources physiques dans les systèmes optiques pourraient être contrôlés en ajustant les paramètres de décalage de Stark. Aussi, ce dernier peut également modifier les autres propriétés des ressources physiques nécessaires à la mise en œuvre des tâches de traitement de l'information quantique.

\subsection{Contenu de la publication 5:}



\chaptertoc{Conclusion générale et perspectives}
La théorie de l'information quantique est devenue l'une des nouvelles interfaces entre la science et la technologie au cours des dernières décennies. Contrairement à la théorie de l'information classique qui porte sur le bit classique, la théorie de l'information quantique est centrée sur le concept de qubit qui peut exister dans une superposition de deux états en même temps. L'importance des ressources quantiques est évidente du fait que certaines tâches de traitement de l'information, comme la téléportation et la sécurité de l'information, ne peuvent être accomplies avec les seules ressources classiques et que leurs résultats sont grandement améliorés lorsque les ressources classiques sont combinées aux ressources quantiques. Alors que la superposition et les corrélations quantiques sont des ressources quantiques bien établies. Des travaux de recherche importants ont récemment été menés pour explorer la possibilité d'utiliser la cohérence quantique comme ressource pour des applications de traitement de l'information. D'autre part, les progrès récents dans le domaine de l'estimation quantique des paramètres ont été stimulés par la perspective d'une deuxième révolution quantique, visant à exploiter les phénomènes quantiques pour améliorer les performances de plusieurs tâches relevant de la théorie de l'information.\par

L'essentiel de cette thèse, et le principal objectif de l'auteur, est de trouver les méthodes analytiques pour résoudre plusieurs problèmes ouverts en théorie de l'information quantique. Nous nous sommes principalement intéressés à trouver les nombreuses connexions entre les théories des ressources quantiques et la théorie de l'estimation quantique, où certains résultats intéressants ont été trouvés et discutés tout au long de cette thèse. Dans cette partie, nous fournissons un résumé technique du contenu de la thèse, discutons de l'avenir de ce travail, et tirons des conclusions générales basées sur les résultats que nous avons obtenus. A travers ce manuscrit, nous avons essayé de présenter notre travail de manière conviviale et concise. Celui-ci est réparti en quatre chapitres plus d'une introduction générale constituant les piliers de cette thèse.\par

Le premier chapitre sert de révision des concepts d'introduction importants de la théorie de l'information quantique, en commençant par expliquer l'information d'un point de vue physique. Ensuite, nous explorons le lien entre le caractère aléatoire et le manque d'information, puis nous étudions la différence entre l'information classique et l'information quantique, ainsi qu'un bref aperçu des mesures correspondantes. Nous avons également évoqué les outils mathématiques de base utilisés dans la description des états quantiques, le formalisme de l'opérateur de densité, des qubits, des mesures quantiques, des types d'entropies et du phénomène d'intrication quantique. Enfin, le chapitre se termine par quelques critères et mesures de l'intrication dans les systèmes quantiques avec leurs expressions analytiques en basse démension.\par

De nombreuses études ont porté sur la réalisation de l'intrication dans des états quantiques et beaucoup d'efforts ont été consacrés à l'étude de la dynamique de l'intrication, de sa décohérence et de son contrôle dans différents systèmes. C'est malgré tous ces progrès que l'absence d'une théorie complète de l'intrication persiste et que la détermination de l'intrication d'un état donné reste un problème ouvert. Pour les systèmes bipartites, à deux et trois niveaux, le problème peut être abordé à l'aide de la concurrence ou de la négativité. Pour les systèmes multipartites, même ce progrès limité est resté insaisissable. Par ailleurs, il a été démontré que même les états séparables peuvent jouer un rôle important dans l'exécution de protocoles de communication et d'information plus performants que les protocoles classiques. Il est donc intéressant d'étudier les avantages fondamentaux des corrélations quantiques au-delà de l'intrication.\par 

Dans le deuxième chapitre de cette thèse, nous avons développé des méthodes analytiques qui permettent la détection et la quantification des corrélations de type discorde avec des opérations locales sur un seul sous-système, et qui ont promis de représenter une approche évolutive de la caractérisation des corrélations à la lumière de l'augmentation de la dimension de l'espace de Hilbert avec le nombre de particules en interaction. Tout d'abord, nous fournissons une prescription générale pour calculer la discorde quantique des états à deux qubits. Ensuite, nous étudions l'interaction entre les mesures d'intrication et la discorde quantique. Le problème d'optimisation pour l'entropie conditionnelle, et de manière équivalente pour la discorde quantique, est reformulé en utilisant l'entropie linéaire au lien de l'entropie de von Neumann dans le calcul de corrélation classique induit par la mesure locale. Ce processus permet d'obtenir deux expressions analytiques de la discorde quantique, l'une pour les états qubit-qubit arbitraire, et l'autre pour des états bipartites de type qudit-qubit. Nous construisons également un point de vue alternatif sur la corrélation quantique à partir du concept d'incertitude quantique, en prouvant un lien frappant entre deux caractéristiques quantiques, à savoir la corrélation quantique des états et l'incertitude sur les observables locales. En effet, chaque fois que nous effectuons une mesure sur un sous-système d'un système bipartite dont l'état possède une corrélation quantique, l'incertitude quantique sur cet état est garantie d'apparaître. L'incertitude quantique minimale sur une seule mesure locale est elle-même un quantificateur de corrélation quantique. À la fin de ce chapitre, nous avons étudié un ensemble de lois conservatrices qui régissent la façon dont l'intrication et la discorde quantique sont distribuées dans les systèmes multipartites. Ces égalités relient les contraintes de l'intrication distribuée à la discorde distribuée et vice versa.\par

En revanche, l'interaction inévitable entre un système quantique et son environnement détruit généralement la cohérence et le caractère quantique du système. Cependant, une dynamique cohérente et les survivances des corrélations quantiques sont des conditions nécessaires au traitement de l'information quantique. Une description détaillée du mécanisme de décohérence est donc d'une grande importance pour le développement des technologies quantiques et la caractérisation précise du bruit agissant sur un système quantique est le principal outil pour concevoir des protocoles robustes contre les effets néfastes de la décohérence. L'approche habituelle pour étudier la décohérence due à l'interaction entre un système quantique et son environnement consiste à décrire ce dernier comme un réservoir quantique. Mais il existe des situations où cette description peut être difficile ou inappropriée. Dans cette inspiration, le troisième chapitre fournit une description complète de la théorie des systèmes quantiques ouverts, où nous avons discuté en détail le concept d'application dynamique universelle et sa relation avec l'évolution du système quantique ouvert. Nous avons montré que la dynamique du système réduit est obtenue à partir de l'équation de Liouville-Von Neumann, puis nous avons présenté l'équation maîtresse de Lindblad que nous avons dérivée sous des approximations de Born-Markov. Les approches et techniques markoviennes les plus couramment utilisées ont également été présentées. Dans la deuxième partie de ce chapitre, nous explorons le fait que la cohérence et les corrélations quantiques sont intimement liées et peuvent être transformées l'une en l'autre, bien que la première notion a été définie par rapport à un système unique, tandis que la seconde concerne les systèmes bipartites et multipartites. Concernant la caractérisation et la quantification de la cohérence quantique, il existe de nombreuses méthodes différentes qui s'appliquent à des situations différentes. Elles peuvent également être quantifiées de manière similaire aux corrélations quantiques, par exemple en utilisant la distance géométrique de deux états. Ce cadre unifié permet de comprendre les liens intrinsèques entre ces deux notions fondamentales.\par

Dans le quatrième chapitre, nous avons revu les grandes idées sur la métrologie quantique. Notre grand intérêt fut porté principalement sur le rôle des corrélations quantiques en métrologie quantique, afin de fournir une meilleure presicion dans l'estimation d'un paramètre inconnu. En effet, la théorie de l'estimation quantique permet de trouver le paramètre optimal pour estimer efficacement la valeur d'un paramètre. La limite ultime de la précision de la variance associée à un estimateur est bornée selon le théorème de Cramér-Rao quantique et elle est proportionnelle à l'inverse de l'information de Fisher quantique. Les bases de la théorie de l'estimation classique et de celle de l'estimation quantique sont discutées. Par ailleurs, des méthodes analytiques permettant de calculer l'information de Fisher quantique dans plusieurs scénarios ont été obtenues. De surcroît, nous avons lié l'information quantique de Fisher à l'analyse des corrélations quantiques. Nous avons étudié la caractérisation des corrélations quantiques du point de vue de l'information quantique de Fisher locale, qui constitue une mesure raisonnable des corrélations quantiques. Cela signifie que nous pouvons quantifier les corrélations quantiques en termes de l'information quantique de Fisher. Ce lien est devenu apparent lorsque nous avons examiné la relation entre l'information quantique de Fisher locale et l'incertitude quantique locale, qui sont toutes des quantificateurs de corrélations quantiques de type discorde via l'incertitude quantique. Cela donne des avantages significatifs pour examiner le rôle des corrélations quantiques dans la détermination de la haute précision du paramètre estimé.\par

Dans le cinquième chapitre, nous avons inséré nos contributions consacrées à ces différents axes majeurs. Le premier travail portait principalement sur le problème des mesures de corrélations quantiques dans les systèmes bipartites à deux qubits. Nous nous sommes attachés à étudier, via quelques modèles spécifiques, la dynamique des corrélations quantiques sous les effets induisant la disparition de la cohérence. Le premier modèle est un système à deux qubits couplé à deux réservoirs bosoniques indépendants, dont trois types de réservoirs: sub-ohmique, ohmique et super-ohmique. Le second concerne deux atomes à deux niveaux interagissant avec les modes d'un champ de radiation quantifié. Dans ce sens, nous avons dérivé l'expression analytique de l'incertitude quantique locale pour les états $X$ à deux qubits. En analysant la dynamique de ce quantificateur, et en le comparant avec la discorde quantique de trace et la concurrence dans chaque modèle considéré. Ces trois quantificateurs des corrélations quantiques se comportent différemment, et chaque mesure présente des avantages par rapport à l'autre qui dépendent de chaque système considéré.\par
Dans la seconde contribution, nous nous intéressons à la création de corrélations quantiques entre deux atomes identiques couplés à un environnement markovien commun en présence de modes de champ électromagnétique. Pour cela, nous analysons la dynamique des corrélations quantiques, avec une attention particulière à la concurrence, à la discorde quantique de trace et à l'incertitude quantique locale, pour différents états initiaux de Dicke. Nos résultats montrent que l'effet d'amortissement collectif et l'interaction dipôle-dipôle jouent un rôle essentiel dans l'amélioration des corrélations non classiques au cours du processus de décohérence intrinsèque.\par

Le troisième publication présentait l'expression analytique de l'incertitude quantique locale pour des états $X$ à trois qubits. À titre d'illustration, nous avons calculé la corrélation non-classique, dans les cas particuliers des états mixtes GHZ et des états de type Bell, en utilisant ce quantificateur de corrélation quantique. Les résultats obtenus sont comparés à ceux obtenus par la discorde quantique entropique et de la négativité. La quantité de corrélations quantiques quantifiées par l'incertitude quantique locale est presque similaire à celle mesurée par la discorde quantique entropique. Cela indique que l'incertitude quantique locale constitue un quantificateur approprié pour traiter la corrélation quantique dans les systèmes multi-qubit. Nous examinons également la dynamique de ces quantificateurs sous les effets des canaux d'amortissement de phase, de dépolarisation et d'inversion de phase. Nous constatons que l'incertitude quantique locale montre plus de robustesse de corrélation quantique. Enfin, nous montrons que l'incertitude quantique locale satisfait le principe de monogamie pour les états à trois qubits.\par

Dans le quatrième papier, nous avons examiné le rôle des corrélations quantiques dans les protocoles d'estimation de phase en métrologie quantique. Nous avons donné la relation entre l'incertitude quantique locale et l'information quantique locale de Fisher, et nous montrons que ces deux mesures sont essentielles pour déterminer la précision dans les protocoles métrologiques. Surtout, la présente étude suggère que l'information quantique de Fisher locale et l'incertitude quantique locale présentent une variation similaire. Ces résultats soulignent l'importance et l'interaction entre ces deux types particuliers de corrélations quantiques.\par

La cinquième publication est consacrée à la compréhension des liens entre la corrélation et la cohérence quantique dans les systèmes quantiques. En particulier, nous avons étudié la dynamique de la cohérence quantique (mesurée à l'aide de la divergence quantique de Jensen-Shannon) et la dynamique de corrélation non-classique (capturée par la discorde quantique) dans deux atomes à deux niveaux interagissant avec un champ de cavité électromagnétique monomode. Une attention particulière est consacrée à la présence des décalages de Stark. Nous avons montré que le décalage de Stark et le nombre de photons de l'état cohérent jouent un rôle clé dans le renforcement ou la disparition de la cohérence et de la discorde quantique au cours du processus de décohérence intrinsèque. En outre, ces deux concepts différents présentent un comportement similaire dans notre modèle, avec l'observation intuitive qui affirme que la quantité de la cohérence quantique est toujours plus grande et va au-delà de la discorde quantique.\par

Plusieurs obstacles doivent encore être surmontés, certaines clés restent perdues et de nombreuses questions sans réponse apparaissent, qui pourront être étudiées dans le cadre de développements ultérieurs. Ceci nous a motivé à présenter les perspectives d'avenir. En effet, nous envisageons d'aller plus loin dans ce terrain, et nous synthétisons ici quelques points de vue:
\begin{itemize}
	\item Durant cette thèse, nous avons discuté de la manière dont l'information quantique peut être quantifiée et caractérisée en utilisant le concept de qubit dans l'espace de Hilbert fini. D'autre part, une autre approche de la construction des qubits a fait l'objet de plusieurs études en utilisant des systèmes à variables continues, où l'information est codée dans la position et l’impulsion des particules. Nous sommes convaincus que nos résultats serviront de base aux futures études à la fois de la corrélation et de la cohérence quantique dans l'espace de phase et devraient être étendus et validés par de grandes classes d'états des systèmes à variables continues.

	\item De nombreux schémas d'estimation multiparamétrique quantique ont été proposés et discutés en ce qui concerne divers systèmes quantiques, et certains d'entre eux ont montré des avancées théoriques par rapport aux schémas à paramètre unique. Cependant, de nombreux problèmes restent ouverts, comme la conception d'une mesure optimale, une mesure particulièrement simple et pratique qui est indépendante des paramètres inconnus, les méthodes analytiques pour saturer la borne de Cramér-Rao, ainsi que le rôle des corrélations quantiques dans la stratégie d'estimation multiparamétrique quantique. Nous pensons que certains de ces problèmes seront résolus dans un avenir proche. 
	\item Très récemment, deux extensions de la thermodynamique (développée à l'origine comme une théorie visant à optimiser l'efficacité des machines thermiques) ont fait progresser la théorie au point où la mécanique quantique devrait être incorporée; Premièrement, le rôle de l'information en thermodynamique, Deuxièmement, les extensions de la thermodynamique au domaine des systèmes microscopiques dans lesquels les fluctuations sont importantes. Ces deux extensions conduisent la thermodynamique comme une théorie de ressource quantique qui nous permet de reformuler ses lois d'une manière naturelle \cite{Deffner2019,Goold2016}. Il est donc très intéressant d'étudier les liens entre cette nouvelle ressource quantique et les autres ressources que nous avons abordées dans cette thèse.	
\end{itemize}
\begin{center}
\begin{tabular}{|>{\columncolor{gray!40}\color{white}\bfseries}p{3cm}||p{3cm}|p{3.5cm}|p{3.5cm}|}
	\hline
	\multicolumn{4}{|c|}{\cellcolor{green!25}Théories des ressources quantiques} \\
	\hline\rowcolor{lightgray}
	Ressources quantique& Intrication quantique &Cohérence quantique &Thermodynamique quantique\\
	\hline
	\'Etats des ressources& \'Etats intriqués & \'Etats cohérents &   \'Etats hors-équilibre\\
	\hline
	Opérations libres& LOCC & Opérations incohérents&Opérations thermodynamiques\\
	\hline
	\'Etats libres & \'Etats séparables & \'Etats incohérents&  États thermiques\\
	\hline
\end{tabular}
\end{center}
Nous espérons que les résultats décrits dans cette thèse seront utilisés pour mieux décrire et comprendre les ressources quantiques et leurs applications en théorie de l'information quantique. Si vous, le lecteur, avez des questions relatives à ces résultats, l'auteur de la thèse est plus que ravi de vous répondre par e-mail. Veuillez contacter l'auteur à l'adresse électronique suivante: \colorbox{red}{abdallah.slaoui@um5s.net.ma}

\includepdf[pages=-]{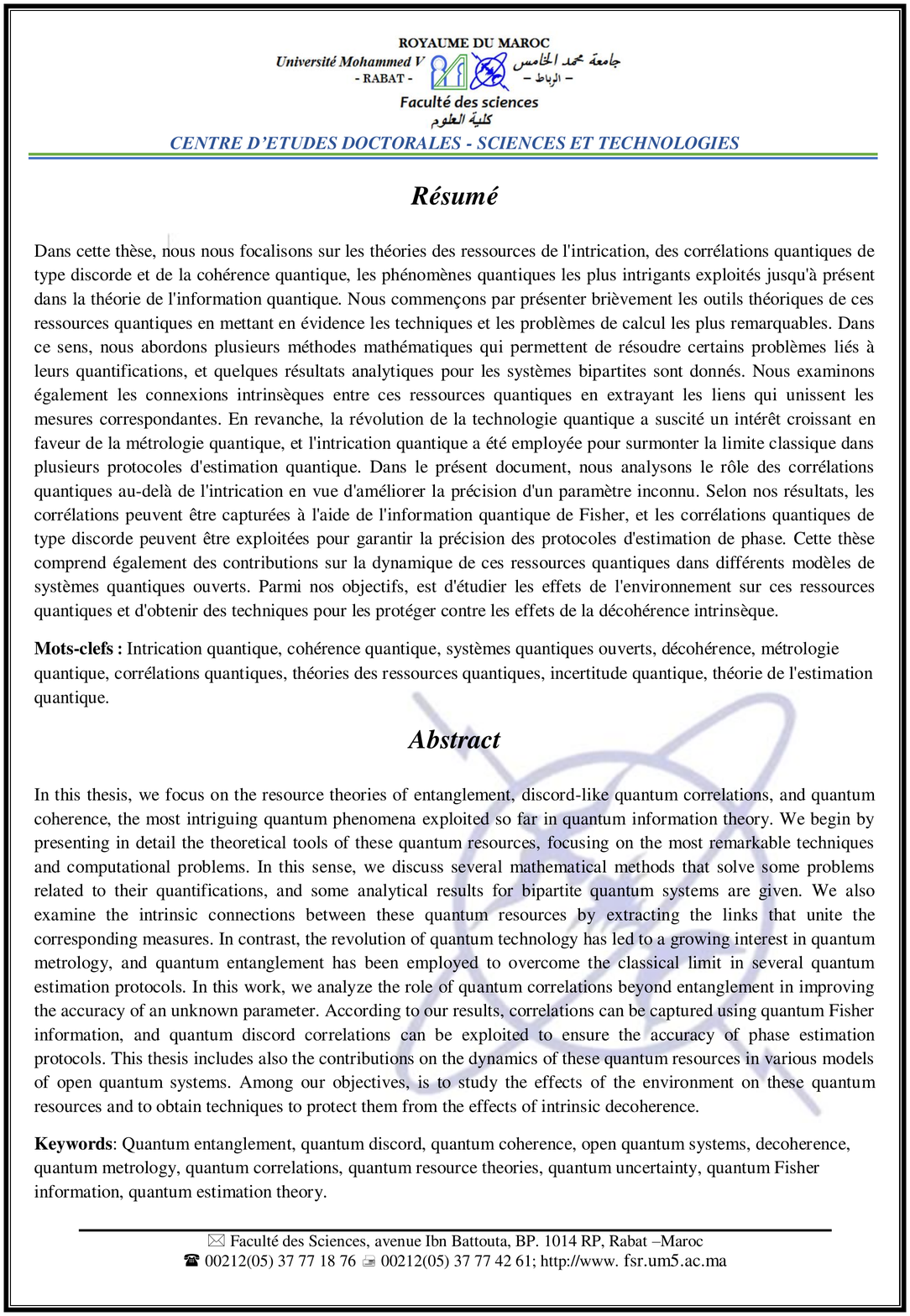}
\end{document}